\documentclass[a4paper,twoside,12pt
]{report}

\usepackage[pdftex]{graphicx}
\usepackage[
draft=false,             
bookmarks=true,          
colorlinks=true,        
anchorcolor=black,       
citecolor=green,         
filecolor=magenta,       
linkcolor=blue,           
urlcolor=magenta,        
pdfborder={0 0 0},       
]{hyperref}

\usepackage{fancyhdr}
\usepackage{framed}

\usepackage{amsmath,amssymb,amsthm}
\usepackage{bm} 
\usepackage{mathtools}
\usepackage{braket}
\usepackage{cases}
\usepackage{subcaption}
\usepackage{booktabs}
\usepackage{cite}

\usepackage[
svgnames]{xcolor} 
\usepackage{tikz} 
\usetikzlibrary{patterns} 
\usepackage[most]{tcolorbox}
\tcbuselibrary{raster,skins}
\tcbuselibrary{skins,breakable,xparse} 

\usepackage{lipsum}
\newcommand{\note}[1]{\textcolor{red}{[#1]}}
\usepackage{comment}

\makeatletter
  \def\@seccntformat#1{%
    \@nameuse{@seccnt@prefix@#1}%
    \@nameuse{the#1}%
    \@nameuse{@seccnt@postfix@#1}%
    \@nameuse{@seccnt@afterskip@#1}}
  \def\@seccnt@prefix@section{}
  \def\@seccnt@postfix@section{}
  \def\@seccnt@afterskip@section{\hskip1em}
  \def\@seccnt@prefix@subsection{}
  \def\@seccnt@postfix@subsection{}
  \def\@seccnt@afterskip@subsection{\hskip1em}
  \def\@seccnt@prefix@subsubsection{$\blacksquare$}
  \def\@seccnt@postfix@subsubsection{}
  \def\@seccnt@afterskip@subsubsection{\hskip1em}
  
\makeatother

\theoremstyle{plain}
\newtheorem{theorem}{Theorem}[chapter]

\theoremstyle{definition}
\newtheorem{example}{Example}[chapter]
\AtBeginEnvironment{example}{%
	\pushQED{\qed}%
}
\AtEndEnvironment{example}{\popQED\endexample}

\theoremstyle{definition}
\newtheorem{remark}{Remark}[chapter]

\makeatletter
    
    \@addtoreset{equation}{chapter}
\makeatother

\usepackage{caption}
\DeclareCaptionLabelFormat{mylabel}{#1 #2:\hspace{1em}}
\renewcommand{\bibname}{References}

\newcommand{\sgn}{\mathop{\mathrm{sgn}}\nolimits}

\begin{document}

\begin{titlepage}
~

\vfill
\begin{center}
{\LARGE\bf Study on a Quantization Condition and} \\[1em]
{\LARGE\bf  the Solvability of Schr\"{o}dinger-type Equations} 
\end{center}


\vspace{.4\textheight}
\begin{center}
{\Large Yuta Nasuda} \\[2ex]
{\large Department of Physics and Astronomy} \\[1ex]
{\large Tokyo University of Science}
\end{center}

\vspace{.05\textheight}
\begin{center}
{\large A thesis submitted for the degree of} \\[1ex]
{\large\it Doctor of Science}
\end{center}

\vspace{.02\textheight}
\begin{center}
{\large March 2024}
\end{center}
\end{titlepage}

\clearpage
\thispagestyle{empty}
~


\clearpage
\def\thepage{\roman{page}}
\setcounter{page}{1}

\chapter*{Abstract}
\addcontentsline{toc}{chapter}{Abstract}
~

In this thesis, we study a quantization condition in relation to the solvability of Schr\"{o}dinger equations.
This quantization condition is called the SWKB (supersymmetric Wentzel--Kramers--Brillouin) quantization condition and has been known in the context of supersymmetric quantum mechanics for decades.
Supersymmetric quantum mechanics has been revealing various aspects of exactly solvable problems of non-relativistic quantum mechanics, such as shape invariance.
It has recently turned out that previous studies of the SWKB condition fail to provide what the condition generally implies.
Moreover, the existing literature on the condition mostly regarded the SWKB condition as a quantization condition of the energy, and few attempts have been made to apply the condition for different purposes.

The main contents of this thesis are recapitulated as follows: the foundation and the application of the SWKB quantization condition.
The first half of this thesis aims to understand the fundamental implications of this condition based on extensive case studies.
We carry out our analyses for the conventional shape-invariant potentials, the exceptional/multi-indexed systems, the Krein--Adler systems, the conventional exactly solvable systems by Junker and Roy, and also classical-orthogonal-polynomially solvable systems with position-dependent effective mass.
It turns out that the exactness of the SWKB quantization condition indicates the exact solvability of a system via the classical orthogonal polynomials.

The SWKB quantization condition provides quantizations of energy, which we call the direct problem of the SWKB.
We formulate the \textit{inverse} problem of the SWKB: the problem of determining the superpotential from a given energy spectrum.
An assumption on the shape of the superpotential is required to make the inverse problem well-posed.
The formulation successfully reconstructs all conventional shape-invariant potentials from the given energy spectra.
We further construct novel solvable potentials, which are classical-orthogonal-polynomially \textit{quasi}-exactly solvable, by this formulation. 
The term ``quasi'' refers to the situation in which only a part of the eigenstate spectrum is obtained exactly by analytic expressions.
We have identified the rule concerning which eigenstates are classical-orthogonal-polynomially solvable and which are not.

We further demonstrate several explicit solutions of the Schr\"{o}dinger equations with the classical-orthogonal-polynomially quasi-exactly solvable potentials, whose family is referred to as a harmonic oscillator with singularity functions in this thesis.
In one case, the energy spectra become isospectral, with several additional eigenstates, to the ordinary harmonic oscillator for special choices of a parameter.
By virtue of this, we formulate a systematic way of constructing infinitely many potentials that are strictly isospectral to the ordinary harmonic oscillator.

The basic knowledge and reviews of previous studies in this field are also provided in this thesis.

\subsection*{Highlights:}
\begin{itemize}\setlength{\leftskip}{-1em}
\item The exactness of the SWKB quantization condition indicates that the system is exactly solvable via the classical orthogonal polynomials.
\item The inverse problem of the SWKB is formulated to construct (novel) classical-orthogonal-polynomially solvable superpotentials.
\item The exact solutions of the Schr\"{o}dinger equations with a new entry of the classical-orthogonal-polynomially (quasi-)exactly solvable potentials defined by piecewise analytic functions: harmonic oscillators with singularity functions, are obtained.
\item Infinitely many potentials that are \textit{strictly} isospectral to the ordinary harmonic oscillator are constructed methodically.
\end{itemize}

\bigskip
\paragraph{Keywords:}
Schr\"{o}dinger equation, supersymmetric quantum mechanics, SWKB quantization condition, exactly solvable problems, classical orthogonal polynomials, inverse problem, piecewise analytic functions, isospectral transformations, harmonic oscillator


\clearpage
\vspace*{\stretch{1}}
\section*{Acknowledgement}
\addcontentsline{toc}{chapter}{Acknowledgement}

\bigskip
I thank my supervisor, Prof. Nobuyuki Sawado for his invaluable supervision, continuous support and patience during my Ph.D. study, in fact, the last nine years as his student.
I also thank Prof. Ryu Sasaki for his useful advice regarding exactly solvable quantum mechanics.
I would also like to express my gratitude to Prof. Naruhiko Aizawa, Prof. Atsushi Nakamula and Prof. Kouichi Toda for their constant support.

I appreciate Prof. Luiz Agostinho Ferreira for his kind hospitality in Instituto de F\'{i}sica de S\~{a}o Carlos of Universidade de S\~{a}o Paulo (IFSC/USP) in 2022.
I would like to extend my thanks to the following individuals for their kind invitations to their institutes during my stay in Brazil: Prof. Pawe\l~Klimas (Universidade Federal de Santa Catarina), Prof. Zhanna Kuznetsova (Universidade Federal do ABC), Prof. Gabriel Luchini (Universidade Federal do Esp\'{i}rito Santo), Dr. Madhusudhan Raman (Instituto de F\'{i}sica Te\'{o}rica of Universidade Estadual Paulista), Prof. Francesco Toppan (Centro Brasileiro de Pesquisas F\'{i}sicas).

My sincere thanks also go to Dr. Yuki Amari, Prof. Ulysses Camara da Silva, Prof. Elso Drigo Filho, Prof. Elchin Jafarov, Prof. Regina Maria Ricotta, Dr. Shota Yanai, and all the others who have helped me to arrive at this point.
The discussions with those people are quite fruitful.

I am equally grateful to my friends who have encouraged me.
Last but not least, I would like to thank my parents always for supporting me.

\bigskip
The project was financially supported in part by JST SPRING Grant Number JPMJSP2151 and the Sasakawa Scientific Research Grant from The Japan Science Society (No. 2022-2011).

\vspace*{\stretch{3}}

\clearpage

\addcontentsline{toc}{chapter}{Contents}
\tableofcontents

\clearpage
\addcontentsline{toc}{chapter}{List of Figures}
\listoffigures

\clearpage
\addcontentsline{toc}{chapter}{List of Tables}
\listoftables

\clearpage
\thispagestyle{empty}
~

\vfill
\hfill{\large\it Alles Gescheite ist schon gedacht worden,~~~~~~~~~~~~~}

\hfill{\large\it man mu\ss nur versuchen, es noch einmal zu denken.~}

\bigskip\medskip
\hfill{\large\it All intelligent thoughts have already been thought;~~~}

\hfill{\large\it what is necessary is only to try to think them again.}

\bigskip
\hfill{\large\rm --- Johann Wolfgang von Goethe}



\clearpage
\pagestyle{fancy}
\renewcommand{\chaptermark}[1]{\markboth{\sc\thechapter.~~#1~}{}}
\renewcommand{\sectionmark}[1]{\markright{\it~\thesection\quad #1}}
\fancyhead{}
\fancyhead[RE]{\leftmark}
\fancyhead[LO]{\rightmark}
\setcounter{page}{1}
\def\thepage{\arabic{page}}
\fancyhead[LE]{\quad\thepage}
\fancyhead[RO]{\thepage\quad}
\fancyfoot{}

\chapter{Introduction}
\label{sec:1}

\section{Backgrounds/Brief History}
\label{sec:1-1}
In modern science, many researchers employ mathematical models to analyze and understand the phenomena they are concerned about.
Such models are expressed as partial differential equations with frequency.
Separation of variables is a powerful tool for solving partial differential equations, and consequently, one often arrives at the eigenvalue problem for an ordinary differential equation.
Strum--Liouville problem is a major, well-known example of this.
In this thesis, we consider the Schr\"{o}dinger equation, which is an example of such problems in non-relativistic quantum mechanics. 
A similar eigenvalue problem appears not only in quantum physics but also in many branches of science.

Exactly solvable problems in non-relativistic quantum mechanics have been attracting scientists' interest in understanding the behavior of particles in various potential energy landscapes.
Since the early days of quantum physics, various methods for the analytical solutions of the Schr\"{o}dinger equation, especially time-independent, one-dimensional ones, have been extensively studied. 
One methodology is based on the factorization of Hamiltonian. 
This factorization method was originally developed by E. Schr\"{o}dinger, who studied an algebraic method to obtain the energy spectrum of the hydrogen atom~\cite{schrodinger1940method} and was later generalized by L. Infeld and T. Hull~\cite{infeld1951factorization}. 
In fact, the Schr\"{o}dinger's work is said to have been stimulated by the ideas of P. M. Dirac~\cite{dirac1981principles} and H. Weyl~\cite{weyl1950theory}.

In terms of the theory of ordinary differential equation, it was known before Schr\"{o}dinger that a first-order nonlinear differential equation called the Riccati equation is equivalent to the corresponding second-order linear ordinary differential equation. 
The factorization method in quantum mechanics can be regarded as a translation of this Riccati's idea into Schr\"{o}dinger equation.

In particle physics, on the other hand, supersymmetry was proposed by H. Miyazawa in 1966~\cite{10.1143/PTP.36.1266}.
Supersymmetry is a symmetry between Bosons and Fermions.
Since the symmetry is actually broken in our universe now, we need supersymmetric field theories where supersymmetry is spontaneously broken.
In the early 1980s, E. Witten made an attempt to understand it in a rather simple model; starting from a field theory, he constructed an effective model at low energies in quantum mechanics~\cite{witten1981dynamical,witten1982constraints}. 
These days, the model is often referred to as \textit{supersymmetric quantum mechanics}~\cite{cooper1995supersymmetry,bagchi2000supersymmetry,cooper2001supersymmetry,junker2012supersymmetric,gangopadhyaya2017supersymmetric,fernandez2018trends,10.1088/2053-2563/aae6d5}.

As studies in supersymmetric quantum mechanics go on, it has turned out that supersymmetric quantum mechanics provides insights into the factorization method and the exact solvability of the Schr\"{o}dinger equation. 
For example, the idea of shape invariance~\cite{gendenshtein1983derivation}, which is a sufficient condition of the exact solvability of Schr\"{o}dinger equation, was developed in this context.
Other than this, an algebraic structure underlying the solvability of Schr\"{o}dinger equation has been revealed, and new methods for constructing solvable potentials have been established based on supersymmetric quantum mechanics.
Today, supersymmetric quantum mechanics is not only studied for understanding supersymmetry but also for exploring the solvability of quantum mechanical systems.

Together with this aspect of supersymmetric quantum mechanics, the field of solving Schr\"{o}dinger equations analytically and constructing new solvable systems is sometimes called \textit{exactly solvable quantum mechanics}~\cite{sasaki2014exactly}. 
A system is exactly solvable when all the eigenvalues and the corresponding eigenfunctions are obtained explicitly in analytical closed forms.
Furthermore, in this thesis, we refer to quasi-exactly solvable problems~\cite{turbiner1988quasi,MR1329549,turbiner2016one,ZNOJIL20161414}, which means that only a part of eigenvalues and the eigenfunctions are obtained analytically.

Potentials like the Coulomb potential and the harmonic oscillator have been listed as exactly solvable models since the early days of quantum mechanics in the 1920s.
Those potentials appeared in the analyses of the phenomena that gave rise to quantum physics.
After that, various solvable potentials have been proposed as models describing quantum phenomena. 
The Morse potential is an example, introduced to describe the interatomic interaction of a diatomic molecule~\cite{morse1929diatomic}.
These potentials are typical cases where the Schr\"{o}dinger equations are solved using the factorization method. 

By around 1950, a general formulation of the factorization method had been established for several dozens of solvable potentials~\cite{infeld1951factorization}.
The method was further generalized by Crum~\cite{10.1093/qmath/6.1.121}.
In the Crum's theorem, the general structure of the solution space of the one-dimensional Schr\"{o}dinger equation was manifested.
Nowadays, the theorem is understood as a simple realization of Darboux transformation~\cite{darboux}, which was already known in the 19th century in mathematics.

Then in 1983, the concept of shape invariance was introduced in the context of supersymmetric quantum mechanics~\cite{gendenshtein1983derivation}.
The factorization method came to be understood in this context; those potentials are exactly solvable because they possess shape invariance and the Crum's idea is obviously applicable to these problems. 
The exactly solvable appeared in Ref. \cite{infeld1951factorization} are now called \textit{conventional} shape-invariant potentials.

In the early 1990s, a novel class of solvable potentials with shape invariance was discovered~\cite{barclay1993new}. 
Those potentials are defined by the power series of a parameter, and known as \textit{scaling} shape-invariant potentials.
Later in the 21st century, yet another class of shape-invariant potentials has been constructed~\cite{quesne2008exceptional,ODAKE2009414,ODAKE2010173}, where the eigenfunctions are expressed in terms of the exceptional orthogonal polynomials~\cite{GOMEZULLATE2009352,GOMEZULLATE2010987}.
They are now understood in the following way.
That is, those potentials are Darboux transformations of conventional shape-invariant potentials~\cite{Sasaki_2010}.
Based on this understanding, the class was generalized to the so-called multi-indexed systems~\cite{ODAKE2011164}.

Back in the 20th century, Krein~\cite{krein1957continuous} and Adler~\cite{adler1994modification} extended the Crum's theorem (which is another realization of Darboux transformation) independently. 
By applying this idea to a known exactly solvable system, one can construct yet another novel class of exactly solvable potentials, and such deformation is sometimes referred to as Krein--Adler transformation.

Now one can see that Darboux transformation is a powerful tool for constructing novel exactly solvable systems. 
However, the above-mentioned exactly solvable systems are essentially all possible ones in this context.
There are also other classes of exactly solvable problems that have been constructed in different contexts. 
The so-called Natanzon class of potentials~\cite{natanzon1971study,natanzon1979general,ginocchio1984class,cooper1987relationship} is an example.
Furthermore, Abraham--Moses transformation~\cite{PhysRevA.22.1333} also transforms an exactly solvable system into another one.
We now know a mountain of construction methods for (exactly) solvable potentials.
It remains to be quite a challenge to obtain a unified framework for these methods.

\section{Motivations}
\label{sec:1-2}
When it comes to actually solving the eigenvalue problem for an ordinary differential equation, we have several directions: analytical and algebraic approaches, numerical computations \textit{etc}.
There have been numerous methods that have ever been proposed (including the ones we mentioned in the previous section).

Among them, approaches using \textit{quantization conditions} can be effective, if you seek for the eigenvalues of your problem, which are supposed to characterize your model.
Quantization conditions were originally discussed in quantum physics, where they have played a unique role as `translators' to help us understand quantum phenomena since the very first days of quantum physics. 
While some quantization conditions like Bohr--Sommerfeld quantization condition, the one in the semi-classical regime, are known to reproduce exact bound-state energy spectra for only a limited number of problems, they can still serve as approximate formulae for obtaining the eigenvalue spectra for other problems.

In 1985, a quantization condition was proposed by Comtet, Bandrauk and Campbell~\cite{comtet1985exactness}, which is often referred to as the SWKB quantization condition.
It gives exact bound-state energy spectra for all the conventional shape-invariant potentials~\cite{dutt1986exactness,BARCLAY1991357,hruska1997accuracy,YIN2010528,GANGOPADHYAYA2020126722,Gangopadhyaya_2021}, while for newly discovered solvable systems, this condition is known to provide only approximate formulae for determining the energy spectra~\cite{khare1989shape,delaney1990susy,doi:10.1139/p91-189,bougie2018supersymmetric,Nasuda:2020aqf,NASUDA2023116087}.
Since the conventional shape-invariant potentials play a central role in exactly solvable quantum mechanics, that is, most exactly solvable problems can be seen as some kind of deformations of the conventional shape-invariant systems, this fact is too good to miss.

So far, various discussions regarding the interpretation of the SWKB quantization condition have been delivered.
Some argued in relation to the exact solvability of a system, while others thought it might be somehow equivalent to the shape-invariant condition.
However, it has recently turned out that all the previous studies of the SWKB condition fail to provide what the condition generally implies. 
Thus we say that an appropriate interpretation of the SWKB quantization condition is still absent.

Moreover, the existing literature on the condition mostly regarded the SWKB condition as a quantization condition of the energies, and few attempts have been made to apply the condition for different purposes.
A similar formulation of the WKB scheme, on the other hand, has been applied to many different contexts such as resurgent theories.
The limitation of the SWKB is probably due to the fact that it has poor compatibility with the wavefunctions.

Our aim of the study is to first provide a mathematical or physical interpretation of the SWKB quantization condition.
Subsequently, based on the interpretation, we are to see how the SWKB condition equation is applied to actual problems in physics \textit{etc}.
We propose a way of constructing novel solvable potentials in terms of the SWKB formalism.

\section{Contributions}
\label{sec:1-3}
The importance of studies in exactly solvable quantum mechanics is found in the following context.
Firstly, solvable models, in contrast to models that can only be solved numerically, allow one to carry out analytical discussions and get richer information.
It therefore becomes possible to demonstrate the properties of a model concretely and thereby reveal the mathematical structures underlying natural phenomena.

However, in reality, most models that describe physical phenomena involve problems that are not solved analytically but numerically.
Even in those problems, a deep understanding of the properties of solvable models will help, and approaching those problems based on the knowledge of the solvable problems is an essential strategy for uncovering the mathematical structures behind the phenomena.
A major example is perturbation theory, which is an approach towards an approximate solution starting from the exact solution of a simpler problem.
It has been made great successes in various areas; even in a case where a perturbation series fails to converge, \textit{i.e.}, a non-perturbative problem, there is a way to subtract some information from the divergent series (\textit{cf.} resurgence theory~\cite{ecalle1985fonctions,10.1215/S0012-7094-98-09311-5,ANICETO20191,Sueishi:2020aa}).

Some computational calculations are based on the theory of orthogonal polynomials, and the theory has recently been developed significantly with the knowledge of exactly solvable quantum mechanics as was mentioned earlier.
Those findings can contribute to the development of new computational methods.

Moreover, exactly solvable quantum mechanics, or supersymmetric quantum mechanics, has provided insights into classical integrable systems such as the Korteweg--de-Vries (KdV) equation and the sine-Gordon equation. 
It is well-known in the literature that a construction method of multi-soliton solutions of these equations is closely related to a Schr\"{o}dinger-type equation.
Conversely, such soliton solutions have been used to construct new solvable potentials. 
Also, the algebraic structure of the solution space of the Schr\"{o}dinger equation plays a key role in stability analyses in the context of soliton theories, where the evaluation formula is a Schr\"{o}dinger-type equation.

Moreover, Schr\"{o}dinger-type equations are everywhere.
It is well-known that Fokker--Planck equation, which is applicable to models in various fields such as plasma physics, biology, economics and financial engineering, is a Schr\"{o}dinger equation in imaginary time, and thereby we are to solve the eigenvalue problem of a negative semi-definite Hamiltonian.
Schr\"{o}dinger-type equations or equations with similar algebraic structures also appear in optics as Helmholtz equation, in cosmology, circuit theory, spin system \textit{etc}.
We expect that the present studies explore new aspects of these branches of science.

\section{Structure of Thesis}
This chapter has presented the introduction of this thesis, which consists of five sections: the background of our study is mentioned with a brief history of the field of research in Sect. \ref{sec:1-1}, the motivation of the research project and its contribution to the field are found in Sects. \ref{sec:1-2} and \ref{sec:1-3} respectively, this section provides the structure of this thesis, and in the subsequent section, we shall have a quick review on the Schr\"{o}dinger equation.

The rest of this thesis is organized as follows.
Chap. \ref{sec:2} is a review on the so-called exactly solvable quantum mechanics.
All the knowledge that one needs to discuss in the following chapters is contained here (and in the related Appendix), but several topics that have little relevance to this thesis are not covered to prevent this chapter from getting lengthy.
Tips that are not usually emphasized in textbooks and review papers are also included.

In Chap. \ref{sec:3}, we discuss SWKB formalism, starting by introducing the SWKB quantization condition. 
The condition equation and several notable properties are given in Sect. \ref{sec:3-1}.
We perform several case studies on the condition equation to understand what the SWKB quantization condition actually means in Sect. \ref{sec:3-2}.
Of course, it is impossible to carry out case studies for all potentials ever constructed, but our five examples are sufficient to grasp the implication.
Sect. \ref{sec:3-3} is devoted to the discussions based on the case studies in the previous section, where we arrive at a conjecture on the implication of the SWKB quantization condition.
These parts are based on the author's works~\cite{Nasuda:2020aqf,10.1007/978-981-19-4751-3_29,NASUDA2023116087}.

So far we have focused on one side of SWKB formalism, where we know the superpotential \textit{a priori} and then apply the condition.
In Sect. \ref{sec:3-4}, we formulate a way of determining the superpotential from a given energy spectrum: the inverse problem of SWKB.

In Chap. \ref{sec:4}, we start with the solution of the Schr\"{o}dinger equation with the potential constructed at the end of the previous chapter in Sect. \ref{sec:4-2}.
The potential is classified into the classical-orthogonal-polynomially quasi-exactly solvable potentials.
We provide general remarks on the solution of the Schr\"{o}dinger equation with the potentials of this kind in advance in Sect. \ref{sec:4-1}.
In Sect. \ref{sec:4-3}, we consider yet other potentials in this class, which is based on the author's works~\cite{nasuda2023harmonic,Nasuda_2023,nasuda-lt15}.

Chap. \ref{sec:5} is the conclusion of the thesis with some perspectives of this study.

Explicit examples are provided to help the readers' understand throughout the thesis.
In addition, supplemental materials are found in the Appendices.

\clearpage
\section{Schr\"{o}dinger Equation}
The Schr\"{o}dinger equation (in $(3+1)$ dimension) has the following form in general:
\begin{equation}
\mathrm{i}\hbar\frac{\partial}{\partial t}\varPsi = \mathcal{H}\varPsi ~.
\label{eq:1-SE}
\end{equation}
We consider a Hamiltonian for a single particle:
\begin{equation}
\mathcal{H} = -\frac{\hbar^2}{2m}\bm{\nabla}^2 + V(\bm{r}) ~,
\end{equation}
where $\bm{\nabla}^2$ is the Laplacian in three dimension, $\mathbb{E}^3$.

We solve the partial differential equation \eqref{eq:1-SE} by separation of variables.
We assume that $\varPsi(\bm{r},t)$ is a product of $f(t)$ and $\psi(\bm{r})$, $\varPsi(\bm{r},t) \equiv f(t)\psi(\bm{r})$, where the dependence of $\varPsi$ on $\bm{r}$, $t$ is separated.
Therefore,
\begin{equation}
\frac{\mathrm{i}\hbar\dfrac{df(t)}{dt}}{f(t)} = \frac{\mathcal{H}\psi(\bm{r})}{\psi(\bm{r})} = \mathrm{const.} \equiv E ~,
\end{equation}
that is,
\begin{subnumcases}
{}
~\mathrm{i}\hbar\frac{df(t)}{dt} = Ef(t) ~,
\label{eq:1-SE_timep} \\[5pt]
~\mathcal{H}\psi(\bm{r}) = E\psi(\bm{r}) ~.
\label{eq:1-SE_spacep}
\end{subnumcases}
Eq. \eqref{eq:1-SE_timep} yields
\begin{equation}
f(t) = \mathrm{e}^{-\mathrm{i}\frac{E}{\hbar}t} ~,
\end{equation}
so essentially, our problem is to solve Eq. \eqref{eq:1-SE_spacep}.
Eq. \eqref{eq:1-SE_spacep} is often referred to as time-independent, or more simply static/stationary, Schr\"{o}dinger equation.

If $V(\bm{r}) \equiv V_x(x)+V_y(y)+V_z(z)$, then Eq. \eqref{eq:1-SE_spacep} is further reduced to three ordinary differential equations by separating the variables, $\psi(\bm{r}) \equiv X(x)Y(y)Z(z)$,
\begin{subequations}
\begin{align}
&\left[ -\frac{\hbar^2}{2m}\frac{d^2}{dx^2} + V_x(x) \right] X(x) = \mathcal{E}^{(x)}X(x) ~, \\
&\left[ -\frac{\hbar^2}{2m}\frac{d^2}{dy^2} + V_y(y) \right] Y(y) = \mathcal{E}^{(y)}Y(y) ~, \\
&\left[ -\frac{\hbar^2}{2m}\frac{d^2}{dz^2} + V_z(z) \right] Z(z) = \mathcal{E}^{(z)}Z(z) ~.
\end{align} 
\end{subequations}
Or, if the potential is spherically symmetric, $V(\bm{r}) \equiv V(r)$, then by $\psi(\bm{r}) \equiv R(r)Y(\theta,\varphi)$, Eq. \eqref{eq:1-SE_spacep} becomes 
\begin{subequations}
\begin{align}
\left[ -\frac{\hbar^2}{2m} \left( \frac{d^2}{dr^2} + \frac{2}{r}\frac{d}{dr} - \frac{\ell (\ell+1)}{r^2} \right) + V(r) \right] R(r) &= \mathcal{E}R(r) ~, \\
-\frac{1}{\sin\theta}\frac{\partial}{\partial\theta}\left( \sin\theta\frac{\partial Y(\theta,\varphi)}{\partial\theta} \right) - \frac{1}{\sin^2\theta}\frac{\partial^2Y(\theta,\varphi)}{\partial\theta^2} &= \ell (\ell+1) Y(\theta,\varphi) ~,
\end{align} 
\end{subequations}
where $Y(\theta,\varphi) = Y_{\ell}^{m}(\theta,\varphi)$ turns out to be a spherical harmonics.


\chapter{Exactly Solvable Quantum Mechanics}
\label{sec:2}

{\small
\begin{leftbar}
\noindent\textbf{Introduction.}\hspace{1em}
In this chapter, we have a review on the so-called exactly solvable quantum mechanics.
We focus exclusively on bound-state problems.
Tips that are not usually emphasized in textbooks and review papers are also included.
For basic materials of non-relativistic quantum mechanics, see, \textit{e.g.}, Refs. \cite{schiff,lifshitz1977quantum,jjsakurai}.
\end{leftbar}
}

\section{Problem Setting}
Throughout this thesis, we consider Hamiltonians for one-dimensional quantum mechanical systems of the following form:
\begin{equation}
\mathcal{H} = \frac{\hat{p}^2}{2m} + V(\hat{x})
= -\frac{\hbar^2}{2m}\frac{d^2}{dx^2} + V(x) ~,~~~
x \in (x_1,x_2) ~,
\label{eq:2-Ham}
\end{equation}
where $x_1$ and/or $x_2$ can be infinite, and $V(x)$ is the potential.
In the formulation in Sect. \ref{sec:2-2}, the potential is a $C$-infinity function, $V(x) \in C^{\infty}$\footnote{In the context of exactly solvable quantum mechanics, we usually deal with the case where $V(x)$ is an analytic function, or a piecewise analytic function.}.
We assume that our Hamiltonian is bounded from below.
For finite $x_1$ and/or $x_2$, we require 
\[
\lim_{x\to x_{1,2}}V(x) = +\infty ~.
\]

The time-independent Schr\"{o}dinger equation:
\begin{equation}
\mathcal{H}\psi(x) = E\psi(x)
\label{eq:2-SE}
\end{equation}
with a constant $E$, is a second-order differential equation.
Especially, we are interested in the eigenvalue problem, namely the problem of determining all the discrete eigenvalues $\{ E_n \}$ and the corresponding eigenfunctions $\{ \psi_n(x) \}$ of the Hamiltonian \eqref{eq:2-Ham},
\begin{equation}
\mathcal{H}\psi_n(x) = E_n\psi_n(x) ~.
\label{eq:2-SEn}
\end{equation}
The numbering of the eigenvalues is monotonically increasing,
\[
E_0 < E_1 < E_2 < \cdots ~,
\]
and the eigenfunctions, $\psi_n(x) \in L^2(\mathbb{R}) \cap C^1$, are orthogonal,
\begin{equation}
( \psi_n(x), \psi_m(x) ) \coloneqq \int_{x_1}^{x_2} \psi_n^{\ast} (x)\psi_m (x) \,dx = h_n\delta_{n,m} ~,~~~~ 0 < h_n < \infty ~.
\end{equation}
In quantum mechanics with one degree of freedom, one can always take the eigenfunctions to be real, $\psi_n(x) \in \mathbb{R}$.
Also, we choose $\psi_0(x) > 0$.
The oscillation theorem states that the $n$-th eigenfunction $\psi_n(x)$ has $n$ nodes in the domain.

A potential is a confining potential, if 
\begin{equation}
\lim_{x\to x_1}V(x) = \lim_{x\to x_2}V(x) = +\infty ~.
\end{equation}
Confining potentials have infinitely many discrete eigenvalues.
Otherwise, potentials are said to be non-confining, and they have finitely or infinitely many discrete eigenvalues.

In the following, we set $2m=1$ for simplicity otherwise noted but retain $\hbar$ for rigorous discussions on the semi-classical regime.

\subsection{Classification of Problems}
Here, we present three ways of classifying problems \eqref{eq:2-SEn} in terms of the potentials.

\subsubsection{Analyticity of a potential}
Every standard textbook on quantum mechanics deals with the square potential well and the harmonic oscillator.
They are not only two of the most significant examples that show up in a wide range of fields in quantum physics, but they represent two of the major categories of exactly solvable potentials.

The one represented by the square potential well is the piecewise constant interactions (See, \textit{e.g.}, Ref.\,\cite{Magyari}), including so-called point interaction models~\cite{PointInteraction}.
Another typical example in this class is the Kronig--Penney model~\cite{Kronig1931}. 
The domain $(x_1,x_2)$ is divided into several subintervals,
\[
(x_1,x_2) = (x_1,a_1) \cup (a_1,a_2) \cup\cdots\cup (a_K,x_2) ~,
\]
and when it comes to constructing the wavefunctions, we require the matching conditions at $x=a_1\ldots,a_K$.
The other major class with the harmonic oscillator is analytical potentials whose eigenfunctions are written in closed analytical form.
A substantial feature of these examples is that they are solved by the classical orthogonal polynomials or the exceptional/multi-indexed orthogonal polynomials~\cite{GOMEZULLATE2009352,GOMEZULLATE2010987,ODAKE2009414,ODAKE2010173,Sasaki_2010,ODAKE2011164}.
Several deformations of this class of potentials have been considered by employing Darboux transformation~\cite{darboux}, Abraham--Moses transformation~\cite{PhysRevA.22.1333} and other techniques (See Fig. \ref{fig:2-SolvablePots} in Sect. \ref{sec:2-4} and the references therein).

The problems lying in the intersection between these two classes, that is, potentials defined by piecewise analytic functions, have also attracted attention.
For instance, the Coulomb plus square-well potential~\cite{PhysRev.71.865}, a finite parabolic quantum well potential~\cite{PhysRevB.48.17316} and the harmonic oscillator potential embedded in an infinite square-well~\cite{10.1119/1.15738} are discussed in atomic and nuclear physics.
Moreover, the sheared harmonic oscillator~\cite{doi:10.1021/j100356a004}, the inverse square root potential~\cite{Ishkhanyan_2015}, the symmetrized Morse-potential short-range interaction~\cite{doi:10.1142/S0217732316500887}, and other potentials such as $-g^2\exp(-|x|)$-potential~\cite{Sasaki_2016}, a double well potential $\min[(x+d)^2,(x-d)^2]$ and its dual single well potential $\max[(x+d)^2,(x-d)^2]$~\cite{quantum4030022,10.1063/5.0127371} are all defined piecewise and solved by the matching of wavefunctions.

\subsubsection{Solvability of a potential}
For a given potential, sometimes the whole spectra and the corresponding wavefunctions are obtained in closed analytic forms, while many other times the eigenvalue problem \eqref{eq:2-SEn} is not solved analytically.
We say a potential is exactly solvable when all the discrete eigenvalues and the corresponding eigenfunctions are obtained in closed analytical forms. 

As was mentioned above, potentials such as the harmonic oscillator and the Coulomb potential are exactly solvable, thanks to the classical orthogonal polynomials.
Most of the well-known solvable potentials are in the class of conventional shape-invariant potentials, and thus they are solvable by the classical orthogonal polynomials.
A major subclass in the exactly solvable potentials is polynomially solvable potentials.
Other examples in this subclass are potentials that are solved by the exceptional/multi-indexed potentials~\cite{GOMEZULLATE2009352,GOMEZULLATE2010987,ODAKE2009414,ODAKE2010173,Sasaki_2010,ODAKE2011164}.

Even when you cannot solve Schr\"{o}dinger equations via orthogonal polynomials, you still have chances that your problems are exactly solvable through well-known special functions such as Bessel functions~\cite{Sasaki_2016}. 
Sometimes, such potentials are said to be non-polynomially solvable.

Between the solvable and the non-solvable lie yet two other classes.
One is called quasi-exactly solvable potentials~\cite{turbiner1988quasi,MR1329549,turbiner2016one,ZNOJIL20161414}.
A potential is quasi-exactly solvable, when only several eigenstates are solved exactly in closed analytical form while other states are not.
In most cases, the quasi-exact solvable states are related by $\mathfrak{sl}(2,\mathbb{R})$ algebra~\cite{turbiner1988quasi}.
Note however that this is not the case with the quasi-exact solvable states for piecewise analytic potentials.

The other one is conditionally exactly solvable potentials, where they are exactly solvable if model parameters satisfy some condition(s).
The concept was originally introduced by Dutra in Ref. \cite{PhysRevA.47.R2435}, and the most well-known examples are the ones by G. Junker and P. Roy, in which they have employed supersymmetric quantum mechanics in their construction~\cite{junker1998conditionally,junker1998supersymmetric}.

\section{Formulation}
\label{sec:2-2}
\subsection{Factorized Hamiltonian}
\label{sec:2-2-1}
We consider the eigenvalue problem of a Hamiltonian \eqref{eq:2-Ham}.
In this formulation, we choose the lowest energy eigenvalue to be zero without loss of generality.
We use the calligraphic ``E'' to emphasize this instead of the Roman ``E'', $\mathcal{E}_n \equiv E_n - E_0$.
Thus, the Hamiltonian is positive semi-definite, $\mathcal{H} \succeq 0$.
In linear algebra, the following theorem is known:

\begin{theorem}[Linear algebra]
Any positive semi-definite hermitian matrix $\mathsf{A}$ can be factorized as a product of a certain matrix, say $\mathsf{P}$ and its hermitian conjugate $\mathsf{P}^{\dag}$, $\mathsf{A} = \mathsf{P}^{\dag}\mathsf{P}$.
\end{theorem}

Although our Hamiltonian is not a matrix, we will apply the idea of factorization to our problem.
That is, we consider that our Hamiltonian has the following factorized form:
\begin{equation}
\mathcal{H} \coloneqq \mathcal{A}^{\dag}\mathcal{A} ~,
\label{eq:2-factorized}
\end{equation}
with the two operators $\mathcal{A}$ and $\mathcal{A}^{\dag}$ being
\begin{gather}
\mathcal{A} \coloneqq \hbar\frac{d}{dx} + W(x) ~,~~~
\mathcal{A}^{\dag} = -\hbar\frac{d}{dx} + W(x) ~, \\
V(x) \equiv W(x)^2 - \hbar\frac{dW(x)}{dx} ~,~~~
\mathcal{H} = -\hbar^2\frac{d^2}{dx^2} + W(x)^2 - \hbar\frac{dW(x)}{dx} ~.
\label{eq:2-RiccatiEq-W}
\end{gather}
Here, a real function $W(x)$ is often referred to as \textit{superpotential}.
Note that the Hamiltonian is invariant under unitary transformations: $\mathcal{A}\to U\mathcal{A}$, $UU^{\dag}=U^{\dag}U=I$ ($I$ is the identity operator).
For example, one can also factorize the Hamiltonian in the following manner:
\[
\mathcal{H} = \left( -\mathrm{i}\hbar\frac{d}{dx} + \mathrm{i}\,W(x) \right) \left( -\mathrm{i}\hbar\frac{d}{dx} - \mathrm{i}\,W(x) \right) ~.
\]

The zero mode of $\mathcal{A}$ gives the ground-state wavefunction,
\begin{equation}
\mathcal{A}\phi_0(x) = 0
\quad\Longrightarrow\quad
\mathcal{A}^{\dag}\mathcal{A}\phi_0(x) = \mathcal{H}\phi_0 = 0
\end{equation}
The first equation is a first-order differential equation, and its formal solution is given as follows:
\begin{equation}
\phi_0(x) \propto \exp\left[ -\frac{1}{\hbar}\int^x W(\bar{x}) \,d\bar{x} \right] ~.
\end{equation}
Inversely, with the knowledge of $\phi_0(x)$, one can always construct the superpotential and the potential whose ground-state wavefunction is $\phi_0(x)$:
\begin{align}
W(x) &= -\hbar\frac{\partial_x\phi_0(x)}{\phi_0(x)} = -\hbar\frac{d}{dx}\ln\phi_0(x) ~, \\
V(x) &= \hbar^2\left[ \left( \frac{d}{dx}\ln\phi_0(x) \right)^2 + \frac{d^2}{dx^2}\ln\phi_0(x) \right] = \hbar^2\frac{\partial_x^2\phi_0(x)}{\phi_0(x)} ~,
\end{align}
where $\partial_x \equiv \dfrac{d}{dx}$ and $\partial_x^2 \equiv \dfrac{d^2}{dx^2}$.
On the other hand, the zero mode of $\mathcal{A}^{\dag}$ is the inverse of the zero mode of $\mathcal{A}$, 
\begin{equation}
\mathcal{A}^{\dag}\phi_0^{-1}(x) = 0 ~.
\label{eq:2-Adag_zeromode}
\end{equation}

\begin{remark}[Multi-degrees of freedom]
For the case of $D$ degrees of freedom, the Hamiltonian is factorized as
\begin{equation}
\mathcal{H} \coloneqq \sum_{j=1}^{D}\mathcal{A}_j^{\dag}\mathcal{A}_j ~,
\end{equation}
and the ground-state wavefunction is given by
\begin{equation}
\mathcal{A}_j\phi_0(x) = 0 
\qquad\text{for $j=1,2,\ldots, D$} ~.
\end{equation}

Also, one can deal with the reduced `radial' equation in our formulation for one degree of freedom.
\begin{enumerate}\setlength{\leftskip}{-1em}
\item \textit{$d$ spacial dimensions ($D=d$)}.\quad
Suppose that the $d$-dimensional Schr\"{o}dinger equation in the Cartesian coordinate $(x_1,x_2,\ldots,x_d)$ is reduced to the following `radial' equation:
\begin{equation}
\left[ -\frac{\partial^2}{\partial r^2} - \frac{d-1}{r}\frac{\partial}{\partial r} + \frac{ \ell(\ell + d - 2)}{r^2} + V(r) \right] R(r) = \mathcal{E}R(r) ~,
\end{equation}
in which $r\coloneqq\sqrt{x_1^2+x_2^2+\cdots+x_d^2}$ and $\ell$ is the angular momentum quantum number.
Also, $R(r)$ denotes the radial wavefunction.
By writing $R(r) \equiv r^{-(d-1)/2}\phi(r)$, the equation becomes
\begin{equation}
\left[ -\frac{\partial^2}{\partial r^2} + \frac{\left( \ell + \frac{d-3}{2} \right)\left( \ell + \frac{d-3}{2} +1 \right)}{r^2} + V(r) \right] \phi(r) = \mathcal{E}\phi(r) ~,
\end{equation}
and the `Hamiltonian' $[\bullet]$ is factorized as
\begin{multline}
-\frac{\partial^2}{\partial r^2} + \frac{\left( \ell + \frac{d-1}{2} \right)\left( \ell + \frac{d-3}{2} \right)}{r^2} + V(r) \\
= \left( -\frac{\partial}{\partial r} + \frac{\ell + \frac{d-3}{2}}{r} + \widetilde{W}(r) \right) \left( \frac{\partial}{\partial r} + \frac{\ell + \frac{d-3}{2}}{r} + \widetilde{W}(r) \right) ~.
\end{multline}

\item \textit{$N$-body problem ($D=N$)}.\quad
Again, suppose that the Schr\"{o}dinger equation for an $N$-body problem is reduced to the following `radial' equation in the hypercentral formalism~\cite{delves1958tertiary,delves1960tertiary,ballot1980application,ferraris1995three,rosati2002hyperspherical}:
\begin{equation}
\left[ -\frac{\partial^2}{\partial x^2} - \frac{3N-4}{x}\frac{\partial}{\partial x} + \frac{ \gamma(\gamma + 3N - 5)}{x^2} + V(x) \right] X(x) = \mathcal{E}X(x) ~,
\end{equation}
where $x$ is the hyper radius and $\gamma$ is the hyper angular momentum quantum number.
Also, $X(x)$ denotes the hyper-radial wavefunction.
By writing $X(x) \equiv x^{-(3N-4)/2}\phi(x)$, the equation becomes
\begin{equation}
\left[ -\frac{\partial^2}{\partial x^2} + \frac{\gamma(\gamma + 3N - 5) + \frac{(3N-4)(3N-6)}{4}}{x^2} + V(x) \right] \phi(x) = \mathcal{E}\phi(x) ~,
\end{equation}
and the `Hamiltonian' $[\bullet]$ is factorized as
\begin{multline}
-\frac{\partial^2}{\partial x^2} + \frac{\gamma(\gamma + 3N - 5) + \frac{(3N-4)(3N-6)}{4}}{x^2} + V(x) \\
= \left( -\frac{\partial}{\partial x} + \frac{\gamma + \frac{3N-6}{2}}{x} + \widetilde{W}(x) \right)\left( \frac{\partial}{\partial x} + \frac{\gamma + \frac{3N-6}{2}}{x} + \widetilde{W}(x) \right) ~.
\end{multline}
\end{enumerate}
\end{remark}

\begin{remark}[Can you always factorize your Hamiltonian?]
One might wonder if a Hamiltonian is always expressed in a factorized form \eqref{eq:2-factorized}.
The answer is \textit{no}, for the first differential equation in Eq. \eqref{eq:2-RiccatiEq-W} is the Riccati equation, which does not always have a solution.

In the following, we consider a simple case where Eq. \eqref{eq:2-RiccatiEq-W} is solvable \textit{i.e.}, the harmonic oscillator:
\begin{equation}
W^2(x) - \frac{dW(x)}{dx} = x^2 - 1 ~.
\label{eq:2-RiccatiEq_HO}
\end{equation}
Here, we set $\hbar = \omega = 1$.
Apparently, $W_1(x) = x$ is a solution of Eq. \eqref{eq:2-RiccatiEq_HO}.
Now we write $W(x) \equiv W_1(x) + \widetilde{W}(x) = x + \widetilde{W}(x)$.
Then, Eq. \eqref{eq:2-RiccatiEq_HO} is reduced to 
\begin{equation}
\frac{d\widetilde{W}(x)}{dx} - 2x\widetilde{W}(x) = \widetilde{W}(x)^2 \qquad \text{: Bernoulli eq.}
\end{equation} 
By putting $\widetilde{W}^{-1}(x) \equiv w(x)$, this becomes a first-order ordinary differential equation:
\begin{equation}
\frac{dw(x)}{dx} = -2xw(x) - 1.
\end{equation}
Therefore, 
\begin{align}
w(x) &= \mathrm{e}^{-\int^x 2x\,dx}\left( -\int^x\mathrm{e}^{\int^x 2x\,dx} \,dx + \mathrm{const.} \right) \nonumber \\
&\equiv \mathrm{e}^{-x^2}\left( C - \int^x\mathrm{e}^{x^2} \,dx \right)
\end{align}
\begin{equation}
\therefore \quad \widetilde{W}(x) = \frac{\mathrm{e}^{x^2}}{C - \displaystyle\int^x\mathrm{e}^{x^2} \,dx} \neq 0 ~.
\end{equation}
Hence, we wind up with the general solution for Eq. \eqref{eq:2-RiccatiEq_HO}:
\begin{equation}
W(x) = x + \frac{\mathrm{e}^{x^2}}{C - \displaystyle\int^x\mathrm{e}^{x^2} \,dx} ~.
\end{equation}
Note that this superpotential has a singularity and so does the ground-state wavefunction $\phi_0(x)=\mathrm{e}^{-\frac{x^2}{2}}\mathrm{e}^{\int^x\widetilde{W} \,dx}$, and this is not a proper quantum mechanical system.
In order to obtain an appropriate ground-state wavefunction, we have to choose a singular solution for Eq. \eqref{eq:2-RiccatiEq-W}.

On the other hand, the Riccati equation with the potential \eqref{eq:2-partner_pot1}:
\begin{equation}
W(x)^2 + \frac{dW(x)}{dx} = x^2 - 1 ~,
\end{equation}
has solutions that have no singularities other than the singular solution $W(x)=x$.
This was already pointed out in 1980s.
For more details, see Refs. \cite{mielnik1984factorization,fernandez1984new,leach1985exact,mitra1989nonuniqueness}.
\end{remark}

\subsection{Associated Hamiltonians, Intertwining Relations, Crum's Theorem}
\subsubsection{An associated Hamiltonian}
In this subsection, we write
\begin{equation}
\mathcal{H}\equiv
\mathcal{H}^{[0]} 
= \mathcal{A}^{[0]\dag}\mathcal{A}^{[0]} 
= -\hbar^2\frac{d^2}{dx^2} + V^{[0]}(x) 
= \left( -\hbar\frac{d}{dx} + W^{[0]}(x) \right) \left( \hbar\frac{d}{dx} + W^{[0]}(x) \right) ~,
\end{equation}
and accordingly, its eigenvalues and the eigenfunctions are denoted by $\mathcal{E}^{[0]}_n$, $\phi^{[0]}_n$,
\begin{equation}
\mathcal{H}^{[0]}\phi^{[0]}_n = \mathcal{E}^{[0]}_n \phi^{[0]}_n ~.
\label{eq:2-SE0}
\end{equation}

Here, we define a new Hamiltonian $\mathcal{H}^{[1]}$ by
\begin{equation}
\mathcal{H}^{[1]} \coloneqq \mathcal{A}^{[0]}\mathcal{A}^{[0]\dag}
=  -\hbar^2\frac{d^2}{dx^2} + V^{[1]}(x) ~,~~~
V^{[1]}(x) \equiv W(x)^2 + \hbar\frac{dW(x)}{dx} ~.
\label{eq:2-partner_pot1}
\end{equation}
This is an associated Hamiltonian of $\mathcal{H}^{[0]}$.
Or sometimes, the two Hamiltonians $\mathcal{H}^{[0]}$, $\mathcal{H}^{[1]}$ are said to be (supersymmetric) partners.
The eigenvalue equation for $\mathcal{H}^{[1]}$ is
\begin{equation}
\mathcal{H}^{[1]}\phi^{[1]}_n = \mathcal{E}^{[1]}_n \phi^{[1]}_n ~.
\label{eq:2-SE1}
\end{equation}

Note that they are related by the following equation:
\begin{equation}
\mathcal{H}^{[1]} = \mathcal{H}^{[0]} + 2\hbar\frac{dW^{[0]}(x)}{dx} 
= \mathcal{H}^{[0]} + 2\hbar\frac{d^2}{dx^2}\ln\phi^{[0]}_0(x) ~.
\end{equation}
This expression is useful to see Eq. \eqref{eq:2-wf_j_Crum} \textit{etc.} are natural extensions of this relation.

\subsubsection{Intertwining relations}
One can see other relations between the two Hamiltonians, that is $\mathcal{H}^{[0]}$ and $\mathcal{H}^{[1]}$ are related via
\begin{gather}
\mathcal{A}^{[0]}\mathcal{H}^{[0]} = \mathcal{A}^{[0]}\mathcal{A}^{[0]\dag}\mathcal{A}^{[0]} = \mathcal{H}^{[1]}\mathcal{A}^{[0]} ~, \\
\mathcal{A}^{[0]\dag}\mathcal{H}^{[1]} = \mathcal{A}^{[0]\dag}\mathcal{A}^{[0]}\mathcal{A}^{[0]\dag} = \mathcal{H}^{[0]}\mathcal{A}^{[0]\dag} ~,
\end{gather}
which is referred to as the \textit{intertwining relations}.

These relations lead to the essentially isospectral property of the two Hamiltonians.
Let $\mathcal{A}^{[0]}$ act on Eq.~\eqref{eq:2-SE0} from the left.
Then we get
\begin{equation}
\mathcal{A}^{[0]}\mathcal{H}^{[0]}\phi^{[0]}_n = \mathcal{E}^{[0]}_n \mathcal{A}^{[0]}\phi^{[0]}_n
= \mathcal{H}^{[1]}\mathcal{A}^{[0]}\phi^{[0]}_n ~,
\end{equation}
which means $\mathcal{A}^{[0]}\phi^{[0]}_n$ is an eigenfunction of $\mathcal{H}^{[1]}$ with the energy $\mathcal{E}^{[0]}_n$.
Note however that, for $n=0$, the state $\phi^{[0]}_0$ is deleted, $\mathcal{A}^{[0]}\phi^{[0]}_0 = 0$. 
Also, by acting $\mathcal{A}^{[0]\dag}$ on Eq.~\eqref{eq:2-SE1} from the left, we get
\begin{equation}
\mathcal{A}^{[0]\dag}\mathcal{H}^{[1]}\phi^{[1]}_n = \mathcal{E}^{[1]}_n \mathcal{A}^{[0]\dag}\phi^{[1]}_n
= \mathcal{H}^{[1]}\mathcal{A}^{[0]\dag}\phi^{[1]}_n ~,
\end{equation}
and therefore $\mathcal{A}^{[0]\dag}\phi^{[1]}_n$ is an eigenfunction of $\mathcal{H}^{[0]}$ with the energy $\mathcal{E}^{[1]}_n$.
These imply the following:

\begin{enumerate}\setlength{\leftskip}{-1em}
\item The two Hamiltonians $\mathcal{H}^{[0]}$, $\mathcal{H}^{[1]}$ are essentially isospectral,
	\begin{equation}
	\mathcal{E}_n^{[1]} = \mathcal{E}_{n+1}^{[0]} \equiv \mathcal{E}_{n+1} ~,~~~
	\mathcal{E}_0^{[0]} \equiv \mathcal{E}_0 = 0 ~,~~~
	n=0,1,2,\ldots ~.
	\end{equation}
	Note that the lowest eigenstate of $\mathcal{H}^{[1]}$, \textit{i.e.}, $\mathcal{E}_0^{[1]}=\mathcal{E}_1$, is not zero.
\item The eigenfunctions of $\mathcal{H}^{[1]}$ are
	\begin{equation}
	\phi_n^{[1]}(x) \propto \mathcal{A}^{[0]}\phi_{n+1}^{[0]}(x)
	= \frac{\mathrm{W}\left[\phi_0^{[0]},\phi_{n+1}^{[0]}\right](x)}{\phi_0^{[0]}(x)} ~,~~~
	n=1,2,\ldots ~,
	\end{equation}
	where $\mathrm{W}[f_1,\ldots,f_n](x)$ denotes the Wronskian.
	The operator $\mathcal{A}^{[0]}$ maps an eigenfunction of $\mathcal{H}^{[0]}$ with $n$ nodes to that of $\mathcal{H}^{[1]}$ with $n-1$ nodes, or deletes the lowest eigenstate of $\mathcal{H}^{[0]}$.
	Usually, the coefficient of proportionality is taken to be $1/\sqrt{\mathcal{E}_{n+1}}$.
\item The eigenfunctions of $\mathcal{H}^{[0]}$ are
	\begin{equation}
	\phi_{n+1}^{[0]}(x) \propto \mathcal{A}^{[0]\dag}\phi_{n}^{[1]}(x) ~,~~~
	n=0,1,2,\ldots ~,
	\end{equation}
	which means the operator $\mathcal{A}^{[0]\dag}$ maps an eigenfunction of $\mathcal{H}^{[1]}$ with $n$ nodes to that of $\mathcal{H}^{[0]}$ with $n+1$ nodes.
	Usually, the coefficient of proportionality is taken to be $1/\sqrt{\mathcal{E}_{n+1}}$.
\end{enumerate}

The transformation from $\mathcal{H}^{[0]}$ to $\mathcal{H}^{[1]}$ (or reversely from $\mathcal{H}^{[1]}$ to $\mathcal{H}^{[0]}$) is called Darboux--Crum transformation.
We visualize the transformation in Fig. \ref{fig:2-SUSYtransf}.

\begin{figure}[t]
\centering
	\begin{tikzpicture}
	\draw[->,thick,>=stealth] (0,-1)--(0,9);
	\node[right] at(-1.5,0) {$\mathcal{E}_0=0$};
	\node[right] at(-1.5,2) {$\mathcal{E}_1$};
	\node[right] at(-1.5,4) {$\mathcal{E}_2$};
	\node[right] at(-1.5,6) {$\mathcal{E}_3$};
	\node[right] at(-1.5,8) {$\mathcal{E}_4$};
	
	\draw[ultra thick] (1,0)--(3,0) node[above] at(2.1,0) {{\footnotesize $\phi_0^{[0]}$}};
	\draw[ultra thick] (1,2)--(3,2) node[above] at(2.1,2) {{\footnotesize $\phi_1^{[0]}$}};
	\draw[ultra thick] (1,4)--(3,4) node[above] at(2.1,4) {{\footnotesize $\phi_2^{[0]}$}};
	\draw[ultra thick] (1,6)--(3,6) node[above] at(2.1,6) {{\footnotesize $\phi_3^{[0]}$}};
	\draw[ultra thick] (1,8)--(3,8) node[above] at(2.1,8) {{\footnotesize $\phi_4^{[0]}$}};
	\node at(2.1,-1.5) {$\mathcal{H}^{[0]}$};
	
	\draw[ultra thick] (6,2)--(8,2) node[above] at(7.1,2) {{\footnotesize $\phi_0^{[1]}$}};
	\draw[ultra thick] (6,4)--(8,4) node[above] at(7.1,4) {{\footnotesize $\phi_1^{[1]}$}};
	\draw[ultra thick] (6,6)--(8,6) node[above] at(7.1,6) {{\footnotesize $\phi_2^{[1]}$}};
	\draw[ultra thick] (6,8)--(8,8) node[above] at(7.1,8) {{\footnotesize $\phi_3^{[1]}$}};
	\node at(7.1,-1.5) {$\mathcal{H}^{[1]}$};

	\draw[dashed] (0,0)--(1,0);
	\draw[dashed] (0,2)--(1,2);
	\draw[dashed] (0,4)--(1,4);
	\draw[dashed] (0,6)--(1,6);
	\draw[dashed] (0,8)--(1,8);
	
	\draw[dashed] (3,2)--(6,2);
	\draw[dashed] (3,4)--(6,4);
	\draw[dashed] (3,6)--(6,6);
	\draw[dashed] (3,8)--(6,8);
	
	\draw[->] (3.1,2.1)--(5.9,2.1) node[above] at(4.4,2.1) {$\mathcal{A}^{[0]}$};
	\draw[<-] (3.1,1.9)--(5.9,1.9) node[below] at(4.4,1.9) {$\mathcal{A}^{[0]\dag}$};
	\draw[->] (3.1,4.1)--(5.9,4.1) node[above] at(4.4,4.1) {$\mathcal{A}^{[0]}$};
	\draw[<-] (3.1,3.9)--(5.9,3.9) node[below] at(4.4,3.9) {$\mathcal{A}^{[0]\dag}$};
	\draw[->] (3.1,6.1)--(5.9,6.1) node[above] at(4.4,6.1) {$\mathcal{A}^{[0]}$};
	\draw[<-] (3.1,5.9)--(5.9,5.9) node[below] at(4.4,5.9) {$\mathcal{A}^{[0]\dag}$};
	\draw[->] (3.1,8.1)--(5.9,8.1) node[above] at(4.4,8.1) {$\mathcal{A}^{[0]}$};
	\draw[<-] (3.1,7.9)--(5.9,7.9) node[below] at(4.4,7.9) {$\mathcal{A}^{[0]\dag}$};
	\end{tikzpicture}
\caption[Interrelation between the partner Hamiltonians $\mathcal{H}^{[0]}$, $\mathcal{H}^{[1]}$.]
	{Interrelation between the partner Hamiltonians $\mathcal{H}^{[0]}$, $\mathcal{H}^{[1]}$. 
	They are related to each other by Darboux--Crum transformation.
	The spectral property can be seen as a realization of supersymmetry in quantum mechanics.}
\label{fig:2-SUSYtransf}
\end{figure}

\subsubsection{Crum's theorem}
We define a sequence of Hamiltonians associated with $\mathcal{H}$ as follows.
One can construct associated Hamiltonian systems as many as the number of the discrete eigenvalues of $\mathcal{H}^{[0]}$.
\begin{align}\qquad
\mathcal{H}^{[0]} &= \mathcal{A}^{[0]\dag}\mathcal{A}^{[0]} ~,~~~
\mathcal{H}^{[1]} = \mathcal{A}^{[0]}\mathcal{A}^{[0]\dag} ~, \\
\mathcal{H}^{[1]} &\equiv \mathcal{A}^{[1]\dag}\mathcal{A}^{[1]} + \mathcal{E}_1 ~,~~~
\mathcal{H}^{[2]} \coloneqq \mathcal{A}^{[1]}\mathcal{A}^{[1]\dag} + \mathcal{E}_1 ~, \\
&\vdots \nonumber \\
\mathcal{H}^{[j]} &\equiv \mathcal{A}^{[j]\dag}\mathcal{A}^{[j]} + \mathcal{E}_{j} ~,~~~
\mathcal{H}^{[j+1]} \coloneqq \mathcal{A}^{[j]}\mathcal{A}^{[j]\dag} + \mathcal{E}_{j} ~, \\
&\vdots \nonumber \\
&\qquad\qquad j = 0,1,\ldots \text{~(as many as the \# of discrete eigenvalues of $\mathcal{H}^{[0]}$)} ~,
\nonumber
\end{align}
where the operators $\mathcal{A}^{[j]}$ and $\mathcal{A}^{[j]}$ are 
\begin{equation}
\mathcal{A}^{[j]} \coloneqq \hbar\frac{d}{dx} + W^{[j]}(x) ~,~~~
\mathcal{A}^{{[j]}\dag} = -\hbar\frac{d}{dx} + W^{[j]}(x) ~,
\end{equation}
In what follows, $\phi_n^{[j]}(x)$ denotes the $n$-th eigenfunction of $\mathcal{H}^{[j]}$,
\begin{equation}
\mathcal{H}^{[j]}\phi_n^{[j]}(x) = \mathcal{E}_{n+j}\phi_n^{[j]}(x) ~.
\end{equation}

From the definitions, $\mathcal{H}^{[j]}$ and $\mathcal{H}^{[j+1]}$ are related linearly by
\begin{gather}
\mathcal{A}^{[j]}\mathcal{H}^{[j]} = \mathcal{A}^{[j]}\mathcal{A}^{[j]\dag}\mathcal{A}^{[j]} = \mathcal{H}^{[j+1]}\mathcal{A}^{[j]} ~, \\
\mathcal{A}^{[j]\dag}\mathcal{H}^{[j+1]} = \mathcal{A}^{[j]\dag}\mathcal{A}^{[j]}\mathcal{A}^{[j]\dag} = \mathcal{H}^{[j]}\mathcal{A}^{[j]\dag} ~.
\end{gather}
These equalities lead  
\begin{align}
\phi^{[j+1]}_n(x) &\propto \mathcal{A}^{[j]}\phi^{[j]}_{n+1}(x) ~, \\
\phi^{[j]}_{n+1}(x) &\propto \mathcal{A}^{[j]\dag}\phi^{[j+1]}_n(x) ~,
\end{align}
whose coefficients of proportionality is taken to be $1/\sqrt{\mathcal{E}_{n+j+1}-\mathcal{E}_j}$.

$\mathcal{H}^{[j+1]}$ and $\phi_n^{[j+1]}(x)$ are expressed in terms of in terms of $\mathcal{H}^{[0]}$ and $\phi_n^{[0]}(x)$ by virtue of the Wronskian as follows:
\begin{align}
\mathcal{H}^{[j+1]} &= \mathcal{H}^{[0]} - 2\hbar^2\frac{d^2}{dx^2}\ln\left|\mathrm{W}\left[\phi_0^{[0]},\phi_1^{[0]},\ldots,\phi_{j}^{[0]}\right](x)\right| ~, \\
\phi_n^{[j+1]}(x) &\propto \frac{\mathrm{W}\left[\phi_0^{[0]},\phi_1^{[0]},\ldots,\phi_{j}^{[0]},\phi_{n+j}^{[0]}\right](x)}{\mathrm{W}\left[\phi_0^{[0]},\phi_1^{[0]},\ldots,\phi_{j}^{[0]}\right](x)} 
\qquad\text{with the energy $\mathcal{E}_{n+j}$} ~,
\label{eq:2-wf_j_Crum}
\end{align}
or in terms of $\mathcal{H}^{[j]}$ and $\phi_n^{[j]}(x)$ as
\begin{align}
\mathcal{H}^{[j+1]} &= \mathcal{H}^{[j]} + 2\hbar\frac{d^2}{dx^2}\ln\phi^{[j]}_0(x) ~, \\
\phi_n^{[j+1]}(x) &\propto \frac{\mathrm{W}\left[\phi_0^{[j]},\phi_{n+1}^{[j]}\right](x)}{\phi_0^{[j]}(x)} 
\qquad\text{with the energy $\mathcal{E}_{n+j}$} ~.
\end{align}

We arrive at the following theorem:

\begin{theorem}[Crum, 1955~\cite{10.1093/qmath/6.1.121}]
For a given Hamiltonian system $\mathcal{H} \equiv \mathcal{H}^{[0]}$, there are associated Hamiltonian systems $\mathcal{H}^{[1]}, \mathcal{H}^{[2]}, \ldots$, as many as the total number of discrete eigenvalues of the original system $\mathcal{H}^{[0]}$. 
They share the same eigenvalues $\{\mathcal{E}_n\}$ of the original system and the eigenfunctions of $\mathcal{H}^{[j]}$ and $\mathcal{H}^{[j+1]}$ are related linearly by $\mathcal{A}^{[j]}$ and $\mathcal{A}^{[j]\dag}$.
\end{theorem}

\begin{figure}[t]
\centering
\scalebox{0.75}{
	\begin{tikzpicture}
	\draw[->,thick,>=stealth] (0,-1)--(0,9);
	\node[right] at(-1.5,0) {$\mathcal{E}_0=0$};
	\node[right] at(-1.5,2) {$\mathcal{E}_1$};
	\node[right] at(-1.5,4) {$\mathcal{E}_2$};
	\node[right] at(-1.5,6) {$\mathcal{E}_3$};
	\node[right] at(-1.5,8) {$\mathcal{E}_4$};
	
	\draw[ultra thick] (1,0)--(3,0) node[above] at(2.1,0) {{\footnotesize $\phi_0^{[0]}$}};
	\draw[ultra thick] (1,2)--(3,2) node[above] at(2.1,2) {{\footnotesize $\phi_1^{[0]}$}};
	\draw[ultra thick] (1,4)--(3,4) node[above] at(2.1,4) {{\footnotesize $\phi_2^{[0]}$}};
	\draw[ultra thick] (1,6)--(3,6) node[above] at(2.1,6) {{\footnotesize $\phi_3^{[0]}$}};
	\draw[ultra thick] (1,8)--(3,8) node[above] at(2.1,8) {{\footnotesize $\phi_4^{[0]}$}};
	\node at(2.1,-1.5) {$\mathcal{H}^{[0]}$};
	
	\draw[ultra thick] (6,2)--(8,2) node[above] at(7.1,2) {{\footnotesize $\phi_0^{[1]}$}};
	\draw[ultra thick] (6,4)--(8,4) node[above] at(7.1,4) {{\footnotesize $\phi_1^{[1]}$}};
	\draw[ultra thick] (6,6)--(8,6) node[above] at(7.1,6) {{\footnotesize $\phi_2^{[1]}$}};
	\draw[ultra thick] (6,8)--(8,8) node[above] at(7.1,8) {{\footnotesize $\phi_3^{[1]}$}};
	\node at(7.1,-1.5) {$\mathcal{H}^{[1]}$};
	
	\draw[ultra thick] (11,4)--(13,4) node[above] at(12.1,4) {{\footnotesize $\phi_0^{[2]}$}};
	\draw[ultra thick] (11,6)--(13,6) node[above] at(12.1,6) {{\footnotesize $\phi_1^{[2]}$}};
	\draw[ultra thick] (11,8)--(13,8) node[above] at(12.1,8) {{\footnotesize $\phi_2^{[2]}$}};
	\node at(12.1,-1.5) {$\mathcal{H}^{[2]}$};
	
	\draw[ultra thick] (16,6)--(18,6) node[above] at(17.1,6) {{\footnotesize $\phi_0^{[3]}$}};
	\draw[ultra thick] (16,8)--(18,8) node[above] at(17.1,8) {{\footnotesize $\phi_1^{[3]}$}};
	\node at(17.1,-1.5) {$\mathcal{H}^{[3]}$};

	\draw[dashed] (0,0)--(1,0);
	\draw[dashed] (0,2)--(1,2);
	\draw[dashed] (0,4)--(1,4);
	\draw[dashed] (0,6)--(1,6);
	\draw[dashed] (0,8)--(1,8);

	\draw[dashed] (3,2)--(6,2);
	\draw[dashed] (3,4)--(6,4);
	\draw[dashed] (3,6)--(6,6);
	\draw[dashed] (3,8)--(6,8);
	
	\draw[dashed] (8,4)--(11,4);
	\draw[dashed] (8,6)--(11,6);
	\draw[dashed] (8,8)--(11,8);

	\draw[dashed] (13,6)--(16,6);
	\draw[dashed] (13,8)--(16,8);

	\draw[->] (3.1,2.1)--(5.9,2.1) node[above] at(4.4,2.1) {$\mathcal{A}^{[0]}$};
	\draw[<-] (3.1,1.9)--(5.9,1.9) node[below] at(4.4,1.9) {$\mathcal{A}^{[0]\dag}$};
	\draw[->] (3.1,4.1)--(5.9,4.1) node[above] at(4.4,4.1) {$\mathcal{A}^{[0]}$};
	\draw[<-] (3.1,3.9)--(5.9,3.9) node[below] at(4.4,3.9) {$\mathcal{A}^{[0]\dag}$};
	\draw[->] (3.1,6.1)--(5.9,6.1) node[above] at(4.4,6.1) {$\mathcal{A}^{[0]}$};
	\draw[<-] (3.1,5.9)--(5.9,5.9) node[below] at(4.4,5.9) {$\mathcal{A}^{[0]\dag}$};
	\draw[->] (3.1,8.1)--(5.9,8.1) node[above] at(4.4,8.1) {$\mathcal{A}^{[0]}$};
	\draw[<-] (3.1,7.9)--(5.9,7.9) node[below] at(4.4,7.9) {$\mathcal{A}^{[0]\dag}$};

	\draw[->] (8.1,4.1)--(10.9,4.1) node[above] at(9.4,4.1) {$\mathcal{A}^{[1]}$};
	\draw[<-] (8.1,3.9)--(10.9,3.9) node[below] at(9.4,3.9) {$\mathcal{A}^{[1]\dag}$};
	\draw[->] (8.1,6.1)--(10.9,6.1) node[above] at(9.4,6.1) {$\mathcal{A}^{[1]}$};
	\draw[<-] (8.1,5.9)--(10.9,5.9) node[below] at(9.4,5.9) {$\mathcal{A}^{[1]\dag}$};
	\draw[->] (8.1,8.1)--(10.9,8.1) node[above] at(9.4,8.1) {$\mathcal{A}^{[1]}$};
	\draw[<-] (8.1,7.9)--(10.9,7.9) node[below] at(9.4,7.9) {$\mathcal{A}^{[1]\dag}$};

	\draw[->] (13.1,6.1)--(15.9,6.1) node[above] at(14.4,6.1) {$\mathcal{A}^{[2]}$};
	\draw[<-] (13.1,5.9)--(15.9,5.9) node[below] at(14.4,5.9) {$\mathcal{A}^{[2]\dag}$};
	\draw[->] (13.1,8.1)--(15.9,8.1) node[above] at(14.4,8.1) {$\mathcal{A}^{[2]}$};
	\draw[<-] (13.1,7.9)--(15.9,7.9) node[below] at(14.4,7.9) {$\mathcal{A}^{[2]\dag}$};

	\node at(19,7.2) {$\cdots$};
	\end{tikzpicture}
}
\caption[A visualization of Crum's theorem.]
	{A visualization of Crum's theorem.
	It reveals the general structure of the solution space of the one-dimensional Schr\"{o}dinger equation.}
\label{fig:2-CrumTh}
\end{figure}
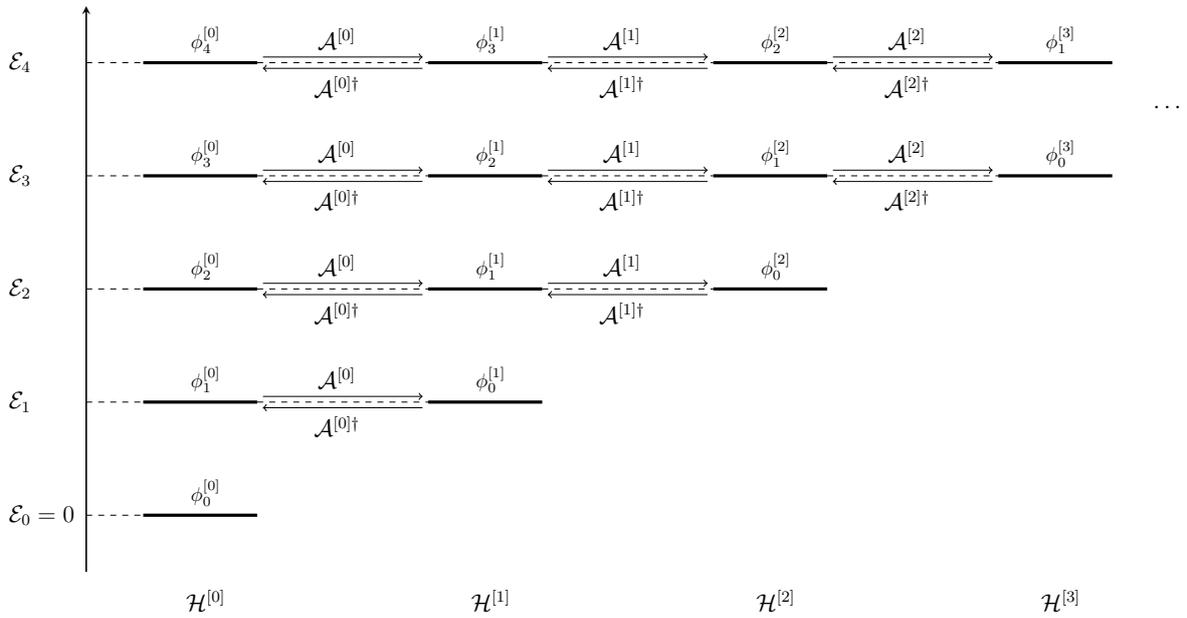

\noindent
We visualize the situation of the Crum's theorem in Fig. \ref{fig:2-CrumTh}.

\begin{remark}[Supersymmetric quantum mechanics]
The formulation above is often referred to as supersymmetric quantum mechanics (SUSY QM).
However, it does not necessarily concern supersymmetry itself.
As one can see, the Crum's theorem (1955, Ref. \cite{10.1093/qmath/6.1.121}) plays an essential role in our formulation, but they are about a decade before the dawn of supersymmetry (in the mid-1960s).
\end{remark}

\section{Shape Invariance}
\subsection{Generalities}
If the partner Hamiltonians $\mathcal{H}^{[0]},\mathcal{H}^{[1]}$ (or generally $\mathcal{H}^{[j]},\mathcal{H}^{[j+1]}$) are similar in `shape', that is, the potentials are in the same functional form but different in the parameter(s) in it, say $\bm{a}=(a_1,a_2,\ldots)$, the Hamiltonians are said to be \textit{shape invariant}~\cite{gendenshtein1983derivation}.
More precisely, the two Hamiltonians are shape invariant, if
\begin{equation}
-\hbar^2\frac{d^2}{dx^2} + W(x;\bm{a})^2 + \hbar\frac{dW(x;\bm{a})}{dx}
= -\hbar^2\frac{d^2}{dx^2} + W(x;f(\bm{a}))^2 - \hbar\frac{dW(x;f(\bm{a}))}{dx} + R(\bm{a}) ~,
\end{equation}
where we write the parameter dependency on the superpotential explicitly, $W(x)\equiv W(x;\bm{a})$, and $f(\bm{a})$ is some function of $\bm{a}$, \textit{e.g.}, $f(a_1,a_2) = (a_1+1,a_2-1)$.
$R(\bm{a})$ is a constant depending on the parameter(s) $\bm{a}$.
Note that if $\mathcal{H}^{[0]},\mathcal{H}^{[1]}$ are shape invariant, then $\mathcal{H}^{[j]}$ and $\mathcal{H}^{[j+1]}$ for any $j$ are shape invariant,
\begin{multline}
-\hbar^2\frac{d^2}{dx^2} + W(x;f^j(\bm{a}))^2 + \hbar\frac{dW(x;f^j(\bm{a}))}{dx} \\
= -\hbar^2\frac{d^2}{dx^2} + W(x;f^{j+1}(\bm{a}))^2 - \hbar\frac{dW(x;f^{j+1}(\bm{a}))}{dx} + \sum_{k=0}^jR(f^k(\bm{a})) 
\protect\footnotemark .
\end{multline}
\footnotetext{For a function $f(x)$ and a positive integer $m$, $f^m(x)$ is a shorthand for $\underbrace{f\circ f\circ\cdots\circ f}_{\text{$m$ times}}(x)$, while $f(x)^m$ means $[f(x)]^m$.}

The shape invariance is a sufficient condition of the exact solvability of the Schr\"{o}dinger equation.
The only knowledge of the ground state $\phi^{[0]}_0(x)$ solves the whole spectra in Fig. \ref{fig:2-CrumTh}.
For any Hamiltonian $\mathcal{H}^{[j]}$, the ground-state wavefunction is 
\begin{multline}
\quad
\mathcal{A}^{[j]}\phi^{[j]}_0(x;\bm{a}) = \left( \frac{d}{dx} + W(x;f^j(\bm{a})) \right)\phi^{[j]}_0(x;\bm{a}) = 0 \\
\Longrightarrow\quad
\phi^{[j]}_0(x;\bm{a}) \propto \phi^{[0]}_0(x;f^j(\bm{a})) ~.
\quad
\end{multline}
Then, it is almost obvious from Fig. \ref{fig:2-CrumTh} that the eigenfunction $\phi^{[j]}_n(x)$ is
\begin{equation}
\phi^{[j]}_n(x) \propto \mathcal{A}^{[j]\dag}\mathcal{A}^{[j+1]\dag}\cdots\mathcal{A}^{[j+n-1]\dag} \phi^{[0]}_0(x;f^{j+n}(\bm{a})) ~.
\end{equation}
Note that in the context of the Crum's theorem, the shape invariance is understood as the relation between two eigenstates by a simple parameter change, $\phi^{[j+1]}_n(x;f^{j+1}(\bm{a}))\propto\phi^{[j]}_n(x;f^j(\bm{a}))$.

As for the energy eigenvalues, $R$ plays a central role.
Since
\begin{equation}
\mathcal{H}^{[j]} + \mathcal{E}_{j} 
= -\hbar^2\frac{d^2}{dx^2} + W(x;f^{j}(\bm{a}))^2 - \hbar\frac{dW(x;f^{j}(\bm{a}))}{dx} + \sum_{k=0}^{j-1}R(f^k(\bm{a})) ~,
\end{equation}
one can identify 
\begin{equation}
\mathcal{E}_{j} = \sum_{k=0}^{j-1}R(f^k(\bm{a})) ~.
\end{equation}
Actually, $R(f^j(\bm{a}))$ means the energy gap between the ground-state energies of $\mathcal{H}^{[j]}$ and $\mathcal{H}^{[j+1]}$, \textit{i.e.}, $R(f^j(\bm{a})) = \mathcal{E}_{j+1}-\mathcal{E}_{j}$.

In summary, the $n$-th eigenfunction of the original Hamiltonian $\mathcal{H}^{[0]}$ and its eigenvalue are
\begin{align}
&\text{Eigenfunction: }& &\phi^{[0]}_n(x;\bm{a}) \propto \mathcal{A}^{[0]\dag}\mathcal{A}^{[1]\dag} \cdots \mathcal{A}^{[n-1]\dag} \phi^{[0]}_0(x;f^{n}(\bm{a})) ~,\quad 
\label{eq:2-SI_Rodrigues} \\
&\text{Eigenvalue: }& &\mathcal{E}_n = \sum_{k=0}^{n-1}R(f^k(\bm{a})) ~.
\label{eq:2-SI_Energies}
\end{align}
Eq. \eqref{eq:2-SI_Rodrigues} can be seen as the generalization of Rodrigues' formula.

\subsection{Conventional Shape-invariant Potentials}
\label{sec:2-3-2}
\subsubsection{1-dim. harmonic oscillator (H)}
The ground-state wavefunction of the one-dimensional (1-dim.) harmonic oscillator $\phi_0(x)$ is
\begin{equation}
\phi_0(x) = \mathrm{e}^{-\frac{\omega x^2}{2\hbar}} ~,~~~ x \in (-\infty,\infty) ~,
\end{equation}
with the angular frequency $\omega >0$.
Starting from the ground-state wavefunction, the superpotential $W(x)$ and the potential $V^{[0]}(x)$ are constructed as follows:
\begin{align}
W(x) &\coloneqq -\hbar\frac{d}{dx}\ln\phi_0(x) = \omega x ~, \\
V^{[0]}(x) &\coloneqq W(x)^2 - \hbar\frac{dW(x)}{dx} = \omega^2x^2 - \hbar\omega ~.
\end{align}

Now the partner potential $V^{[1]}(x)$ is
\begin{equation}
V^{[1]}(x) = W(x)^2 + \hbar\frac{dW(x)}{dx} = \omega^2x^2 + \hbar\omega ~.
\end{equation}
Clearly, the 1-dim. harmonic oscillator is shape invariant, and no parameter changes during the shape-invariant transformation.
From \eqref{eq:2-SI_Energies}, the energy eigenvalue $\mathcal{E}_n$ is
\begin{equation}
\mathcal{E}_n = 2n\hbar\omega ~,~~~
n=0,1,2,\ldots ~.
\end{equation}

The Schr\"{o}dinger equation is 
\begin{equation}
-\hbar^2\frac{d^2\phi_n(x)}{dx^2} + \left( \omega^2x^2 - \hbar\omega \right) \phi_n(x) = 2n\hbar\omega\phi_n(x) ~,
\end{equation}
which is reduced to 
\begin{equation}
-\frac{d^2\check{\phi}_n(\xi)}{d\xi^2} + \left( \xi^2 - 1 \right) \check{\phi}_n(\xi) = 2n\check{\phi}_n(\xi)
\end{equation}
under $x\to\xi\equiv\sqrt{\omega/\hbar}\,x$ and $\phi_n(x)\equiv\check{\phi}_n(\xi)$.
The solution of this equation, \textit{i.e.}, the eigenfunction turns out to be
\begin{equation}
\check{\phi}_n(\xi) = \mathrm{e}^{-\frac{\xi^2}{2}}H_n(\xi)
= \phi_n(x) = \mathrm{e}^{-\frac{\omega x^2}{2\hbar}}H_n\left( \sqrt{\frac{\omega}{\hbar}}x \right) 
\equiv \phi_n^{\rm (H)}(x) ~,
\end{equation}
in which $H_n(x)$ is the Hermite polynomial of degree $n$.

\subsubsection{Radial oscillator (L)}
The ground-state wavefunction of the radial oscillator $\phi_0(x)$ is
\begin{equation}
\phi_0(x) = \mathrm{e}^{-\frac{\omega x^2}{2\hbar}}x^g ~,~~~ x \in (0,\infty) ~,~~~ g > \frac{1}{2} ~,
\end{equation}
with the angular frequency $\omega >0$.
Starting from the ground-state wavefunction, the superpotential $W(x)$ and the potential $V^{[0]}(x)$ are constructed as follows:
\begin{align}
W(x) &\coloneqq -\hbar\frac{d}{dx}\ln\phi_0(x) = \omega x - \frac{\hbar g}{x} ~, 
\label{eq:2-superpot_L} \\
V^{[0]}(x) &\coloneqq W(x)^2 - \hbar\frac{dW(x)}{dx} = \omega^2x^2 + \frac{\hbar^2g(g-1)}{x^2} - \hbar\omega(2g+1) ~.
\end{align}

Now the partner potential $V^{[1]}(x)$ is
\begin{equation}
V^{[1]}(x) = W(x)^2 + \hbar\frac{dW(x)}{dx} = \omega^2x^2 + \frac{\hbar^2g(g+1)}{x^2} - \hbar\omega(2g-1) ~.
\end{equation}
The radial oscillator is shape invariant under the following change of a parameter:  $g\to g+1$.
From \eqref{eq:2-SI_Energies}, the energy eigenvalue $\mathcal{E}_n$ is
\begin{equation}
\mathcal{E}_n = 4n\hbar\omega ~,~~~
n=0,1,2,\ldots ~.
\end{equation}

The Schr\"{o}dinger equation is 
\begin{equation}
-\hbar^2\frac{d^2\phi_n(x)}{dx^2} + \left[ \omega^2x^2 + \frac{\hbar^2g(g-1)}{x^2} - \hbar\omega(2g+1) \right] \phi_n(x) = 4n\hbar\omega\phi_n(x) ~,
\end{equation}
which is reduced to 
\begin{equation}
-\frac{d^2\check{\phi}_n(\xi)}{d\xi^2} + \left[ \xi^2 + \frac{g(g-1)}{\xi^2} - 2g - 1 \right] \check{\phi}_n(\xi) = 4n\check{\phi}_n(\xi)
\end{equation}
under $x\to\xi\equiv\sqrt{\omega/\hbar}\,x$ and $\phi_n(x)\equiv\check{\phi}_n(\xi)$.
The solution of this equation, \textit{i.e.}, the eigenfunction turns out to be
\begin{equation}
\check{\phi}_n(\xi) = \mathrm{e}^{-\frac{\xi^2}{2}}\xi^gL_n^{(g-\frac{1}{2})}(\xi^2)
= \phi_n(x) = \mathrm{e}^{-\frac{\omega x^2}{2\hbar}}\left( \sqrt{\frac{\omega}{\hbar}}x \right)^gL_n^{(g-\frac{1}{2})}\left( \frac{\omega}{\hbar}x^2 \right) 
\equiv \phi_n^{\rm (L)}(x) ~,
\end{equation}
in which $L_n^{(\alpha)}(x)$ is the Laguerre polynomial of degree $n$.

\subsubsection{P\"{o}schl--Teller potential (J)}
The ground-state wavefunction of the P\"{o}schl--Teller potential $\phi_0(x)$ is
\begin{equation}
\phi_0(x) = (\sin x)^g(\cos x)^h ~,~~~ x \in \left( 0,\frac{\pi}{2} \right) ~,~~~ g,h > \frac{1}{2} ~.
\end{equation}
Starting from the ground-state wavefunction, the superpotential $W(x)$ and the potential $V^{[0]}(x)$ are constructed as follows:
\begin{align}
W(x) &\coloneqq -\hbar\frac{d}{dx}\ln\phi_0(x) = -\hbar (g\cot x - h\tan x) ~,
\label{eq:2-superpot_J} \\
V^{[0]}(x) &\coloneqq W(x)^2 - \hbar\frac{dW(x)}{dx} = \frac{\hbar^2 g(g+1)}{\sin^2x} + \frac{\hbar^2 h(h+1)}{\cos^2x} - \hbar^2(g+h)^2 ~.
\end{align}

Now the partner potential $V^{[1]}(x)$ is
\begin{equation}
V^{[1]}(x) = W(x)^2 + \hbar\frac{dW(x)}{dx} = \frac{\hbar^2 g(g+1)}{\sin^2x} + \frac{\hbar^2 h(h+1)}{\cos^2x} - \hbar^2(g+h)^2  ~.
\end{equation}
The P\"{o}schl--Teller potential is shape invariant under the following change of parameters:  $g\to g+1$, $h\to h+1$.
From \eqref{eq:2-SI_Energies}, the energy eigenvalue $\mathcal{E}_n$ is
\begin{equation}
\mathcal{E}_n = 4\hbar^2n(n+g+h) ~,~~~
n=0,1,2,\ldots ~.
\end{equation}

The Schr\"{o}dinger equation is 
\begin{equation}
-\hbar^2\frac{d^2\phi_n(x)}{dx^2} + \left[ \frac{\hbar^2 g(g-1)}{\sin^2x} + \frac{\hbar^2 h(h-1)}{\cos^2x} - \hbar^2(g+h)^2 \right] \phi_n(x) = 4\hbar^2n(n+g+h)\phi_n(x) ~,
\end{equation}
which is reduced to 
\begin{equation}
-\frac{d^2\phi_n(x)}{dx^2} + \left[ \frac{g(g-1)}{\sin^2x} + \frac{h(h-1)}{\cos^2x} - (g+h)^2 \right] \phi_n(x) = 4n(n+g+h)\phi_n(x) ~.
\end{equation}
The solution of this equation, \textit{i.e.}, the eigenfunction turns out to be
\begin{equation}
\phi_n(x) = (\sin x)^g(\cos x)^h P_n^{(g-\frac{1}{2},h-\frac{1}{2})}(\cos 2x) 
\equiv \phi_n^{\rm (J)}(x) ~,
\end{equation}
in which $P_n^{(\alpha,\beta)}(x)$ is the Jacobi polynomial of degree $n$.

\subsubsection{Other conventional shape-invariant potentials}
\begin{itemize}\setlength{\leftskip}{-1em}\bf

\item $\dfrac{1}{\sin^2x}$-potential\quad{\rm (*A special case of the P\"{o}schl--Teller potenial with $g=h$.)}
\begin{itemize}\setlength{\leftskip}{-2em}
\item[] $\displaystyle\phi_0(x) = (\sin x)^g ~,~~~ x \in (0,\pi) ~,~~~ g > \frac{1}{2} ~,~~~ g \to g+1 ~.$
\item[] $\displaystyle W(x) = -g\cot x ~,~~~ V(x) = \frac{\hbar^2g(g-1)}{\sin^2x} - \hbar^2g^2 ~.$
\item[] $\displaystyle\mathcal{E}_n = \hbar^2n(n+2g) ~,~~~ \phi_n(x) = (\sin x)^g P_n^{(g-\frac{1}{2},g-\frac{1}{2})}(\cos x) ~,~~~ n=0,1,2,\ldots ~.$
\end{itemize}

\item Coulomb potential
\begin{itemize}\setlength{\leftskip}{-2em}\rm
\item[] $\displaystyle\phi_0(x) = \mathrm{e}^{-\frac{e^2x}{2\hbar^2g}}x^g ~,~~~ x \in (0,\infty) ~,~~~ g > \frac{1}{2} ~,~~~ \text{$e$: electric charge} ~,~~~ g \to g+1 ~.$
\item[] $\displaystyle W(x) = \frac{e^2}{2\hbar g} - \frac{\hbar g}{x} ~,~~~ V(x) = \frac{\hbar^2g(g-1)}{x^2} - \frac{e^2}{x} + \frac{e^4}{4\hbar^2g^2} ~.$
\item[] $\displaystyle\mathcal{E}_n = \frac{e^4}{4\hbar^2g^2} - \frac{e^4}{4\hbar^2(g+n)^2} ~,~~~ \phi_n(x) = \mathrm{e}^{-\frac{e^2x}{2\hbar^2(g+n)}}\left( \frac{e^2}{2\hbar^2}x \right)^g L_n^{(2g-1)}\left( \frac{e^2}{\hbar^2(g+n)}x \right) ~,$

	\hfill{$n=0,1,2,\ldots ~.$}
\end{itemize}

\item Kepler problem on a hypersphere
\begin{itemize}\setlength{\leftskip}{-2em}\rm
\item[] $\displaystyle\mathrm{e}^{-\frac{\mu}{g}x}(\sin x)^g ~,~~~ x \in (0,\pi) ~,~~~ g > \frac{3}{2} ~,~~~ \mu > 0 ~,~~~ g \to g+1 ~.$
\item[] $\displaystyle W(x) = \frac{\hbar\mu}{g} - \hbar g\cot x ~,~~~ V(x) = \frac{\hbar^2g(g-1)}{\sin^2x} - 2\hbar^2\mu\cot x + \frac{\hbar^2\mu^2}{g^2} - \hbar^2g^2 ~.$
\item[] $\displaystyle\mathcal{E}_n = \frac{\hbar^2\mu^2}{g^2} - \frac{\hbar^2\mu^2}{(g+n)^2} - \hbar^2g^2 + \hbar^2(g+n)^2 \,,~~ \phi_n(x) = \mathrm{e}^{-\frac{\mu}{g+n}x}(\sin x)^{g+n} (\mathrm{i}^{-n})P_n^{(\alpha_n,\beta_n)}(\mathrm{i}\cot x) ~,$

	\hfill{with\quad $\displaystyle\alpha_n \equiv -g - n + \mathrm{i}\frac{\mu}{g+n} ~,~~~ \beta_n \equiv -g - n - \mathrm{i}\frac{\mu}{g+n} ~,~~~ n=0,1,2,\ldots ~.$}
\end{itemize}

\item Morse potential
\begin{itemize}\setlength{\leftskip}{-2em}\rm
\item[] $\displaystyle\phi_0(x) = \mathrm{e}^{hx-\mu\mathrm{e}^x} ~,~~~ x \in (-\infty,\infty) ~,~~~ h, \mu > 0 ~,~~~ h \to h-1 ~.$
\item[] $\displaystyle W(x) = \hbar (\mu\mathrm{e}^x - h) ~,~~~ V(x) = \hbar^2\mu^2\mathrm{e}^{2x} - \hbar^2\mu(2h+1)\mathrm{e}^x + \hbar^2h^2 ~.$
\item[] $\displaystyle\mathcal{E}_n = \hbar^2n(n+2g) ~,~~~ \phi_n(x) = \mathrm{e}^{hx-\mu\mathrm{e}^x}(2\mu\mathrm{e}^x)^{-n} L_n^{(2h-2n)}(2\mu\mathrm{e}^x) ~,~~~ n = 0,1,\ldots, \lfloor h \rfloor ~.$
\end{itemize}

\item $\dfrac{1}{\cosh^2x}$-potential\quad{\rm (*A special case of the Rosen--Morse potenial with $\mu\to 0$.)}
\begin{itemize}\setlength{\leftskip}{-2em}\rm
\item[] $\displaystyle\phi_0(x) = (\cosh x)^{-h} ~,~~~ x \in (-\infty,\infty) ~,~~~ h > \frac{1}{2} ~,~~~ h \to h-1 ~.$
\item[] $\displaystyle W(x) = \hbar h\tanh x ~,~~~ V(x) = -\frac{\hbar^2h(h+1)}{\cosh^2x} + \hbar^2h^2 ~.$
\item[] $\displaystyle\mathcal{E}_n = 2\hbar^2nh - \hbar^2n^2 ~,~~~ \phi_n(x) = (\cosh x)^{n-h} P_n^{(h-n,h-n)}(\tanh x) ~,~~~ n=0,1,\ldots, \lfloor h \rfloor ~.$
\end{itemize}

\item Rosen--Morse potential
\begin{itemize}\setlength{\leftskip}{-1.5em}\rm
\item[] $\displaystyle\phi_0(x) = \mathrm{e}^{-\frac{\mu}{h}x}(\cosh x)^{-h}  ~,~~~ x \in (-\infty,\infty) ~,~~~ h > \sqrt{\mu} > 0 ~,~~~ h \to h-1 ~.$
\item[] $\displaystyle W(x) = \frac{\hbar\mu}{h} + \hbar h\tanh x ~,~~~ V(x) = -\frac{\hbar^2h(h+1)}{\cosh^2x} + 2\hbar^2\mu\tanh x + \hbar^2h^2 + \frac{\hbar^2\mu^2}{h^2} ~.$
\item[] $\displaystyle\mathcal{E}_n = \hbar^2h^2 - \hbar^2(h-n)^2 + \frac{\hbar^2\mu^2}{h^2} - \frac{\hbar^2\mu^2}{(h-n)^2} \,,~~ \phi_n(x) = \mathrm{e}^{-\frac{\mu}{h-n}x}(\cosh x)^{n-h} P_n^{(\alpha_n,\beta_n)}(\tanh x) ~,$

	\hfill{with\quad $\displaystyle\alpha_n \equiv h - n + \frac{\mu}{h-n} ~,~~~ \beta_n \equiv h - n -\frac{\mu}{h-n} ~,~~~ n=0,1,\ldots, \lfloor h \rfloor ~.$}
\end{itemize}

\item Hyperbolic symmetric top II
\begin{itemize}\setlength{\leftskip}{-1.5em}\rm
\item[] $\displaystyle\phi_0(x) = \mathrm{e}^{-\mu\tan^{-1}\sinh x}(\cosh x)^{-h} ~,~~~ x \in (-\infty,\infty) ~,~~~ h,\mu > 0 ~,~~~ h \to h-1 ~.$
\item[] $\displaystyle W(x) = \frac{\hbar\mu}{\cosh x} + \hbar h\tanh x ~,$

	 \hfill{$\displaystyle V(x) = \frac{-\hbar^2h(h+1) + \hbar^2\mu^2 + \hbar^2\mu(2h+1)\sinh x}{\cosh^2x} + \hbar^2h^2 ~.$}
\item[] $\displaystyle\mathcal{E}_n = 2\hbar^2nh - \hbar^2n^2 ~,~~~ \phi_n(x) = \mathrm{e}^{-\mu\tan^{-1}\sinh x}(\cosh x)^{-h}(\mathrm{i}^{-n})P_n^{(\alpha,\beta)}(\mathrm{i}\sinh x) ~,$

	\hfill{with\quad $\displaystyle\alpha \equiv -h - \frac{1}{2} - \mathrm{i}\mu ~,~~~ \beta \equiv -h - \frac{1}{2} + \mathrm{i}\mu ~,~~~ n=0,1,\ldots, \lfloor h \rfloor ~.$}
\end{itemize}

\item Eckart potential
\begin{itemize}\setlength{\leftskip}{-1.5em}\rm
\item[] $\displaystyle\phi_0(x) = \mathrm{e}^{-\frac{\mu}{g}x}(\sinh x)^g ~,~~~ x \in (0,\infty) ~,~~~ \sqrt{\mu} > g > \frac{1}{2} ~,~~~ g \to g+1 ~.$
\item[] $\displaystyle W(x) = \frac{\hbar\mu}{g} + \hbar g\coth x ~,~~~ V(x) = \frac{\hbar^2g(g-1)}{\sinh^2x} - 2\hbar^2\mu\coth x + \hbar^2g^2 + \frac{\hbar^2\mu^2}{g^2} ~.$
\item[] $\displaystyle\mathcal{E}_n = \hbar^2g^2 - \hbar^2(g+n)^2 + \frac{\hbar^2\mu^2}{g^2} - \frac{\hbar^2\mu^2}{(g+n)^2} ~,~~~ \phi_n(x) = \mathrm{e}^{-\frac{\mu}{g+n}x}(\sinh x)^{g+n} P_n^{(\alpha_n,\beta_n)}(\coth x) ~,$

	\hfill{with\quad $\displaystyle\alpha_n \equiv -g - n + \frac{\mu}{g+n} ~,~~~ \beta_n \equiv -g - n -\frac{\mu}{g+n} ~,~~~ n=0,1,\ldots, \lfloor \sqrt{\mu}-g \rfloor ~.$}
\end{itemize}

\item Hyperbolic P\"{o}schl--Teller potential
\begin{itemize}\setlength{\leftskip}{-1.5em}\rm
\item[] $\displaystyle(\sinh x)^g(\cosh x)^{-h} ~,~~~ x \in (0,\infty) ~,~~~ h > g > \frac{1}{2} ~,~~~ g \to g+1 ~,~~~ h \to h-1 ~.$
\item[] $\displaystyle W(x) = -\hbar (g\coth x - h\tanh x) ~,~~~ V(x) = \frac{\hbar^2g(g-1)}{\sinh^2x} - \frac{\hbar^2h(h+1)}{\cosh^2x} + \hbar^2(h-g)^2 ~.$
\item[] $\displaystyle\mathcal{E}_n = 4\hbar^2n(h-g-n) ~,~~~ \phi_n(x) = (\sinh x)^g(\cosh x)^{-h} P_n^{(g-\frac{1}{2},-h-\frac{1}{2})}(\cosh 2x) ~,$

	\hfill{$\displaystyle n=0,1,\ldots, \left\lfloor \frac{h-g}{2} \right\rfloor ~.$}
\end{itemize}

\end{itemize}

\subsubsection{On the solvability via classical orthogonal polynomials}
Let us consider a Schr\"{o}dinger equation for a conventional shape-invariant potential:
\begin{equation}
\mathcal{H}\phi_n(x) = \left[ -\hbar^2\frac{d^2}{dx^2} + V(x) \right] \phi_n(x) = \mathcal{E}_n\phi_n(x) ~.
\label{eq:2-SE_cSI}
\end{equation}
We perform the similarity transformation of this Hamiltonian using the ground-state wavefunction,
\begin{equation}
\mathcal{H} = -\hbar^2\frac{d^2}{dx^2} + V(x) 
\quad\to\quad
\widetilde{\mathcal{H}} \coloneqq \phi_0(x)^{-1}\circ\mathcal{H}\circ\phi_0(x) = -\hbar^2\frac{d^2}{dx^2} + 2\hbar W(x)\frac{d}{dx} ~.
\end{equation}
Assuming that the wave function is of the following factorized form: $\phi_n(x) \equiv \phi_0(x)P_n(x)$, the Schr\"{o}dinger equation \eqref{eq:2-SE_cSI} becomes
\begin{equation}
\widetilde{\mathcal{H}}P_n(x) = \left[ -\hbar^2\frac{d^2}{dx^2} + 2\hbar W(x)\frac{d}{dx} \right] P_n(x) = \mathcal{E}_nP_n(x) ~,
\end{equation}
which turns out to be one of the Hermite, Laguerre or Jacobi differential equations.

\begin{example}[H]
For the 1-dim. harmonic oscillator, the Hamiltonian with the unit $\hbar=\omega=1$:
\begin{equation}
\mathcal{H} = -\frac{d^2}{dx^2} + x^2 - 1 ~,
\end{equation}
is transformed into
\begin{equation}
\widetilde{\mathcal{H}} = -\frac{d^2}{dx^2} + 2x\frac{d}{dx} 
\end{equation}
under the similarity transformation via $\phi_0(x)=\mathrm{e}^{-x^2/2}$.
The resulting differential equation is
\begin{equation}
-\frac{d^2P_n(x)}{dx^2} + 2x\frac{dP_n(x)}{dx} = \mathcal{E}_nP_n(x) ~,
\end{equation}
which is equivalent to the Hermite equation \eqref{eq:H_deq} with the identification $\mathcal{E}_n=2n$.
\end{example}

\begin{example}[L]
For the radial oscillator, the Hamiltonian with the unit $\hbar=\omega=1$:
\begin{equation}
\mathcal{H} = -\frac{d^2}{dx^2} + x^2 + \frac{g(g-1)}{x^2} - (2g+1) ~,
\end{equation}
is transformed into
\begin{equation}
\widetilde{\mathcal{H}} = -4z\frac{d^2}{dz^2} + 4 \left( z - g - \frac{1}{2} \right) \frac{d}{dz} 
\end{equation}
in terms of $z = x^2$ under the similarity transformation via $\phi_0(x)=\mathrm{e}^{-x^2/2}x^{-g}$.
The resulting differential equation is
\begin{equation}
-4z\frac{d^2P_n(z)}{dz^2} + 4 \left( z - g - \frac{1}{2} \right) \frac{dP_n(z)}{dz} = \mathcal{E}_nP_n(z) ~,
\end{equation}
which is equivalent to the Laguerre equation \eqref{eq:L_deq} with the identification $\mathcal{E}_n=4n$.
\end{example}

\begin{example}[J]
For the P\"{o}schl--Teller potential, the Hamiltonian with the unit $\hbar=1$:
\begin{equation}
\mathcal{H} = -\frac{d^2}{dx^2} + \frac{g(g-1)}{\sin^2x} + \frac{h(h-1)}{\cos^2x} - (g+h)^2 ~,
\end{equation}
is transformed into
\begin{equation}
\widetilde{\mathcal{H}} = -4(1-y^2)\frac{d^2}{dy^2} + 4\left[ h-\frac{1}{2} - \left( g-\frac{1}{2} \right) (g+h+1) y \right] \frac{d}{dy} 
\end{equation}
in terms of $y = \cos 2x$ under the similarity transformation via $\phi_0(x)=(\sin x)^g(\cos x)^h$.
The resulting differential equation is
\begin{equation}
-4(1-y^2)\frac{d^2P_n(y)}{dy^2} + 4\left[ h-\frac{1}{2} - \left( g-\frac{1}{2} \right) (g+h+1) y \right] \frac{dP_n(y)}{dy} = \mathcal{E}_nP_n(y) ~,
\end{equation}
which is equivalent to the Jacobi equation \eqref{eq:J_deq} with the identification $\mathcal{E}_n=4n(n+g+h)$.
\end{example}

\subsubsection{Interrelation among conventional shape-invariant potentials}
It has been discussed that all the Schr\"{o}dinger equations for the conventional shape-invariant potentials are mapped to the Schr\"{o}dinger equation for either 1-d harmonic oscillator (H), radial oscillator (L) or P\"{o}schl--Teller potential (J) under certain changes of variables: $x\mapsto z=z(x)$.
Formally, the Schr\"{o}dinger equation for a conventional shape-invariant potential:
\begin{equation}
\left[ -\hbar^2\frac{d^2}{dx^2} + V(x) \right]\phi_n(x) = \mathcal{E}_n\phi_n(x) ~,
\end{equation}
changes into
\begin{equation}
\left[ -\hbar^2\frac{d^2}{dz^2} + v(z) \right]\varphi_n(z) = \varepsilon_n\varphi_n(z) ~,
\end{equation}
with some reparametrizations (See the following example and Tab. \ref{tab:2-Interrel_cSI-SE} for the details).

\begin{example}[Coulomb potential]
For the Coulomb potential, the Schr\"{o}dinger equation is
\begin{equation}
\left[ -\hbar^2\frac{d^2}{dx^2} + \frac{\hbar^2g(g-1)}{x^2} - \frac{e^2}{x} + \frac{e^4}{4\hbar^2g^2} \right]\phi_n(x) = \left( \frac{e^4}{4\hbar^2g^2} - \frac{e^4}{4\hbar^2(g+n)^2} \right) \phi_n(x) ~,
\end{equation}
with
\begin{equation}
\phi_n(x) = \mathrm{e}^{-\frac{e^2x}{2\hbar^2(g+n)}}\left( \frac{e^2}{2\hbar^2}x \right)^g L_n^{(2g-1)}\left( \frac{e^2}{\hbar^2(g+n)}x \right) ~.
\end{equation}
This transforms into the Schr\"{o}dinger equation for the radial oscillator under the change of variables 
\begin{equation}
x \mapsto z = \sqrt{x}
\qquad\text{\it i.e.}\quad
x = z^2 ~.
\end{equation}
In the meanwhile, we write the wavefunction $\phi_n(x) \equiv \sqrt{2z}\,\varphi_n(z)$ and also the paramters
\[
\frac{e^2}{\hbar (g+n)} \equiv \omega ~,~~~ 2g - \frac{1}{2} \equiv \bar{g} ~.
\]
The resulting differential equation is
\begin{equation}
\left[ -\hbar^2\frac{d^2}{dz^2} + \frac{\hbar^2\bar{g}(\bar{g}-1)}{z^2} + \omega^2z^2 - \hbar\omega (2\bar{g}+1) \right]\varphi_n(z) = 4n\hbar\omega \varphi_n(z) ~,
\end{equation}
with
\begin{align}
\varphi_n(x) &= \frac{1}{\sqrt{2z}}\,\phi_n(x(z)) 
= \frac{1}{\sqrt{2z}} \cdot \mathrm{e}^{-\frac{e^2z^2}{2\hbar^2(g+n)}}\left( \frac{e^2}{2\hbar^2}z^2 \right)^g L_n^{(2g-1)}\left( \frac{e^2}{\hbar^2(g+n)}z^2 \right) \nonumber \\
&\propto \mathrm{e}^{-\frac{\omega z^2}{2\hbar}}\left( \sqrt{\frac{\omega}{\hbar}}\,z \right)^{\bar{g}} L_n^{( \bar{g}-\frac{1}{2} )}\left( \frac{\omega}{\hbar}z^2 \right) ~.
\end{align}
\end{example}

\begin{table}[p]
\centering
\caption{Interrelation among the Schr\"{o}dinger equations of conventional shape-invariant potentials.}
\label{tab:2-Interrel_cSI-SE}
\scalebox{0.65}{
	\begin{tabular}{ccccll}
	\toprule
	Class & Superpotential & Energy &~~& \begin{tabular}{c}
			Change of variables~~ \\
			$x \mapsto z=z(x)$~~
		\end{tabular} & Relation among parameters \\
	\midrule
	H & \begin{tabular}{c}
			1-dim. HO \\
			$\omega z$ 
		\end{tabular} & $2n\hbar\omega$ && --- & --- \\
	\midrule
	L & \begin{tabular}{c}
			Radial osc. \\
			$\omega z - \dfrac{\hbar g}{z}$ 
		\end{tabular} & $4n\hbar\omega$ && --- & --- \\
	  & \begin{tabular}{c}
	  		Coulomb pot. \\
			$\dfrac{e^2}{2\hbar\bar{g}} - \dfrac{\hbar\bar{g}}{x}$ 
		\end{tabular} & $\dfrac{e^4}{4\hbar^2\bar{g}^2} - \dfrac{e^4}{4\hbar^2(\bar{g}+n)^2}$ && $z=\sqrt{x}$ & $\omega = \dfrac{e^2}{\hbar(g+n)}$, $g = 2\bar{g}-\dfrac{1}{2}$ \\
	  & \begin{tabular}{c}
	  		Morse pot. \\
			$\hbar(\mu\mathrm{e}^x - h)$ 
		\end{tabular} & $\hbar^2(2nh-n^2)$ && $z=\mathrm{e}^{x/2}$ & $\omega = 2\hbar\mu$, $g = 2(h-n)+\dfrac{1}{2}$ \\
	\midrule
	J & \begin{tabular}{c}
			P\"{o}schl--Teller pot. \\
			$\hbar(h\tan z - g\cot z)$ 
		 \end{tabular} & $4\hbar^2n(n+g+h)$ && --- & --- \\
	   & \begin{tabular}{c}
			Hyperbolic symmetric top II \\
			$\hbar\left( \dfrac{\mu}{\cosh x} + \bar{h}\tanh x \right)$ 
		 \end{tabular} & $\hbar^2(2n\bar{h}-n^2)$ && $z = \dfrac{1}{2}\arccos (\mathrm{i}\sinh x)$ & $g = -\bar{h}-\mu\mathrm{i}$, $h = -\bar{h}+\mu\mathrm{i}$ \\
	   & \begin{tabular}{c}
			Rosen--Morse pot. \\
			$\hbar\left( \dfrac{\mu}{\bar{h}} + \bar{h}\tanh x \right)$ 
		 \end{tabular} & $\hbar^2\left( 2n\bar{h}-n^2 + \dfrac{\mu^2}{\bar{h}^2} - \dfrac{\mu^2}{(\bar{h}-n)^2} \right)$ && $z = \dfrac{1}{2}\arccos (\tanh x)$ & \begin{tabular}{l} 
		 	$g = \bar{h}-n+\dfrac{\mu}{\bar{h}-n}+\dfrac{1}{2}$, \\
			$h = \bar{h}-n-\dfrac{\mu}{\bar{h}-n}+\dfrac{1}{2}$
		 \end{tabular} \\
	   & \begin{tabular}{c}
			Eckart pot. \\
			$\hbar\left( \dfrac{\mu}{\bar{g}} - \bar{g}\coth x \right)$ 
		 \end{tabular} & $\hbar^2\left( 2n\bar{g}-n^2 + \dfrac{\mu^2}{\bar{g}^2} - \dfrac{\mu^2}{(\bar{g}+n)^2} \right)$ && $z = \dfrac{1}{2}\arccos (\coth x)$ & \begin{tabular}{l} 
		 	$g = -\bar{g}-n+\dfrac{\mu}{\bar{g}+n}+\dfrac{1}{2}$, \\
			$h = -\bar{g}-n-\dfrac{\mu}{\bar{g}+n}+\dfrac{1}{2}$
		 \end{tabular} \\
	   & \begin{tabular}{c}
			Hyperbolic P\"{o}schl--Teller pot. \\
			$\hbar(\bar{h}\tanh x - g\coth x)$ 
		 \end{tabular} & $4\hbar^2n(\bar{h}-g-n)$ && $z = \arcsin (-\mathrm{i}\sinh x)$ & $h = -\bar{h}$ \\
	\bottomrule
	\end{tabular}
}
\end{table}



\clearpage
\section{Other Exactly Solvable Potentials: an Overview}
\label{sec:2-4}
Exactly solvable quantum mechanical potentials have been studied since the early days of quantum mechanics.
Among them is the factorization method~\cite{infeld1951factorization}, which is now understood in connection with shape invariance.
This class of exactly solvable potentials is referred to as the conventional shape-invariant systems.
The term ``conventional'' reflects the fact that these systems were already known in the 1950's.

\begin{figure}[t]
\centering
\includegraphics[scale=1]{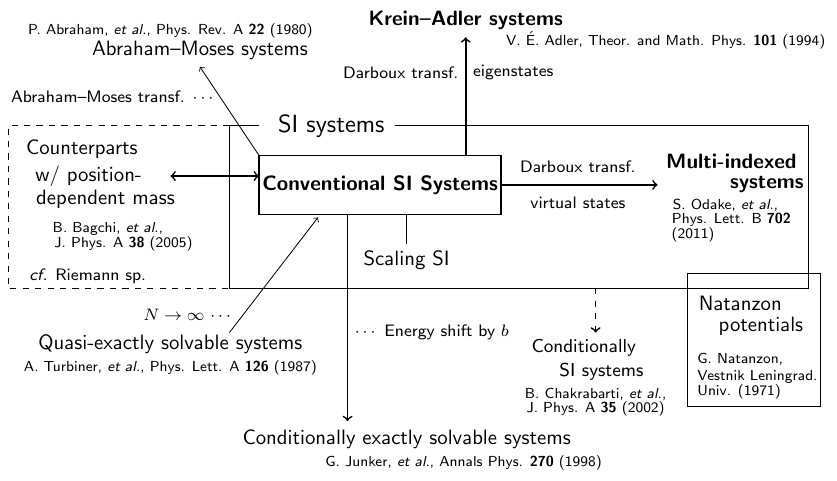}
\caption[Several classes of exactly solvable quantum mechanical systems in connection with the conventional shape-invariant systems.]
	{Several classes of exactly solvable quantum mechanical systems in connection with the conventional shape-invariant systems.}
\label{fig:2-SolvablePots}
\end{figure}

After that, a number of classes of exactly solvable systems have been constructed in relation to the conventional shape-invariant systems.
A part of those exactly solvable quantum mechanical systems (references are given in the figure) and how they are related to the conventional shape-invariant systems are summarized in Fig. \ref{fig:2-SolvablePots}.

\chapter{SWKB Formalism}
\label{sec:3}

{\small
\begin{leftbar}
\noindent\textbf{Introduction.}\hspace{1em}
In this chapter, we discuss SWKB formalism, starting with introducing the SWKB quantization condition. 
The condition equation and several notable properties are given in Sect. \ref{sec:3-1}.
We perform, in Sect. \ref{sec:3-2}, several case studies on the condition equation were conducted to understand what the SWKB quantization condition actually means.
Sect. \ref{sec:3-3} is devoted to the discussions based on the case studies in the previous section, where we arrive at a conjecture on the implication of the SWKB quantization condition.

The SWKB condition equation can be used not only to evaluate the energy spectrum from a given superpotential but to determine the superpotential from a given energy spectrum.
In Sect. \ref{sec:3-4}, we formulate a way of doing so: the inverse problem of the SWKB.

\bigskip\noindent
$\blacktriangleright$\hspace{1em}\bf
This chapter is based in part on the author's works: Refs. \cite{Nasuda:2020aqf,10.1007/978-981-19-4751-3_29,NASUDA2023116087}.
\end{leftbar}
}

\section{SWKB Quantization Condition}
\label{sec:3-1}
In 1985, A. Comtet, A. D. Bandrauk and D. K. Campbell proposed a quantization condition in the context of supersymmetric quantum mechanics.
This condition is often referred to as \textit{supersymmetric WKB}, or in short \textit{SWKB}, quantization condition.
In Ref. \cite{comtet1985exactness}, it is also called the CBC formula, which is an initialism for Comtet--Bandrauk--Campbell.

\subsection{Condition Equation}
The condition equation is
\begin{equation}
\int_{a_{\rm L}}^{a_{\rm R}} \sqrt{\mathcal{E}_n - W(x)^2} \,dx = n\pi\hbar ~,~~~
n=0,1,2,\ldots ~,
\label{eq:3-SWKB}
\end{equation}
where $a_{\rm L}$, $a_{\rm R}$ are the two roots of the equation $\mathcal{E}_n - W(x)^2 =0$ with $a_{\rm L}<a_{\rm R}$.
For the case where the equation has more than two roots, see Sect. \ref{sec:3-2-4}.
Let $I_{\rm SWKB}$ denote the left hand side of Eq. \eqref{eq:3-SWKB}, and call it SWKB integral in the following,
\begin{equation}
I_{\rm SWKB} \equiv \int_{a_{\rm L}}^{a_{\rm R}} \sqrt{\mathcal{E}_n - W(x)^2} \,dx ~.
\label{eq:3-SWKBint}
\end{equation}

\subsubsection{Properties}
Here, we list two well-known properties of this condition:

\begin{enumerate}\setlength{\leftskip}{-1em}
\item The ground state, $n=0$, satisfies the condition exactly for any potentials by construction. 
	This is demonstrated as follows. 
	We have taken the ground-state energy eigenvalue $\mathcal{E}_0$ to be zero (See Sect. \ref{sec:2-2-1}).
	Also, the superpotential $W(x)$ has at least one zero, at the point where the ground-state wavefunction $\phi_0(x)$ takes its extremum, because $W(x) = -\hbar\partial_x\phi_0(x)/\phi_0(x)$.
	We write this point $x=a$.
	Then the SWKB integral is
	\begin{equation}
	\int_{a}^{a} \sqrt{0 - W(x)^2} \,dx \equiv 0 ~.
	\end{equation}
\item As for the excited states, the condition is not exactly satisfied in general, but the following exceptions.
	That is, the conventional shape-invariant potentials.
	We will give direct proofs in Sects. \ref{sec:3-2-2}. 
	In this thesis, we present yet other exceptions by considering a natural extension of the condition equation (See Sect. \ref{sec:3-2-6}).
\end{enumerate}

Furthermore, we have shown the following three properties of the condition equation in this thesis:

\begin{enumerate}\setlength{\leftskip}{-1em}
\setcounter{enumi}{2}
\item $\hbar$ can always be factored out from the condition equation. 
	Thus, SWKB formalism should be formulated independently from the (reduced) Planck constant $\hbar$~\cite{Nasuda:2020aqf}.
	\hfill{$\Rightarrow$ Sect. \ref{sec:3-2}.}
\item Although the condition equation is not exactly satisfied in most cases, the SWKB integrals remain around $n\pi\hbar$ in many cases.
	We say that in those cases the condition is \textit{approximately} satisfied~\cite{Nasuda:2020aqf,NASUDA2023116087}.
	\hfill{$\Rightarrow$ Sect. \ref{sec:3-2}.}
\item The exactness of the SWKB quantization condition is closely related to the exact solvability of a potential via classical orthogonal polynomials.
	\hfill{$\Rightarrow$ Sect. \ref{sec:3-3-4}.}
\end{enumerate}

\subsubsection{Formal derivation of the condition equation}
Historically, the SWKB quantization condition was `derived' from Bohr--Sommerfeld quantization condition, or the quantization condition in the context of the WKB approximation:
\begin{equation}
\int_{x_{\rm L}}^{x_{\rm R}} \sqrt{\mathcal{E}_n - V(x)} \,dx = \left( n+\frac{1}{2} \right)\pi\hbar ~,~~~
n=0,1,2,\ldots ~,
\label{eq:3-WKB}
\end{equation}
where $x_{\rm L}$, $x_{\rm R}$ are the classical turning points, that are the two roots of the equation $\mathcal{E}_n - V(x) =0$ with $x_{\rm L}<x_{\rm R}$.

The idea of the derivation is rather simple; we expand the left-hand side of Eq. \eqref{eq:3-WKB} in $\hbar$ by assuming that $\mathcal{E}_n$ and $W(x)$ are independent on $\hbar$.
That is,
\begin{align}
\int_{x_{\rm L}}^{x_{\rm R}} \sqrt{\mathcal{E}_n - V(x)} \,dx 
&= \int_{x_{\rm L}}^{x_{\rm R}} \sqrt{\mathcal{E}_n - W(x)^2 + \hbar\frac{dW(x)}{dx}} \,dx \nonumber \\
&\cong \int_{a_{\rm L}}^{a_{\rm R}} \sqrt{\mathcal{E}_n - W(x)^2} \,dx 
	+ \frac{\hbar}{2} \left[ \int_{a_{\rm L}}^{a_{\rm R}} \frac{1}{\sqrt{\mathcal{E}_n - W(x)^2}}\frac{dW(x)}{dx}\,dx \right. \nonumber \\
	&\qquad \left. + \sqrt{\mathcal{E}_n-W(a_{\rm R})^2}\left.\frac{da_{\rm R}}{d\hbar}\right|_{\hbar\to 0} 
	- \sqrt{\mathcal{E}_n-W(a_{\rm L})^2}\left.\frac{da_{\rm L}}{d\hbar}\right|_{\hbar\to 0} \right]
	+ \mathcal{O}(\hbar^2) \nonumber \\
&= \int_{a_{\rm L}}^{a_{\rm R}} \sqrt{\mathcal{E}_n - W(x)^2} \,dx 
	+ \frac{\hbar}{2} \int_{a_{\rm L}}^{a_{\rm R}} \frac{1}{\sqrt{\mathcal{E}_n - W(x)^2}}\frac{dW(x)}{dx} \,dx 
	+ \mathcal{O}(\hbar^2) ~.
\end{align}
Here, 
\begin{equation}
\frac{\hbar}{2} \int_{a_{\rm L}}^{a_{\rm R}} \frac{1}{\sqrt{\mathcal{E}_n - W(x)^2}}\frac{dW(x)}{dx} \,dx 
= \left. \frac{\hbar}{2}\sin^{-1}\left[ \frac{W(x)}{\sqrt{\mathcal{E}_n}} \right] \right|_{a_{\rm L}}^{a_{\rm R}}
= \frac{1}{2}\pi\hbar ~,
\end{equation}
and thus, if we drop $\mathcal{O}(\hbar^2)$, we get
\begin{equation}
\int_{a_{\rm L}}^{a_{\rm R}} \sqrt{\mathcal{E}_n - W(x)^2} \,dx + \frac{1}{2}\pi\hbar
= \left( n+\frac{1}{2} \right)\pi\hbar ~,
\end{equation}
which is equivalent to Eq. \eqref{eq:3-SWKB}.

However, the above derivation is only a \textit{formal} one, for $\mathcal{E}_n$ and $W(x)$ are usually dependent on $\hbar$ (See, \textit{e.g.}, Sect. \ref{sec:2-3-2}) in various ways.
Therefore the expansion in $\hbar$ is not mathematically justified.

\begin{remark}
The results that come out from the assumption of the $\hbar$-independency on $\mathcal{E}_n$ and $W(x)$ sometimes agree with that from mathematically-justified methods, presumably for the following facts.

The parameters always appear in the form of $\hbar g \equiv G$, which behaves like an adiabatic invariant in the classical limit $\hbar\to 0$.
However, these facts cannot be derived from the shape-invariant condition or any other definitions theorems, \textit{etc.}

Moreover, it is illogical when the classical limit of WKB goes to SWKB and comes to reproduce exact bound-state spectra, even for \textit{quantum} levels (lower $n$'s).
\end{remark}

\subsection{Quantization of Energy}
\label{sec:3-1-2}
The SWKB quantization condition is usually seen as a quantization condition for the energy.
In this subsection, we present how the condition equation quantizes the energy.

\subsubsection{Conventional shape-invariant potentials}
As was mentioned above, the SWKB quantization condition reproduces exact bound-state spectra for any conventional shape-invariant potential.
One can deduce the energy spectra from the condition equation analytically in these cases.
In the calculations, we refer to, \textit{e.g.}, Ref. \cite{gradshteyn2014tables}.

\begin{example}[H]
The SWKB integral is 
\begin{equation}
I_{\rm SWKB} = \int_{-\sqrt{\mathcal{E}}/\omega}^{\sqrt{\mathcal{E}}/\omega} \sqrt{\mathcal{E} - \omega^2x^2} \,dx 
= \frac{\pi\mathcal{E}}{2\omega} ~.
\label{eq:3-QuantizationOfEnergy_H}
\end{equation}
Therefore, the quantization condition $I_{\rm SWKB} = n\pi\hbar$ yields $\mathcal{E}_n = 2n\hbar\omega$.
\end{example}

\begin{example}[L]
The SWKB integral is 
\begin{equation}
I_{\rm SWKB} = \int_{a_-}^{a_+} \sqrt{\mathcal{E} - \left( \omega x - \frac{\hbar g}{x} \right)^2} \,dx 
= \frac{\pi\mathcal{E}}{4\omega}  ~,~~~
{a_{\pm}}^2 \equiv \frac{\mathcal{E}+2\hbar\omega g \pm \sqrt{\mathcal{E}(\mathcal{E}+4\hbar\omega g)}}{2\omega^2} ~.
\label{eq:3-QuantizationOfEnergy_L}
\end{equation}
Therefore, the quantization condition $I_{\rm SWKB} = n\pi\hbar$ yields $\mathcal{E}_n = 4n\hbar\omega$.
\end{example}

\begin{example}[J]
The SWKB integral is 
\begin{align}
I_{\rm SWKB} &= \int_{a_-}^{a_+} \sqrt{\mathcal{E} - \hbar^2( g\cot x - h\tan x )^2} \,dx 
= \frac{\pi}{2}\left[ \sqrt{\mathcal{E}+\hbar^2(g+h)^2} - \hbar(g+h) \right] ~, \nonumber \\
{a_{\pm}}^2 &\equiv \arctan\frac{\mathcal{E}+2\hbar^2gh \pm \sqrt{\mathcal{E}(\mathcal{E}+4\hbar^2gh)}}{2\hbar^2h^2} ~.
\label{eq:3-QuantizationOfEnergy_J}
\end{align}
Therefore, the quantization condition $I_{\rm SWKB} = n\pi\hbar$ yields $\mathcal{E}_n = 4\hbar^2n(n+g+h)$.
\end{example}

\subsubsection{Other potentials}
On the other hand, it is well-known that the SWKB quantization condition does not reproduce exact bound-state spectra for other potentials, but can provide approximate values.
The quantization of energy for such potentials has already been carried out by several authors~\cite{khare1989shape,delaney1990susy,doi:10.1139/p91-189,bougie2018supersymmetric}.

\subsection{Relation to Quantum Hamilton--Jacobi Theory}
An interesting correspondence between the SWKB quantization condition and the quantum Hamilton--Jacobi theory~\cite{leacock1983hamilton_action,leacock1983hamilton} was first revealed by Bhalla \textit{et al.} in Ref. \cite{bhalla1996exactness,bhalla1997quantum}.
This subsection is a brief review on this discussion.

\subsubsection{Exact quantization condition in quantum Hamilton--Jacobi theory}
In quantum Hamilton--Jacobi theory, an \textit{exact} quantization condition has been discussed.
The condition equation is
\begin{equation}
J_{\rm QHJ} \coloneqq \frac{1}{2\pi} \oint_C p(x;\mathcal{E}_n) \,dx ~,~~~
J_{\rm QHJ} = n\hbar ~,~~~
n=0,1,2,\ldots ~,
\label{eq:3-QHJ-QC}
\end{equation}
where $p(x;\mathcal{E})$ is the quantum momentum function, satisfying the following Riccati-type equation called the quantum Hamilton--Jacobi equation:
\begin{equation}
p(x;\mathcal{E})^2 - \mathrm{i}\hbar\frac{dp(x;\mathcal{E})}{dx} = \mathcal{E} - V(x) ~,~~~
p(x;\mathcal{E}) \equiv -\mathrm{i}\hbar\frac{\partial_x\phi(x)}{\phi(x)} ~,
\label{eq:3-QHJ-eq}
\end{equation}
and $C$ is a counterclockwise contour in the complex $x$-plane enclosing the classical turning points $x_{\rm L}$ and $x_{\rm R}$.
This equation means the quantization of the quantum action variable $J_{\rm QHJ}$.
This quantization condition holds exactly for any quantum-mechanical potentials $V(x)$.

The exactness of the condition is guaranteed by the fact that an $n$-th eigenfunction has $n$ nodes between the two classical turning points (this is known as the oscillation theorem~\cite{lifshitz1977quantum}), which produce the poles of $p(x;\mathcal{E})$ having residue $-\mathrm{i}\hbar$ along with the real axis.
The Cauchy's argument principle gives Eq. \eqref{eq:3-QHJ-QC}.
We note that Gozzi~\cite{gozzi1986nodal} discovered the same equation independently of Leacock~\cite{leacock1983hamilton_action,leacock1983hamilton}.

\subsubsection{The correspondence to SWKB}
In the following, we show the correspondence between Eq. \eqref{eq:3-QHJ-QC} and the SWKB condition equation.
In order to compare the two quantization conditions, we first extend our SWKB integral to a contour integral on a complex $x$-plane,
\begin{equation}
I_{\rm SWKB} \to
J_{\rm SWKB} \coloneqq \frac{1}{2\pi} \oint_{C'} \sqrt{\mathcal{E}_n - W(x)^2}\,dx ~,
\label{eq:3-SWKBonCC}
\end{equation}
where $C'$ is a counterclockwise contour enclosing the branch cut of $\sqrt{\mathcal{E}_n - W(x)^2}$ from $a_{\rm L}$ to $a_{\rm R}$.
This $J_{\rm SWKB}$ is to be quantized as $J_{\rm SWKB}=n\hbar$.

Now let us assume that 
\begin{equation}
p(x;\mathcal{E}_n) \equiv -\mathrm{i}\hbar \left( \frac{\partial_x\phi_0(x)}{\phi_0(x)} + \frac{\partial_x\mathcal{P}_n(x)}{\mathcal{P}_n(x)} \right)
= \mathrm{i}\left( W(x) - \hbar\frac{\partial_x\mathcal{P}_n(x)}{\mathcal{P}_n(x)} \right) ~,
\end{equation}
or equivalently $\phi_n(x) \equiv \phi_0(x)\mathcal{P}_n(x)$, where $\mathcal{P}_n(x)$ is such a function that $\partial_x\mathcal{P}_n(x)/\mathcal{P}_n(x)$ has the $n$ poles of residue $1$ at the same points as the nodes of $\phi_n(x)$ along with the real axis, but $\partial_x\mathcal{P}_n(x)/\mathcal{P}_n(x)$ can have other poles off the real axis. 
Then, the quantum action variable becomes
\begin{equation}
J_{\rm QHJ} = \frac{1}{2\pi}\oint_C \sqrt{-\left( W(x) - \hbar\frac{\partial_x\mathcal{P}_n(x)}{\mathcal{P}_n(x)} \right)^2} \,dx
= \frac{1}{2\pi}\oint_C \sqrt{\mathcal{E}_n - W(x)^2 + \hbar^2\frac{d}{dx}\left( \frac{\partial_x\mathcal{P}_n(x)}{\mathcal{P}_n(x)} \right)}\,dx ~.
\end{equation}

Comparing Eqs. \eqref{eq:3-QHJ-QC} and \eqref{eq:3-SWKBonCC}, one can see that $J_{\rm SWKB} = J_{\rm QHJ}$ when $\partial_x\bigl( \partial_x\mathcal{P}_n(x)/\mathcal{P}_n(x) \bigr)$ does not have any singularity outside the contour $C$, which is to be realized for all conventional shape-invariant potentials.
On the contrary, for other potentials, where the quantization of $I_{\mathrm{SWKB}}$ is not exact, it is easy to guess that $\partial_x\bigl( \partial_x\mathcal{P}_n(x)/\mathcal{P}_n(x) \bigr)$ has singularities outside the contour.
In summary, the SWKB is exact quantization condition when the pole structure of the quantum momentum function and that of the SWKB integrand coincide outside the contours, \textit{i.e.}, $\partial_x\bigl( \partial_x\mathcal{P}_n(x)/\mathcal{P}_n(x) \bigr)$ has no singularity outside the contour $C$.

\begin{example}[H]
We show that the pole structures of the SWKB integrand and the quantum action variable coincide outside the contours $C$ and $C'$ for the case of 1-dim. harmonic oscillator (H).
We put $\hbar = \omega = 1$ for simplicity without loss of generality.

The SWKB integral with the complex variable $x$ has a fixed pole at infinity, and no other singularities outside the contour $C'$.
The integrand is Laurent expanded as
\begin{equation}
\sqrt{2n - x^2}
\cong \mathrm{i} \left( x - \frac{n}{x} - \frac{n}{2x^3} - \cdots \right) ~,
\end{equation}
and therefore, the residue of the fixed pole is $-\mathrm{i}n$. 


On the other hand, we assume that the quantum action variable in terms of a complex variable $y\equiv 1/x$ has the following expansion form:
\begin{equation}
p(y^{-1};2n) \cong \sum_{j=0}^{\infty} a_jy^j + \sum_{k=0}^{\infty} \frac{b_k}{y^k} ~.
\end{equation}
We require that the quantum action variable satisfies the quantum Hamilton--Jacobi equation:
\[
p(y^{-1};2n)^2 + \mathrm{i}y^2\frac{dp(y^{-1};2n)}{dy} = 2n - \frac{1}{y^2} + 1 ~,
\]
which determines the expansion coefficient $\{ a_j \}$, $\{ b_k \}$ as 
\begin{equation}
a_1 = -\mathrm{i}n ~,~~~
b_1 = \mathrm{i} ~,~~~
\text{otherwise}~~ a_j = b_k = 0 ~. 
\end{equation}
One can see from this that the quantum action variable also has a fixed pole at infinity with residue $-\mathrm{i}n$.
The singularity structures of the quantum momentum function and the SWKB integral outside the contours $C$ and $C'$ are exactly the same, and thereby the SWKB condition is also an exact quantization condition by Bhalla \textit{et al.}'s argument~\cite{bhalla1996exactness,bhalla1997quantum}.
\end{example}

Note however that usually the contour integral for $J_{\mathrm{SWKB}}$ is not so straightforward other than in the cases of conventional shape-invariant potentials; the contour integrations for the singularities cannot be performed analytically. 
Sometimes infinitely many number of poles and branch cuts appear on the complex $x$-plane, which are responsible for the breaking of SWKB condition. 
Instead, we give numerical computations for all of such examples below.
At the end of this section, we introduce a quantity to describe the discrepancy:
\begin{equation}
\Delta \coloneqq \pi (J_{\mathrm{QHJ}} - J_{\mathrm{SWKB}}) = n\pi\hbar - I_{\rm SWKB} ~.
\label{eq:3-Delta}
\end{equation}

\section{(Non-)exactness of SWKB Quantization Condition: Case Studies}
\label{sec:3-2}
\subsection{Overview and Structure of this Section}

\begin{figure}[t]
\centering
	\begin{tikzpicture}
	\node at (-2,2.75) {\underline{\underline{Case Studies}}};
	\node at (-2,2) {\S\,3.2.2};
	\node at (-2,1.4) {\S\,3.2.3};
	\node at (-2,0.8) {\S\,3.2.4};
	\node at (-2,0.2) {\S\,3.2.5};
	\node at (-2,-0.4) {\S\,3.2.6};
	\node at (2,2.75) {\underline{\underline{Discussions}}};
	\node at (2,2) {\S\,3.3.2};
	\node at (2,1.4) {\S\,3.3.3};
	\node at (2,0.8) {\S\,3.3.4};
	\node at (2,0.2) {\S\,3.3.5};
	\draw[thick] (-1,2)--(1,2);
	\draw[thick] (-1,1.4)--(1,2);
	\draw[thick,dashed] (-1,1.4)--(1,1.4);
	\draw[thick] (-1,0.8)--(1,1.4);
	\draw[thick] (-1,0.2)--(1,1.4);
	\draw[thick] (-1,2)--(1,0.8);
	\draw[thick] (-1,-0.4)--(1,0.8);
	\draw[thick] (2.8,1.4)--(3,1.4)--(3,0.8)--(2.8,0.8);
	\draw[thick,->] (3,1.1)--(3.2,1.1)--(3.2,0.2)--(2.8,0.2);
	\end{tikzpicture}
\caption{The relation between Sect. \ref{sec:3-2} and Sect. \ref{sec:3-3}.}
\label{fig:3-2_and_3-3}
\end{figure}
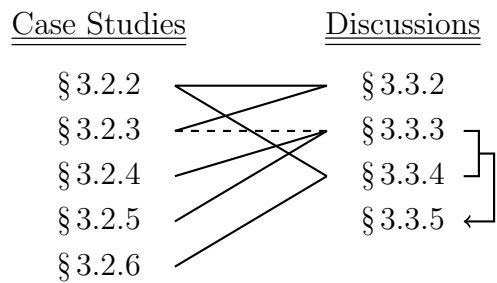

We carry out several case studies on whether a potential satisfies the SWKB condition equation.
As is expected, except in the case with the conventional shape-invariant potentials (Sect. \ref{sec:2-3-2}), the condition equation is not exactly satisfied.
Furthermore, we find the following two aspects:

\begin{enumerate}\setlength{\leftskip}{-1em}
\item The SWKB integrals remain around $n\pi\hbar$, \textit{i.e.}, the condition is approximately satisfied, in many cases~\cite{Nasuda:2020aqf,NASUDA2023116087}.
\item In the cases of position-dependent effective mass, where the Schr\"{o}dinger equations are solved via the classical orthogonal polynomials as in the cases of conventional shape-invariant potentials, the SWKB-exactness is restored by considering a natural extension of the condition equation.
\end{enumerate}

The rest of this section contains five subsections, each of which is devoted to a different class of exactly solvable potentials.
Each subsection starts with a brief introduction of why we study this case.
Then, we present the condition equations and numerical(/analytical) studies on them.
We also provide numerical(/analytical) studies on the related exact quantization condition \eqref{eq:3-QHJ-QC} for comparison.
The discussions based on these case studies are provided in the subsequent section.
Fig. \ref{fig:3-2_and_3-3} shows which case studies provoke which discussion(s).

\subsection{SWKB for Conventional Shape-invariant Potentials}
\label{sec:3-2-2}
In this subsection, we make some comments on the SWKB quantization condition for conventional shape-invariant potentials.

\subsubsection{Brief introduction}
All the examples that were mentioned in the Comtet \textit{et al.}'s original paper~\cite{comtet1985exactness} are now classified as the conventional shape-invariant potentials.
In the next year, Dutt \textit{et al.} first demonstrated that the SWKB quantization condition reproduces the exact bound-state spectra for all conventional shape-invariant potentials~\cite{dutt1986exactness}.

Later, in 1997, Hru\v{s}ka \textit{et al.} also showed the exactness of the SWKB condition by computing the integrations analytically~\cite{hruska1997accuracy}. 
We have also done the same calculation for our list in Sect. \ref{sec:2-3-2} to obtain the same result.
The details are presented in the following paragraph.
Other kinds of proofs have been given by several authors~\cite{BARCLAY1991357,YIN2010528,GANGOPADHYAYA2020126722,Gangopadhyaya_2021}.
We further discuss the interrelation of the SWKB integrals for the conventional shape-invariant potentials in Sect. \ref{sec:3-3-4}.

\subsubsection{Analytical calculations}
The analytical calculations to show the SWKB-exactness are almost identical to the calculations in Sect. \ref{sec:3-1-2} (See Eqs. \eqref{eq:3-QuantizationOfEnergy_H}--\eqref{eq:3-QuantizationOfEnergy_J}).
By replacing $\mathcal{E}$ in Eqs. \eqref{eq:3-QuantizationOfEnergy_H}--\eqref{eq:3-QuantizationOfEnergy_J} to $\mathcal{E}_n$ given in Sect. \ref{sec:2-3-2}, we obtain $I_{\rm SWKB} = n\pi\hbar$.

\begin{example}[H]\quad
$\displaystyle I_{\rm SWKB} = \frac{\pi\cdot 2n\hbar\omega}{2\omega} = n\pi\hbar$.
\end{example}

\begin{example}[L]\quad
$\displaystyle I_{\rm SWKB} = \frac{\pi\cdot 4n\hbar\omega}{4\omega} = n\pi\hbar$.
\end{example}

\begin{example}[J]\quad
$\displaystyle I_{\rm SWKB} = \frac{\pi}{2}\left[ \sqrt{4\hbar^2 n(n+g+h) + \hbar^2(g+h)^2} - \hbar (g+h) \right] = n\pi\hbar$.
\end{example}

\subsection{SWKB for Exceptional/Multi-indexed Systems}
\label{sec:3-2-3}
\subsubsection{Brief introduction}
It has long been considered that the SWKB quantization condition reproduces the exact bound-state spectra for all (additive) shape-invariant potentials since Dutt \textit{et al.} showed in 1986 that the SWKB is an exact quantization condition for all the shape-invariant potentials known at that time.
Starting in 1993, exactly solvable potentials with a different type of shape invariance, \textit{i.e.}, scaling shape invariance, have been constructed~\cite{barclay1993new}.
It was soon confirmed that the statement did not hold for this class of shape invariance~\cite{BARCLAY1993263}.
After that, it was important to mention ``all \textit{additive} shape-invariant potentials'' when stating the proposition.

After two and a half decades, however, Bougie \textit{et al.} argued that a newly constructed type of additive shape-invariant potentials, that is, the so-called exceptional systems (which are special cases of the multi-indexed systems) may not satisfy the SWKB condition equation exactly~\cite{bougie2018supersymmetric}.
It was striking, but we saw it skeptically.
This was mainly because we found a mathematically `awkward' way of introducing the superpotential in their paper\footnote{They employ the same idea in Refs. \cite{bougie2010generation,bougie2012supersymmetric,bougie2015generation,mallow2020inter}.}.
Also, our numerical experiments showed that the condition equation \eqref{eq:3-SWKB} held in good approximation, and we thought we could attribute the small discrepancy to computational precision (See Tab. \ref{tab:3-num_ex}).
However, it turned out that their statement still held after introducing the superpotential in a mathematically-justified way, and the small discrepancy was not due to computational errors~\cite{Nasuda:2020aqf}.

In the rest of this subsection, we verify Bougie \textit{et al.}'s argument in our formulation, and also extend it for the other families of exceptional systems and also for the multi-indexed systems.

\subsubsection{The condition equation}
For the exceptional/multi-indexed systems, the SWKB condition \eqref{eq:3-SWKB} reads
\begin{equation}
\int_{a_{\rm L}}^{a_{\rm R}} \sqrt{\mathcal{E}_{\mathcal{D};n}^{\rm (M,\ast)} - \left( \hbar\frac{d}{dx}\ln\left|\phi_{\mathcal{D};0}^{\rm (M,\ast)}(x)\right| \right)^2} \,dx = n\pi\hbar ~,~~~
n = 0,1,2,\ldots ~,~~~
\ast = \mathrm{L}, \mathrm{J} ~.
\end{equation}
This equation is reduced to  
\begin{align}
\mathrm{L:}\qquad
&\int_{a_{\rm L}'}^{a_{\rm R}'} \sqrt{n - z\left( \frac{d}{dz}\ln\left|\check{\phi}_{\mathcal{D};0}^{\rm (M,L)}(z)\right| \right)^2} \frac{dz}{\sqrt{z}} = n\pi ~, 
\label{eq:3-SWKB_ML} \\
\mathrm{J:}\qquad
&\int_{a_{\rm L}'}^{a_{\rm R}'} \sqrt{n(n+g+h) - (1-y^2)\left( \frac{d}{dy}\ln\left|\check{\phi}_{\mathcal{D};0}^{\rm (M,J)}(y)\right| \right)^2} \frac{dy}{\sqrt{1-y^2}} = n\pi ~,
\label{eq:3-SWKB_MJ}
\end{align}
with $a_{\rm L}'$ and $a_{\rm R}'$ being the roots of the equations where inside the square roots are put to zero.
Here, $z\equiv\omega x^2/\hbar$, $y=\cos 2x$, and also $\phi_{\mathcal{D};0}^{\rm (M,L)}(x)\equiv \check{\phi}_{\mathcal{D};0}^{\rm (M,L)}(z)$, $\phi_{\mathcal{D};0}^{\rm (M,J)}(x)\equiv \check{\phi}_{\mathcal{D};0}^{\rm (M,J)}(y)$.
Note that these formulae depend on $g$ and $h$, but are independent of $\hbar$ and $\omega$.
Thus one can safely put $\hbar=\omega=1$ without loss of generality.

\subsubsection{Numerical study}
So far, we have no way of carrying out the SWKB integrations \eqref{eq:3-SWKB_ML} and \eqref{eq:3-SWKB_MJ} for $n\geqslant 1$ analytically.
After a few numerical experiments (See Tab. \ref{tab:3-num_ex}), it turns out that the condition equation is never exactly (but approximately) satisfied.
This agrees with Bougie \textit{et al.}'s point \cite{bougie2018supersymmetric}.

\begin{table}[t]
\centering
\caption[The results of several numerical experiment on the SWKB integrals \eqref{eq:3-SWKB_ML} for the type II $X_1$-Laguerre system.]
	{The results of several numerical experiment on the SWKB integrals \eqref{eq:3-SWKB_ML} for the type II $X_1$-Laguerre system.
	The integrals divided by $\pi$ are displayed with six digits.}
\label{tab:3-num_ex}
\begin{tabular}{c|cccc}
$g~\backslash~n$ & $1$ & $2$ & $10$ & $100$ \\
\hline
$3$ & $0.997674$ & $1.99781$ & $~9.99930$ & $100.000$ \\
$10$ & $0.999989$ & $1.99998$ & $~9.99998$ & $100.000$ \\
$100$ & $1.00000~\,$ & $2.00000$ & $10.0000~~$ & $100.000$ \\
\end{tabular}
\end{table}

In what follows, we calculate the left-hand sides of Eqs. \eqref{eq:3-SWKB_ML} and \eqref{eq:3-SWKB_MJ} numerically to see how accurate the SWKB condition equations hold for this class of additive shape-invariant systems.
In order to evaluate the accuracy, we introduce the following quantity, the `relative error':
\begin{equation}
\mathrm{Err}(n) \coloneqq \frac{I_{\rm SWKB} - n\pi\hbar}{I_{\rm SWKB}} ~,~~~
n = 1,2,\ldots ~.
\label{eq:3-Err}
\end{equation}
For the case of $n=0$, where the condition equation is always exact by construction, $I_{\rm SWKB}=0$, we define $\mathrm{Err}=0$.

\begin{figure}[t]
\centering
\begin{minipage}[t]{.48\linewidth}
\includegraphics[width=\linewidth]{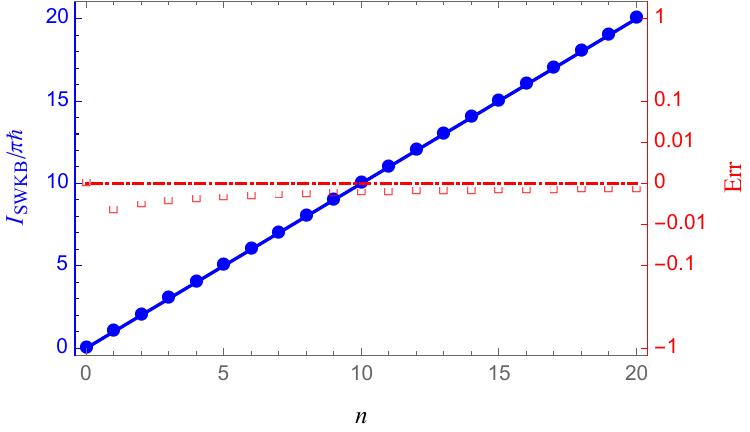}
\subcaption{$g=3$.}
\end{minipage}
\quad
\begin{minipage}[t]{.48\linewidth}
\includegraphics[width=\linewidth]{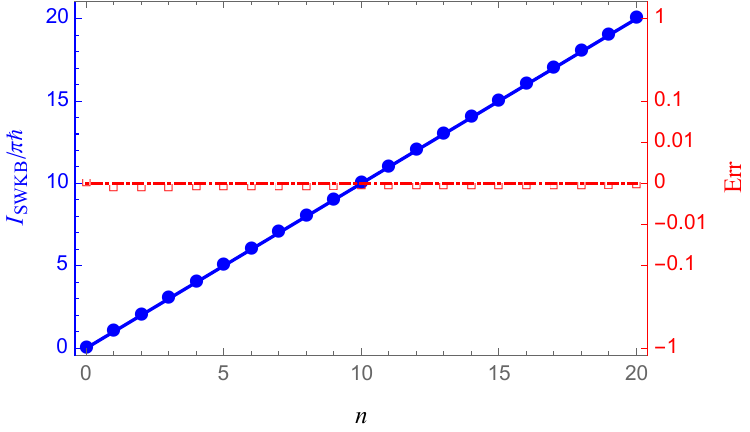}
\subcaption{$g=10$.}
\end{minipage}
\caption[The SWKB integrals for the type II $X_1$-Laguerre systems \eqref{eq:3-SWKB_ML} with several choices of the parameter.]
	{The SWKB integrals for the type II $X_1$-Laguerre systems \eqref{eq:3-SWKB_ML} with several choices of the parameter.
	The blue dots are the values of the SWKB integrals, and the red squares are the corresponding relative errors defined by Eq. \eqref{eq:3-Err}, while the blue solid line and the red chain line mean that the SWKB condition is exact, $\mathrm{Err}(n)= 0$.}
\label{fig:3-Ex-SWKB}
\end{figure}

\begin{example}[Type II $X_1$-Laguerre systems]
We take type II $X_1$-Laguerre systems with the parameters (a) $g=3$, which corresponds to the analysis of Ref. \cite{bougie2018supersymmetric}, and also (b) $g=10$ as examples.
In Fig. \ref{fig:3-Ex-SWKB}, we show the results of our numerical analysis of the SWKB integrals \eqref{eq:3-SWKB_ML}.

Note that we have employed the following rescaling for the plot of $\mathrm{Err}(n)$:
\begin{equation}
\mathrm{Err} \quad\to\quad
\sgn (\mathrm{Err})\times 2^{\log_{10}|\mathrm{Err}|} ~.
\end{equation}
We employ the same rescaling for all the plots of $\mathrm{Err}(n)$ hereafter.
\end{example}

\begin{figure}[p]
\centering
\begin{minipage}[t]{.48\linewidth}
\includegraphics[width=\linewidth]{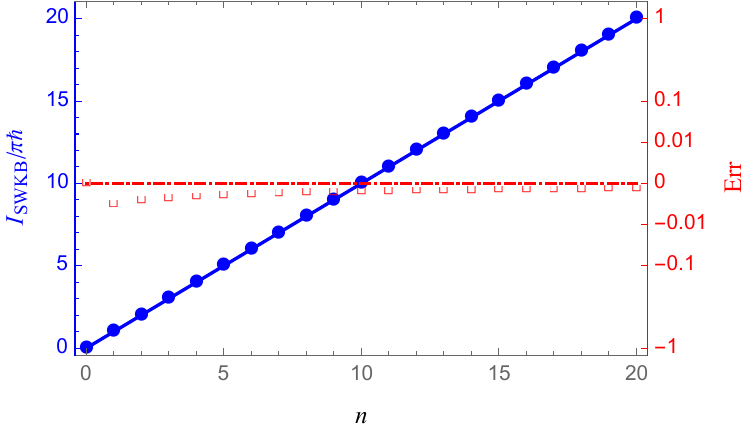}
\subcaption{$\mathcal{D} = \{ 1 \} \cap \{ 2 \}$.}
\label{fig:3-MIL-SWKB_1}
\end{minipage}
\quad
\begin{minipage}[t]{.48\linewidth}
\includegraphics[width=\linewidth]{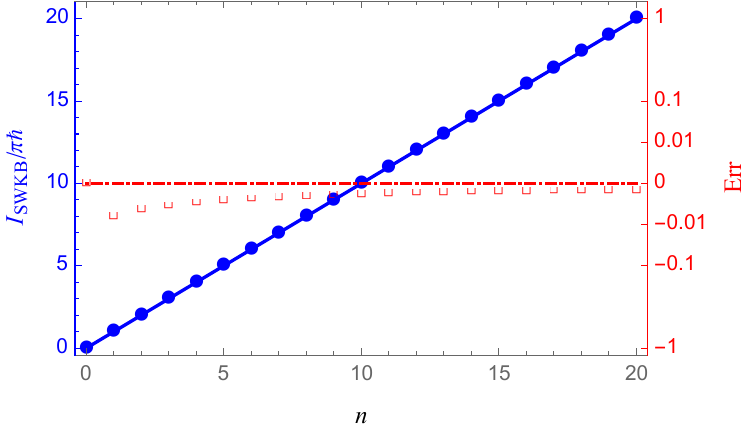}
\subcaption{$\mathcal{D} = \{ 1,2 \} \cap \{ 2,3 \}$.}
\label{fig:3-MIL-SWKB_2}
\end{minipage}
\caption[The SWKB integrals for the multi-indexed Laguerre systems \eqref{eq:3-SWKB_ML} with $g=5$.]
	{The SWKB integrals for the multi-indexed Laguerre systems \eqref{eq:3-SWKB_ML} with $g=5$.
	The blue dots are the values of the SWKB integrals, and the red squares are the corresponding relative errors defined by Eq. \eqref{eq:3-Err}, while the blue solid line and the red chain line mean that the SWKB condition is exact, $\mathrm{Err}(n)= 0$.}
\label{fig:3-MIL-SWKB}
\end{figure}

\begin{example}[Multi-indexed Laguerre systems]
Our next examples are more general cases of the multi-indexed Laguerre system.
We fix $g=5$ here.
Fig. \ref{fig:3-MIL-SWKB_1} is the results with the choice of $\mathcal{D} = \mathcal{D}^{\rm I}\cap\mathcal{D}^{\rm II} = \{ 1 \} \cap \{ 2 \}$, and Fig. \ref{fig:3-MIL-SWKB_2} is those with $\mathcal{D} = \mathcal{D}^{\rm I}\cap\mathcal{D}^{\rm II} = \{ 1,2 \} \cap \{ 2,3 \}$.
\end{example}

\begin{figure}[p]
\centering
\begin{minipage}[t]{.48\linewidth}
\includegraphics[width=\linewidth]{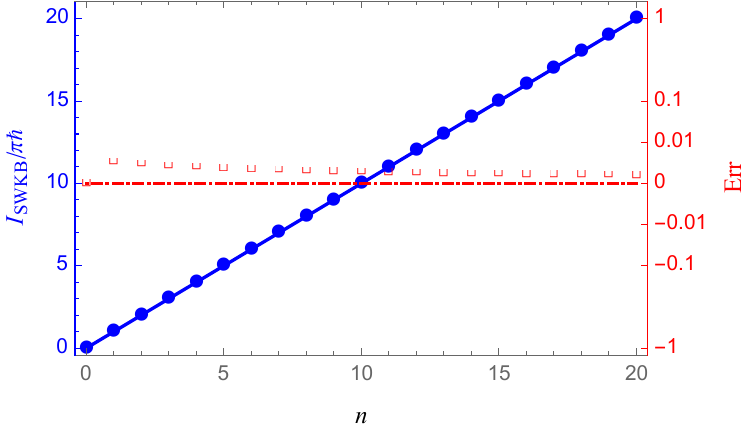}
\subcaption{$\mathcal{D} = \{ 1 \} \cap \{ 2 \}$.}
\label{fig:3-MIJ-SWKB_1}
\end{minipage}
\quad
\begin{minipage}[t]{.48\linewidth}
\includegraphics[width=\linewidth]{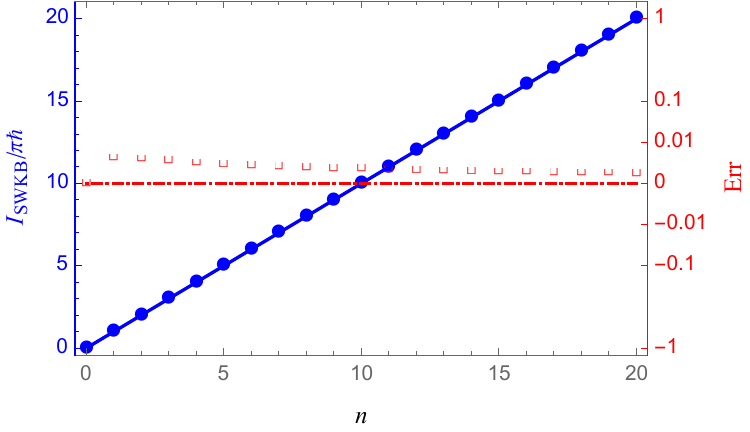}
\subcaption{$\mathcal{D} = \{ 1,2 \} \cap \{ 2,3 \}$.}
\label{fig:3-MIJ-SWKB_2}
\end{minipage}
\caption[The SWKB integrals for the multi-indexed Jacobi systems \eqref{eq:3-SWKB_MJ} with $(g,h)=(5,6)$.]
	{The SWKB integrals for the multi-indexed Jacobi systems \eqref{eq:3-SWKB_MJ} with $(g,h)=(5,6)$.
	The blue dots are the values of the SWKB integrals, and the red squares are the corresponding relative errors defined by Eq. \eqref{eq:3-Err}, while the blue solid line and the red chain line mean that the SWKB condition is exact, $\mathrm{Err}(n)= 0$.}
\label{fig:3-MIJ-SWKB}
\end{figure}

\begin{example}[Multi-indexed Jacobi systems]
For the multi-indexed-Jacobi system, we choose $(g,h)=(5,6)$ here.
We show the results for the choices of $\mathcal{D} = \mathcal{D}^{\rm I}\cap\mathcal{D}^{\rm II} = \{ 1 \} \cap \{ 2 \}$ in Fig. \ref{fig:3-MIJ-SWKB_1} and $\mathcal{D} = \mathcal{D}^{\rm I}\cap\mathcal{D}^{\rm II} = \{ 1,2 \} \cap \{ 2,3 \}$ in Fig. \ref{fig:3-MIJ-SWKB_2}.
\end{example}

One can immediately see from the numerical analyses that the condition equations are never exactly satisfied for $n\geqslant 1$, but it is notable that the relative errors are always $|\mathrm{Err}(n)| \lesssim 10^{-3}$. 
The behaviors look similar in all cases; the maximal errors occur at $n=1$, and as $n$ grows, $|\mathrm{Err}(n)|$ gradually reduces, and in the limit $n\to\infty$, the SWKB condition will be restored. 
A similar thing can be said for the parameter $g$.
For larger $g$, the relative errors get smaller with the same $\mathcal{D}$. 
Another feature revealed by our numerical analyses is that the SWKB integrals are always underestimated for the multi-indexed Laguerre systems, while it is always overestimated for the multi-indexed Jacobi systems.

\subsection{SWKB for Krein--Adler Systems}
\label{sec:3-2-4}
In the previous subsection, we have evaluated the SWKB integrals for the exceptional/multi-indexed systems, which are constructed from the conventional shape-invariant potentials by Darboux transformations.
In this subsection, we will examine the SWKB integral for Krein--Adler systems, which are another class of exactly solvable systems constructed from the conventional shape-invariant potentials by Darboux transformation.

Krein--Adler transformation corresponds to a deletion of eigenstates, so the energy spectra are deformed radically during the transformation.
It would be interesting to consider how this deformation affects the SWKB integrals. 

\subsubsection{The condition equation}
For the Krein--Adler systems, the SWKB condition \eqref{eq:3-SWKB} becomes
\begin{equation}
\int_{a_{\rm L}}^{a_{\rm R}} \sqrt{\mathcal{E}_{\mathcal{D};n}^{\rm (K,\ast)} - \left( \hbar\frac{d}{dx}\ln\left|\phi_{\mathcal{D};0}^{\rm (K,\ast)}(x)\right| \right)^2} \,dx = n\pi\hbar ~,~~~
n = 0,1,2,\ldots ~,~~~
\ast = \mathrm{H}, \mathrm{L}, \mathrm{J} ~.
\end{equation}
This equation is reduced to 
\begin{align}
\mathrm{H:}\qquad
&\int_{a_{\rm L}'}^{a_{\rm R}'} \sqrt{2\breve{n} - \left( \frac{d}{d\xi}\ln\left|\check{\phi}_{\mathcal{D};0}^{\rm (K,H)}(\xi)\right| \right)^2} \,d\xi = n\pi ~, 
\label{eq:3-SWKB_KAH} \\
\mathrm{L:}\qquad
&\int_{a_{\rm L}'}^{a_{\rm R}'} \sqrt{\breve{n} - z\left( \frac{d}{dz}\ln\left|\check{\phi}_{\mathcal{D};0}^{\rm (K,L)}(z)\right| \right)^2} \frac{dz}{\sqrt{z}} = n\pi ~, 
\label{eq:3-SWKB_KAL} \\
\mathrm{J:}\qquad
&\int_{a_{\rm L}'}^{a_{\rm R}'} \sqrt{\breve{n}(\breve{n}+g+h) - (1-y^2)\left( \frac{d}{dy}\ln\left|\check{\phi}_{\mathcal{D};0}^{\rm (K,J)}(y)\right| \right)^2} \frac{dy}{\sqrt{1-y^2}} = n\pi ~,
\label{eq:3-SWKB_KAJ} \\[1ex]
&\qquad\text{with}\quad
\breve{n} \in \{ n \} \backslash \mathcal{D} ~,
\end{align}
in which $a_{\rm L}'$ and $a_{\rm R}'$ are the roots of the equations where inside the square roots equal zero.
Also, $\xi\equiv\sqrt{\omega/\hbar}\,x$, $z\equiv\xi^2$, $y=\cos 2x$, and $\phi_{\mathcal{D};0}^{\rm (K,H)}(x)\equiv \check{\phi}_{\mathcal{D};0}^{\rm (K,H)}(\xi)$, $\phi_{\mathcal{D};0}^{\rm (K,L)}(x)\equiv \check{\phi}_{\mathcal{D};0}^{\rm (K,L)}(z)$, $\phi_{\mathcal{D};0}^{\rm (K,J)}(x)\equiv \check{\phi}_{\mathcal{D};0}^{\rm (K,J)}(y)$.
Note that these formulae depend on $g$ and $h$, but are totally independent of $\hbar$ and $\omega$.
Therefore we set $\hbar=\omega=1$ hereafter.

\subsubsection{Prescription for cases with more than two `turning points'}
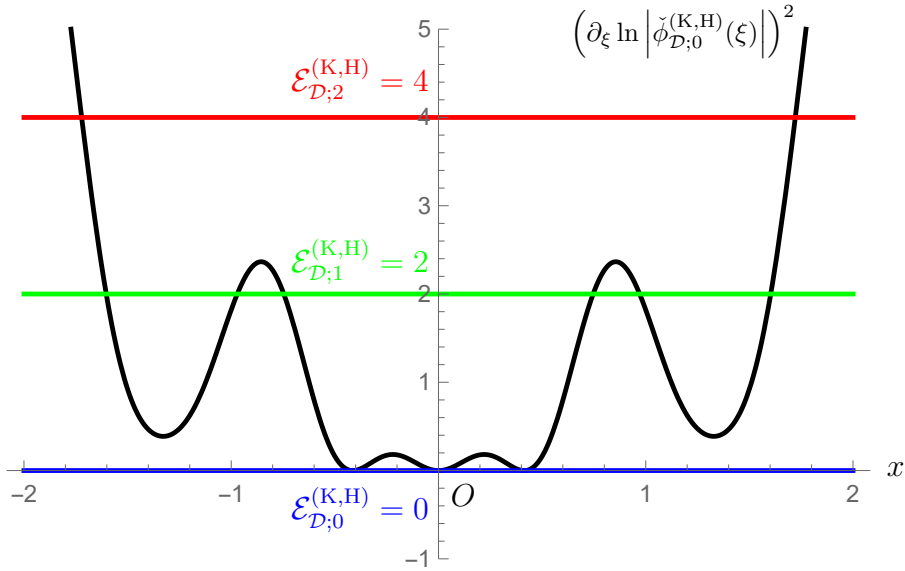
\begin{figure}[t]
\centering
	\begin{tikzpicture}
	\node at (0,0) {\includegraphics[scale=1]{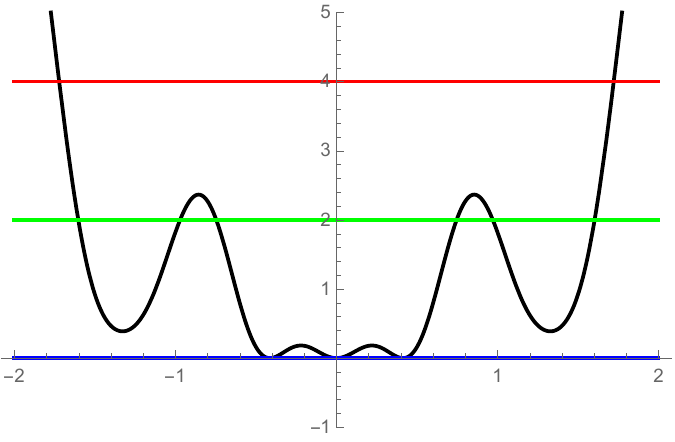}};
	\node[right] at (0,-2.65) {$O$};
	\node at (6,-2.3) {$x$};
	\node at (3.2,3.5) {\footnotesize$\left( \partial_{\xi}\ln\left|\check{\phi}_{\mathcal{D};0}^{\rm (K,H)}(\xi)\right| \right)^2$};
	\node[below left] at (0,-2.4) {\textcolor{blue}{$\mathcal{E}_{\mathcal{D};0}^{(\mathrm{K,H})}=0$}};
	\node[above left] at (0,0) {\textcolor{green}{$\mathcal{E}_{\mathcal{D};1}^{(\mathrm{K,H})}=2$}};
	\node[above left] at (0,2.4) {\textcolor{red}{$\mathcal{E}_{\mathcal{D};2}^{(\mathrm{K,H})}=4$}};
	\end{tikzpicture}
\caption[The plot of the square of the superpotential $\left( \partial_{\xi}\ln\left|\check{\phi}_{\mathcal{D};0}^{\rm (K,H)}(\xi)\right| \right)^2$ for $d=4$.]
	{The plot of the square of the superpotential $\left( \partial_{\xi}\ln\left|\check{\phi}_{\mathcal{D};0}^{\rm (K,H)}(\xi)\right| \right)^2$ for $d=4$.
	When $n=1$, this system has two sets of turning points.}
\label{fig:3-SeveralTurningPoints}
\end{figure}

In some cases, the equation `$\mathcal{E}_n - W(x)^2 = 0$' has more than two roots, $\{ (a_i,a_{i+1})\;;\; i=1,2,\ldots \}$.
For example, if we take $\ast = \mathrm{H}$ and choose $\mathcal{D}=\{ 4,5 \}$, the equation for the first excited state has four roots (See Fig. \ref{fig:3-SeveralTurningPoints}).
In such cases, we employ a prescription of replacing the left-hand side of Eq. (\ref{eq:3-SWKB}) as follows~\cite{Nasuda:2020aqf}:
\begin{equation}
\int_{a_{\rm L}}^{a_{\rm R}} \sqrt{\mathcal{E}_n - W(x)^2} \,dx
\quad\to\quad
\sum_i \int_{a_i}^{a_{i+1}} \sqrt{\mathcal{E}_n - W(x)^2} \,dx ~.
\label{eq:3-prescription}
\end{equation}

In the following of this subsection, we restrict ourselves to the cases $\mathcal{D}=\{ d,d+1 \}$ with $d=1,2,\cdots$ to simplify our discussions.

\subsubsection{In quantum-Hamilton--Jacobi point of view}
We compare the pole structures of the SWKB integrands and those of the quantum momentum functions on the complex $x$-plane for Krein--Adler systems.
One can presume as follows.
They coincide, except for the branch cut(s) on the real axis in the former, if the SWKB is also an exact condition; when they do not coincide, we expect that it is not exact.

\begin{example}[H]
Take the case with $\ast=\mathrm{H}$ as an example.
The quantum momentum function of the system is 
\begin{equation}
p(x,\mathcal{E}_n) = -\mathrm{i} \left( -x - \frac{\partial_x\mathrm{W}[H_d,H_{d+1}](x)}{\mathrm{W}[H_d,H_{d+1}](x)} + \frac{\partial_x\mathrm{W}[H_d,H_{d+1},H_{\breve{n}}](x)}{\mathrm{W}[H_d,H_{d+1},H_{\breve{n}}](x)} \right) ~.
\end{equation}

The plots of the singularity structures for both the quantum momentum function and the SWKB integrand for the first excited state with $d = 1$ are displayed in Fig. \ref{fig:3-KAH-QMF-SingStruct}.
These figures are the results of the analytic calculation; the position of each singularity is obtained analytically.
Apparently, they do not coincide with each other and the quantization of the SWKB integral is not exact.

\begin{figure}[p]
\centering
\begin{minipage}[t]{.48\linewidth}
\centering
\includegraphics[scale=1.5]{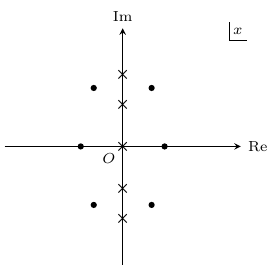}
\subcaption{The QMF.}
\end{minipage}
\quad
\begin{minipage}[t]{.48\linewidth}
\centering
\includegraphics[scale=1.5]{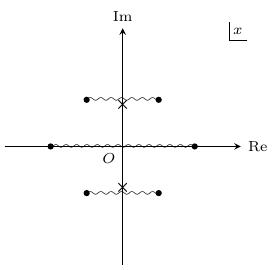}
\subcaption{The SWKB integral.}
\end{minipage}
\caption[Poles (x-marks) and branch cuts (wavy lines) of the quantum momentum function (QMF) $p(x;\mathcal{E}_1)$ and the SWKB integrand $\sqrt{\mathcal{E}_1-W(x)^2}$ for a Krein--Adler Hermite system.]
	{Poles (x-marks) and branch cuts (wavy lines) on a complex $x$-plane are displayed for (a) the quantum momentum function (QMF) $p(x;\mathcal{E}_1)$ for a Krein--Adler system, and (b) the SWKB integrand $\sqrt{\mathcal{E}_1-W(x)^2}$.
	We also plot the zeros with closed dots on both figures.
	The poles are at (a) $x=0,~\pm\mathrm{i}/\sqrt{2},~\pm\mathrm{i}\sqrt{3/2}$, (b) $x=\pm\mathrm{i}/\sqrt{2}$ and the nodes exist at (a) $x=\pm1/\sqrt{2},~\left( \pm\sqrt[4]{33-12\sqrt{6}} \pm\sqrt[4]{33+12\sqrt{6}} \right) \Big/ 2\sqrt{2},
	~\left( \pm\sqrt[4]{33-12\sqrt{6}} \mp\sqrt[4]{33+12\sqrt{6}} \right) \Big/ 2\sqrt{2}$, (b) $x=\pm\sqrt{3/2},~\left( \pm\sqrt{3}\pm\mathrm{i}\sqrt{5} \right) \big/ 2\sqrt{2}$.}
\label{fig:3-KAH-QMF-SingStruct}
\end{figure}

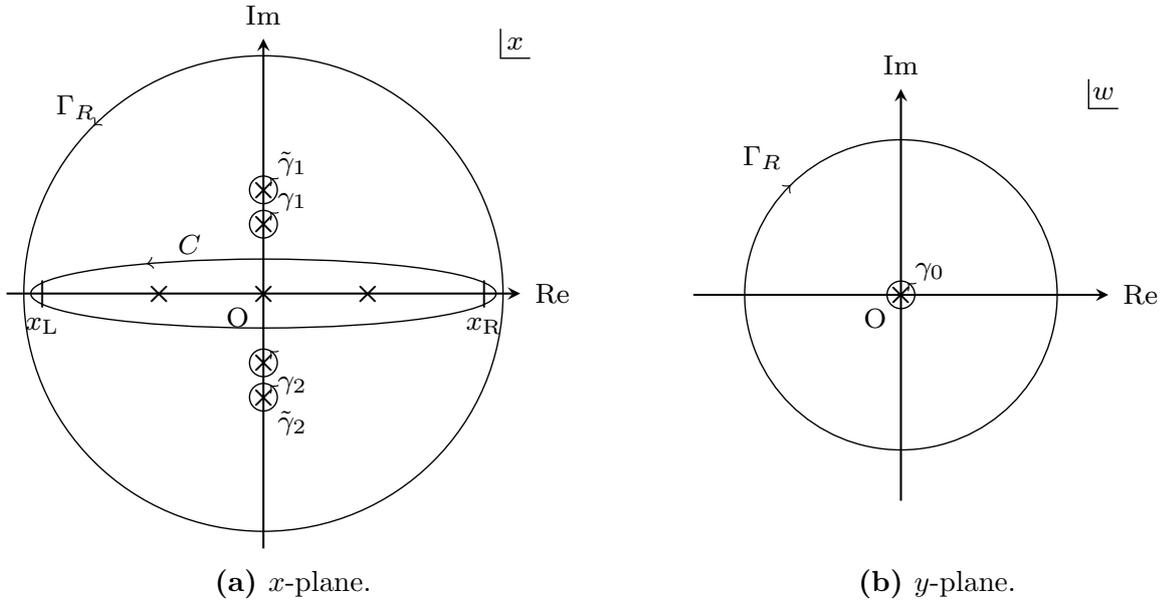
\begin{figure}[p]
\centering
\begin{minipage}[b]{.48\linewidth}
\centering\scalebox{1.3}{
	\begin{tikzpicture}
	\draw[semithick,->,>=stealth] (0,-2.6)--(0,2.6) node[above] {\scriptsize $\mathrm{Im}$};
	\draw[semithick,->,>=stealth] (-2.6,0)--(2.6,0) node[right] {\scriptsize $\mathrm{Re}$};
	\draw (2.4,2.7)--(2.4,2.4)--(2.7,2.4) node at (2.55,2.55) {\scriptsize $x$};
	\node [below left] at(0,0) {\scriptsize $\mathrm{O}$};
	
	\node at (0,0) {\scriptsize $\bm{\times}$};
	\node at (1.05737,0) {\scriptsize $\bm{\times}$};
	\node at (-1.05737,0) {\scriptsize $\bm{\times}$};
	\node at (0,0.707107) {\scriptsize $\bm{\times}$};
	\node at (0,-0.707107) {\scriptsize $\bm{\times}$};
	\node at (0,1.05737) {\scriptsize $\bm{\times}$};
	\node at (0,-1.05737) {\scriptsize $\bm{\times}$};
	
	\node at (2.23607,0) {\scriptsize $\bm{|}$} node[below] at (2.23607,-0.1) {\scriptsize $x_{\mathrm{R}}$};
	\node at (-2.23607,0) {\scriptsize $\bm{|}$} node[below] at (-2.23607,-0.1) {\scriptsize $x_{\mathrm{L}}$};
	
	\draw (0,0) circle (69pt);
	\draw[->] (-48.79pt,48.79pt)--(-48.8pt,48.78pt);
	\node at (-1.9,1.9) {\scriptsize $\Gamma_R$};
	
	\draw (0,0) circle (67pt and 10pt);
	\draw[->] (-33.5pt,8.66pt)--(-33.56pt,8.65pt);
	\node at (-0.75,0.5) {\scriptsize $C$};
	
	\draw (0,0.707107) circle (4pt) node[above right] {\scriptsize $\gamma_1$};
	\draw[->] (2pt,3.46pt+0.707107cm)--(1.99pt,3.47pt+0.707107cm);
	\draw (0,-0.707107) circle (4pt) node[below right] {\scriptsize $\gamma_2$};
	\draw[->] (2pt,3.46pt-0.707107cm)--(1.99pt,3.47pt-0.707107cm);
	\draw (0,1.05737) circle (4pt) node[above right] {\scriptsize $\tilde{\gamma}_1$};
	\draw[->] (2pt,3.46pt+1.05737cm)--(1.99pt,3.47pt+1.05737cm);
	\draw (0,-1.05737) circle (4pt) node[below right] {\scriptsize $\tilde{\gamma}_2$};
	\draw[->] (2pt,3.46pt-1.05737cm)--(1.99pt,3.47pt-1.05737cm);
	\end{tikzpicture}
}
\subcaption{$x$-plane.}
\end{minipage}
\quad
\begin{minipage}[b]{.48\linewidth}
\centering\scalebox{1.3}{
	\begin{tikzpicture}
	\draw[semithick,->,>=stealth] (0,-2.1)--(0,2.1) node[above] {\scriptsize $\mathrm{Im}$};
	\draw[semithick,->,>=stealth] (-2.1,0)--(2.1,0) node[right] {\scriptsize $\mathrm{Re}$};
	\draw (1.9,2.2)--(1.9,1.9)--(2.2,1.9) node at (2.05,2.05) {\scriptsize $w$};
	\node [below left] at(0,0) {\scriptsize $\mathrm{O}$};
	
	\node at (0,0) {\scriptsize $\bm{\times}$};
	
	\draw (0,0) circle (45pt);
	\draw[->] (-31.82pt,31.82pt)--(-31.81pt,31.83pt);
	\node at (-1.4,1.4) {\scriptsize $\Gamma_R$};
	
	\draw (0,0) circle (4pt) node[above right] {\scriptsize $\gamma_0$};
	\draw[->] (2pt,3.46pt)--(1.99pt,3.47pt);
	\end{tikzpicture}
}
\bigskip\medskip
\subcaption{$y$-plane.}
\end{minipage}
\caption{The contours of the integrations in Eq. \eqref{eq:QHJ_ContourInt_KAH}.}
\label{fig:3-KAH-QMF-SingStruct-Contours}
\end{figure}

In what follows we obtain analytically the exact bound-state spectrum from the quantization condition for the quantum action variable \eqref{eq:3-QHJ-QC}.
As we mentioned, $C$ in Eq. \eqref{eq:3-QHJ-QC} is the counterclockwise contour enclosing the two classical turning points $x_{\mathrm{L,R}}$.
For the Krein--Adler systems, in general, the quantum momentum functions have an isolated pole at $x\to\infty$, $4d-2$ fixed poles other than that and $\breve{n}$ moving poles, including $n$ moving poles on the real axis.
For the names of the contours enclosing these poles counterclockwise, see Fig. \ref{fig:3-KAH-QMF-SingStruct-Contours}.
Hence, the following equation holds:
\begin{equation}
J_{\Gamma_R} = J_{\mathrm{QHJ}} + \sum_{j=1}^{4d-2}J_{\gamma_j} + \sum_{j=1}^{\breve{n}-n}J_{\tilde{\gamma}_j} ~,~~~
J_{\bullet} \coloneqq \frac{1}{2\pi} \oint_{\bullet} p(x;\mathcal{E}) \,dx ~.
\label{eq:QHJ_ContourInt_KAH}
\end{equation}
Here, taking into account that $\mathrm{W}[H_d,H_{d+1}](x)$, $\mathrm{W}[H_d,H_{d+1},1](x)$ and $\mathrm{W}[H_d,H_{d+1},H_{\breve{n}}](x)$ are polynomials of degree $2d$, $2d-2$ and $2d-2+\breve{n}$ respectively, the second and the third terms of the right-hand side of the first equation in Eq. \eqref{eq:QHJ_ContourInt_KAH} are 
\begin{align}
\sum_{i=1}^{4d-2}J_{\gamma_i} &= (2d-2) - 2d = -2 ~, \\
\sum_{j=1}^{\breve{n}-n}J_{\tilde{\gamma}_j} &= -(2d-2) + (2d-2+\breve{n}-n) = \breve{n}-n ~.
\end{align}

We evaluate $J_{\Gamma_R}$ by changing variables $x\to y\equiv 1/x$ and then employing the Laurant expansion of the quantum momentum function and the quantum Hamilton--Jacobi equation.
Then we obtain $J_{\Gamma_R} = \mathcal{E}/2 - 2$.
Therefore, the quantization condition yields $\mathcal{E} = 2\breve{n}$.
\end{example}

\subsubsection{Numerical study}
In the case of the Krein--Adler systems, we again have no way of executing the SWKB integrals \eqref{eq:3-SWKB_KAH}--\eqref{eq:3-SWKB_KAJ} for $n\geqslant 1$ analytically.
Thus we numerically compute the left-hand side of Eqs. \eqref{eq:3-SWKB_KAH}--\eqref{eq:3-SWKB_KAJ}, together with the relative error \eqref{eq:3-Err} to show the accuracy of the SWKB condition equation.

\begin{figure}[p]
\centering
\begin{minipage}[t]{.48\linewidth}
\includegraphics[width=\linewidth]{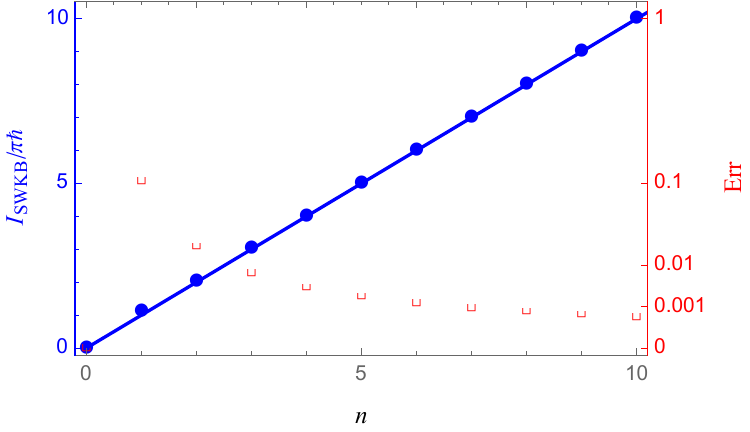}
\subcaption{$d=1$.}
\end{minipage}
\quad
\begin{minipage}[t]{.48\linewidth}
\includegraphics[width=\linewidth]{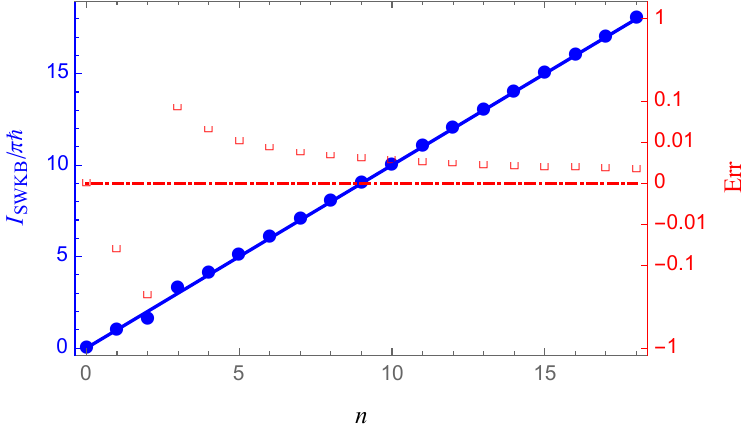}
\subcaption{$d=3$.}
\end{minipage}

\medskip
\begin{minipage}[t]{.48\linewidth}
\includegraphics[width=\linewidth]{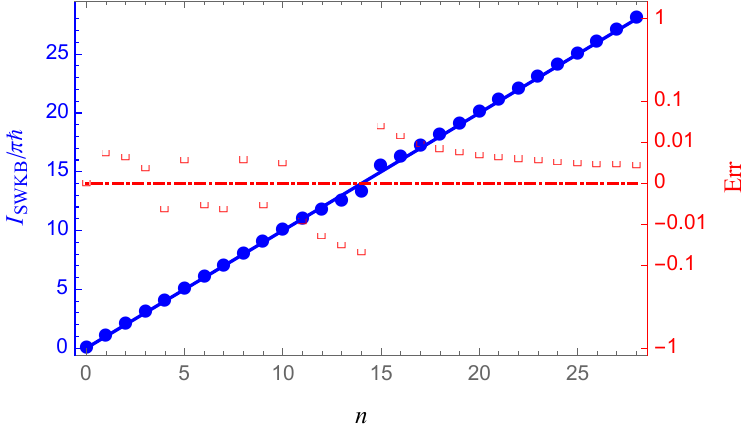}
\subcaption{$d=15$.}
\end{minipage}
\quad
\begin{minipage}[t]{.48\linewidth}
~
\end{minipage}
\caption[The SWKB integrals for the Krein--Adler Hermite systems \eqref{eq:3-SWKB_KAH} with several choices of the parameter.]
	{The SWKB integrals for the Krein--Adler Hermite systems \eqref{eq:3-SWKB_KAH} with several choices of the parameter.
	The blue dots are the values of the SWKB integrals, and the red squares are the corresponding relative errors defined by Eq. \eqref{eq:3-Err}, while the blue solid line and the red chain line mean that the SWKB condition is exact, $\mathrm{Err}(n)= 0$.}
\label{fig:3-KAH-SWKB}
\end{figure}

\begin{example}[H]
Our first examples displayed in Fig. \ref{fig:3-KAH-SWKB} are the Krein--Adler system with $\ast=\mathrm{H}$ and (a) $d=1$, (b)$d=3$, (c) $d=15$.
\end{example}

\begin{figure}[p]
\centering
\begin{minipage}[t]{.48\linewidth}
\includegraphics[width=\linewidth]{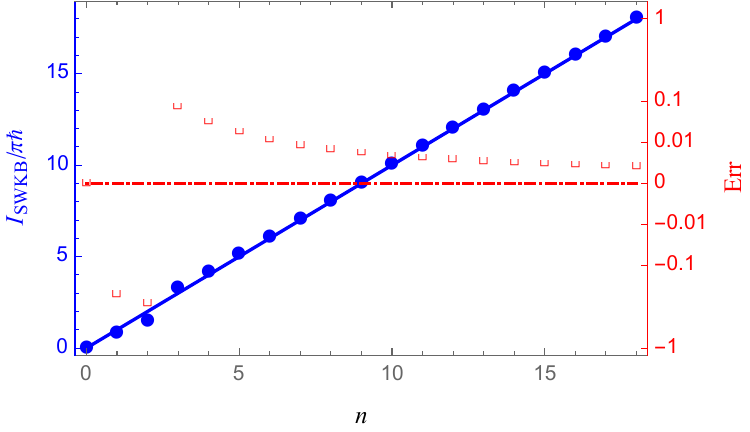}
\subcaption{$d=3$ and $g=3$.}
\end{minipage}
\quad
\begin{minipage}[t]{.48\linewidth}
\includegraphics[width=\linewidth]{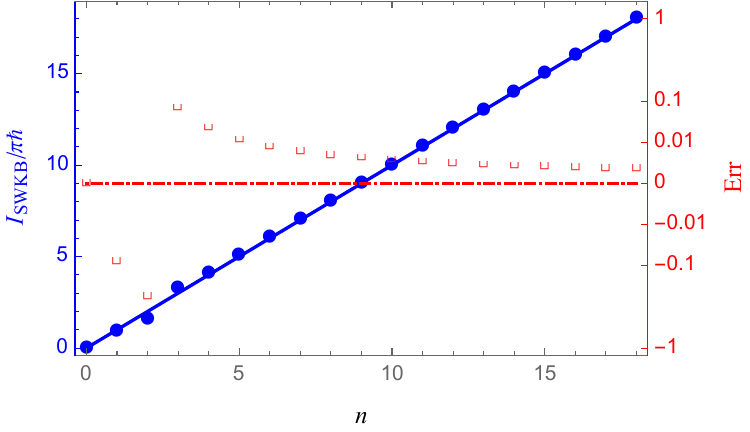}
\subcaption{$d=3$ and $g=30$.}
\end{minipage}

\medskip
\begin{minipage}[t]{.48\linewidth}
\includegraphics[width=\linewidth]{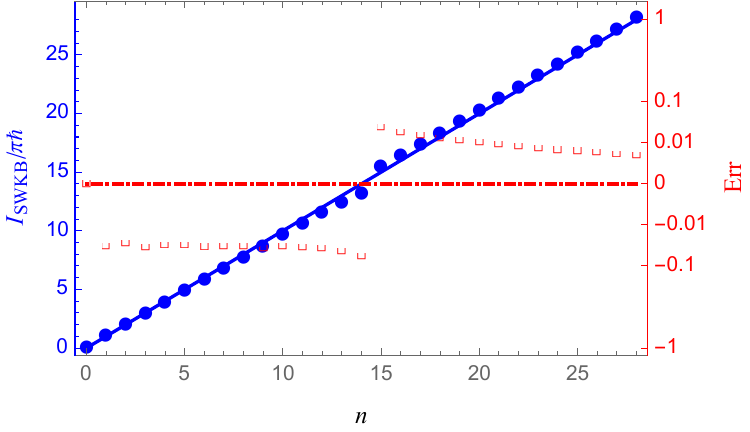}
\subcaption{$d=15$ and $g=3$.}
\end{minipage}
\quad
\begin{minipage}[t]{.48\linewidth}
~
\end{minipage}
\caption[The SWKB integrals for the Krein--Adler Laguerre systems \eqref{eq:3-SWKB_KAL} with several choices of the parameters.]
	{The SWKB integrals for the Krein--Adler Laguerre systems \eqref{eq:3-SWKB_KAL} with several choices of the parameters.
	The blue dots are the values of the SWKB integrals, and the red squares are the corresponding relative errors defined by Eq. \eqref{eq:3-Err}, while the blue solid line and the red chain line mean that the SWKB condition is exact, $\mathrm{Err}(n)= 0$.}
\label{fig:3-KAL-SWKB}
\end{figure}

\begin{example}[L]
For $\ast=\mathrm{L}$, we set (a) $d=3$ and $g=3$, (b) $d=3$ and $g=30$, (c) $d=15$ and $g=3$ as illustrations.
See Fig. \ref{fig:3-KAL-SWKB}.
\end{example}

\begin{figure}[p]
\centering
\begin{minipage}[t]{.48\linewidth}
\includegraphics[width=\linewidth]{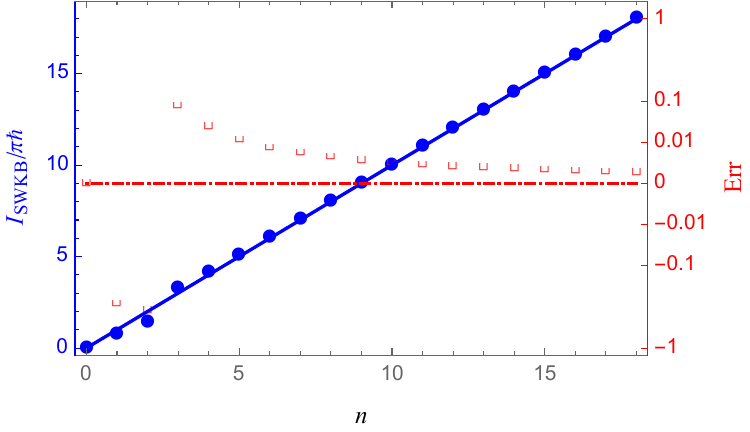}
\subcaption{$d=3$ and $(g,h)=(3,4)$.}
\end{minipage}
\quad
\begin{minipage}[t]{.48\linewidth}
\includegraphics[width=\linewidth]{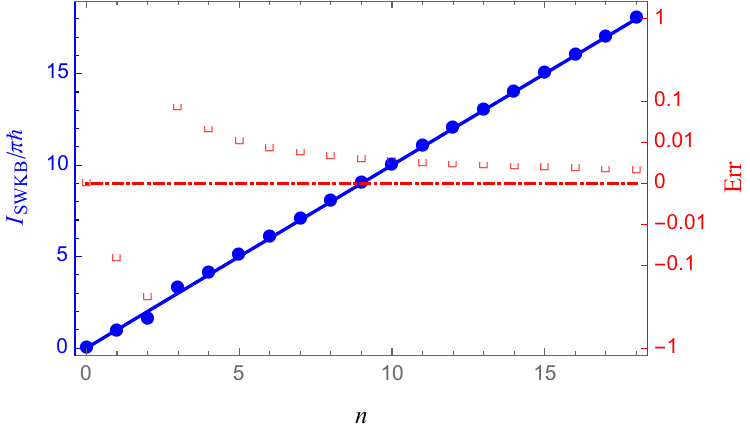}
\subcaption{$d=3$ and $(g,h)=(30,40)$.}
\end{minipage}

\medskip
\begin{minipage}[t]{.48\linewidth}
\includegraphics[width=\linewidth]{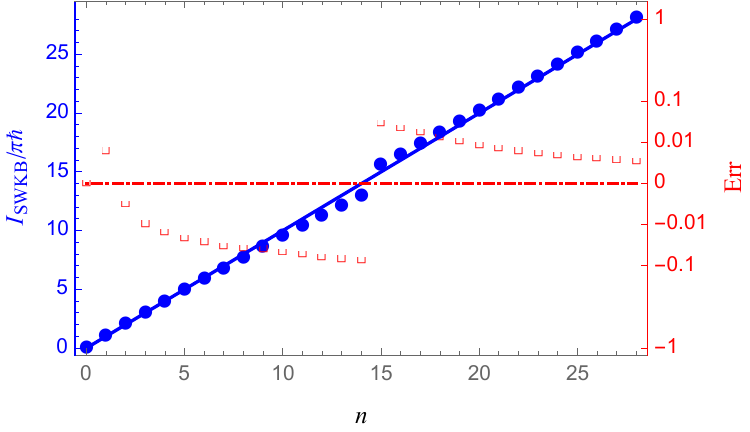}
\subcaption{$d=15$ and $(g,h)=(3,4)$.}
\end{minipage}
\quad
\begin{minipage}[t]{.48\linewidth}
~
\end{minipage}
\caption[The SWKB integrals for the Krein--Adler Jacobi systems \eqref{eq:3-SWKB_KAJ} with several choices of the parameters.]
	{The SWKB integrals for the Krein--Adler Jacobi systems \eqref{eq:3-SWKB_KAJ} with several choices of the parameters.
	The blue dots are the values of the SWKB integrals, and the red squares are the corresponding relative errors defined by Eq. \eqref{eq:3-Err}, while the blue solid line and the red chain line mean that the SWKB condition is exact, $\mathrm{Err}(n)= 0$.}
\label{fig:3-KAJ-SWKB}
\end{figure}

\begin{example}[J]
Next, for $\ast=\mathrm{J}$, (a) $d=3$, $(g,h)=(3,4)$,  (b) $d=3$, $(g,h)=(30,40)$, (c) $d=15$, $(g,h)=(3,4)$.
The numerical results are shown in Fig. \ref{fig:3-KAJ-SWKB}.
\end{example}

One can immediately see from our numerical calculations that the condition equations do not hold.
What is different from the previous case is that the relative error $\mathrm{Err}(n)$ does not decrease monotonically as growing $n$ in each case here.
The maximum of the error occurs in the vicinity of the deleted levels, and also the errors tend to be of opposite sign between the below and the above of the deleted levels. 
Note that the behavior at $n\to 0$ and $n\to \infty$ is not symmetrical, \textit{i.e.}, for the smaller $n$, the value still decreases but seems not to go to the exact condition. 
For the cases of Laguerre and Jacobi, we numerically confirm that the relative errors decrease as the larger value of the parameters $g,h$.
A notable feature in the case of Hermite is that when we delete higher levels, the integral value exhibits oscillating behavior around the exact one below the deleted levels, which is likely to be caused by our prescription \eqref{eq:3-prescription}.

\bigskip
In the following, we exclusively concentrate on the cases of the deformations/transformations of the 1-dim. harmonic oscillator for simplicity.

\subsection{SWKB for Conditionally Exactly Solvable Potentials}
\label{sec:3-2-5}
In the previous subsection, we have seen that the discrepancy $|\mathrm{Err}|(n)$ takes its maximum value around the deleted levels, where the energy distribution is radically different from the rest of the spectrum.
We now dig deeper into the relation between the SWKB-(non)exactness and the level structure.
Here, we would like to know what happens between the harmonic oscillator and its Krein--Adler transformation.
Since Darboux transformation (and thereby Krein--Adler transformation) is a discrete transformation, we have no idea what is going on between them.
Now we need an exactly solvable system that smoothly connects the harmonic oscillator and its Krein--Adler transformation with a continuous parameter.

The conditionally exactly solvable systems by Junker and Roy meet our needs, and we shall employ them.
In addition, these systems contain two continuous parameters, say $b$ and $\beta$. 
The former describes the modification of level structures, while changing the latter means the isospectral deformation of a potential.

\subsubsection{The condition equation}
For the conditionally exactly solvable potential with $\ast=\mathrm{H}$ by Junker and Roy, the SWKB condition \eqref{eq:3-SWKB} is
\begin{equation}
\int_{a_{\rm L}}^{a_{\rm R}} \sqrt{\mathcal{E}_n^{\rm (C,H)} - \left( \hbar\frac{d}{dx}\ln\left|\phi_0^{\rm (C,H)}(x)\right| \right)^2} \,dx = n\pi\hbar ~,~~~
n = 0,1,2,\ldots ~.
\end{equation}
This equation is reduced to 
\begin{equation}
\int_{a_{\rm L}'}^{a_{\rm R}'} \sqrt{2n+b - \left( \frac{d}{d\xi}\ln\left|\check{\phi}_0^{\rm (C,H)}(\xi)\right| \right)^2} \,d\xi = n\pi ~,
\label{eq:3-SWKB_CES}
\end{equation}
where $a_{\rm L}'$ and $a_{\rm R}'$ are the roots of the equation obtained by setting the inside of the square root equal to zero.
Again, $\xi\equiv\sqrt{\omega/\hbar}\,x$ and $\phi_0^{\rm (C,H)}(x)\equiv \check{\phi}_0^{\rm (C,H)}(\xi)$.
This formula is also totally independent of $\hbar$ and $\omega$, and thereby we fix $\hbar=\omega=1$.

\subsubsection{In quantum-Hamilton--Jacobi point of view}
We investigate the pole structure of the quantum momentum function:
\begin{equation}
p(x;\mathcal{E}_n) = -\mathrm{i} \left\{ \frac{\partial_x\phi_0^{\rm (C,H)}(x)}{\phi_0^{\rm (C,H)}(x)} + \frac{u^{\rm (H)}(x)\partial_x\phi_n^{\rm (C,H)}(x)}{\left[ Cu^{\rm (H)}(x) + \partial_xu^{\rm (H)}(x) \right] \phi_n^{\rm (H)}(x)} + \frac{\partial_xu^{\rm (H)}(x)}{u^{\rm (H)}(x)} \right\} ~,
\end{equation}
with some constant $C$.

\begin{figure}[p]
\centering
	\begin{minipage}[t]{0.48\linewidth}
	\centering
	\includegraphics[width=.8\linewidth]{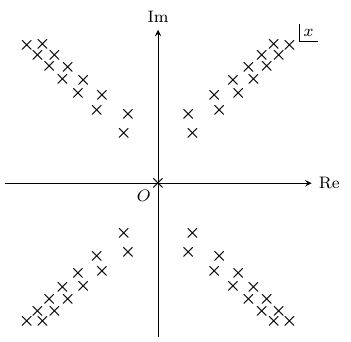}
	\subcaption{$b=-0.5$.}
	\label{fig:3-CES-QMF-SingSturct_a}
	\end{minipage}
	\quad
	\begin{minipage}[t]{0.48\linewidth}
	\centering
	\includegraphics[width=.8\linewidth]{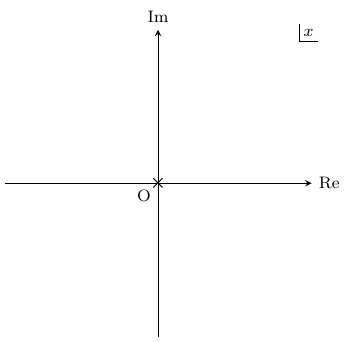}
	\subcaption{$b=0$.}
	\label{fig:3-CES-QMF-SingSturct_b}
	\end{minipage}
	
	\medskip
	\begin{minipage}[t]{0.48\linewidth}
	\centering
	\includegraphics[width=.8\linewidth]{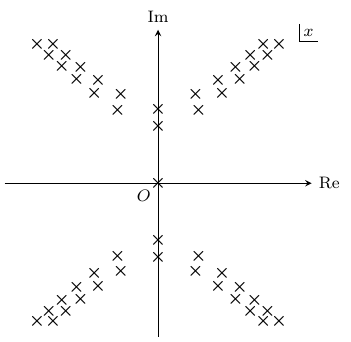}
	\subcaption{$b=0.1$.}
	\label{fig:3-CES-QMF-SingSturct_c}
	\end{minipage}
	\quad
	\begin{minipage}[t]{0.48\linewidth}
	\centering
	\includegraphics[width=.8\linewidth]{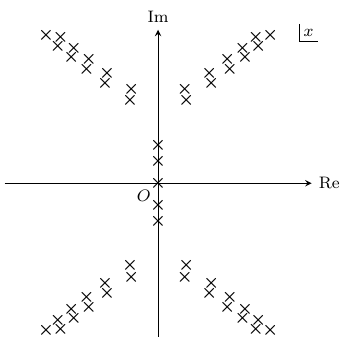}
	\subcaption{$b=3.5$.}
	\label{fig:3-CES-QMF-SingSturct_d}
	\end{minipage}
	
	\medskip
	\begin{minipage}[t]{0.48\linewidth}
	\centering
	\includegraphics[width=.8\linewidth]{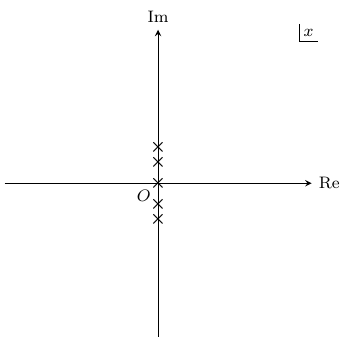}
	\subcaption{$b=4$.}
	\label{fig:3-CES-QMF-SingSturct_e}
	\end{minipage}
	\quad
	\begin{minipage}[t]{0.48\linewidth}
	\centering
	\includegraphics[width=.8\linewidth]{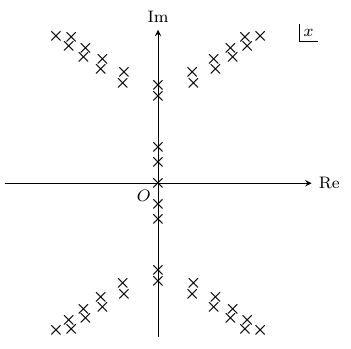}
	\subcaption{$b=4.1$.}
	\label{fig:3-CES-QMF-SingSturct_f}
	\end{minipage}
\caption[The singularity structures of the quantum momentum functions for the first excited states of the conditionally exactly solvable system with various $b$'s.]
	{The singularity structures of the quantum momentum functions for the first excited states of the conditionally exactly solvable system with various $b$'s.
	Poles are plotted by x-marks.
	The location of each pole is calculated numerically.
	Note that (b) and (e) are identical to that of the 1-dim. harmonic oscillator and the Krein--Adler Hermite system with $d=1$ respectively.}
\label{fig:3-CES-QMF-SingSturct}
\end{figure}

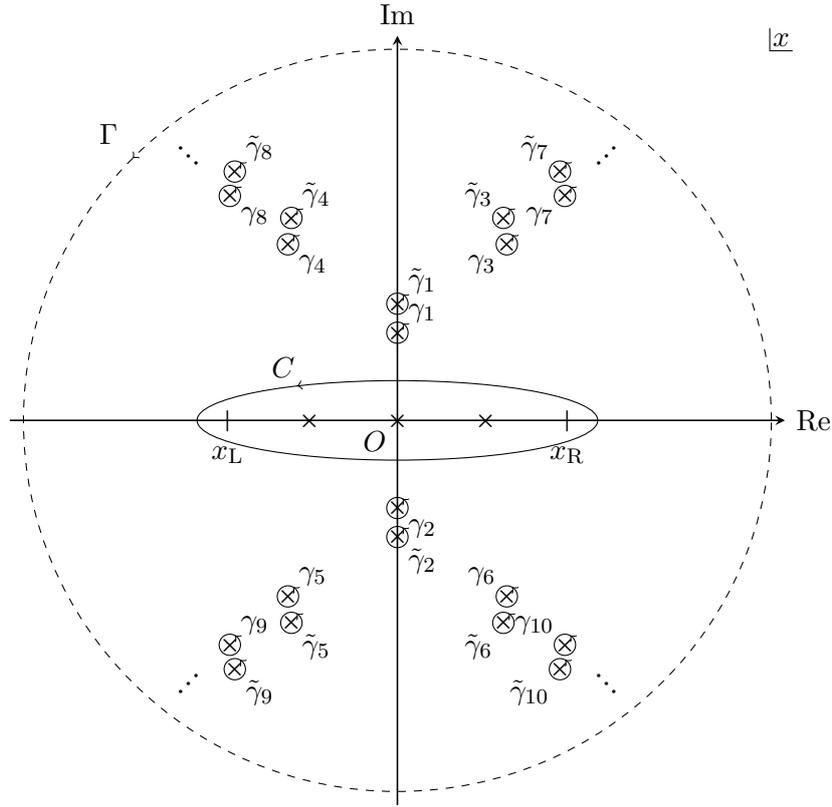
\begin{figure}[t]
\centering
	\begin{tikzpicture}
	\draw[semithick,->,>=stealth] (0,-5.1)--(0,5.1) node[above] {\small $\mathrm{Im}$};
	\draw[semithick,->,>=stealth] (-5.1,0)--(5.1,0) node[right] {\small $\mathrm{Re}$};
	\draw (4.9,5.2)--(4.9,4.9)--(5.2,4.9) node at (5.05,5.05) {\small $x$};
	\node [below left] at(0,0) {\small $O$};
	
	\node at (0,0) {\scriptsize $\bm{\times}$};
	\node at (1.159,0) {\scriptsize $\bm{\times}$};
	\node at (-1.159,0) {\scriptsize $\bm{\times}$};
	\node at (0,1.16234) {\scriptsize $\bm{\times}$};
	\node at (0,-1.16234) {\scriptsize $\bm{\times}$};
	\node at (0,1.54664) {\scriptsize $\bm{\times}$};
	\node at (0,-1.54664) {\scriptsize $\bm{\times}$};
	\node at (1.43986,2.33398) {\scriptsize $\bm{\times}$};
	\node at (-1.43986,2.33398) {\scriptsize $\bm{\times}$};
	\node at (1.43986,-2.33398) {\scriptsize $\bm{\times}$};
	\node at (-1.43986,-2.33398) {\scriptsize $\bm{\times}$};
	\node at (1.3946,2.68424) {\scriptsize $\bm{\times}$};
	\node at (-1.3946,2.68424) {\scriptsize $\bm{\times}$};
	\node at (1.3946,-2.68424) {\scriptsize $\bm{\times}$};
	\node at (-1.3946,-2.68424) {\scriptsize $\bm{\times}$};
	\node at (2.20467,2.97556) {\scriptsize $\bm{\times}$};
	\node at (-2.20467,2.97556) {\scriptsize $\bm{\times}$};
	\node at (2.20467,-2.97556) {\scriptsize $\bm{\times}$};
	\node at (-2.20467,-2.97556) {\scriptsize $\bm{\times}$};
	\node at (2.13929,3.29892) {\scriptsize $\bm{\times}$};
	\node at (-2.13929,3.29892) {\scriptsize $\bm{\times}$};
	\node at (2.13929,-3.29892) {\scriptsize $\bm{\times}$};
	\node at (-2.13929,-3.29892) {\scriptsize $\bm{\times}$};
	
	\node at (2.23607,0) {\scriptsize $\bm{|}$} node[below] at (2.23607,-0.2) {\small $x_{\mathrm{R}}$};
	\node at (-2.23607,0) {\scriptsize $\bm{|}$} node[below] at (-2.23607,-0.2) {\small $x_{\mathrm{L}}$};
	
	\draw[dashed] (0,0) circle (140pt);
	\draw[->] (-98.99pt,98.99pt)--(-99pt,98.98pt);
	\node at (-3.8,3.8) {\small $\Gamma$};
	
	\draw (0,0) circle (75pt and 15pt);
	\draw[->] (-37.5pt,12.99pt)--(-37.515pt,12.987pt);
	\node at (-1.5,0.7) {\small $C$};
	
	\draw (0,1.16234) circle (4pt) node[above right] {\small $\gamma_1$};
	\draw[->] (2pt,3.46pt+1.16234cm)--(1.99pt,3.47pt+1.16234cm);
	\draw (0,-1.16234) circle (4pt) node[below right] {\small $\gamma_2$};
	\draw[->] (2pt,3.46pt-1.16234cm)--(1.99pt,3.47pt-1.16234cm);
	\draw (0,1.54664) circle (4pt) node[above right] {\small $\tilde{\gamma}_1$};
	\draw[->] (2pt,3.46pt+1.54664cm)--(1.99pt,3.47pt+1.54664cm);
	\draw (0,-1.54664) circle (4pt) node[below right] {\small $\tilde{\gamma}_2$};
	\draw[->] (2pt,3.46pt-1.54664cm)--(1.99pt,3.47pt-1.54664cm);
	\draw (1.43986,2.33398) circle (4pt) node[below left] {\small $\gamma_3$};
	\draw[->] (2pt+1.43986cm,3.46pt+2.33398cm)--(1.99pt+1.43986cm,3.47pt+2.33398cm);
	\draw (1.3946,2.68424) circle (4pt) node[above left] {\small $\tilde{\gamma}_3$};
	\draw[->] (2pt+1.3946cm,3.46pt+2.68424cm)--(1.99pt+1.3946cm,3.47pt+2.68424cm);
	\draw (-1.43986,2.33398) circle (4pt) node[below right] {\small $\gamma_4$};
	\draw[->] (2pt-1.43986cm,3.46pt+2.33398cm)--(1.99pt-1.43986cm,3.47pt+2.33398cm);
	\draw (-1.3946,2.68424) circle (4pt) node[above right] {\small $\tilde{\gamma}_4$};
	\draw[->] (2pt-1.3946cm,3.46pt+2.68424cm)--(1.99pt-1.3946cm,3.47pt+2.68424cm);
	\draw (-1.43986,-2.33398) circle (4pt) node[above right] {\small $\gamma_5$};
	\draw[->] (2pt-1.43986cm,3.46pt-2.33398cm)--(1.99pt-1.43986cm,3.47pt-2.33398cm);
	\draw (-1.3946,-2.68424) circle (4pt) node[below right] {\small $\tilde{\gamma}_5$};
	\draw[->] (2pt-1.3946cm,3.46pt-2.68424cm)--(1.99pt-1.3946cm,3.47pt-2.68424cm);
	\draw (1.43986,-2.33398) circle (4pt) node[above left] {\small $\gamma_6$};
	\draw[->] (2pt+1.43986cm,3.46pt-2.33398cm)--(1.99pt+1.43986cm,3.47pt-2.33398cm);
	\draw (1.3946,-2.68424) circle (4pt) node[below left] {\small $\tilde{\gamma}_6$};
	\draw[->] (2pt+1.3946cm,3.46pt-2.68424cm)--(1.99pt+1.3946cm,3.47pt-2.68424cm);
	\draw (2.20467,2.97556) circle (4pt) node[below left] {\small $\gamma_7$};
	\draw[->] (2pt+2.20467cm,3.46pt+2.97556cm)--(1.99pt+2.20467cm,3.47pt+2.97556cm);
	\draw (2.13929,3.29892) circle (4pt) node[above left] {\small $\tilde{\gamma}_7$};
	\draw[->] (2pt+2.13929cm,3.46pt+3.29892cm)--(1.99pt+2.13929cm,3.47pt+3.29892cm);
	\draw (-2.20467,2.97556) circle (4pt) node[below right] {\small $\gamma_8$};
	\draw[->] (2pt-2.20467cm,3.46pt+2.97556cm)--(1.99pt-2.20467cm,3.47pt+2.97556cm);
	\draw (-2.13929,3.29892) circle (4pt) node[above right] {\small $\tilde{\gamma}_8$};
	\draw[->] (2pt-2.13929cm,3.46pt+3.29892cm)--(1.99pt-2.13929cm,3.47pt+3.29892cm);
	\draw (-2.20467,-2.97556) circle (4pt) node[above right] {\small $\gamma_9$};
	\draw[->] (2pt-2.20467cm,3.46pt-2.97556cm)--(1.99pt-2.20467cm,3.47pt-2.97556cm);
	\draw (-2.13929,-3.29892) circle (4pt) node[below right] {\small $\tilde{\gamma}_9$};
	\draw[->] (2pt-2.13929cm,3.46pt-3.29892cm)--(1.99pt-2.13929cm,3.47pt-3.29892cm);
	\draw (2.20467,-2.97556) circle (4pt) node[above left] {\small $\gamma_{10}$};
	\draw[->] (2pt+2.20467cm,3.46pt-2.97556cm)--(1.99pt+2.20467cm,3.47pt-2.97556cm);
	\draw (2.13929,-3.29892) circle (4pt) node[below left] {\small $\tilde{\gamma}_{10}$};
	\draw[->] (2pt+2.13929cm,3.46pt-3.29892cm)--(1.99pt+2.13929cm,3.47pt-3.29892cm);
	
	\node at (2.65,3.4) {$\cdot$};
	\node at (2.75,3.5) {$\cdot$};
	\node at (2.85,3.6) {$\cdot$};
	\node at (-2.65,3.4) {$\cdot$};
	\node at (-2.75,3.5) {$\cdot$};
	\node at (-2.85,3.6) {$\cdot$};
	\node at (2.65,-3.4) {$\cdot$};
	\node at (2.75,-3.5) {$\cdot$};
	\node at (2.85,-3.6) {$\cdot$};
	\node at (-2.65,-3.4) {$\cdot$};
	\node at (-2.75,-3.5) {$\cdot$};
	\node at (-2.85,-3.6) {$\cdot$};
	\end{tikzpicture}
\caption[The contours of the integrations in Eq. \eqref{eq:QHJ_ContourInt_CES}.]
	{The contours of the integrations in Eq. \eqref{eq:QHJ_ContourInt_CES}.
	The dots $\cdots$ show that the infinite number of pairwise poles lie in the plane.
	The dashed contour $\Gamma$ is a virtual contour that would enclose all the poles except for the one at infinity.}
\label{fig:3-CES-QMF-SingStruct-Contours}
\end{figure}

In Fig. \ref{fig:3-CES-QMF-SingSturct}, we display our numerical results with several $b$'s, where we set $\beta=0$ and $n=1$.
They reveal notable features of the quantum momentum function of the conditionally exactly solvable system.
Except for $b=0$ (1-dim. harmonic oscillator) and $b=4N$ (Krein--Adler), the quantum momentum function has an infinite number of poles in the complex plane.
At $b=0$, there is just one pole at the origin $x=0$ (Fig. \ref{fig:3-CES-QMF-SingSturct_b}).
For $b\neq 0$, an infinite number of poles appear in the complex plane (Fig. \ref{fig:3-CES-QMF-SingSturct_c}) and also 
$4N+1$ poles on the imaginary axis for $4(N-1) < b \leqslant 4N$.
The poles on the imaginary axis approach the origin $x=0$ as $b$ grows, while the other poles remain almost the same locations (Fig. \ref{fig:3-CES-QMF-SingSturct_d}).
When $b$ reaches $4N$, all the poles except the ones on the imaginary axis disappear (Fig. \ref{fig:3-CES-QMF-SingSturct_e}).
Again, as $b$ grows further, infinite poles appear in the complex plane (Fig. \ref{fig:3-CES-QMF-SingSturct_f}). 
A notable feature is that these poles except for the origin $x=0$ (and the one at $x\to\infty$) are pairwise 
with the residues $1$ and $-1$, respectively.
Therefore, for the contour integral of $J_{\mathrm{QHJ}}$, these contributions exactly vanish and only the 
residue at $x\to\infty$ contributes to the integral;
\begin{equation}
J_{\Gamma} = J_{\mathrm{QHJ}} + \sum_{i=1}^{\infty} J_{\gamma_i} + \sum_{j=1}^{\infty} J_{\tilde{\gamma}_j} 
= J_{\mathrm{QHJ}} ~,
\label{eq:QHJ_ContourInt_CES}
\end{equation}
which we have numerically verified.
For the definitions of the contours, see Fig. \ref{fig:3-CES-QMF-SingStruct-Contours}.
Note that this is just the quantization of the quantum action variable, not the quantization of the energy, and then there is no direct method for calculating the energy $\mathcal{E}$ from the quantization condition.

\begin{figure}[p]
\centering
	\begin{minipage}[t]{0.48\linewidth}
	\centering
	\includegraphics[width=.8\linewidth]{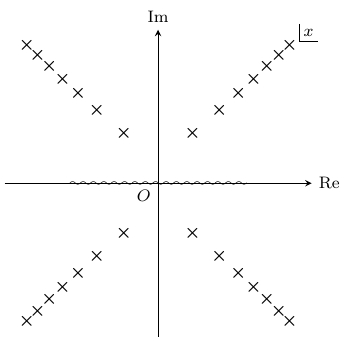}
	\subcaption{$b=-0.5$.}
	\label{}
	\end{minipage}
	\quad
	\begin{minipage}[t]{0.48\linewidth}
	\centering
	\includegraphics[width=.8\linewidth]{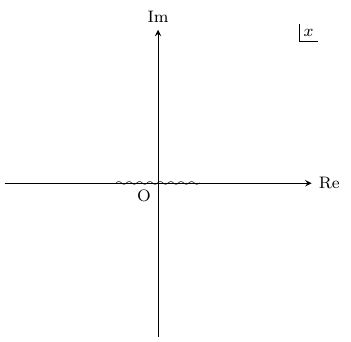}
	\subcaption{$b=0$.}
	\label{}
	\end{minipage}
	
	\medskip
	\begin{minipage}[t]{0.48\linewidth}
	\centering
	\includegraphics[width=.8\linewidth]{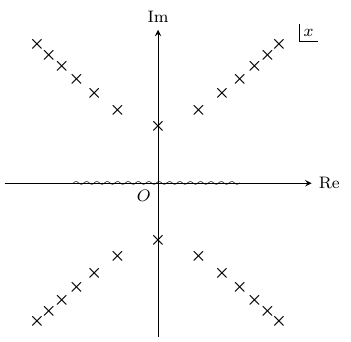}
	\subcaption{$b=0.1$.}
	\label{}
	\end{minipage}
	\quad
	\begin{minipage}[t]{0.48\linewidth}
	\centering
	\includegraphics[width=.8\linewidth]{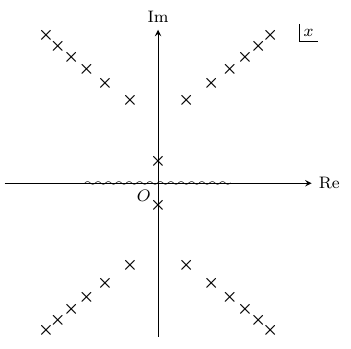}
	\subcaption{$b=3.5$.}
	\label{}
	\end{minipage}
	
	\medskip
	\begin{minipage}[t]{0.48\linewidth}
	\centering
	\includegraphics[width=.8\linewidth]{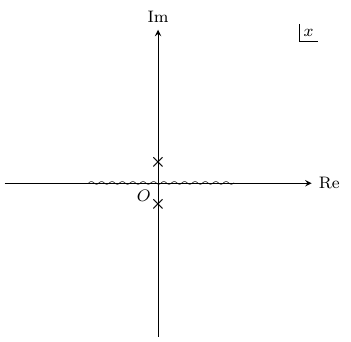}
	\subcaption{$b=4$.}
	\label{}
	\end{minipage}
	\quad
	\begin{minipage}[t]{0.48\linewidth}
	\centering
	\includegraphics[width=.8\linewidth]{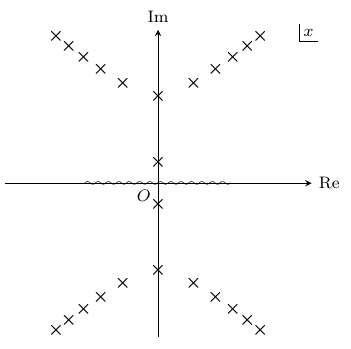}
	\subcaption{$b=4.1$.}
	\label{}
	\end{minipage}
\caption[The singularity structures of the SWKB integrands for the first excited states of the conditionally exactly solvable system with various $b$'s.]
	{The singularity structures of the SWKB integrands for the first excited states of the conditionally exactly solvable system with various $b$'s.
	The poles are plotted by x-marks and the branch cut on the real axis is shown by wavy lines.
	Other branch cuts are removed from these cartoons.
	The location of each pole is calculated numerically.
	Note that (b) and (e) are identical to that of the 1-dim. harmonic oscillator and the Krein--Adler Hermite system with $d=1$ respectively.}
\label{fig:3-CES-SWKB-SingSturct}
\end{figure}

On the other hand, for the SWKB integration, the situation is worse. 
An infinite number of the poles appeared in the complex plane are not pairwise and then, no cancellation of the residue of the poles occurs (See Fig. \ref{fig:3-CES-SWKB-SingSturct}). 
Also, there appear branch cuts other than the one on the real axis (which are sometimes referred to as ``other branch cuts'').
They have nonzero contributions on the contour integral $J_{\mathrm{SWKB}}$.
They spread all over the complex plane, but we do not plot in Fig. \ref{fig:3-CES-SWKB-SingSturct} to make it easier to see.
These are origins of the non-exactness of the SWKB conditions and also the essential difficulty 
for the explicit calculation of $J_{\mathrm{SWKB}}$ and let alone the quantization of energies in this formalism.
This gives us an intuition that we have to rely on a perturbative treatment to analyze the condition further.

\subsubsection{Numerical study}
The comparison of Fig. \ref{fig:3-CES-SWKB-SingSturct} with Fig. \ref{fig:3-CES-QMF-SingSturct} indicates that the SWKB condition \eqref{eq:3-SWKB_CES} breaks, which we demonstrate numerically.

\begin{figure}[t]
\centering
\begin{minipage}[t]{.48\linewidth}
\includegraphics[width=\linewidth]{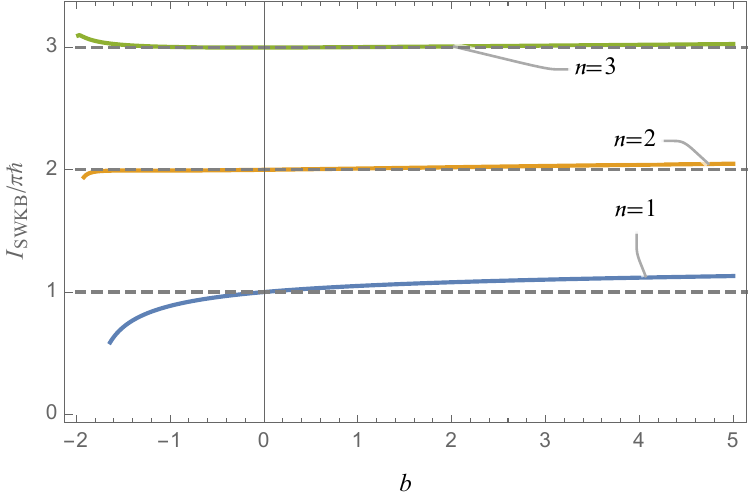}
\subcaption{$\beta = 0$.}
\label{fig:3-CES-SWKB_b}
\end{minipage}
\quad
\begin{minipage}[t]{.48\linewidth}
\includegraphics[width=\linewidth]{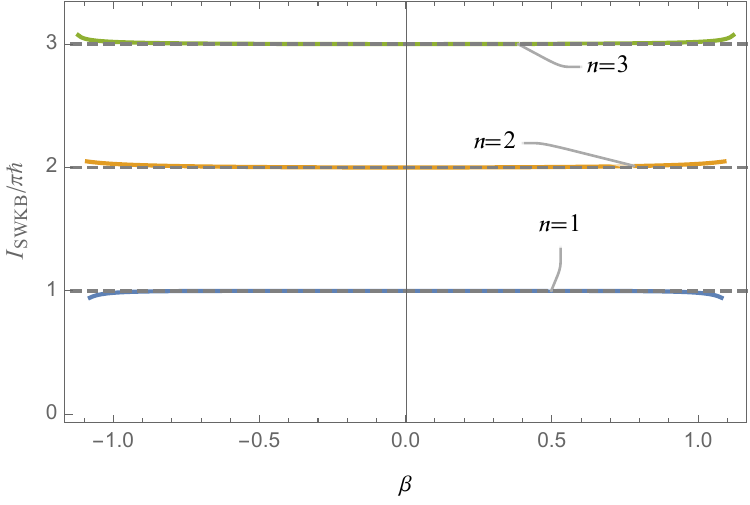}
\subcaption{$b = 0$.}
\label{fig:3-CES-SWKB_beta}
\end{minipage}
\caption[The values of the SWKB integral with the conditionally exactly solvable potential for $n=1,2,3$.]
	{The values of the SWKB integral with the conditionally exactly solvable potential for $n=1,2,3$.
	Range of each plot is limited so that the systems have no more than two turning points.
	The parametric conditions yield $b>-2$ and $|\beta| < 2/\sqrt{\pi}$ respectively.}
\label{fig:3-CES-SWKB}
\end{figure}

Fig. \ref{fig:3-CES-SWKB_b} shows the $b$-dependency of the SWKB integral with $\beta=0$, and Fig. \ref{fig:3-CES-SWKB_beta} is the $\beta$-dependency of the SWKB integral with $b=0$.
The SWKB integral grows with the parameter $b$ around $b=0$, while it exhibits plateau behavior (but the condition is never exactly satisfied) around $\beta=0$.
Different behaviors are seen as the parameters approach their boundaries:
\begin{equation}
\beta = \pm \frac{2\varGamma\left( \frac{b}{4} + 1 \right)}{\varGamma\left( \frac{b}{4} + \frac{1}{2} \right)} ~.
\end{equation}
These statements hold for general $b\neq 0$ and $\beta \neq 0$ cases.
The numerical calculations (Fig. \ref{fig:3-CES-SWKB}) support our conjecture \cite{Nasuda:2020aqf} that the level structure guarantees approximate satisfaction of the SWKB condition.

At the end, we would like to point out that in the cases of $\ast=\mathrm{L,J}$, one can obtain similar results.

\subsection{SWKB with Position-dependent Effective Mass}
\label{sec:3-2-6}
Our previous case studies show that the SWKB-exactness occurs if and only if the potential is one of the conventional shape-invariant ones, whose solvability is guaranteed by the classical orthogonal polynomials.
Since those potentials can be mapped into either H, L or J, there is a chance that other classes of exactly solvable problems whose potentials are also mapped to either H, L or J.
Here, we test the condition by employing classical-orthogonal-polynomially exactly solvable problems systems with position-dependent effective mass~\cite{PhysRev.152.683,PhysRevB.27.7547,Bagchi_2005,quesne-SIGMA2007,sym12111853,Jafarov-2021}.

\subsubsection{Naive application of SWKB}
First, we analytically show that a naive application of SWKB condition \eqref{eq:3-SWKB} fails to reproduce exact bound-state spectra for this class of exactly solvable problems.
Take a deformed harmonic oscillator:
\begin{equation}
W(x) = \omega x ~,~~~
\mathcal{E}_n = 2n\hbar\omega + \hbar^2\alpha n^2
\end{equation}
as an example.
The SWKB integral yields 
\begin{equation}
\int_{-a}^{a}\sqrt{2n\hbar\omega + \hbar^2\alpha n^2 - \omega^2x^2} \,dx
= n\pi\hbar\left( 1 + \frac{n\hbar\alpha}{2\omega} \right)
\neq n\pi\hbar ~,
\end{equation}
where $a\equiv\sqrt{2n\hbar\omega + \hbar^2\alpha n^2}/\omega$.
Clearly, the condition equation does not hold except for $\alpha=0$, which corresponds to the ordinary harmonic oscillator.

\subsubsection{Extension of SWKB formula}
Here, remember that we take the unit $2m=1$ throughout the thesis.
Let us now take back the $m$-dependency into account, for this class of exactly solvable potentials concerns `mass' in the first place.
The SWKB condition equation reads
\begin{equation}
\int_{a_{\rm L}}^{a_{\rm R}} \sqrt{2m \left[ \mathcal{E}_n - W(x)^2 \right]} \,dx = n\pi\hbar
~,~~~ n = 0,1,2,\ldots ~.
\end{equation}
Our systems have a mass depending on the coordinate, \textit{i.e.}, we use $m$ instead of $m(x)\equiv ( 2\eta(x)^2 )^{-1}$.
Therefore, a simple guessing tells the following extended version of the quantization condition equation:
\begin{equation}
\int_{a_{\rm L}}^{a_{\rm R}} \frac{\sqrt{\mathcal{E}_n - W(x)^2}}{\eta(x)} \,dx = n\pi\hbar
~,~~~ n = 0,1,2,\ldots ~,
\end{equation}
in which $a_{\rm L}$, $a_{\rm R}$ are the two roots of the equation $\mathcal{E}_n - W(x)^2=0$.

Here, we have extended the SWKB condition equation by replacing the mass $m$ by $( 2\eta(x)^2 )^{-1}$, which is in accordance with the construction of the potentials.

\begin{example}[Deformed harmonic oscillator]
The simplest example of the deformed shape-invariant systems would be the deformed harmonic oscillator, where $\eta(x) = 1 + \alpha x^2$.
The extended SWKB integral is
\begin{align}
\int_{-a}^{a}\frac{\sqrt{2n\hbar\omega + \hbar^2\alpha n^2 - \omega^2x^2}}{1+\alpha x^2}\,dx 
&= 2\omega\int_0^a \frac{\sqrt{a^2-x^2}}{1+\alpha x^2} \,dx \nonumber \\
&= 2\omega \cdot \frac{1}{\alpha}\frac{\hbar n\alpha}{2\omega}\pi \nonumber \\
&= n\pi\hbar ~.
\end{align}
\end{example}

The extended SWKB quantization condition also holds for another type of quantum mechanical system with position-dependent effective mass, which is solvable via the classical orthogonal polynomials, but the Hamiltonian is not deformed shape invariant~\cite{Bagchi_2005,quesne-SIGMA2007,sym12111853}.
Here, we provide an example.

\begin{example}[Semi-confined harmonic oscillator \cite{Jafarov-2021}]
The superpotential, the position-dependent effective mass and the energy eigenvalues of this system are given by
\begin{equation}
W(x) = \begin{cases}
	\omega x\sqrt{\dfrac{a}{x+a}} & x>-a \\
	-\infty & x\leqslant -a
\end{cases} ~,~~~ 
\eta(x) =  \begin{cases}
	\sqrt{\dfrac{x+a}{a}} & x>-a \\
	\infty & x\leqslant -a
\end{cases} ~,~~~
\mathcal{E}_n = 2n\hbar\omega ~.
\end{equation}
Then,
\begin{align}
\int_{a_-}^{a_+} \sqrt{\frac{a}{x+a}\left( 2n\hbar\omega - \frac{a\omega^2x^2}{x+a} \right)} \,dx
&= \omega a\int_{a'_-}^{a'_+} \sqrt{(y-a'_-)(a'_+-y)} \,\frac{dy}{y} \nonumber \\
&= \omega a\left[ -a\cdot\frac{\pi}{2} + \left( \frac{\hbar n}{\omega a}+a \right)\frac{\pi}{2} - a\cdot\frac{\pi}{2} + \left( \frac{\hbar n}{\omega a}+a \right)\frac{\pi}{2} \right] \nonumber \\[3pt]
&= n\pi\hbar ~,
\end{align}
with
\begin{equation}
x+a \equiv y ~,~~~
a'_{\pm} = \frac{\hbar n \pm \sqrt{\hbar^2n^2+2\hbar\omega a^2n}}{\omega a} - a ~.
\end{equation}
\end{example}

\section{(Non-)exactness of SWKB Quantization Condition: Discussions}
\label{sec:3-3}
\subsection{Overview and Structure of this Section}
In the previous section, we have carried out the five case studies.
In this section, we discuss the implication of the SWKB quantization condition based on them.

This section is organized as follows.
First, the subsequent three subsections are devoted to the exploration of the implication of SWKB quantization condition.
In Sect. \ref{sec:3-3-2}, we trace the history with our new examples given in Sects. \ref{sec:3-2-2}--\ref{sec:3-2-5}.
Next, in Sect. \ref{sec:3-3-3}, we discuss SWKB quantization condition in relation to the level distributions, which is indicated by our calculations in Sects. \ref{sec:3-2-4} and \ref{sec:3-2-5}.
Then in Sect. \ref{sec:3-3-4}, we also investigate the close relation between SWKB quantization condition and the classical orthogonal polynomials. 
See also Fig. \ref{fig:3-2_and_3-3}.
The last subsection is on an application of SWKB condition equation, other than the quantization of energy.
The arguments provided here connect our case studies and what are discussed in the rest of the thesis.

\subsection{Interpretation of the SWKB: Historical Arguments}
\label{sec:3-3-2}
It has long been explored what the SWKB implies.
First, we shall have a quick review on what have been considered together with our case studies in the previous section.

\subsubsection{SWKB-exactness and exact solvability}
At the time of 1985, when the SWKB quantization condition was originally proposed, researchers were interested in the relation between the exactness of the quantization condition and the exact solvability of potentials.
They thought that the exact solvability of potentials perhaps explained the exactness of SWKB.
Some even believed that the SWKB was exact for all exactly solvable potentials, which situation DeLaney and Nieto expressed by the word ``folklore''~\cite{delaney1990susy}.

In 1989, Khare and Varshni gave counter-examples of this folklore~\cite{khare1989shape}.
The authors discussed the SWKB conditions for Ginocchio potential~\cite{ginocchio1984class} and also for a potential that is isospectral to the harmonic oscillator, both of which are exactly solvable but are not shape invariant.
Their statement is that the shape invariance may be a necessary condition for the exactness of the SWKB condition.

Since any proof of this conjecture is absent, it is worth examining other exactly solvable, but not conventional shape-invariant, potentials. 
A year after Ref. \cite{khare1989shape}, D. DeLaney and M. M. Nieto did for the Abraham--Moses systems~\cite{PhysRevA.22.1333}, which is also SWKB-nonexact.
They also concluded that the SWKB is neither exact nor never worse than WKB for this class of exactly solvable potentials.
Years later, yet another demonstration has been carried out by Bougie \textit{et al.} for a new type of additional shape-invariant potential~\cite{bougie2018supersymmetric}.
The examples in the previous three subsections (Sects. \ref{sec:3-2-3}--\ref{sec:3-2-5}) also disprove it.
Now we know that many exactly solvable potentials are not SWKB-exact, and the mere fact that a potential is exactly solvable never explains the exactness of the SWKB quantization condition.

\subsubsection{SWKB-exactness and shape invariance}
As was mentioned earlier, it has long been considered that the SWKB quantization condition reproduces the exact bound-state spectra for all additive shape-invariant potentials since Dutt \textit{et al.} showed in 1986 that the SWKB is an exact quantization condition for all the known shape-invariant potentials at that time.
In 2018, Bougie \textit{et al.} argued that the newly constructed type of additive shape-invariant potentials may not satisfy the SWKB condition equation exactly~\cite{bougie2018supersymmetric}.
Although it contains dubious discussions, we have verified them and extended them to more general cases in Sect. \ref{sec:3-2-3}.
Now one can safely say that the SWKB condition is not exact for all additive shape-invariant potentials, and one could give infinitely many examples for this statement.

\subsection{SWKB and Level Structures}
\label{sec:3-3-3}
We would like to discuss from our numerical results what guarantees the exact/approximate satisfaction of the SWKB condition.
From our numerical observations on the Krein--Adler systems in which the maximal errors $|\mathrm{Err}|$ are seen around the deleted levels $\mathcal{D}$ and get smaller as $n$ steps away from $\mathcal{D}$, it may be the whole distribution of the energy eigenvalues that is responsible for the approximate satisfaction of the SWKB condition.
That is, the modifications of the conventional shape-invariant systems change the level structures of the systems, and so do the values of the SWKB integral.

\begin{remark}[]
In the cases of the multi-indexed systems, the maximal errors are seen at $n=1$.
This can be explained as follows.
The multi-indexed systems are obtained through the Darboux transformations with the  so-called virtual-state wavefunctions as seed solutions.
This can be regarded as the deletion of `eigenstates with negative eigenvalues'.
A similar thing to the case with the Krein--Adler systems also happens here, \textit{i.e.}, the maximal error $|\mathrm{Err}|$ is seen around the deleted levels.
Since the condition equation for $n=0$ always holds exactly by construction, thus for $n\geqslant 1$, the maximum error should appear at $n=1$, the closest to the deleted levels.
\end{remark}

In Sect. \ref{sec:3-2-5}, we have demonstrated numerically how the exactness of the condition equation breaks by employing the conditionally exactly solvable potential.
Also, we have succeeded that for the isospectral deformation, the SWKB integral basically remains approximately the same values as the change of continuous deformation parameter.
Those results support our statement that the approximate satisfaction of the SWKB condition is guaranteed by the whole distribution of the energy spectrum of a system.
Or the approximate satisfaction indicates that the level structure is quite similar to that of the corresponding conventional shape-invariant potential.

\subsubsection{Series expansion of the discrepancy}
Our demonstration with the conditionally exactly solvable systems allows us a perturbative treatment of the discrepancy $\Delta$ defined in Eq. \eqref{eq:3-Delta}.
By series expanding the discrepancy, we further analyze the behavior shown in Fig. \ref{fig:3-CES-SWKB}, \textit{i.e.}, how the condition equation breaks as the parameters $b,\beta$ grow.

The basic idea is as follows. 
We consider small perturbations from the exact case: $b=\beta=0$, where the condition becomes exact, since the systems are equivalent to the original conventional shape-invariant ones.
Then we employ Taylor expansion for the SWKB integrand around the point where the SWKB condition is exact.

However, we are not sure with which parameter the integral is supposed to be expanded (Remember we have factored out $\hbar$, and the integral has no $\hbar$-dependency).
It is notable that, for the exact cases the main part of the SWKB integral is of the form $\sqrt{(x-a_{\mathrm{L}})(a_{\mathrm{R}}-x)}$.
Here, we employ the following trick:
\begin{align}
I_{\rm SWKB}
&= \int_{a_{\mathrm{L}}}^{a_{\mathrm{R}}} \sqrt{(x-a_{\mathrm{L}})(a_{\mathrm{R}}-x)}\sqrt{1+\frac{2n + b - W^{(\mathrm{C,H})}(x)^2 - (x-a_{\mathrm{L}})(a_{\mathrm{R}}-x)}{(x-a_{\mathrm{L}})(a_{\mathrm{R}}-x)}} \,dx \nonumber \\
&\cong \frac{(a_{\mathrm{R}} - a_{\mathrm{L}})^2}{8}\pi + \sum_{k=1}^{\infty}\frac{(-1)^k(2k)!}{(1-2k)(k!)^24^k} \int_{a_{\mathrm{L}}}^{a_{\mathrm{R}}} \frac{ \left[ 2n + b - W^{(\mathrm{C,H})}(x)^2 - (x-a_{\mathrm{L}})(a_{\mathrm{R}}-x) \right]^k}{\left[ (x-a_{\mathrm{L}})(a_{\mathrm{R}}-x) \right]^{k-\frac{1}{2}}} \,dx ~,
\label{eq:3-Exp_CES}
\end{align}
where $W^{(\mathrm{C,H})}(x)$ denotes
\[
W^{(\mathrm{C,H})}(x) \coloneqq -\frac{d}{dx}\ln\phi_0^{\rm (C,H)}(x) ~.
\]
After the series expansion, we use the fact to obtain Eq. \eqref{eq:3-Exp_CES} that the integrand converges uniformly where one can swap the orders of the integration and the limit to infinity.
Note that 
\begin{equation}
\frac{2n + b - W^{(\mathrm{C,H})}(x)^2 - (x-a_{\mathrm{L}})(a_{\mathrm{R}}-x)}{(x-a_{\mathrm{L}})(a_{\mathrm{R}}-x)} 
\end{equation}
equals zero if and only if $b=\beta=0$.
The radius of convergence for the expansion is thus
\begin{equation}
\left| \frac{2n + b - W^{(\mathrm{C,H})}(x)^2 - (x-a_{\mathrm{L}})(a_{\mathrm{R}}-x)}{(x-a_{\mathrm{L}})(a_{\mathrm{R}}-x)} \right| = 1 ~.
\label{eq:3-ConvR_CES}
\end{equation}

\begin{figure}[t]
\centering
\includegraphics[scale=1]{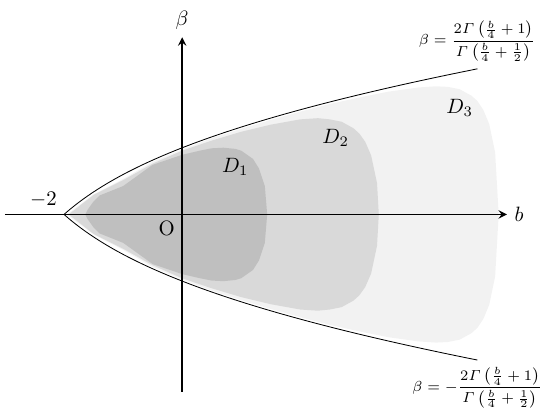}
\caption[The domains $D_n$ for $n=1,2,3$, where the series \eqref{eq:3-Exp_CES} converge.]
	{The domains $D_n$ for $n=1,2,3$, where the series \eqref{eq:3-Exp_CES} converge.
	Here, one can see that $D_1\subset D_2\subset D_3$.}
\label{fig:3-CES-SWKB-ExpDomain}
\end{figure}

In terms of Eq. \eqref{eq:3-Exp_CES} a choice of parameters $(b,\beta)$ within this radius of convergence corresponds to a conditionally exactly solvable system which  is connected to the original conventional shape-invariant potential.
When a choice of parameters $(b,\beta)$ is outside the radius, such a system simply does not relate to the original conventional shape-invariant potential in terms of those series.
We plot the domains where the series converges for $n=1,2,3$ on $(b,\beta)$-plane in Fig. \ref{fig:3-CES-SWKB-ExpDomain}.
One can obtain the domains by solving Eq. \eqref{eq:3-ConvR_CES} numerically.
Let us call each domain $D_n$ respectively.
We have checked numerically that as $n$ grows, the radius of convergence is enlarged; $D_1 \subset D_2 \subset D_3 \subset\cdots$.
A quantitative argument of the inclusion relation of domains $D_n$ supports this result.
Thus we conclude that there always exist sets of model parameters where the expansion \eqref{eq:3-Exp_CES} is possible for any $n$, and it is enough to consider the domain $D_1$ so that the expansion formula \eqref{eq:3-Exp_CES} holds for any $n$.

\begin{figure}[p]
\centering
\begin{minipage}[t]{.48\linewidth}
\includegraphics[width=\linewidth]{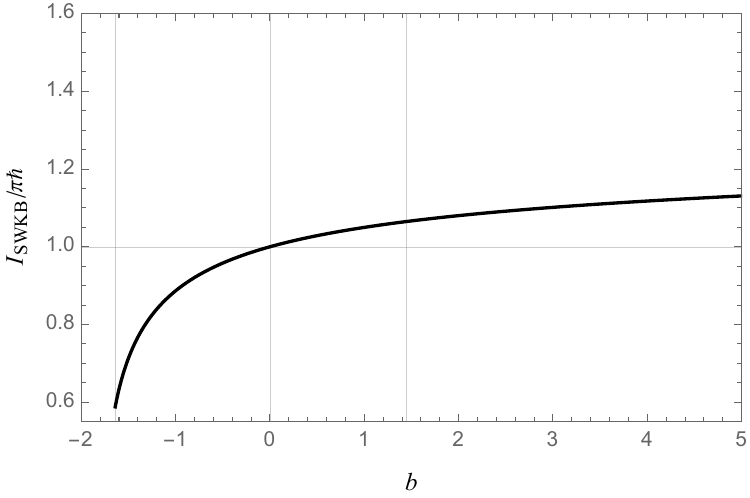}
\subcaption{$\beta = 0$.}
\end{minipage}
\quad
\begin{minipage}[t]{.48\linewidth}
\includegraphics[width=\linewidth]{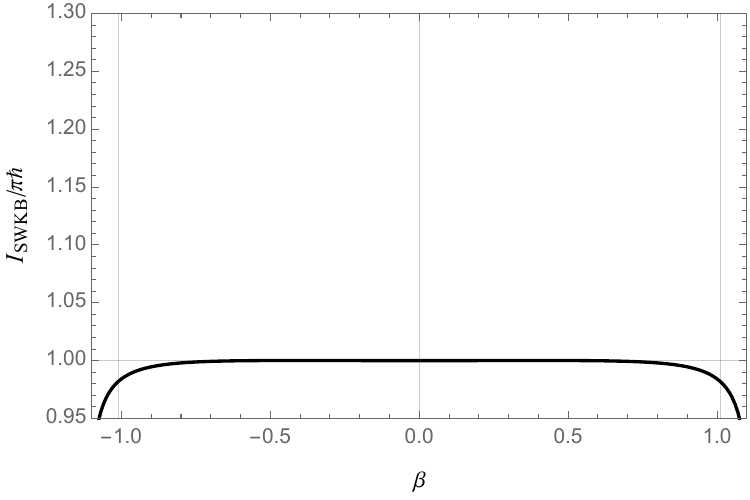}
\subcaption{$b = 0$.}
\end{minipage}

\medskip
\begin{minipage}[t]{.48\linewidth}
\includegraphics[width=\linewidth]{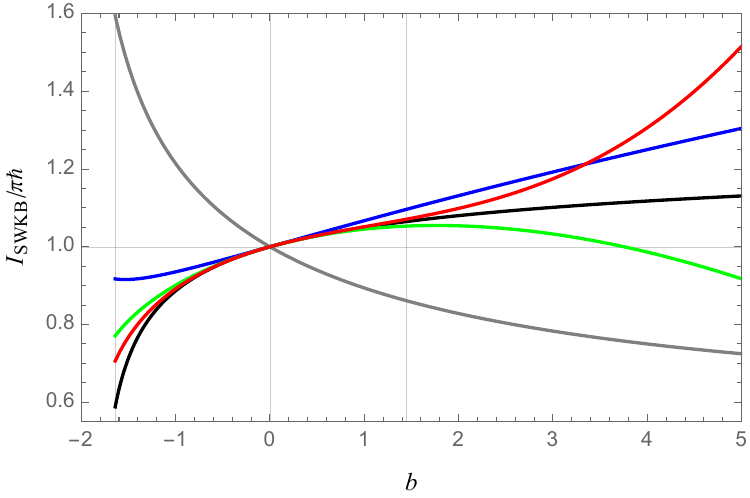}
\subcaption{$\beta = 0$.}
\end{minipage}
\quad
\begin{minipage}[t]{.48\linewidth}
\includegraphics[width=\linewidth]{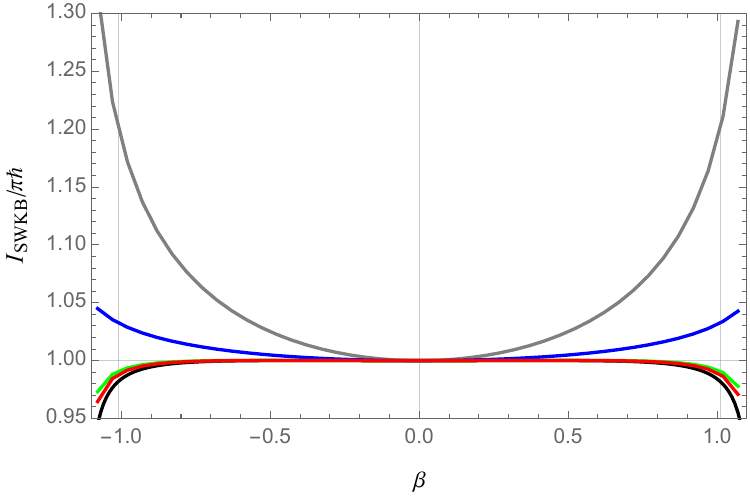}
\subcaption{$b = 0$.}
\end{minipage}
\caption[Plots of the SWKB integral \eqref{eq:3-SWKB_CES} for $n=1$ with (a) $\beta=0$, (b) $b=0$, and its the power series approximations in different orders with (c) $\beta=0$, (d) $b=0$.]
	{Plots of the SWKB integral \eqref{eq:3-SWKB_CES} for $n=1$ with (a) $\beta=0$, (b) $b=0$, and its the power series approximations in different orders with (c) $\beta=0$, (d) $b=0$.
	We plot the $0$th order of the expansion in gray, and the expansion up to first order in blue, up to second order in green and up to third order in red, while the black curve shows the numerical results.
	The range of each plot is determined so that the systems will not have more than two turning points.
	The parametric conditions yield $b>-2$ and $|\beta| < 2/\sqrt{\pi}$ respectively.
	The light gray lines at (a,c) $b\approx -1.64, 1.44$, (b,d) $|\beta|\approx 1.01$ show the radius of convergence \eqref{eq:3-ConvR_CES}.}
\label{fig:3-CES-SWKB-Exp}
\end{figure}

In Fig. \ref{fig:3-CES-SWKB-Exp}, we plot the SWKB integral \eqref{eq:3-SWKB_CES} with different orders of the power series approximation.
For the glancing behavior of the SWKB integral \eqref{eq:3-SWKB_CES}, the first few orders of the expansion formula are sufficient.

\clearpage
\subsection{SWKB and Solvability via Classical Orthogonal Polynomials}
\label{sec:3-3-4}
In the previous subsection, we expand the SWKB integral of the conditional exactly solvable systems around that of the harmonic oscillator.
Also, our discussions above are always the comparisons of the SWKB integral of some potential with that of a conventional shape-invariant potential.
Now, we return to the SWKB condition for the conventional shape-invariant potentials itself to get a mathematical insight.

\subsubsection{SWKB for conventional shape-invariant potentials revisited}
All the conventional shape-invariant potentials are exactly solvable by virtue of the classical orthogonal polynomials.
Also, we have already mentioned that all the Schr\"{o}dinger equations of the conventional shape-invariant potentials are mapped to that of either H, L, J by changing variable (See Sect. \ref{sec:2-3-2}).
Now, one might wonder if this is the case with the SWKB condition equations.
It actually is, and we have confirmed it for all the cases listed in Sect. \ref{sec:2-3-2}.
Formally,
\begin{equation}
\int_{a_{\rm L}}^{a_{\rm R}} \sqrt{\mathcal{E}_n -W(x)^2} \,dx
= \int_{\alpha_{\rm L}}^{\alpha_{\rm R}} \sqrt{\varepsilon_n -w(z)^2} \,dz
= n\pi\hbar ~.
\end{equation}
We provide an illustrative example below (See also Tab. \ref{tab:3-Interrel_cSI-SWKB}).

\begin{example}[Coulomb potential]
The SWKB integral for the Coulomb potential:
\begin{equation}
\int_{a_{\rm L}}^{a_{\rm R}} \sqrt{\frac{e^4}{4\hbar^2\bar{g}^2} - \frac{e^4}{4\hbar^2(\bar{g}+n)^2} - \left( \frac{e^2}{2\hbar \bar{g}} - \frac{\hbar \bar{g}}{x} \right)^2} \,dx 
= \int_{a_{\rm L}}^{a_{\rm R}} \sqrt{-\frac{e^4}{4\hbar^2(\bar{g}+n)^2} + \frac{e^2}{x} - \frac{\hbar^2\bar{g}^2}{x^2}} \,dx  ~,
\end{equation}
maps to 
\begin{equation}
\int_{a'_{\rm L}}^{a'_{\rm R}} \sqrt{-\frac{e^4}{\hbar^2(\bar{g}+n)^2}z^2 + 4e^2 - \frac{4\hbar^2\bar{g}^2}{z^2}} \,dz  ~,
\end{equation}
under $x\mapsto z=\sqrt{x}$.
With the identification 
\[
\frac{e^2}{\hbar (\bar{g}+n)} \equiv \omega ~,~~~ 2\bar{g} \equiv g ~,
\]
the integral becomes
\begin{equation}
\int_{a''_{\rm L}}^{a''_{\rm R}} \sqrt{-\omega^2z^2 + 2\hbar\omega g + 4n\hbar\omega - \frac{\hbar^2g^2}{z^2}} \,dz 
= \int_{a''_{\rm L}}^{a''_{\rm R}} \sqrt{4n\hbar\omega - \left( \omega z - \frac{\hbar g}{z} \right)^2} \,dz  ~,
\end{equation}
which is exactly the SWKB integral for the radial oscillator (L), and equivalent to $n\pi\hbar$.
\end{example}

\begin{table}[t]
\centering
\caption[Interrelation among the SWKB integrals of conventional shape-invariant potentials.]
	{Interrelation among the SWKB integrals of conventional shape-invariant potentials.
	Note that the relations among parameters differ from those in Tab. \ref{tab:2-Interrel_cSI-SE}.}
\label{tab:3-Interrel_cSI-SWKB}
\scalebox{0.65}{
	\begin{tabular}{ccccll}
	\toprule
	Class & Superpotential & Energy &~~& \begin{tabular}{c}
			Change of variables~~ \\
			$x \mapsto z$~~
		\end{tabular} & Relation among parameters \\
	\midrule
	H & \begin{tabular}{c}
			1-dim. HO \\
			$\omega z$ 
		\end{tabular} & $2n\hbar\omega$ && --- & --- \\
	\midrule
	L & \begin{tabular}{c}
			Radial osc. \\
			$\omega z - \dfrac{\hbar g}{z}$ 
		\end{tabular} & $4n\hbar\omega$ && --- & --- \\
	  & \begin{tabular}{c}
	  		Coulomb pot. \\
			$\dfrac{e^2}{2\hbar\bar{g}} - \dfrac{\hbar\bar{g}}{x}$ 
		\end{tabular} & $\dfrac{e^4}{4\hbar^2\bar{g}^2} - \dfrac{e^4}{4\hbar^2(\bar{g}+n)^2}$ && $z=\sqrt{x}$ & $\omega = \dfrac{e^2}{\hbar(g+n)}$, $g = 2\bar{g}$ \\
	  & \begin{tabular}{c}
	  		Morse pot. \\
			$\hbar(\mu\mathrm{e}^x - h)$ 
		\end{tabular} & $\hbar^2(2nh-n^2)$ && $z=\mathrm{e}^{x/2}$ & $\omega = 2\hbar\mu$, $g = 2(h-n)$ \\
	\midrule
	J & \begin{tabular}{c}
			P\"{o}schl--Teller pot. \\
			$\hbar(h\tan z - g\cot z)$ 
		 \end{tabular} & $4\hbar^2n(n+g+h)$ && --- & --- \\
	   & \begin{tabular}{c}
			Hyperbolic symmetric top II \\
			$\hbar\left( \dfrac{\mu}{\cosh x} + \bar{h}\tanh x \right)$ 
		 \end{tabular} & $\hbar^2(2n\bar{h}-n^2)$ && $z = \dfrac{1}{2}\arccos (\mathrm{i}\sinh x)$ & $g = -\bar{h}-\mu\mathrm{i}$, $h = -\bar{h}+\mu\mathrm{i}$ \\
	   & \begin{tabular}{c}
			Rosen--Morse pot. \\
			$\hbar\left( \dfrac{\mu}{\bar{h}} + \bar{h}\tanh x \right)$ 
		 \end{tabular} & $\hbar^2\left( 2n\bar{h}-n^2 + \dfrac{\mu^2}{\bar{h}^2} - \dfrac{\mu^2}{(\bar{h}-n)^2} \right)$ && $z = \dfrac{1}{2}\arccos (\tanh x)$ & \begin{tabular}{l} 
		 	$g = \bar{h}-n+\dfrac{\mu}{\bar{h}-n}$, \\
			$h = \bar{h}-n-\dfrac{\mu}{\bar{h}-n}$
		 \end{tabular} \\
	   & \begin{tabular}{c}
			Eckart pot. \\
			$\hbar\left( \dfrac{\mu}{\bar{g}} - \bar{g}\coth x \right)$ 
		 \end{tabular} & $\hbar^2\left( 2n\bar{g}-n^2 + \dfrac{\mu^2}{\bar{g}^2} - \dfrac{\mu^2}{(\bar{g}+n)^2} \right)$ && $z = \dfrac{1}{2}\arccos (\coth x)$ & \begin{tabular}{l} 
		 	$g = -\bar{g}-n+\dfrac{\mu}{\bar{g}+n}$, \\
			$h = -\bar{g}-n-\dfrac{\mu}{\bar{g}+n}$
		 \end{tabular} \\
	   & \begin{tabular}{c}
			Hyperbolic P\"{o}schl--Teller pot. \\
			$\hbar(\bar{h}\tanh x - g\coth x)$ 
		 \end{tabular} & $4\hbar^2n(\bar{h}-g-n)$ && $z = \arcsin (-\mathrm{i}\sinh x)$ & $h = -\bar{h}$ \\
	\bottomrule
	\end{tabular}
}
\end{table}

\subsubsection{Extension of SWKB formula}
The extended version of the SWKB quantization condition:
\begin{equation}
\int_{a_{\rm L}}^{a_{\rm R}} \sqrt{2m(x) [\mathcal{E}_n - W(x)^2]} \,dx
= \int_{a_{\rm L}}^{a_{\rm R}} \frac{\sqrt{\mathcal{E}_n - W(x)^2}}{\eta(x)} \,dx 
= n\pi\hbar
~,~~~ n = 0,1,2,\ldots ~,
\end{equation}
can also be understood in the following manner, that is, in this extension $1/\eta(x)$ means a Jacobian of the change of variables in the following sense.
As is discussed in Ref. \cite{Bagchi_2005}, the Schr\"{o}dinger equation with position-dependent effective mass:
\begin{equation}
\left[ -\hbar^2\left( \sqrt{\eta(x)}\frac{d}{dx}\sqrt{\eta(x)} \right)^2 + V(x) \right]\phi(x) = \mathcal{E}\phi(x)
\end{equation}
is mapped to the following Schr\"{o}dinger-type equation under the change of variable $x\mapsto z=z(x)$:
\begin{equation}
z(x) \coloneqq \kappa \int^x \frac{dx}{\eta(x)} ~,~~~
\left( -\hbar^2\frac{d^2}{dz^2} + v(z) \right)\varphi(z) = \varepsilon\varphi(z) ~,
\end{equation}
with $\kappa$ being a constant, and
\begin{equation}
v(z) \equiv \frac{1}{\kappa^2}V(x) ~,~~~
\varepsilon \equiv \frac{1}{\kappa^2}\mathcal{E} ~,~~~
\varphi(z) \equiv \sqrt{\eta(x)}\,\phi(x) ~.
\label{eq:3-vV_rel_etc}
\end{equation}
If $v(z)$ is a conventional shape-invariant potential, the following SWKB condition holds:
\begin{equation}
\int_{\alpha_{\rm L}}^{\alpha_{\rm R}} \sqrt{\varepsilon_n - \left( \hbar\frac{d}{dz}\ln\varphi_0(z) \right)^2} \,dz = n\pi\hbar ~.
\end{equation}
Now, let us change the variable $z$ to $x$ in the SWKB integral, and using \eqref{eq:3-vV_rel_etc}, we get
\begin{align}
\int_{\alpha_{\rm L}}^{\alpha_{\rm R}} \sqrt{\varepsilon_n - \left( \hbar\frac{d}{dz}\ln\varphi_0(z) \right)^2} \,dz 
&= \int_{a_{\rm L}}^{a_{\rm R}} \sqrt{\frac{1}{\kappa^2}\mathcal{E}_n - \left( \hbar\frac{\eta(x)}{\kappa}\frac{d}{dx}\ln\sqrt{\eta(x)}\,\phi_0(x) \right)^2} \frac{\kappa}{\eta(x)} \,dx \nonumber \\
&= \int_{a_{\rm L}}^{a_{\rm R}} \sqrt{\mathcal{E}_n - W(x)^2} \,\frac{dx}{\eta(x)} ~.
\end{align}
This computation also proves that the extended SWKB condition always reproduces exact bound-state spectra for the deformed shape-invariant systems~\cite{Bagchi_2005,quesne-SIGMA2007,sym12111853}.

\subsubsection{The conjecture}
In this thesis, we aim to figure out what the SWKB quantization condition actually implies.
We have already realized that when you see the SWKB quantization condition as a tool for estimating the energy spectrum for a given superpotential, the integral roughly means how the spectrum is close to that for the conventional shape-invariant potentials.
However, this does not account fully for the SWKB-(non)exactness.
From our discussions in the present subsection, we arrive at the following conjecture: 
\begin{quotation}\noindent\it
There exists $n \in \mathbb{Z}_{\geqslant 0}$ such that the SWKB quantization condition holds exactly, if and only if the main part of the wavefunction $\phi_n$ with the same $n$, as well as the ground-state wavefunction $\phi_0$, are expressed in terms of the classical orthogonal polynomials with different orders.
\end{quotation}

At the end, we note that mathematical proof of it is still absent, but we have checked that the conjecture holds for all the solvable systems in the author's knowledge.

\subsection{Application of SWKB Quantization Condition}
As in the extension of exactly solvable potentials to \textit{quasi-}exactly solvable problems, one might wonder whether it is possible to realize that the SWKB condition is \textit{quasi} exact: the SWKB condition reproduces exact bound-state spectra only for several (could be infinitely many, but not the whole spectrum) eigenstates, say $n=n_1,n_2,\ldots$.
A simple guess tells that for $n=n_j$ the energy eigenvalues are exactly $2n_j$ as the SWKB estimates, and the corresponding eigenfunctions are expressed in terms of the classical orthogonal polynomials.

So far, we have no such systems, and our next move is to construct such potentials.
Before moving on, we summarize the basic idea of what we have done throughout the current chapter in a single sentence: the SWKB quantization condition estimates the energy spectrum for a given superpotential.
We call a ``direct problem'' this problem of obtaining spectrum for a given superpotential.
In the subsequent section, we are going to formulate a construction method of superpotentials from a given energy spectrum in the context of the SWKB formalism.

\clearpage
\section{Inverse Problem in SWKB Formalism}
\label{sec:3-4}
In the previous several sections, we have dealt with problems of determining energy spectra from given superpotentials,
\[
W(x) \text{: given}
\quad\longrightarrow\quad
\{ \mathcal{E}_n \} ~.
\]
Now we discuss the problem in literally the inverted direction:
\[
\{ \mathcal{E}_n \} \text{: given}
\quad\longrightarrow\quad
W(x) ~.
\]


\subsection{Formulation}
We assume that $W(x)^2$ has only one minimum at $x=x_0$.
Also, let $x_{\pm}$ ($x_- < x_+$) denote the values of $x$ where $W(x)^2=\mathcal{E}$ (See the figure below, Fig. \ref{fig:3-xpm}).

\begin{figure}[h]
\centering
	\begin{tikzpicture}
	\draw[semithick,->,>=stealth] (-3,0)--(3,0) node[right] {$x$};
	\draw[very thick,domain=-2.3:2.3] plot(\x,{0.4*pow(\x,2)}) node[above right] {$W(x)^2$};
	%
	\draw (2.75,1.5)--(-2.5,1.5) node[left] {$\mathcal{E}$};
	\draw[dashed] (-1.9365,1.5)--(-1.9365,0) node[below] {$x_-$};
	\draw[dashed] (1.9365,1.5)--(1.9365,0) node[below] {$x_+$};
	\draw (0,0.075)--(0,-0.075);
	\node[below] at (0,0) {$x_0$};
	\end{tikzpicture}
\caption{The definition of $x_{\pm}$.}
\label{fig:3-xpm}
\end{figure}
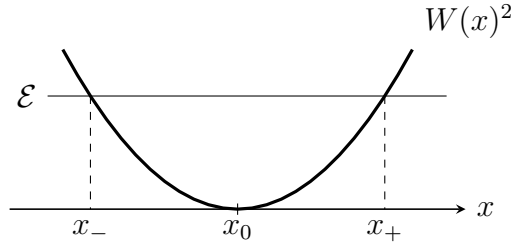

First, we differentiate the SWKB integral with respect to $\mathcal{E}$, 
\begin{equation}
\frac{dI}{d\mathcal{E}} = I'(\mathcal{E}) = \int_{x_-}^{x_+} \frac{dx}{2\sqrt{\mathcal{E}-W(x)^2}} ~.
\end{equation}
\begin{figure}[t]
\centering
	\begin{minipage}{0.48\linewidth}
	\centering
	\begin{tikzpicture}
	\draw[semithick,->,>=stealth] (0,0)--(5,0) node[right] {$W^2$};
	\draw[semithick,->,>=stealth] (0,0)--(0,5) node[above] {$\mathcal{E}$};
	\node[below left] at (0,0) {$O$};
	\draw[dotted] (0,0)--(4.75,4.75) node[above right] {\footnotesize $\mathcal{E}=W^2$};
	\draw[ultra thick] (3,3)--(3,0);
	\draw[dashed] (3,3)--(0,3) node[left] {$\mathcal{E}$};
	\draw[pattern=north west lines] (4,0)--(0,0)--(4,4)--cycle node[below] {$\alpha$};
	\end{tikzpicture}
	\subcaption{}
	\end{minipage}
	\begin{minipage}{0.48\linewidth}
	\centering
	\begin{tikzpicture}
	\draw[semithick,->,>=stealth] (0,0)--(5,0) node[right] {$W^2$};
	\draw[semithick,->,>=stealth] (0,0)--(0,5) node[above] {$\mathcal{E}$};
	\node[below left] at (0,0) {$O$};
	\draw[dotted] (0,0)--(4.75,4.75) node[above right] {\footnotesize $\mathcal{E}=W^2$};
	\draw[dashed] (0,2.5)--(2.5,2.5);
	\draw[ultra thick] (2.5,2.5)--(4,2.5);
	\draw[dashed] (2.5,2.5)--(2.5,0) node[below] {$W^2$};
	\draw[dashed] (4,4)--(0,4) node[left] {$\alpha$};
	\draw[pattern=north west lines] (4,0)--(0,0)--(4,4)--cycle node[below] {$\alpha$};
	\end{tikzpicture}
	\subcaption{}
	\end{minipage}
\caption{The domain of integration in \eqref{eq:3-a_quantity}.}
\label{fig:3-dom_int}
\end{figure}
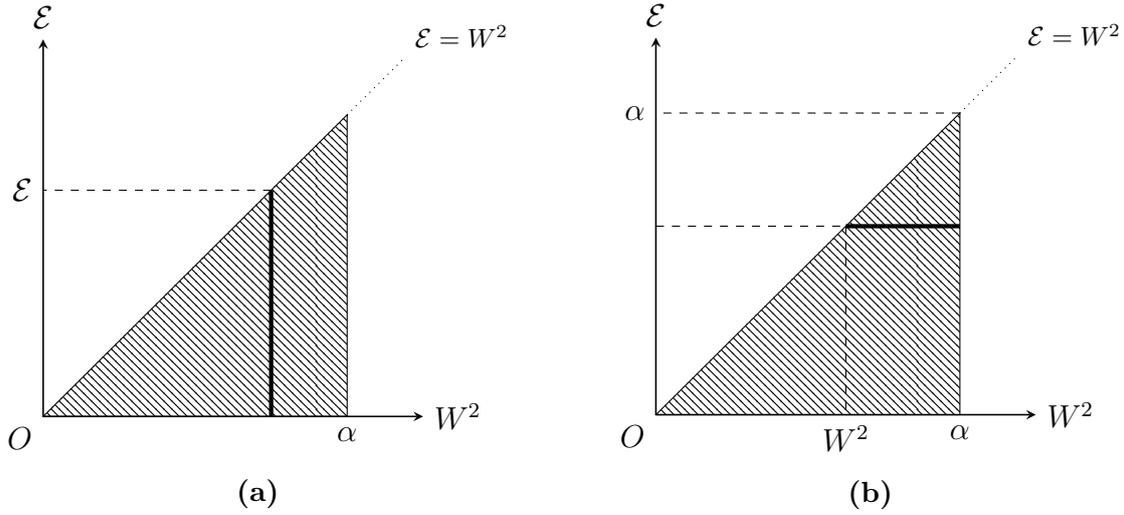
After some algebra, we obtain
\begin{equation}
\frac{dI}{d\mathcal{E}}
= \int_{x_-}^{x_+} \frac{dx}{2\sqrt{\mathcal{E}-W(x)^2}} 
= \frac{1}{2} \int_{0}^{\mathcal{E}} \left( \frac{dx_2}{d(W^2)} - \frac{dx_1}{d(W^2)} \right) \frac{d(W^2) }{\sqrt{\mathcal{E}-W^2}} ~.
\end{equation}
Now, we calculate a quantity $\displaystyle\int_{0}^{\alpha} \frac{I'(\mathcal{E})}{\sqrt{\alpha-\mathcal{E}}} \,d\mathcal{E}$,
\begin{align}
\int_{0}^{\alpha} \frac{I'(\mathcal{E})}{\sqrt{\alpha-\mathcal{E}}} \,d\mathcal{E}
&= \frac{1}{2} \int_{0}^{\alpha} \left[ \int_{0}^{\mathcal{E}} \left( \frac{dx_2}{d(W^2)} - \frac{dx_1}{d(W^2)} \right) \frac{d(W^2) }{\sqrt{(\alpha-\mathcal{E})(\mathcal{E}-W^2)}} \right] d\mathcal{E} \nonumber \\
&= \frac{1}{2} \int_{0}^{\alpha} \left[ \left( \frac{dx_2}{d(W^2)} - \frac{dx_1}{d(W^2)} \right) \int_{W^2}^{\alpha} \frac{d\mathcal{E}}{\sqrt{(\alpha-\mathcal{E})(\mathcal{E}-W^2)}} \right] d(W^2) \nonumber \\
&= \frac{\pi}{2} \int_{0}^{\alpha} \left( \frac{dx_2}{d(W^2)} - \frac{dx_1}{d(W^2)} \right) d(W^2) \nonumber \\
&= \frac{\pi}{2} \Bigl[ x_+(\alpha) - x_-(\alpha) \Bigr] ~.
\label{eq:3-a_quantity}
\end{align}
From the first line to the second, see Fig. \ref{fig:3-dom_int}, and to reduce the second line in Eq.\,\eqref{eq:3-a_quantity}, we have used
\begin{equation}
\int_{W^2}^{\alpha} \frac{d\mathcal{E}}{\sqrt{(\alpha-\mathcal{E})(\mathcal{E}-W^2)}} 
= \left. -\arcsin\left( \frac{-2\mathcal{E}+\alpha+W^2}{\alpha-W^2} \right) \right|_{W^2}^{\alpha}
= \pi ~.
\end{equation}
On the other hand, the SWKB condition \eqref{eq:3-SWKB} gives
\begin{equation}
\frac{dI}{d\mathcal{E}} = \pi\hbar\frac{dn}{d\mathcal{E}} = \pi\hbar \left( \frac{d\mathcal{E}}{dn} \right)^{-1} ~,
\end{equation}
which leads to
\begin{equation}
\int_{0}^{\alpha} \frac{I'(\mathcal{E})}{\sqrt{\alpha-\mathcal{E}}} \,d\mathcal{E}
= \pi\hbar\int_{0}^{\alpha} \frac{1}{\sqrt{\alpha-\mathcal{E}}} \left( \frac{d\mathcal{E}}{dn} \right)^{-1} d\mathcal{E} ~.
\label{eq:3-a_quantity-another}
\end{equation}
Using Eqs. \eqref{eq:3-a_quantity} and \eqref{eq:3-a_quantity-another} with the identification $\alpha=W^2$, we get
\begin{equation}
x_+(W^2) - x_-(W^2) 
= 2\hbar \int_{0}^{W^2} \frac{1}{\sqrt{W^2-\mathcal{E}}} \left( \frac{d\mathcal{E}}{dn} \right)^{-1} d\mathcal{E} ~. 
\label{eq:3-formula_inverseSWKB}
\end{equation}

Here, we require $x_-$ can be expressed in terms of $x_+$, say
\begin{equation}
x_- = f(x_+) ~,
\end{equation}
which is closely related to the symmetry of $W^2$.
The right-hand side of Eq. \eqref{eq:3-formula_inverseSWKB} is a function of $W^2$, which is denoted symbolically by $2\hbar\times F(W^2)$. 
Then, the square of superpotential $W^2$ (or $|W|$) is obtained formally as
\begin{align}
W^2 &= W(x_+)^2 = \frac{1}{2\hbar}\,F^{-1}\left( x_+ - f(x_+) \right) \\
\text{or}\qquad
|W| &= |W(x_+)| = \sqrt{\frac{1}{2\hbar}\,F^{-1}\left( x_+ - f(x_+) \right)} ~,\quad
\end{align}
where $F^{-1}$ denotes the inverse function of $F$.

\subsection{Construction of Conventional Shape-invariant Potentials}
Before we move on to our problem of constructing new potentials, here we briefly show that our formulation does actually replicate the conventional shape-invariant potentials with the three explicit examples.

\begin{example}[H]
\label{ex:3-inverseProb_H}
For a given energy spectrum:
\begin{equation}
\mathcal{E} = \mathcal{E}(n) = 2n\hbar\omega ~,
\end{equation}
the SWKB condition reproduces the superpotential of 1-dim. harmonic oscillator.
We choose $x_0=0$ for simplicity, and assume that $W^2$ is symmetric, \textit{i.e.},
\begin{equation}
x_+(W^2) = -x_-(W^2) \equiv x(W^2) ~.
\label{eq:3-nverseProb_H-assumption}
\end{equation}
Then Eq. \eqref{eq:3-formula_inverseSWKB} reads
\begin{equation}
2x(W^2) 
= \frac{1}{\omega} \int_{0}^{W^2} \frac{d\mathcal{E}}{\sqrt{W^2-\mathcal{E}}}
= \frac{2}{\omega} \,|W|
\end{equation}
therefore
\begin{equation}
W(x) = \omega x ~.
\end{equation}
\end{example}

\begin{example}[L]
In our next example, we choose
\begin{equation}
\mathcal{E} = \mathcal{E}(n) = 4n\hbar\omega ~,
\end{equation}
and also assume 
\begin{equation}
x_-(W^2) \,x_+(W^2) = {x_0}^2
\quad\text{\it i.e.}\quad
x_-(W^2) = \frac{{x_0}^2}{x_+(W^2)} ~,
\end{equation}
to obtain the radial oscillator.
Then Eq. \eqref{eq:3-formula_inverseSWKB} is
\begin{equation}
x_+(W^2) - \frac{{x_0}^2}{x_+(W^2)} = \frac{1}{2\omega} \int_{0}^{W^2} \frac{d\mathcal{E}}{\sqrt{W^2-\mathcal{E}}}
= \frac{1}{\omega} \,|W| ~.
\end{equation}
Hence,
\begin{gather} 
|W(x_-)| = -\omega x_- + \frac{\omega {x_0}^2}{x_-} ~,~~~
|W(x_+)| = \omega x_+ - \frac{\omega {x_0}^2}{x_+} \nonumber \\
\therefore\quad
W(x) = \omega x - \frac{\omega {x_0}^2}{x} ~.
\end{gather}
If we write $x_0\equiv \sqrt{g/\omega}$, the resulting superpotential is exactly the same as Eq. \eqref{eq:2-superpot_L}.
\end{example}

\begin{example}[J]
An energy spectrum
\begin{equation}
\mathcal{E} = \mathcal{E}(n) = 4n(n+g+h)
\end{equation}
reproduces the P\"{o}schl--Teller superpotential, together with
\begin{equation}
\tan x_-(W^2) \tan x_+(W^2) = \tan^2 x_0
\quad\text{\it i.e.}\quad
x_-(W^2) = \arctan\left( \frac{\tan^2 x_0}{\tan x_+(W^2)} \right) ~.
\end{equation}
Eq. \eqref{eq:3-formula_inverseSWKB} is
\begin{align}
x_+(W^2) - x_-(W^2) &= 2 \int_{0}^{W^2} \frac{dn}{\sqrt{W^2-\mathcal{E}(n)}} \nonumber \\
\therefore\quad
x_+(W^2) - \arctan\left( \frac{\tan^2 x_0}{\tan x_+(W^2)} \right) 
&= 2\hbar\int\limits_{\mathcal{E}(n)=0}^{\mathcal{E}(n)=W^2} \frac{dn}{\sqrt{W^2-4\hbar^2(g+h)n-4\hbar^2n^2}} \nonumber \\
&= \arctan\left( \frac{|W|}{\hbar g+\hbar h} \right) ~,
\end{align}
where $\mathcal{E}(n)=0$ corresponds to $n=0$ and $\mathcal{E}(n)=W^2$ corresponds to
\begin{equation}
n = \frac{\sqrt{W^2+(g+h)^2}-(g+h)}{2\hbar} ~.
\end{equation}
Hence,
\begin{equation}
\begin{aligned} 
|W(x)| &= \hbar(g+h)\sin^2x_0\cot x_- - \hbar(g+h)\cos^2x_0\tan x_- ~, \\
|W(x)| &= \hbar(g+h)\cos^2x_0\tan x_+ - \hbar(g+h)\sin^2x_0\cot x_+ ~,
\end{aligned}
\end{equation}
\begin{equation}
\therefore\quad
W(x) = -\hbar(g+h)\sin^2x_0\cot x + \hbar(g+h)\cos^2x_0\tan x ~, \quad
\end{equation}
where the identification $x_0\equiv \arccos\left( \sqrt{\dfrac{h}{g+h}}\,\right)$ yields Eq. \eqref{eq:2-superpot_J}.
\end{example}

\begin{remark}
Since all the conventional shape-invariant potentials are reduced into one of the above three examples by a coordinate transformation, it is sufficient to consider the three examples to check our formulation.
\end{remark}

\subsection{Construction of a Novel Solvable Potentials}
\label{sec:3-4-4}
We have formulated the inverse problem.
Now we construct superpotentials that satisfy the SWKB condition equation exactly for several $n$'s but do not for other $n$'s.
Here, we restrict ourselves to the modulation\footnote{I personally prefer the word ``modulation'' to ``deformation'' or ``transformation'' when describing the situation here, for the `angular velocity' changes during the `modulation' indeed.} of the ordinary harmonic oscillator (H).
Recall Example \ref{ex:3-inverseProb_H}.

\subsubsection{Construction of superpotentials}
Here, instead of Eq. \eqref{eq:3-nverseProb_H-assumption}, we take 
\begin{equation}
x_+(W^2) = -\gamma x_-(W^2) ~,
\end{equation}
where a positive constant $\gamma$ describes the modulation.
One can take $0<\gamma\leqslant 1$ without loss of generality.
$\gamma =1$ corresponds to the ordinary harmonic oscillator (Example \ref{ex:3-inverseProb_H}).
Thereby, for the given $\mathcal{E}=\mathcal{E}(n)=2n\hbar\omega$, we get
\begin{align}
(1+\gamma) x_2(W^2) &= \frac{1}{\omega}\int_0^{W^2} \frac{d\mathcal{E}}{\sqrt{W^2-\mathcal{E}}} \nonumber \\
&= \frac{2}{\omega}|W| 
\qquad\qquad\therefore\quad
W(x) = \begin{cases}
	\dfrac{1+\gamma}{2}\omega x & (x<0) \\[2ex]
	\dfrac{1+\gamma}{2\gamma}\omega x & (x>0)
\end{cases} ~,
\end{align}
where the corresponding potential is
\begin{equation}
V(x) = \begin{cases}
	\left( \dfrac{1+\gamma}{2} \right)^2\omega^2x^2 - \dfrac{1+\gamma}{2}\omega & (x<0) \\[2ex]
	\left( \dfrac{1+\gamma}{2\gamma} \right)^2\omega^2x^2 - \dfrac{1+\gamma}{2\gamma}\omega & (x>0)
\end{cases} ~.
\label{eq:3-SWKB_osc}
\end{equation}
In Sect. \ref{sec:4-2}, we shall solve the Schr\"{o}dinger equation for this potential.

\subsubsection{On SWKB-(non)exactness}
SWKB quantization condition \eqref{eq:3-SWKB} predicts the energy spectrum of this system as
\begin{equation}
\mathcal{E} = \mathcal{E}^{\rm (SWKB)}_n = 2n\hbar\omega ~.
\end{equation}
Here, the SWKB integral is
\begin{align}
I_{\rm SWKB} 
&= \int_{a_{\rm L}}^0 \sqrt{\mathcal{E} - \left( \frac{1+\gamma}{2} \right)^2\omega^2x^2} \,dx
+ \int_0^{a_{\rm R}} \sqrt{\mathcal{E} - \left( \frac{1+\gamma}{2\gamma} \right)^2\omega^2x^2} \,dx  
= \frac{\pi\mathcal{E}}{2\omega} ~, \nonumber \\
a_{\rm L} &= \frac{2\sqrt{\mathcal{E}}}{(1+\gamma)\omega} ~,~~~
a_{\rm R} = \frac{2\gamma\sqrt{\mathcal{E}}}{(1+\gamma)\omega}  ~.
\end{align}

This prediction does not agree with the exact bound-state spectrum for $n\geqslant 1$ in general.
However, when $\gamma$ is a rational number, say $\gamma=p/q$ with $p$ and $q$ being relatively prime, the quantization is `quasi-exact'.
If $p+q$ is an even number, then it gives an exact value only for
\[
n = \frac{p+q}{2}k ~,~~~
k = 0,1,2,\ldots ~. 
\]
On the other hand, if $p+q$ is an odd number, then 
\[
n = (p+q)k ~,~~~
k = 0,1,2,\ldots ~. 
\]
Note that the corresponding eigenfunctions with such $n$'s are expressed in terms of the Hermite polynomials.

\subsection{Classical Analogue}
\subsubsection{Determination of the potential energy from the period of oscillation}
In classical mechanics, there is a classical problem of determining the potential energy from the period of oscillation.

Let us consider a particle (mass $1/2$) under a potential $U(x)$ moving back and forth repeatedly between $x_-$ and $x_+$ (turning points) with the period $T$.
When the particle possesses the energy $E$, the period of oscillation is
\begin{equation}
T = T(E) = \int_{x_-}^{x_+} \frac{dx}{\sqrt{E-U(x)}} ~.
\label{eq:3-cl_period}
\end{equation} 
One is to identify the potential energy $U(x)$ from this equation with the knowledge of the period of oscillation $T(E)$.
Mathematically, this is a problem of solving the integral equation \eqref{eq:3-cl_period} about an unknown function $U(x)$.
Suppose $U(x)$ has only one minimum. 
$x=x(U)$ is two-valued, and let $x_-$ denote the smaller value and $x_+$ the larger one.
Then, Eq. \eqref{eq:3-cl_period} becomes
\begin{equation}
x_+ -x_- = \frac{1}{\pi}\int_{0}^{U}\frac{T(E)\,dE}{\sqrt{U-E}}
\end{equation}
The simplest example would be the determination of a harmonic potential $U(x) \propto x^2$ from a constant period $T=\mathrm{const.}$ 
Here, one needs to assume that $U(x)$ is symmetric, \textit{i.e.}, $x_- = -x_+$ to obtain the harmonic potential.
Therefore, 
\begin{equation}
x = \frac{T}{2\pi}\int_{0}^{U}\frac{dE}{\sqrt{U-E}} = \frac{T}{\pi}\sqrt{U}
\quad\Longrightarrow\quad
U(x) = \frac{\pi^2}{T^2}x^2 
\end{equation}
(For more details, see, \textit{e.g.}, Ref. \cite{lifshitz1976mechanics}).

\subsubsection{Construction of isochronous potentials}
In the early modern period, Galilei discovered that the period of oscillating pendulums of equal length is constant, or \textit{isochronous}, regardless of the amplitude, which turned out to be wrong.
Huygens proved that oscillators moving along cycloidal curves are actually isochronous.
Since then, scientists have asked how to construct isochronous oscillators.

One method for constructing potentials that realize isochronous motions of a particle is to solve the integral equation \eqref{eq:3-cl_period} with $T=\mathrm{const.}$ under different anzats.
For example, suppose $U(x)$ has only one minimum at $x=0$ and call the coordinate on the left $x_1<0$ and right $x_2>0$. 
Then the ansatz $x_1(U) = -\gamma x_2(U)$ where $\gamma$ is a positive parameter realizes isochronous potential energy $U(x)$ other than the harmonic potential:
\begin{equation}
U(x) = \begin{cases}
	\dfrac{\pi^2(1+\gamma)^2}{T^2}x^2 & (x<0) \\[2ex]
	\dfrac{\pi^2(1+\gamma)^2}{T^2\gamma^2}x^2 & (x>0)
\end{cases}
\end{equation}
They are all isochronous, regardless of the value of $\gamma$. 
For further discussions, see, \textit{e.g.}, Refs. \cite{10.1119/1.15063,10.1119/1.2839560,10.1119/1.4963770,10.1119/1.5019025}.

\subsubsection{Inverse problem in WKB formalism}
We can also consider this kind of inverse problem, that is, the problem of determining a potential $V(x)$ from a given energy spectrum $\{ E_n \}$ in semi-classical theories, using the Bohr--Sommerfeld quantization condition \eqref{eq:3-WKB}.
One can find simple cases in exercises of a basic quantum mechanics course (See, \textit{e.g.}, Ref. \cite{Magyari}).
Note that the energy spectrum $\{ E_n \}$ coincides exactly with the energy eigenvalues of the resulting $V(x)$, only in the cases where Bohr--Sommerfeld quantization condition reproduces exact bound-state spectra.

In solving the inverse problem, first we differentiate the sides of Eq. \eqref{eq:3-WKB} in $E$,
\begin{equation}
\frac{d}{dE}\int_{x_1}^{x_2} \sqrt{E-V(x)} \,dx 
= \int_{x_1}^{x_2} \frac{dx}{2\sqrt{E-V(x)}} 
= \pi\hbar\frac{dn}{dE} ~,
\end{equation}
where the integral 
\[
\int_{x_1}^{x_2} \frac{dx}{\sqrt{E-V(x)}} 
\]
can be regarded as a quantum analogue of the period of oscillation.
The quantization condition specifies `the period' as $\pi\hbar\,dn/dE$, which is constant if $E$ is linear in $n$.
Thus, in quantum mechanics, the concept of isochronism is translated into the equidistance of energy spectra.

Similar to the case of isochronous deformation of potentials in classical mechanics, one can construct `isochronous' potentials:
\begin{equation}
V(x) = \begin{cases}
	\left( \dfrac{1+\gamma}{2} \right)^2\omega^2x^2 & (x<0) \\[2ex]
	\left( \dfrac{1+\gamma}{2\gamma} \right)^2\omega^2x^2 & (x>0)
\end{cases} ~
\label{eq:3-sheared_osc}
\end{equation}
in the context of WKB approximation.
Here, we have assumed that $E=2n\hbar\omega$.
However, the energy spectrum of this potential is not exactly equidistant, which is considered as a quantum effect.~\cite{ASOREY20071444,Dorignac_2005}.
When 
\[
\gamma = \frac{4k+1}{4\ell+1} \equiv \frac{p}{q}
\quad\text{or}\quad
\gamma = \frac{4k+3}{4\ell+3} \equiv \frac{p}{q} ~,~~~
k,\ell \in \mathbb{Z}_{\geqslant 0} ~,
\]
we have such energy spectrum that is equidistant every $\dfrac{p+q}{2}$ states from $n=\dfrac{p+q-2}{4}$.
For further discussions, see Refs. \cite{Stillinger:1989aa,Chadzitaskos:2022kql}.

\clearpage
\section{Summary of Chapter 3}
In this chapter, we have thoroughly investigated SWKB quantization condition \eqref{eq:3-SWKB}, aiming mainly to understand what the condition actually implies.
We first introduced the condition equation with its several properties in Sect. \ref{sec:3-1}.
 
Our study began in Sect. \ref{sec:3-2} with extensive case studies: the SWKB integrals of the conventional shape-invariant potentials, the exceptional/multi-indexed systems, the Krein--Adler systems, the conventional exactly solvable systems by Junker and Roy, and classical-orthogonal-polynomially solvable systems with position dependent effective mass are analyzed.
For the classical-orthogonal-polynomially solvable systems with position-dependent effective mass, the condition is naturally extended to reflect the position-dependency of mass.
During the calculations, we learned the following two notable properties of the condition: (1) $\hbar$ is always factored out of the condition equation and the SWKB can be discussed outside the context of semi-classical approximation, (2) the SWKB condition holds in good approximation in many examples, which seems to be guaranteed by the distribution of the whole energy spectrum.
After several discussions, we have concluded in Sect. \ref{sec:3-3} that the exactness of the SWKB quantization condition indicates that the system is exactly solvable via the classical orthogonal polynomials. 

The SWKB quantization condition provides quantizations of energy, which we call the direct problem of the SWKB.
Our study was in this course so far.
Sometimes, modeling a phenomenon from the experimental data of a spectrum is a central issue in physics and other sciences.
It is quite natural to consider the \textit{inverse} problem of the SWKB, that is, the problem of determining the superpotential from a given energy spectrum, which we have formulated in Sect. \ref{sec:3-4}.
An assumption on the shape of superpotential is required to make the inverse problem well-posed.
Since we have revealed that SWKB is related to the classical-orthogonal-polynomial solvability of a system, one expects that this formulation allows one to construct novel classical-orthogonal-polynomially solvable superpotentials.
First, we reconstructed conventional shape-invariant potentials from a given energy spectra as a test of our formulation.
Then we have constructed novel superpotentials, which are classical-orthogonal-polynomially \textit{quasi}-exactly solvable and satisfy the SWKB quantization condition quasi-exactly.
The solution is discussed in the subsequent chapter.

\subsubsection{Main statements of this chapter}
\begin{itemize}\setlength{\leftskip}{-1em}
\item The exactness of the SWKB quantization condition indicates that the system is exactly solvable via the classical orthogonal polynomials (Sect. \ref{sec:3-3}).
\item The inverse problem of the SWKB is formulated to construct (novel) classical-orthogonal-polynomially solvable superpotentials (Sect. \ref{sec:3-4}).
\end{itemize}

\clearpage
\thispagestyle{empty}
~

\chapter{Modulations of Harmonic Oscillator: Novel Solvable Potentials}
\label{sec:4}

{\small
\begin{leftbar}
\noindent\textbf{Introduction.}\hspace{1em}
In this chapter, we start with solving the Schr\"{o}dinger equation with the potential constructed at the end of the previous chapter in Sect. \ref{sec:4-2}.
The potential is classified into the classical-orthogonal-polynomially quasi-exactly solvable potentials.
We provide general remarks on the solution of the Schr\"{o}dinger equation with the potentials of this kind in advance in Sect. \ref{sec:4-1}.
In Sect. \ref{sec:4-3}, we consider yet other potentials in this class, which is abstracted as ``harmonic oscillator with singularity functions''.

\bigskip\noindent
$\blacktriangleright$\hspace{1em}\bf
This chapter is chiefly based on the author's works: Refs. \cite{nasuda2023harmonic,Nasuda_2023,nasuda-lt15}.
\end{leftbar}
}

\section{Solution Method: General Remarks}
\label{sec:4-1}
All the potentials we consider in this chapter are confining potentials on the real line, $x\in (-\infty,\infty)$, 
\[
\lim_{x\to +\infty} V(x) = \lim_{x\to -\infty} V(x) = +\infty ~,
\]
and thereby have infinitely many discrete eigenstates and no scattering solutions.
Also, the eigenfunctions are square-integrable: $\psi_n(x) \in L^2(\mathbb{R})\cap C^1$, which leads to the following boundary conditions:
\begin{equation}
\lim_{x\to +\infty}\psi(x) = \lim_{x\to -\infty}\psi(x) = 0 ~.
\label{eq:4-BC_inf}
\end{equation}

Moreover, our potentials are not analytic at $x=0$, so in solving the Schr\"{o}dinger equations one must be careful about the matching condition there:
\begin{equation}
\lim_{x\to 0^{+}}\psi(x) = \lim_{x\to 0^{-}}\psi(x) ~,~~~
\lim_{x\to 0^{+}}\frac{d\psi(x)}{dx} = \lim_{x\to 0^{-}}\frac{d\psi(x)}{dx} ~,
\label{eq:4-BC_origin}
\end{equation}
or simply
\begin{equation}
\lim_{x\to 0^{+}}\frac{d}{dx}\ln|\psi(x)| = \lim_{x\to 0^{-}}\frac{d}{dx}\ln|\psi(x)| ~.
\end{equation}

In addition, it is convenient for us to note that the general solution of the following second-order ordinary differential equation: 
\begin{equation}
-\frac{d^2\psi(x)}{dx^2} + \left( x^2-1 \right)\psi(x) = E\psi(x) ~,
\end{equation}
is
\begin{equation}
\psi(x) = \mathrm{e}^{-\frac{x^2}{2}} \left[ \alpha \,{}_1F_1\left( -\frac{E}{4};\frac{1}{2};x^2 \right) + \beta x \,{}_1F_1\left( -\frac{E-2}{4};\dfrac{3}{2};x^2 \right) \right] ~,
\label{eq:4-general_sol}
\end{equation} 
where $\alpha$ and $\beta$ are constants, and ${}_1F_1(a;c;x)$ denotes the Kummer's confluent hypergeometric function.
Of course, in the case of the ordinary harmonic oscillator, we require the square-integrability of $\psi(x)$ in $(-\infty,\infty)$, and get
\begin{equation}
E = E_n = 2n ~,~~~
\psi(x) = \phi^{\rm (H)}_n(x) = \mathrm{e}^{-\frac{x^2}{2}}H_n(x) ~,~~~
n = 0,1,2,\ldots ~,
\end{equation}
where $H_n(x)$ is the $n$-th order Hermite polynomial.

\section{SWKB-induced Quadratic Oscillator}
\label{sec:4-2}
\subsection{The Potential}
In this section, we consider the potential \eqref{eq:3-SWKB_osc}, constructed in Sect. \ref{sec:3-4-4}.
Here, we set $\omega =1$ without loss of generality,
\begin{equation}
V(x) = \begin{cases}
	\left( \dfrac{1+\gamma}{2} \right)^2x^2 - \dfrac{1+\gamma}{2} & (x<0) \\[2ex]
	\left( \dfrac{1+\gamma}{2\gamma} \right)^2x^2 - \dfrac{1+\gamma}{2\gamma} & (x>0)
\end{cases} ~.
\label{eq:4-SWKB_osc}
\end{equation}
We should emphasize here that this potential has a finite gap at $x=0$, which plays a key role in its isospectral property.

Moreover, we are allowed to restrict ourselves to $0<\gamma\leqslant 1$, since the potential is invariant under the parity transformation and the discrete transformation of the parameter, $\gamma \leftrightarrow 1/\gamma$.
At $\gamma=1$, this potential reproduces the ordinary harmonic oscillator.
On the other hand, the limit $\gamma\to 0$ makes the potential a harmonic oscillator on the negative half line.


\subsection{The Solutions}
We solve the Schr\"{o}dinger equation with the potential \eqref{eq:4-SWKB_osc}:
\begin{subequations}
\begin{align}
&\left[ -\dfrac{d^2}{dx^2} + \left( \dfrac{1+\gamma}{2} \right)^2x^2 - \dfrac{1+\gamma}{2} \right] \phi(x) = E\phi(x) & &(x<0) ~, \\[1ex]
&\left[ -\dfrac{d^2}{dx^2} + \left( \dfrac{1+\gamma}{2\gamma} \right)^2x^2 - \dfrac{1+\gamma}{2\gamma} \right] \phi(x) = E\phi(x) & &(x>0) ~,
\end{align}
\end{subequations}
which is piecewise solvable and the wavefunction is
\begin{equation}\small
\phi(x) = \begin{cases}
	\mathrm{e}^{-\frac{1+\gamma}{2}\frac{x^2}{2}}\left[ \alpha_- \,{}_1F_1\left( -\dfrac{\frac{E}{1+\gamma}}{2};\dfrac{1}{2};\dfrac{1+\gamma}{2}x^2 \right) + \beta_- \sqrt{\dfrac{1+\gamma}{2}}\, x\, {}_1F_1\left( -\dfrac{\frac{E}{1+\gamma}-1}{2};\dfrac{3}{2};\dfrac{1+\gamma}{2}x^2 \right)  \right] & ~ \\
	& \hspace{-2.5em} (x<0) \\
	\mathrm{e}^{-\frac{1+\gamma}{2\gamma}\frac{x^2}{2}}\left[ \alpha_+ \,{}_1F_1\left( -\dfrac{\frac{\gamma E}{1+\gamma}}{2};\dfrac{1}{2};\dfrac{1+\gamma}{2\gamma}x^2 \right) + \beta_+ \sqrt{\dfrac{1+\gamma}{2\gamma}}\, x\, {}_1F_1\left( -\dfrac{\frac{\gamma E}{1+\gamma}-1}{2};\dfrac{3}{2};\dfrac{1+\gamma}{2\gamma}x^2 \right)  \right] & ~ \\
	& \hspace{-2.5em} (x>0)
\end{cases} ~,
\end{equation}
with $\alpha_{\pm}$, $\beta_{\pm}$ being constants to be determined from the connecting condition \eqref{eq:4-BC_origin}.
The condition yields the following relations among the constants:
\begin{equation}
\alpha_+ = \alpha_- ~,~~~
\beta_+ = \sqrt{\gamma}\,\beta_- ~.
\end{equation}

Next, we determine the energy eigenvalues from the boundary condition at $x\to\pm\infty$, Eq. \eqref{eq:4-BC_inf}.
Since the asymptotic forms of $\phi(x)$ are 
\begin{subequations}
\begin{align}
&\phi(x) \overset{\tiny x\to -\infty}{\sim} 
\left[ \alpha_{-}\frac{\varGamma\left( \frac{1}{2} \right)}{\varGamma\left( -\frac{\frac{E}{1+\gamma}}{2} \right)} - \beta_{-}\frac{\varGamma\left( \frac{3}{2} \right)}{\varGamma\left( -\frac{\frac{E}{1+\gamma}-1}{2} \right)} \right] \mathrm{e}^{-\frac{1+\gamma}{2}\frac{x^2}{2}}\left( \sqrt{\frac{1+\gamma}{2}}\,x \right)^{-\frac{\frac{E}{1+\gamma}}{2}-\frac{1}{2}} ~, \\[1ex]
&\phi(x) \overset{\tiny x\to +\infty}{\sim} 
\left[ \alpha_{+}\frac{\varGamma\left( \frac{1}{2} \right)}{\varGamma\left( -\frac{\frac{\gamma E}{1+\gamma}}{2} \right)} + \beta_{+}\frac{\varGamma\left( \frac{3}{2} \right)}{\varGamma\left( -\frac{\frac{\gamma E}{1+\gamma}-1}{2} \right)} \right] \mathrm{e}^{-\frac{1+\gamma}{2\gamma}\frac{x^2}{2}}\left( \sqrt{\frac{1+\gamma}{2\gamma}}\,x \right)^{-\frac{\frac{\gamma E}{1+\gamma}}{2}-\frac{1}{2}} ~,
\end{align}
\end{subequations}
it follows that $E$, $\alpha_{\pm}$, $\beta_{\pm}$ satisfy both of the following two transcendental equations:
\begin{subequations}
\begin{align}
\alpha_{-}\frac{\varGamma\left( \frac{1}{2} \right)}{\varGamma\left( -\frac{\frac{E}{1+\gamma}}{2} \right)} - \beta_{-}\frac{\varGamma\left( \frac{3}{2} \right)}{\varGamma\left( -\frac{\frac{E}{1+\gamma}-1}{2} \right)} &= 0 ~, \\
\alpha_{+}\frac{\varGamma\left( \frac{1}{2} \right)}{\varGamma\left( -\frac{\frac{\gamma E}{1+\gamma}}{2} \right)} + \beta_{+}\frac{\varGamma\left( \frac{3}{2} \right)}{\varGamma\left( -\frac{\frac{\gamma E}{1+\gamma}-1}{2} \right)} &= 0 ~,
\end{align}
\end{subequations}
which lead to
\begin{equation}
\frac{\varGamma\left( -\frac{\frac{\gamma E}{1+\gamma}-1}{2} \right)}{\varGamma\left( -\frac{\frac{\gamma E}{1+\gamma}}{2} \right)} = -\sqrt{\gamma} \frac{\varGamma\left( -\frac{\frac{E}{1+\gamma}-1}{2} \right)}{\varGamma\left( -\frac{\frac{E}{1+\gamma}}{2} \right)} ~,
\label{eq:4-transcendental_SWKB_osc}
\end{equation}
with
\begin{equation}
\beta_- = 2\frac{\varGamma\left( -\frac{\frac{E}{1+\gamma}-1}{2} \right)}{\varGamma\left( -\frac{\frac{E}{1+\gamma}}{2} \right)}\alpha_- ~,~~~
\beta_+ = -2\frac{\varGamma\left( -\frac{\frac{\gamma E}{1+\gamma}-1}{2} \right)}{\varGamma\left( -\frac{\frac{\gamma E}{1+\gamma}}{2} \right)}\alpha_+ ~.
\end{equation}
One is to solve the transcendental equation \eqref{eq:4-transcendental_SWKB_osc} graphically to determine the energy eigenvalues as in the case with the finite square potential well, but with the following two exceptions.

We have two kinds of exact roots of the equation \eqref{eq:4-transcendental_SWKB_osc}.
One is with $E$ such that
\begin{equation}
\frac{\frac{E}{1+\gamma}}{2} \in \mathbb{Z}_{\geqslant 0}
\quad\wedge\quad
\frac{\frac{\gamma E}{1+\gamma}}{2} \in \mathbb{Z}_{\geqslant 0} ~,
\label{eq:4-E1st_SWKB_osc}
\end{equation}
\textit{i.e.}, $E$ is a multiple of $2(1+\gamma)$ and $\dfrac{2(1+\gamma)}{\gamma}$ at the same time.
Here, $\beta_{\pm}=0$ but $\alpha_{\pm}\neq 0$.
The other one is with $E$ such that
\begin{equation}
\frac{\frac{E}{1+\gamma}-1}{2} \in \mathbb{Z}_{\geqslant 0}
\quad\wedge\quad
\frac{\frac{\gamma E}{1+\gamma}-1}{2} \in \mathbb{Z}_{\geqslant 0} ~,
\label{eq:4-E2nd_SWKB_osc}
\end{equation}
which means $E$ is an odd multiple of $1+\gamma$ and $\dfrac{1+\gamma}{\gamma}$ simultaneously, and we set $\alpha_{\pm}=0, \beta_{\pm}\neq 0$ in this case.
Note that these $E$'s do not cover the whole spectrum except for $\gamma=1$.

\subsubsection{Case $\gamma\in\mathbb{Q}$: Hermite-polynomial solutions}
Here, let us assume $\gamma$ is a rational number, $\gamma\in\mathbb{Q}$, which is a necessary condition for $E$'s to exist such that Eqs. \eqref{eq:4-E1st_SWKB_osc} and \eqref{eq:4-E2nd_SWKB_osc}.
We write
\begin{equation}
\gamma \equiv \frac{p}{q} ~,
\end{equation}
where $p$ and $q$ are relatively prime integers or $p=q=1$.
Then, the Schr\"{o}dinger equation becomes
\begin{subequations}
\begin{align}
&\left[ -\dfrac{d^2}{dx^2} + \left( \dfrac{p+q}{2q} \right)^2x^2 - \dfrac{p+q}{2q} \right] \phi(x) = E\phi(x) & &(x<0) \\[1ex]
&\left[ -\dfrac{d^2}{dx^2} + \left( \dfrac{p+q}{2p} \right)^2x^2 - \dfrac{p+q}{2p} \right] \phi(x) = E\phi(x) & &(x>0) ~.
\end{align}
\label{eq:4-SE_pq}
\end{subequations}

In this parametrization, Eqs. \eqref{eq:4-E1st_SWKB_osc} and \eqref{eq:4-E2nd_SWKB_osc} read 
\[
\frac{qE}{2(p+q)} \in \mathbb{Z}_{\geqslant 0}
\quad\wedge\quad
\frac{pE}{2(p+q)} \in \mathbb{Z}_{\geqslant 0} 
\]
and
\[
\frac{qE}{2(p+q)} - \frac{1}{2} \in \mathbb{Z}_{\geqslant 0}
\quad\wedge\quad
\frac{pE}{2(p+q)} - \frac{1}{2} \in \mathbb{Z}_{\geqslant 0} ~,
\]
respectively, meaning that $E$ is a multiple of $p+q$.
If $E$ is an even multiple of $p+q$, then $\alpha_{\pm}\neq 0$ and $\beta_{\pm}=0$; while if $E$ is an odd multiple of $p+q$, then $\alpha_{\pm}=0$ and $\beta_{\pm}\neq 0$.

Taking the fact that $E=0$ is always a root of Eqs. \eqref{eq:4-transcendental_SWKB_osc} with $\alpha_{\pm}\neq 0$ and $\beta_{\pm}=0$ into account, we identify that the energy 
\begin{equation}
E = (p+q)m ~,~~~ m \in \mathbb{Z}_{\geqslant 0}
\end{equation}
is the $(p+q)m/2$-th excited-state energy.
Therefore, we can say that if $p+q$ is even, the spectrum is equidistant for every $(p+q)/2$ states, and if odd, the spectrum is equidistant for every $p+q$ states.

The corresponding eigenfunctions are written in terms of Hermite polynomials of piecewise different orders.
For the case where $p+q$ is even, the eigenfunction with energy $E=(p+q)k$, $k=0,1,2,\ldots$ is
\begin{equation}
\phi_n(x) = \begin{cases}
	\displaystyle \mathcal{N}_k^{(-)} \mathrm{e}^{-\frac{p+q}{4q}x^2}H_{qk}\left( \sqrt{\frac{p+q}{2q}\,x} \right) & (x<0) \\[2ex]
	\displaystyle \mathcal{N}_k^{(+)} \mathrm{e}^{-\frac{p+q}{4p}x^2}H_{pk}\left( \sqrt{\frac{p+q}{2p}\,x} \right) & (x>0)
\end{cases} ~,~~~
n = \frac{p+q}{2}k ~.
\end{equation}
$\mathcal{N}_k^{(\pm)}$ are constants, whose ratio $\mathcal{N}_k^{(-)}/\mathcal{N}_k^{(+)}$ is determined by the matching condition \eqref{eq:4-BC_origin},
\begin{equation}
\frac{\mathcal{N}_k^{(-)}}{\mathcal{N}_k^{(+)}} = \begin{cases}
	\dfrac{(pk)!\frac{qk}{2}!}{(qk)!\frac{pk}{2}!} & (\text{$k$: even}) \\[2ex]
	(-1)^{\frac{p-q}{2}}\sqrt{\dfrac{p}{q}}\cdot\dfrac{(pk-1)!\frac{qk-1}{2}!}{(qk-1)!\frac{pk-1}{2}!} & (\text{$k$: odd})
\end{cases} ~.
\end{equation}
On the other hand, for $p+q$ is odd, the eigenfunction with energy $E=2(p+q)k$, $k=0,1,$ $2,\ldots$ is
\begin{equation}
\phi_n(x) = \begin{cases}
	\displaystyle \mathcal{N}_k^{(-)} \mathrm{e}^{-\frac{p+q}{4q}x^2}H_{2qk}\left( \sqrt{\frac{p+q}{2q}\,x} \right) & (x<0) \\[2ex]
	\displaystyle \mathcal{N}_k^{(+)} \mathrm{e}^{-\frac{p+q}{4p}x^2}H_{2pk}\left( \sqrt{\frac{p+q}{2p}\,x} \right) & (x>0)
\end{cases} ~,~~~
n = (p+q)k ~,
\end{equation}
with
\begin{equation}
\frac{\mathcal{N}_k^{(-)}}{\mathcal{N}_k^{(+)}} = (-1)^{k}\frac{(2pk)!(qk)!}{(2qk)!(pk)!} ~.
\end{equation}

\begin{figure}[p]
\centering
\includegraphics[scale=0.9]{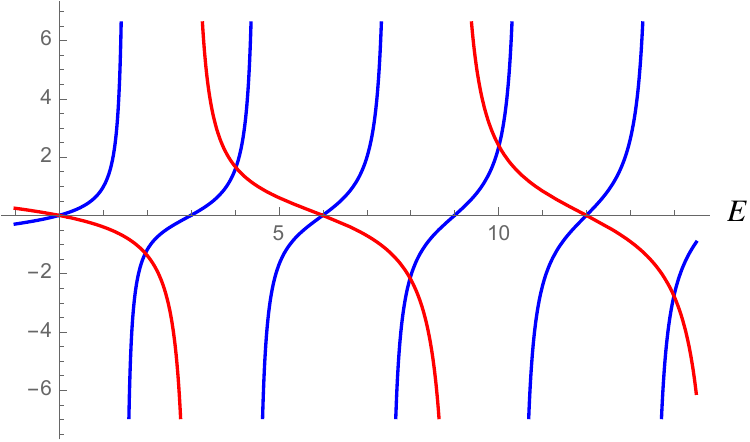}
\caption[Graphical solution of Eq. \eqref{eq:4-transcendental_SWKB_osc} with $\gamma=1/2$.]
	{Graphical solution of Eq. \eqref{eq:4-transcendental_SWKB_osc} with $\gamma=1/2$.
    The blue curves correspond to the left-hand side of the equation, while the red ones are the right-hand side.
    The intersections of these curves determine the energy eigenvalues.
    The numerical solutions are displayed in Tab. \ref{tab:4-Eigenvalues_p1q2}.}
\label{fig:4-GraphicalSol_p1q2}
\end{figure}

\begin{figure}[p]
\centering
\includegraphics[scale=0.8]{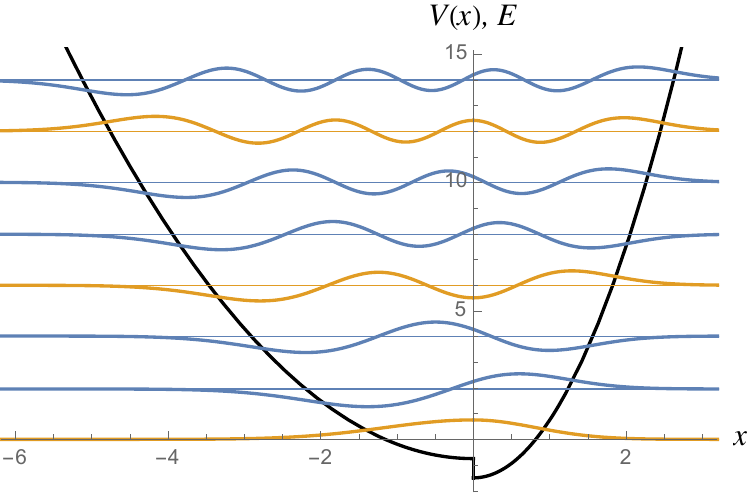}
\caption[The solutions of the eigenvalue problem \eqref{eq:4-SE_pq} with $(p,q)=(1,2)$.]
	{The solutions of the eigenvalue problem \eqref{eq:4-SE_pq} with $(p,q)=(1,2)$.
	 Thin lines show the energy spectrum, and the colored curve on each line is the corresponding eigenfunction.
	The states plotted in yellow possess the Hermite-polynomial solvability, while those colored in blue do not.
    The potential is also plotted in this figure by a black curve.}
\label{fig:4-sol_p1q2}
\end{figure}

\begin{example}
Take $\gamma = p/q=1/2$ as our first example, where $3n'$-th eigenvalues ($n'=0,1,2,\ldots$) are explicitly known, while for other eigenstates we are to solve Eq. \eqref{eq:4-transcendental_SWKB_osc} graphically (See Fig. \ref{fig:4-GraphicalSol_p1q2}).
First several energy eigenvalues are displayed in Tab. \ref{tab:4-Eigenvalues_p1q2}.
The solutions are summarized in Fig. \ref{fig:4-sol_p1q2}.
\end{example}

\begin{figure}[p]
\centering
\includegraphics[scale=0.9]{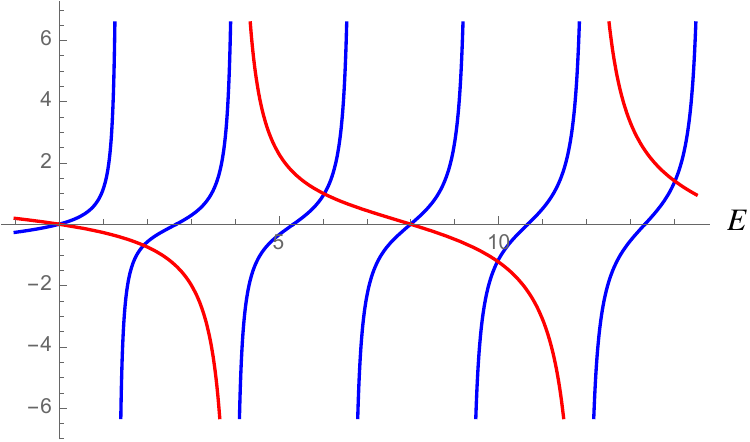}
\caption[Graphical solution of Eq. \eqref{eq:4-transcendental_SWKB_osc} with $\gamma=1/3$.]
	{Graphical solution of Eq. \eqref{eq:4-transcendental_SWKB_osc} with $\gamma=1/3$.
    The blue curves correspond to the left-hand side of the equation, while the red ones are the right-hand side.
    The intersections of these curves determine the energy eigenvalues.
    The numerical solutions are displayed in Tab. \ref{tab:4-Eigenvalues_p1q3}.}
\label{fig:4-GraphicalSol_p1q3}
\end{figure}

\begin{figure}[p]
\centering
\includegraphics[scale=0.8]{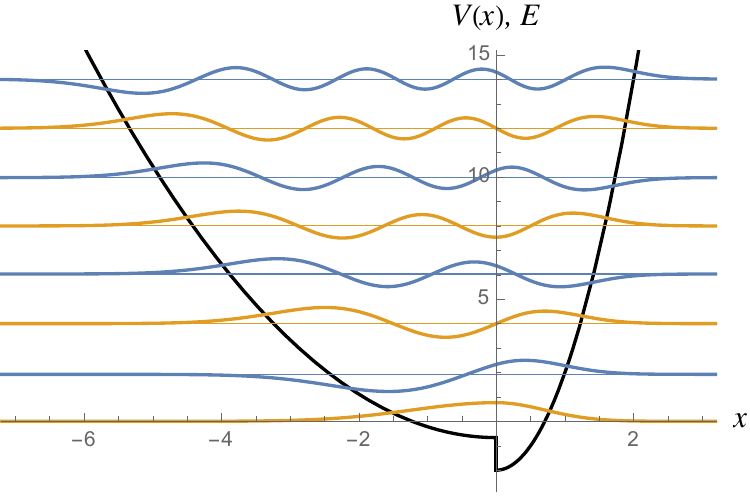}
\caption[The solutions of the eigenvalue problem \eqref{eq:4-SE_pq} with $(p,q)=(1,3)$.]
	{The solutions of the eigenvalue problem \eqref{eq:4-SE_pq} with $(p,q)=(1,3)$.
	 Thin lines show the energy spectrum, and the colored curve on each line is the corresponding eigenfunction.
	The states plotted in yellow possess the Hermite-polynomial solvability, while those colored in blue do not.
    The potential is also plotted in this figure by a black curve.}
\label{fig:4-sol_p1q3}
\end{figure}

\begin{example}
Take $\gamma = p/q=1/3$.
Here, $2n'$-th eigenvalues ($n'=0,1,2,\ldots$) are explicitly known, while for other eigenstates we are to solve Eq. \eqref{eq:4-transcendental_SWKB_osc} graphically (See Fig. \ref{fig:4-GraphicalSol_p1q3}).
First several energy eigenvalues are displayed in Tab. \ref{tab:4-Eigenvalues_p1q3}.
The solutions are summarized in Fig. \ref{fig:4-sol_p1q3}.
\end{example}

\clearpage

\begin{table}[p]
\caption[First several energy eigenvalues for Eq. \eqref{eq:4-SE_pq} with $(p,q)=(1,2)$ with six digits.]
	{First several energy eigenvalues for Eq. \eqref{eq:4-SE_pq} with $(p,q)=(1,2)$ with six digits.
	These values are obtained by solving Eq. \eqref{eq:4-transcendental_SWKB_osc} or finding the intersections in Fig. \ref{fig:4-GraphicalSol_p1q2} numerically.
	The energy gaps between two successive eigenstates are roughly $2$, while those for the harmonic oscillator are exactly $2$ in our unit.}
\label{tab:4-Eigenvalues_p1q2}
\centering
\begin{tabular}{cr}
\toprule
$n$ & $E_n$~~~~ \\
\midrule
$0$ & $0$ (exact) \\
$1$ & $1.96156$ \\
$2$ & $4.02277$ \\
$3$ & $6$ (exact) \\
$4$ & $7.98757$ \\
$5$ & $10.0101$ \\
$6$ & $12$ (exact) \\
\bottomrule
\end{tabular}
\end{table}

\clearpage

\clearpage

\begin{table}[p]
\caption[First several energy eigenvalues for Eq. \eqref{eq:4-SE_pq} with $(p,q)=(1,3)$ with six digits.]
	{First several energy eigenvalues for Eq. \eqref{eq:4-SE_pq} with $(p,q)=(1,3)$ with six digits.
	These values are obtained by solving Eq. \eqref{eq:4-transcendental_SWKB_osc} or finding the intersections in Fig. \ref{fig:4-GraphicalSol_p1q3} numerically.
	The energy gaps between two successive eigenstates are roughly $2$, while those for the harmonic oscillator are exactly $2$ in our unit.}
\label{tab:4-Eigenvalues_p1q3}
\centering
\begin{tabular}{cr}
\toprule
$n$ & $E_n$~~~~ \\
\midrule
$0$ & $0$ (exact) \\
$1$ & $1.92412$ \\
$2$ & $4$ (exact) \\
$3$ & $6.03248$ \\
$4$ & $8$ (exact) \\
$5$ & $9.97945$ \\
$6$ & $12$ (exact) \\
\bottomrule
\end{tabular}
\end{table}

\clearpage

\clearpage
\begin{example}[$\gamma=p'/12$]
We show every how many states the energy spectra are equidistant for the choices of $\gamma=p'/12$ ($p'=(0,)1,2,\ldots,12$) in Tab. \ref{tab:4-p'_N}.
\end{example}

\begin{table}[t]
\centering
\caption{For $\gamma=p'/12$ ($p'=(0,)1,2,\ldots,12$), the energy spectra are equidistant every $N$ states.}
\label{tab:4-p'_N}
\begin{tabular}{cccccccccccccc}
\toprule
$p'$ & ($0$) & $1$ & $2$ & $3$ & $4$ & $5$ & $6$ & $7$ & $8$ & $9$ & $10$ & $11$ & $12$ \\
\midrule
$N$ & ($1$) & $13$ & $7$ & $5$ & $2$ & $17$ & $3$ & $19$ & $5$ & $7$ & $11$ & $23$ & $1$ \\
\bottomrule
\end{tabular}
\end{table}

\subsection{Discussions}

\subsubsection{Equidistance of energy spectra}
We have learned that for a given $\gamma=p/q$, the spectrum is equidistant for every $(p+q)/2$ states for even $p+q$ and every $p+q$ states for odd $p+q$.
It would be interesting to consider the other way around.
That is, suppose you want an energy spectrum that is equidistant for every $N$ states.
Then what value of $\gamma$ should you choose?
We provide the set(s) $(p,q)$ for the first several $N$'s in Tab. \ref{tab:4-N_pq}. 
Note that we have $N'$ sets of $(p,q)$ for $N=2N'-1$ or $2N'$ with $N'=1,2,\ldots$.

\begin{table}[t]
\centering
\caption{The choice(s) of $(p,q)$ for several $N$'s, which means every how many states the energy spectra are equidistant.}
\label{tab:4-N_pq}
\begin{tabular}{ccccc}
\toprule
$N$ & $(p,q)$ \\
\midrule
$1$ & $(1,1)$ & & & \\
$2$ & $(1,3)$ &  & & \\
$3$ & $(1,2)$ & $(1,5)$ & & \\
$4$ & $(1,7)$ & $(3,5)$ & & \\
$5$ & $(1,4)$ & $(1,9)$ & $(2,3)$ & $(3,7)$ \\
\bottomrule
\end{tabular}
\end{table}

\subsubsection{Comments on a finite jump in potentials in physics}
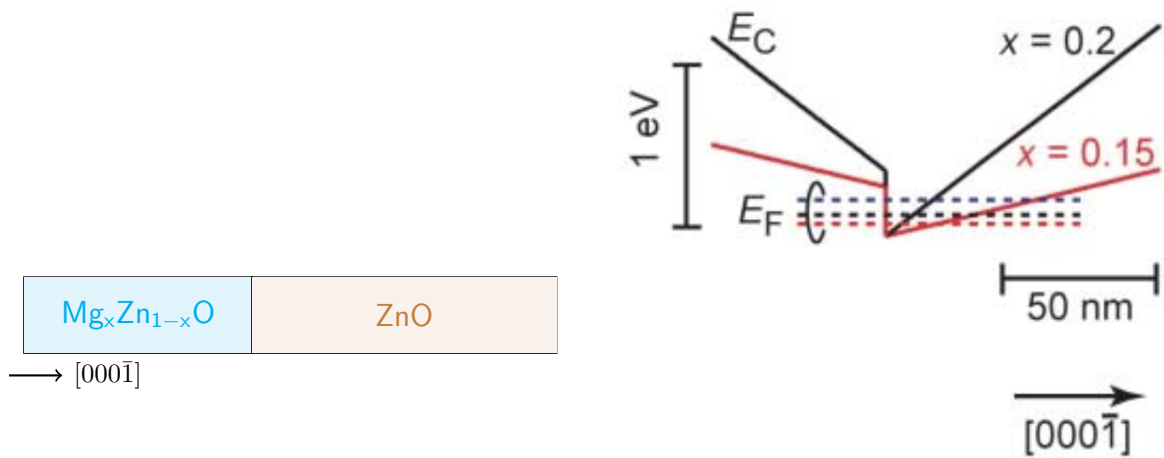
\begin{figure}[h]
\centering
	\begin{minipage}[b]{.45\linewidth}
	\centering
		\begin{tikzpicture}
		\draw[thick] (-3,1)--(-3,0)--(0,0)--(0,1)--cycle;
		\fill[white!90!cyan] (-3,1)--(-3,0)--(0,0)--(0,1)--cycle;
		\node at (-1.5,0.5) {\textcolor{cyan}{$\mathsf{Mg_xZn_{1-x}O}$}};
		\draw[thick,->] (-3.2,-0.3)--(-2.5,-0.3) node[right] {\footnotesize$[000\bar{1}]$};
		\draw[thick] (0,1)--(0,0)--(4,0)--(4,1)--cycle;
		\fill[white!90!brown] (0,1)--(0,0)--(4,0)--(4,1)--cycle;
		\node at (2,0.5) {\textcolor{brown}{$\mathsf{ZnO}$}};
		\end{tikzpicture}
		\vspace{24pt}
	\subcaption[Schematic picture of the $\mathrm{ZnO}/\mathrm{Mg}_{x}\mathrm{Zn}_{1-x}\mathrm{O}$ heterostructure.]{Schematic picture of the $\mathrm{ZnO}/$ $\mathrm{Mg}_{x}\mathrm{Zn}_{1-x}\mathrm{O}$ heterostructure. \\~}
	\end{minipage}
	\quad
	\begin{minipage}[b]{.45\linewidth}
	\centering
	\includegraphics[scale=0.5]{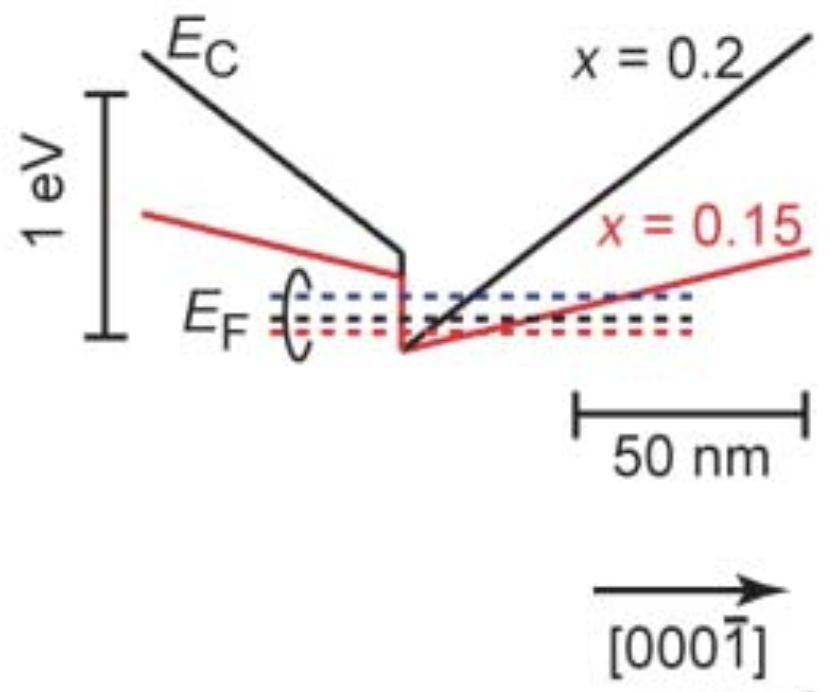}
	\subcaption{Potential diagram near the heterointerfaces. This figure is taken from the middle panel of Fig. 1C in Ref. \cite{doi:10.1126/science.1137430}.}
	\end{minipage}
\caption{A potential with a finite jump in the $\mathrm{ZnO}/\mathrm{Mg}_{x}\mathrm{Zn}_{1-x}\mathrm{O}$ heterostructure.}
\label{fig:offset-pot_1}
\end{figure}

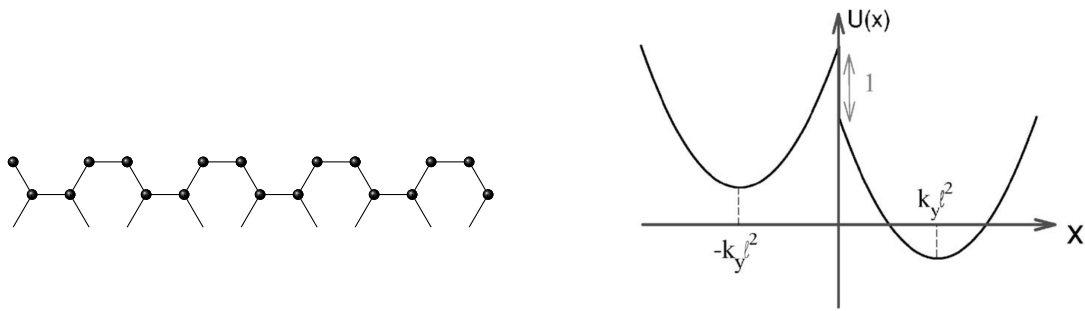
\begin{figure}[h]
\centering
	\begin{minipage}[b]{.45\linewidth}
	\centering
		\begin{tikzpicture}
		\foreach \k in {0,...,3}
		\draw (-0.25+1.5*\k,0.433)--(1.5*\k,0)--(0.5+1.5*\k,0)--(0.75+1.5*\k,0.433)--(1.25+1.5*\k,0.433);
		\draw (-0.25+1.5*4,0.433)--(1.5*4,0);
		\foreach \k in {0,...,3}
		\fill[ball color=black] (-0.25+1.5*\k,0.433) circle (2pt); 
		\foreach \k in {0,...,3}
		\fill[ball color=black] (1.5*\k,0) circle (2pt);
		\foreach \k in {0,...,3}
		\fill[ball color=black] (0.5+1.5*\k,0) circle (2pt); 
		\foreach \k in {0,...,3}
		\fill[ball color=black] (0.75+1.5*\k,0.433) circle (2pt);
		\foreach \k in {0,...,3}
		\fill[ball color=black] (1.25+1.5*\k,0.433) circle (2pt);
		\fill[ball color=black] (1.5*4,0) circle (2pt);
		\foreach \k in {0,...,4}
		\draw (1.5*\k,0)--(-0.25+1.5*\k,-0.433);
		\foreach \k in {0,...,3}
		\draw (0.5+1.5*\k,0)--(0.75+1.5*\k,-0.433);
		\end{tikzpicture}
		\vspace{36pt}
	\subcaption[Schematic picture of the armchair edge.]{Schematic picture of the armchair edge. \\~\\~\\~}
	\end{minipage}
	\quad
	\begin{minipage}[b]{.45\linewidth}
	\centering
	\includegraphics[scale=0.25]{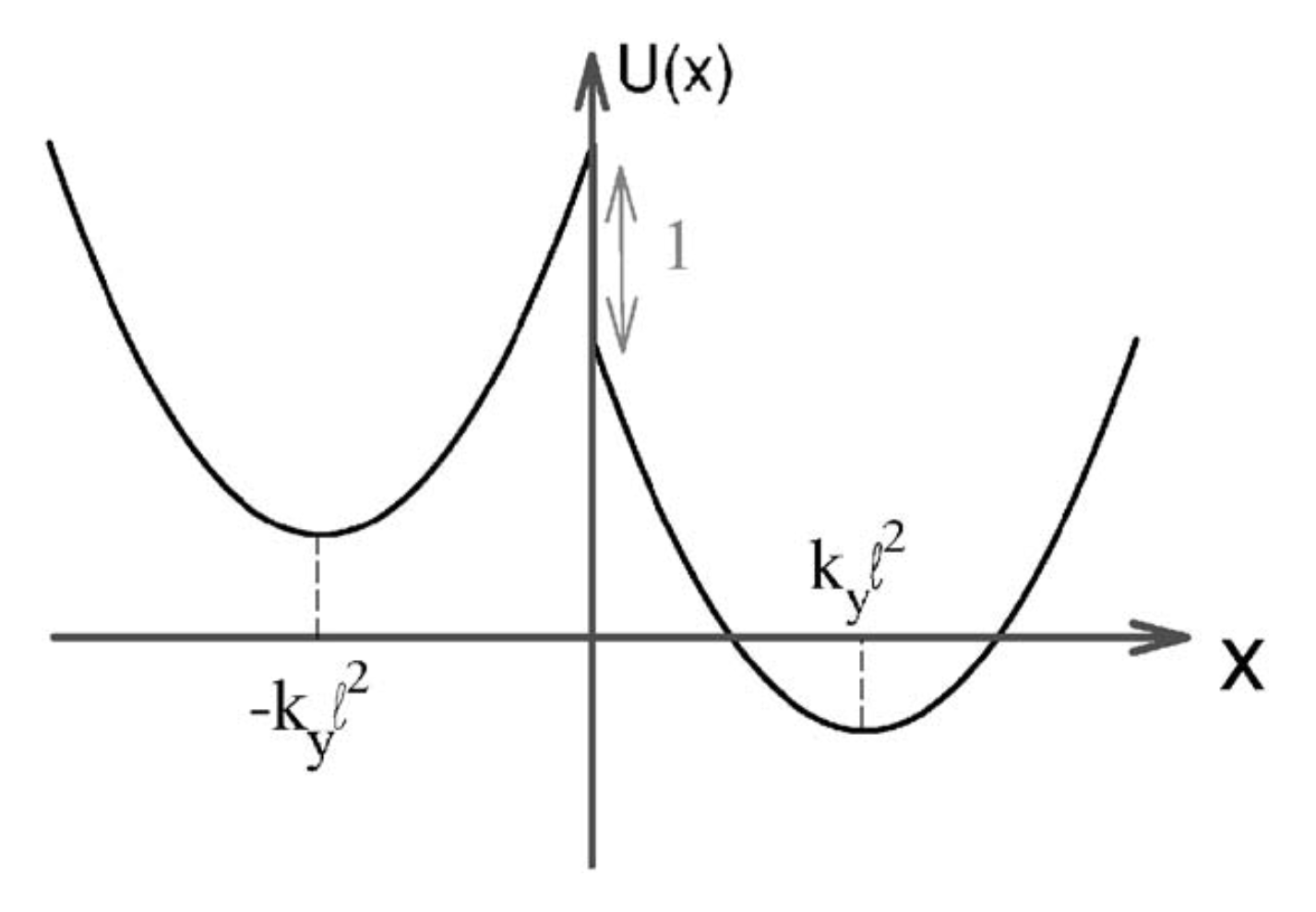}
	\subcaption{A virtual potential for describing the Landau-level structures of graphene ribbons for an armchair edge. This figure is taken from Fig. 4 in Ref. \cite{PhysRevB.73.195408}.}
	\end{minipage}
\caption{A virtual potential with a gap describing the Landau-level structures of graphene ribbons for an armchair edge.}
\label{fig:offset-pot_2}
\end{figure}

One might feel that the potentials with a finite jump, such as \eqref{eq:4-SWKB_osc}, are artificial and have no meaning in reality.
However, motivation for exploring problems with a finite jump in potential is found in the following courses:
\begin{enumerate}\setlength{\leftskip}{-1em}
\item To investigate phenomena that occur at the interface between two different materials such as polar $\mathrm{ZnO}/\mathrm{Mg}_{x}\mathrm{Zn}_{1-x}\mathrm{O}$ heterostructures~\cite{doi:10.1126/science.1137430} (See Fig. \ref{fig:offset-pot_1}).
\item To analyze phenomena by reproducing the doubly degenerated (except for the ground state) energy spectra, as in the case of the Landau levels of the edge states of graphene ribbons for an armchair edge~\cite{PhysRevB.73.195408} (See Fig. \ref{fig:offset-pot_2}).
\end{enumerate}

\clearpage
In summary, one can find a finite jump in potentials when considering interfacial phenomena (\textit{cf}. The Heaviside step function is employed for a simple (rough) model of the free electron in metals).

\section{Harmonic Oscillator with a Step}
\label{sec:4-3}
\subsection{The Potential}
\begin{figure}[h]
\centering
\includegraphics[scale=0.8]{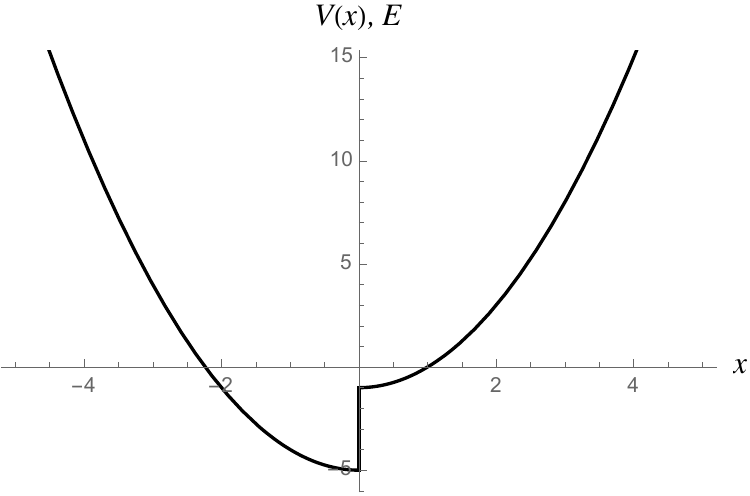}
\caption{The potential \eqref{eq:4-2-pot}.
	Here, the parameter $a$ is set to be $4$ as an example.}
\label{fig:4-2-pot}
\end{figure}

We consider a potential 
\begin{equation}
V(x) = V(x;a) = \begin{cases}
	x^2 - 1 - a & (x<0) \\
	x^2 - 1 & (x>0)
\end{cases} 
\label{eq:4-2-pot}
\end{equation}
(See Fig. \ref{fig:4-2-pot}).
Although the Schr\"{o}dinger equation is not invariant under $a\leftrightarrow -a$, it is sufficient to consider the case where $a$ is a positive constant, $a>0$, by considering the parity transformation plus constant shift of the energy.

\subsection{The Solutions}
Let us first consider the most general case, \textit{i.e.}, arbitrary $a>0$. 
The Schr\"{o}dinger equation:
\begin{subequations}
\begin{align}
&\left[ -\dfrac{d^2}{dx^2} + x^2 - 1 - a \right] \psi(x) = E\psi(x) & &(x<0) \label{eq:4-SEan} \\[1ex]
&\left[ -\dfrac{d^2}{dx^2} + x^2 - 1 \right] \psi(x) = E\psi(x) & &(x>0) ~, \label{eq:4-SEap}
\end{align}
\label{eq:4-SEa}
\end{subequations}
is piecewise solvable, whose solution is
\begin{equation}
\psi(x) =  \begin{cases}
	\mathrm{e}^{-\frac{x^2}{2}}\left[ \alpha \,{}_1F_1\left( -\dfrac{E+a}{4};\dfrac{1}{2};x^2 \right) + \beta x\, {}_1F_1\left( -\dfrac{E+a-2}{4};\dfrac{3}{2};x^2 \right)  \right] 
	& (x<0) \\[2ex]
	\mathrm{e}^{-\frac{x^2}{2}}\left[ \alpha \,{}_1F_1\left( -\dfrac{E}{4};\dfrac{1}{2};x^2 \right) + \beta x\, {}_1F_1\left( -\dfrac{E-2}{4};\dfrac{3}{2};x^2 \right)  \right] 
	& (x>0)
\end{cases} ~,
\end{equation}
with $\alpha$ and $\beta$ being constants.
The energy eigenvalues $E_n$ are determined through the boundary conditions at infinity \eqref{eq:4-BC_inf}, that is,
\begin{equation}
\alpha\frac{\varGamma\left( \frac{1}{2} \right)}{\varGamma\left( -\frac{E+a}{4} \right)} - \beta\frac{\varGamma\left( \frac{3}{2} \right)}{\varGamma\left( -\frac{E+a-2}{4} \right)} = 0 
~~\text{and}~~ 
\alpha\frac{\varGamma\left( \frac{1}{2} \right)}{\varGamma\left( -\frac{E}{4} \right)} + \beta\frac{\varGamma\left( \frac{3}{2} \right)}{\varGamma\left( -\frac{E-2}{4} \right)} = 0 ~.
\label{eq:4-E-det_a}
\end{equation}
From these equations, one obtains the following transcendental equations about $E$, whose roots are the eigenvalues:
\begin{equation}
-\frac{\varGamma\left( -\frac{E}{4} \right)}{\varGamma\left( -\frac{E-2}{4} \right)} = \frac{\varGamma\left( -\frac{E+a}{4} \right)}{\varGamma\left( -\frac{E+a-2}{4} \right)} ~.
\label{eq:4-E-det_a_transcental}
\end{equation}
Note however that, technically, Eq. \eqref{eq:4-E-det_a_transcental} can only be applicable to the case $a\neq 4\ell$ ($\ell=0,1,2,\ldots$), where $\alpha\neq 0$.
For the case where $a=4\ell$, we shall discuss it in detail in the subsequent subsection.

\begin{figure}[p]
\centering
\includegraphics[scale=0.9]{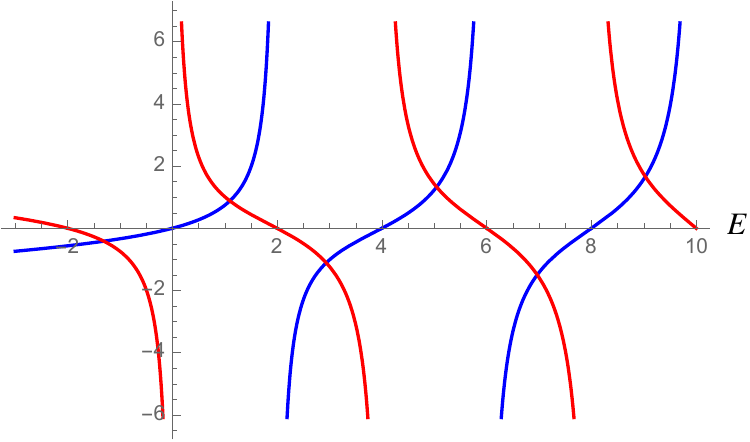}
\caption[Graphical solution of Eq. \eqref{eq:4-E-det_a_transcental}.]
	{Graphical solution of Eq. \eqref{eq:4-E-det_a_transcental}.
	The blue curves correspond to the left-hand side of the equation, while the red ones are the right-hand side.
	The intersections of these curves determine the energy eigenvalues.
	The numerical solutions are displayed in Tab. \ref{tab:4-Eigenvalues_a2}.}
\label{fig:4-GraphicalSol_a2}
\end{figure}

\begin{figure}[p]
\centering
\includegraphics[scale=0.8]{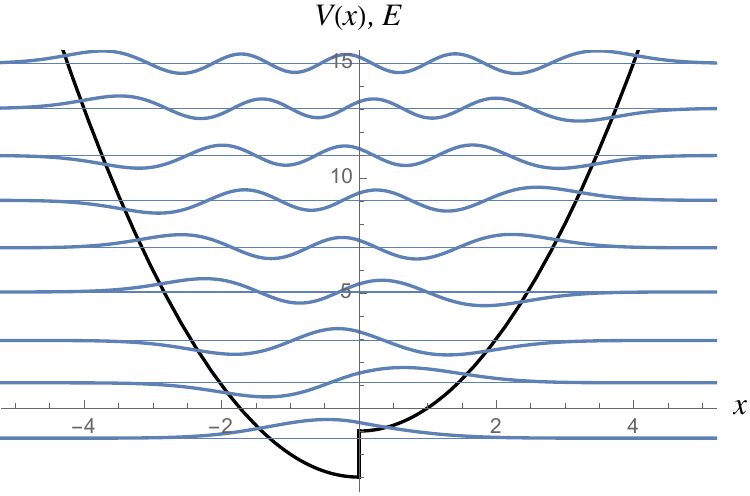}
\caption[The solutions of the eigenvalue problem \eqref{eq:4-SEa} with $a=2$.]
	{The solutions of the eigenvalue problem \eqref{eq:4-SEa} with $a=2$.
    Thin blue lines show the energy spectrum, and the blue curve on each line is the corresponding eigenfunction.
    The potential $V(x;2)$ is also plotted in this figure by a black curve.}
\label{fig:4-sol_a2}
\end{figure}

\begin{example}
We show the case of $a=2$ as an example.
We first solve Eq. \eqref{eq:4-E-det_a_transcental} with $a=2$ to obtain the energy spectrum.
This equation is transcendental, and we are to solve it graphically (See Fig. \ref{fig:4-GraphicalSol_a2}).
The first several energy eigenvalues are displayed in Tab. \ref{tab:4-Eigenvalues_a2} with six digits.
The corresponding eigenfunctions are expressed in terms of $E_n$ as
\begin{equation}
\psi_n(x) \propto \begin{cases}
	\mathrm{e}^{-\frac{x^2}{2}}\left[ {}_1F_1\left( -\dfrac{E_n+2}{4};\dfrac{1}{2};x^2 \right) - 2\dfrac{\varGamma\left( -\frac{E_n-2}{4} \right)}{\varGamma\left( -\frac{E_n}{4} \right)}x\, {}_1F_1\left( -\dfrac{E_n}{4};\dfrac{3}{2};x^2 \right)  \right] 
& (x<0) \\[2.5ex]
	\mathrm{e}^{-\frac{x^2}{2}}\left[ {}_1F_1\left( -\dfrac{E_n}{4};\dfrac{1}{2};x^2 \right) - 2\dfrac{\varGamma\left( -\frac{E_n-2}{4} \right)}{\varGamma\left( -\frac{E_n}{4} \right)}x\, {}_1F_1\left( -\dfrac{E_n-2}{4};\dfrac{3}{2};x^2 \right)  \right]
& (x>0)
\end{cases} ~.
\end{equation}
The solutions are summarized in Fig. \ref{fig:4-sol_a2}.
\end{example}

\subsubsection{Case $a=4\ell$ ($\ell = 1,2,\ldots$): Hermite-polynomial solutions}
Next let us consider the case where $a$ is a multiple of $4$, $a=4\ell$ or $V(x;a) = V(x;4\ell)$ ($\ell=1,2,\ldots$), where all the non-negative eigenfunctions are expressed with the Hermite polynomials as we shall demonstrate below.
The Schr\"{o}dinger equation is
\begin{subequations}
\begin{align}
&\left[ -\dfrac{d^2}{dx^2} + x^2 - 1 - 4\ell \right] \psi(x) = E\psi(x) & &(x<0) \label{eq:4-SE4ln} \\[1ex]
&\left[ -\dfrac{d^2}{dx^2} + x^2 - 1 \right] \psi(x) = E\psi(x) & &(x>0) ~. \label{eq:4-SE4lp}
\end{align}
\label{eq:4-SE4l}
\end{subequations}

\noindent
For this case, the transcendental equations \eqref{eq:4-E-det_a_transcental} reduce to rather simple algebraic equations, and all the eigenfunctions with $E\geqslant 0$ are expressed by the Hermite polynomials $\{H_n(x)\}$.

\begin{table}[p]
\caption[First several energy eigenvalues for $V(x;2)$ with six digits.]{
	First several energy eigenvalues for $V(x;2)$ with six digits.
	These values are obtained by solving Eq. \eqref{eq:4-E-det_a_transcental} or finding the intersections in Fig.\ref{fig:4-GraphicalSol_a2} numerically.
	The energy gaps between two successive eigenstates are roughly $2$, while those for the harmonic oscillator are exactly $2$ in our units.}
\label{tab:4-Eigenvalues_a2}
\centering
\begin{tabular}{cr}
\toprule
$n$ & $E_n$~~~ \\
\midrule
$0$ & $-1.30908$ \\
$1$ & $1.09714$ \\
$2$ & $2.93715$ \\
$3$ & $5.04459$ \\
$4$ & $6.96479$ \\
$5$ & $9.02870$ \\
$6$ & $10.9756$ \\
\bottomrule
\end{tabular}
\end{table}

\clearpage

\clearpage
Before we solve Eq. \eqref{eq:4-SE4l}, let us summarize the solutions of Eqs.\eqref{eq:4-SE4ln} and \eqref{eq:4-SE4lp} solved on the real line, $x\in\mathbb{R}$, respectively.
They are
\begin{subequations}
\begin{align}
&\eqref{eq:4-SE4ln}\Rightarrow &
&E_n^{(-)} = 2n-4\ell ~,~~~ \psi_n^{(-)}(x) = \mathcal{N}_n^{(-)}\mathrm{e}^{-\frac{x^2}{2}}H_n(x) ~,~~~
n = 0,1,2,\ldots ~, \qquad~ \label{eq:4-soln_SE4l_E+} \\
&\eqref{eq:4-SE4lp}\Rightarrow &
&E_n^{(+)} = 2n ~,~~~ \psi_n^{(+)}(x) = \mathcal{N}_n^{(+)}\mathrm{e}^{-\frac{x^2}{2}}H_n(x) ~,~~~
n = 0,1,2,\ldots ~. \qquad ~ \label{eq:4-solp_SE4l_E+}
\end{align}
\label{eq:4-sol_SE4l_E+}
\end{subequations}
Here, they share all non-negative energy eigenvalues.
This suggests that $E_j=2j$ ($j=0,1,2,\ldots$) are the eigenvalues of the system \eqref{eq:4-SE4l} if the coefficients $\mathcal{N}_n^{(\pm)}$ are chosen so that they satisfy the continuity conditions of the wavefunction and its first derivative at the origin \eqref{eq:4-BC_origin}.

For $E=0$, the wavefunctions are 
\begin{equation}
\psi_{2\ell}^{(-)}(x) \propto \mathrm{e}^{-\frac{x^2}{2}}H_{2\ell}(x) ~,~~~
\psi_0^{(+)}(x) \propto \mathrm{e}^{-\frac{x^2}{2}} ~.
\end{equation}
Since
\begin{equation}
\mathrm{e}^{-\frac{0^2}{2}}H_{2\ell}(0) = (-1)^{\ell}\frac{(2\ell)!}{\ell!} ~,~~~
\mathrm{e}^{-\frac{0^2}{2}} = 1 ~,~~~
\frac{d\psi_{2\ell}^{(-)}(0)}{dx} = \frac{d\psi_0^{(+)}(0)}{dx} = 0 ~,
\end{equation}
the continuity conditions at the origin \eqref{eq:4-BC_origin} yields the wavefunction with $E=0$ for the original problem \eqref{eq:4-SE4l} in the following form:
\begin{equation}
\psi_{E=0}(x) = \begin{cases}
	(-1)^{\ell}\dfrac{\ell!}{(2\ell)!}\mathrm{e}^{-\frac{x^2}{2}}H_{2\ell}(x) & (x<0) \\
	\mathrm{e}^{-\frac{x^2}{2}} & (x>0)
\end{cases} ~.
\end{equation}
This is a square-integrable, smooth function of $\ell$ nodes, so from the oscillation theorem, one can safely say that this corresponds to the $\ell$-th excited state, $\psi_{E=0}(x)\equiv\psi_{\ell}(x)$.

Similarly, one can obtain the square-integrable wavefunctions for all the non-negative energies.
The $n$-th excited-state energy eigenvalue for Eq. \eqref{eq:4-SE4l} is $E_n=2(n-\ell)$, where $n$ is greater than or equal to $\ell$, and the corresponding eigenfunction is
\begin{equation}
\psi_n(x) = \begin{cases}
	\mathcal{N}_{n}\,\mathrm{e}^{-\frac{x^2}{2}}H_{n+\ell}(x) & (x<0) \\
	\mathrm{e}^{-\frac{x^2}{2}}H_{n-\ell}(x) & (x>0) 
\end{cases} ~,~~~
n = \ell,\ell+1,\ell+2,\ldots ~,
\end{equation}
where
\begin{subequations}
\begin{align}
\mathcal{N}_{n}
&= (-1)^{\ell}\frac{(n-\ell)!\left( \frac{n+\ell}{2} \right)!}{\left( n+\ell \right)!\left( \frac{n-\ell}{2} \right)!} 
&&\text{if $(n-\ell)$ is even} ~, \\
\mathcal{N}_{n}
&= (-1)^{\ell}\frac{(n-\ell)!\left( \frac{n+\ell-1}{2} \right)!}{\left( n+\ell \right)!\left( \frac{n-\ell-1}{2} \right)!} 
&&\text{if $(n-\ell)$ is odd} ~.
\end{align}
\end{subequations}
Note that this wavefunction satisfies either Neumann or Dirichlet boundary condition at the origin.
For the case where $(n-\ell)$ is even, it complies with
\begin{equation}
\psi_{n}(0) \neq 0 ~,~~~
\frac{d\psi_{n}(0)}{dx} = 0
\qquad\text{: {\it Neumann} boundary condition,}
\end{equation}
while for odd $(n-\ell)$,
\begin{equation}
\psi_{n}(0) =  0 ~,~~~
\frac{d\psi_{n}(0)}{dx} \neq 0
\qquad\text{: {\it Dirichlet} boundary condition.}
\end{equation}
For $E \geqslant 0$, our problem \eqref{eq:4-SE4l} would be rephrased as finding square-integrable solutions of \eqref{eq:4-SEa} under the Neumann/Dirichlet boundary condition at the origin.

For the $\ell$ lowest eigenstates, the eigenfunctions are not expressed by the Hermite polynomials anymore. 
So we need to go back to the problem of solving the piecewise differential equation \eqref{eq:4-SE4l} under the boundary conditions \eqref{eq:4-BC_inf}, \eqref{eq:4-BC_origin}.
Eq.\,\eqref{eq:4-BC_inf} yields 
\begin{equation}
-\frac{\varGamma\left( -\frac{E}{4} \right)}{\varGamma\left( -\frac{E-2}{4} \right)} = \frac{\varGamma\left( -\frac{E+4\ell}{4} \right)}{\varGamma\left( -\frac{E+4\ell-2}{4} \right)} ~.
\end{equation}
This may look a transcendental equation, however, after some algebras,  one finds that it is a degree $\ell$ algebraic equation:
\begin{equation}
-\prod_{k=1}^{\ell}(E+4k-2) = \prod_{k=1}^{\ell}(E+4k) ~.
\label{eq:4-algebraic}
\end{equation}
Here the roots of this equation are denoted by $E_0,E_1,E_2,\ldots, E_{\ell-1}$ from the lowest to higher.
Therefore, for $E<0$, the wavefunctions are 
\begin{multline}
\psi_n(x) = \begin{cases}
	\displaystyle
	\mathrm{e}^{-\frac{x^2}{2}}\left[ {}_1F_1\left( -\frac{E_n+4\ell}{4};\frac{1}{2};x^2 \right) - 2\frac{\varGamma\left( -\frac{E_n-2}{4} \right)}{\varGamma\left( -\frac{E_n}{4} \right)}x\, {}_1F_1\left( -\frac{E_n+4\ell-2}{4};\frac{3}{2};x^2 \right)  \right] & ~ \\
	~ & \hspace{-3em} (x<0) \\
	\displaystyle
	\mathrm{e}^{-\frac{x^2}{2}}\left[ {}_1F_1\left( -\frac{E_n}{4};\frac{1}{2};x^2 \right) - 2\frac{\varGamma\left( -\frac{E_n-2}{4} \right)}{\varGamma\left( -\frac{E_n}{4} \right)}x\, {}_1F_1\left( -\frac{E_n-2}{4};\frac{3}{2};x^2 \right)  \right] & ~ \\
	~ & \hspace{-3em} (x>0) 
\end{cases} ~, \\
n=0,1,\ldots,\ell-1 ~.
\end{multline}

\begin{figure}[t]
\centering
\includegraphics[scale=0.8]{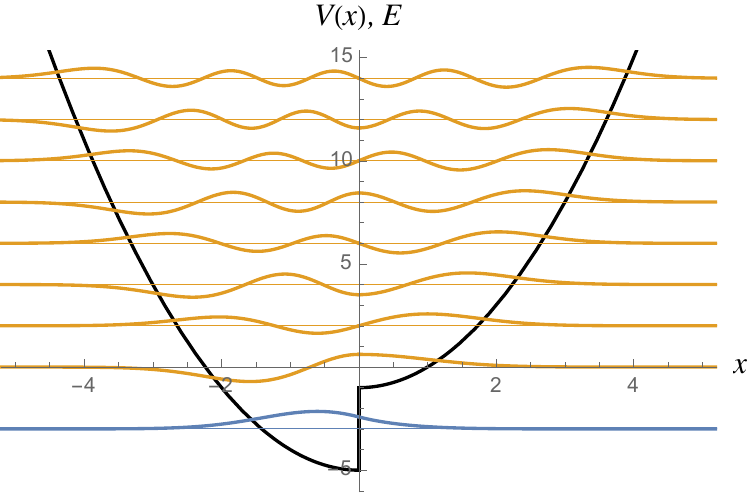}
\caption[The solutions of the eigenvalue problem \eqref{eq:4-SE4l} with $\ell=1$.]
	{The solutions of the eigenvalue problem \eqref{eq:4-SE4l} with $\ell=1$.
	The potential $V(x;4)$ is displayed in this figure by a black curve.
	Thin lines show the energy spectrum, and the colored curve on each line is the corresponding eigenfunction.
	The states plotted in yellow possess the Hermite-polynomial solvability, while that colored in blue does not.}
\label{fig:4-sol_l1}
\end{figure}

\begin{example}
The energy eigenvalues and the corresponding eigenfunctions for $\ell=1$ are
\begin{align}
E_0 &= -3 ~,~~~
\psi_0(x) = \begin{cases}
	\displaystyle
	\mathrm{e}^{-\frac{x^2}{2}}\left[ {}_1F_1\left( -\frac{1}{4};\frac{1}{2};x^2 \right) - 2\frac{\varGamma\left( \frac{5}{4} \right)}{\varGamma\left( \frac{3}{4} \right)}x\, {}_1F_1\left( \frac{1}{4};\frac{3}{2};x^2 \right)  \right] & (x<0)  \\[2.5ex]
	\displaystyle
	\mathrm{e}^{-\frac{x^2}{2}}\left[ {}_1F_1\left( \frac{3}{4};\frac{1}{2};x^2 \right) - 2\frac{\varGamma\left( \frac{5}{4} \right)}{\varGamma\left( \frac{3}{4} \right)}x\, {}_1F_1\left( \frac{5}{4};\frac{3}{2};x^2 \right)  \right] & (x>0)
\end{cases} ~, \\
E_n &= 2(n-1) ~,~~~
\psi_n(x) = \begin{cases}
	\mathcal{N}_{n}\,\mathrm{e}^{-\frac{x^2}{2}}H_{n+1}(x) & (x<0) \\
	\mathrm{e}^{-\frac{x^2}{2}}H_{n-1}(x) & (x>0) 
\end{cases} ~,~~~ 
n = 1,2,3,\ldots ~,
\end{align}
where
\begin{subequations}
\begin{align}
\mathcal{N}_{n}
&= -\frac{1}{2n} 
&&\text{if $n$ is odd} ~, \\
\mathcal{N}_{n}
&= -\frac{1}{2(n+1)}
&&\text{if $n$ is even} ~.
\end{align}
\end{subequations}
They are summarized in Fig. \ref{fig:4-sol_l1}.
\end{example}

\begin{figure}[h]
\centering
\includegraphics[scale=0.8]{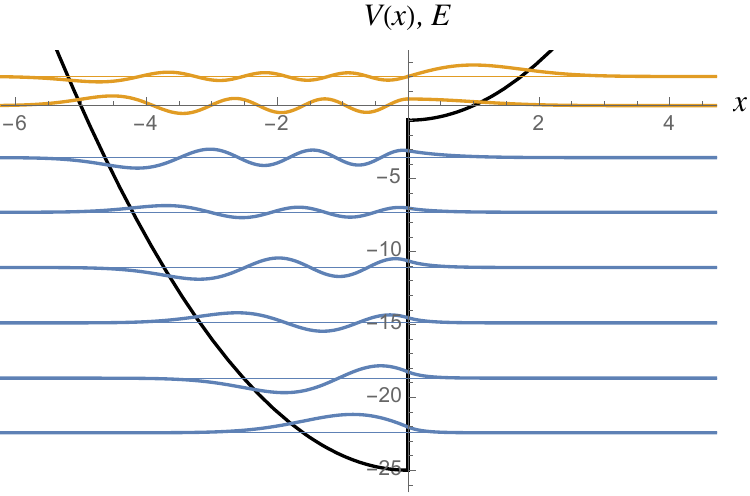}
\caption[The solutions of the eigenvalue problem \eqref{eq:4-SE4l} with $\ell=6$.]
	{The solutions of the eigenvalue problem \eqref{eq:4-SE4l} with $\ell=6$.
	The potential $V(x;24)$ is displayed in this figure by a black curve.
	Thin lines show the energy spectrum, and the colored curve on each line is the corresponding eigenfunction.
	The states plotted in yellow possess the Hermite-polynomial solvability, while those colored in blue do not.}
\label{fig:4-sol_l6}
\end{figure}

\begin{example}
\label{eg:l6}
The energy eigenvalues and the corresponding eigenfunctions for $\ell=6$ are
\begin{align}
\psi_n(x) &= \begin{cases}
	\displaystyle
	\mathrm{e}^{-\frac{x^2}{2}}\left[ {}_1F_1\left( -\frac{E_n+24}{4};\frac{1}{2};x^2 \right) - 2\frac{\varGamma\left( -\frac{E_n-2}{4} \right)}{\varGamma\left( -\frac{E_n}{4} \right)}x\, {}_1F_1\left( -\frac{E_n+22}{4};\frac{3}{2};x^2 \right)  \right] & (x<0) \\[2.5ex]
	\displaystyle
	\mathrm{e}^{-\frac{x^2}{2}}\left[ {}_1F_1\left( -\frac{E_n}{4};\frac{1}{2};x^2 \right) - 2\frac{\varGamma\left( -\frac{E_n-2}{4} \right)}{\varGamma\left( -\frac{E_n}{4} \right)}x\, {}_1F_1\left( -\frac{E_n-2}{4};\frac{3}{2};x^2 \right)  \right] & (x>0) 
\end{cases} ~, \nonumber \\
&\hspace{0.65\textwidth} n=0,1,\ldots, 5 ~, \\
E_n &= 2(n-6) ~,~~~
\psi_n(x) = \begin{cases}
	\mathcal{N}_{n}\,\mathrm{e}^{-\frac{x^2}{2}}H_{n+6}(x) & (x<0) \\
	\mathrm{e}^{-\frac{x^2}{2}}H_{n-6}(x) & (x>0) 
\end{cases} ~,~~~ 
n = 6,7,8,\ldots ~,
\end{align}
where
\begin{subequations}
\begin{align}
\mathcal{N}_{n}
&= \frac{(n-6)!\left( \frac{n+6}{2} \right)!}{\left( n+6 \right)!\left( \frac{n-6}{2} \right)!} 
&&\text{if $n$ is even} ~, \\
\mathcal{N}_{n}
&= \frac{(n-6)!\left( \frac{n+5}{2} \right)!}{\left( n+6 \right)!\left( \frac{n-7}{2} \right)!} 
&&\text{if $n$ is odd} ~.
\end{align}
\end{subequations}

\noindent
They are summarized in Fig.\,\ref{fig:4-sol_l6}, and the negative energy eigenvalues are displayed in Tab.\,\ref{tab:4-Eigenvalue_l6}.

\end{example}

\vfill
\begin{table}[h]
\caption[The negative energy eigenvalues for $V(x;24)$ with six digits.]{
	The negative energy eigenvalues for $V(x;24)$ with six digits.
	These values are obtained by solving Eq. \eqref{eq:4-algebraic} with $\ell=6$.
	The energy gaps between two successive eigenstates are roughly $4$, while those for the harmonic oscillator on the positive half line are exactly $4$ in our units.}
\label{tab:4-Eigenvalue_l6}
\centering
\medskip
\begin{tabular}{cr}
\toprule
$n$ & $E_n$~~~ \\
\midrule
$0$ & $-22.4357$ \\
$1$ & $-18.6885$ \\
$2$ & $-14.8995$ \\
$3$ & $-11.1005$ \\
$4$ & $-7.31152$ \\
$5$ & $-3.56427$ \\
\bottomrule
\end{tabular}
\end{table}

\clearpage
\begin{remark}[On the algebraic equation \eqref{eq:4-algebraic}]
Although we do not have a general formula for solving the $\ell$-th degree algebraic equation \eqref{eq:4-algebraic}, it turns out that the $\ell$ roots have the following property: \textit{if $E=-1-2\ell+\alpha$ is a root of Eq. \eqref{eq:4-algebraic}, $E=-1-2\ell-\alpha$ is also a root.}
Here, we give the proof:

\begin{proof}
We change the variable $E\to \tilde{E}-1-2\ell$.
By this, the equation \eqref{eq:4-algebraic} becomes
\begin{equation}
\prod_{k=1}^{\ell}(\tilde{E}+4k-2\ell-3) + \prod_{k=1}^{\ell}(\tilde{E}+4k-2\ell-1) = 0 ~.
\label{eq:4-algebraic2}
\end{equation}
Now under the transformation: $\tilde{E}\to -\tilde{E}$, the equation transforms
\begin{align*}
&\prod_{k=1}^{\ell}(-\widetilde{E}+4k-2\ell-3) + \prod_{k=1}^{\ell}(-\widetilde{E}+4k-2\ell-1) \\
=\,&(-1)^{\ell}\left( \prod_{k=1}^{\ell}(\widetilde{E}-4k+2\ell+3) + \prod_{k=1}^{\ell}(\widetilde{E}-4k+2\ell+1) \right) \\
=\,&(-1)^{\ell}\left( \prod_{k=1}^{\ell}(\widetilde{E}+4k-2\ell-3) + \prod_{k=1}^{\ell}(\widetilde{E}+4k-2\ell-1) \right) 
= (-1)^{\ell} \times \text{[l.h.s. of Eq. \eqref{eq:4-algebraic2}]} ~.
\end{align*}
Therefore, if $\tilde{E}=\alpha$ satisfies Eq. \eqref{eq:4-algebraic2}, $\tilde{E}=-\alpha$ also satisfies the equation. 
\end{proof}

\bigskip
Note that this property guarantees even numbers of roots of the algebraic equation.
For the odd-$\ell$ case, $E=-1-2\ell$ is also a root of Eq. \eqref{eq:4-algebraic}, which corresponds to $\alpha = 0$ above.
\end{remark}

\subsection{Discussions}
\subsubsection{Wigner distribution function}
We aim to explore the behavior of a particle under the potential \eqref{eq:4-2-pot} by using the analogy of classical dynamics. 
Here, the idea of formulating quantum mechanics in the phase space is a powerful tool to do so, where the Wigner quasiprobability distribution~\cite{PhysRev.40.749} plays a central role (See, \textit{e.g.}, Ref. \cite{curtright2013concise}). 
Numerous analyses of the Wigner function have been carried out in this context (See, \textit{e.g.}, Ref. \cite{nagiyev2023wigner} and references therein).

We compute the Wigner distribution function:
\begin{equation}
\mathcal{W}(p,x) = \frac{1}{4\pi^2}\iint d\lambda d\mu \, \mathrm{e}^{-\mathrm{i}(\lambda p+\mu x)} \bra{\psi} \mathrm{e}^{\mathrm{i}\lambda\hat{p}+\mathrm{i}\mu\hat{x}} \ket{\psi} ~,
\label{eq:4-Wdf_bk}
\end{equation}
of the system~\eqref{eq:4-2-pot}.
In terms of the eigenfunctions, Eq.~\eqref{eq:4-Wdf_bk} is reduced to 
\begin{equation}
\mathcal{W}_n(p,x) = \frac{1}{2\pi}\int_{-\infty}^{\infty} \psi_n^*\left( x-\frac{1}{2}\bar{x} \right)\psi_n\left( x+\frac{1}{2}\bar{x} \right) \mathrm{e}^{-\mathrm{i}p\bar{x}} \,d\bar{x} ~,
\label{eq:4-Wdf}
\end{equation}
where $\psi_n(x)$ is normalized, $\displaystyle\int_{x_1}^{x_2} |\psi_n(x)|^2 \,dx = 1$.
Note that the Wigner function is a real function, $\mathcal{W}_n(p,x) \in \mathbb{R}$.

In our problem, the potential is defined on the whole real line, $x \in (-\infty,\infty)$, and the wavefunctions are piecewise analytic functions, symbolically in the following way:
\begin{equation}
\psi_n (x) = \begin{cases}
	\psi^{(-)}_n(x) & x \leqslant 0 \\
	\psi^{(+)}_n(x) & x \geqslant 0
\end{cases} ~.
\label{eq:pa_eigenf}
\end{equation}
Therefore, the Wigner distribution function is
\begin{align}
\mathcal{W}_n(p,x) =
\frac{1}{2\pi} \bigg[ &\int_{-\infty}^{2x} \psi_n^{(+)}\left( x-\frac{\bar{x}}{2} \right)\psi_n^{(-)}\left( x+\frac{\bar{x}}{2} \right) \mathrm{e}^{-\mathrm{i}p\bar{x}} \,d\bar{x} \nonumber \\
&+ \int_{2x}^{-2x} \psi_n^{(-)}\left( x-\frac{\bar{x}}{2} \right)\psi_n^{(-)}\left( x+\frac{\bar{x}}{2} \right) \mathrm{e}^{-\mathrm{i}p\bar{x}} \,d\bar{x} \nonumber \\
&+ \int_{-2x}^{\infty} \psi_n^{(-)}\left( x-\frac{\bar{x}}{2} \right)\psi_n^{(+)}\left( x+\frac{\bar{x}}{2} \right) \mathrm{e}^{-\mathrm{i}p\bar{x}} \,d\bar{x} \bigg]
\end{align}
for $x \leqslant 0$, and
\begin{align}
\mathcal{W}_n(p,x) =
\frac{1}{2\pi} \bigg[ &\int_{-\infty}^{-2x} \psi_n^{(+)}\left( x-\frac{\bar{x}}{2} \right)\psi_n^{(-)}\left( x+\frac{\bar{x}}{2} \right) \mathrm{e}^{-\mathrm{i}p\bar{x}} \,d\bar{x} \nonumber \\
&+ \int_{-2x}^{2x} \psi_n^{(+)}\left( x-\frac{\bar{x}}{2} \right)\psi_n^{(+)}\left( x+\frac{\bar{x}}{2} \right) \mathrm{e}^{-\mathrm{i}p\bar{x}} \,d\bar{x} \nonumber \\
&+ \int_{2x}^{\infty} \psi_n^{(-)}\left( x-\frac{\bar{x}}{2} \right)\psi_n^{(+)}\left( x+\frac{\bar{x}}{2} \right) \mathrm{e}^{-\mathrm{i}p\bar{x}} \,d\bar{x} \bigg]
\end{align}
for $x \geqslant 0$.

In the case of the ordinary harmonic oscillator, $a\to 0$, it is well-known that the Wigner distribution function~\eqref{eq:4-Wdf} is computed analytically, and one obtains
\begin{equation}
\mathcal{W}_n(p,x) = \frac{(-1)^n}{\pi}\mathrm{e}^{-(p^2+x^2)}L_n^{(0)}\left( 2(p^2+x^2) \right) ~,
\end{equation}
where $L_n^{(\alpha)}(x)$ denotes a Laguerre polynomial of degree $n$.
Here, the Wigner function is circular symmetric.

\begin{figure}[p]
\centering
	\begin{minipage}[t]{0.48\linewidth}
	\centering
	\includegraphics[width=\linewidth]{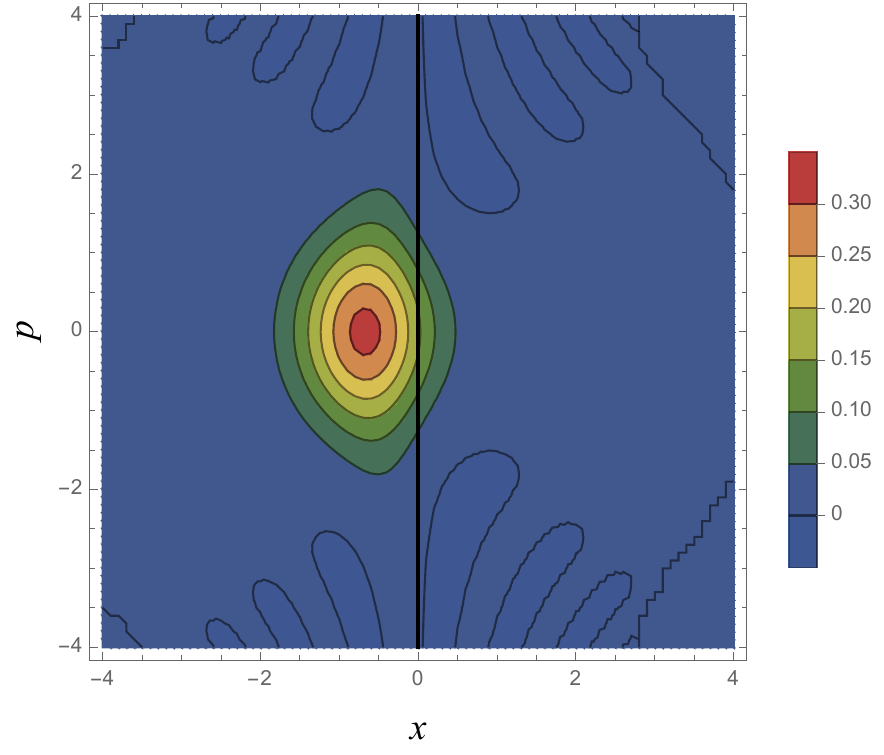}
	\subcaption{$\ell=1$.}
	\label{}
	\end{minipage}
	\quad
	\begin{minipage}[t]{0.48\linewidth}
	\centering
	\includegraphics[width=\linewidth]{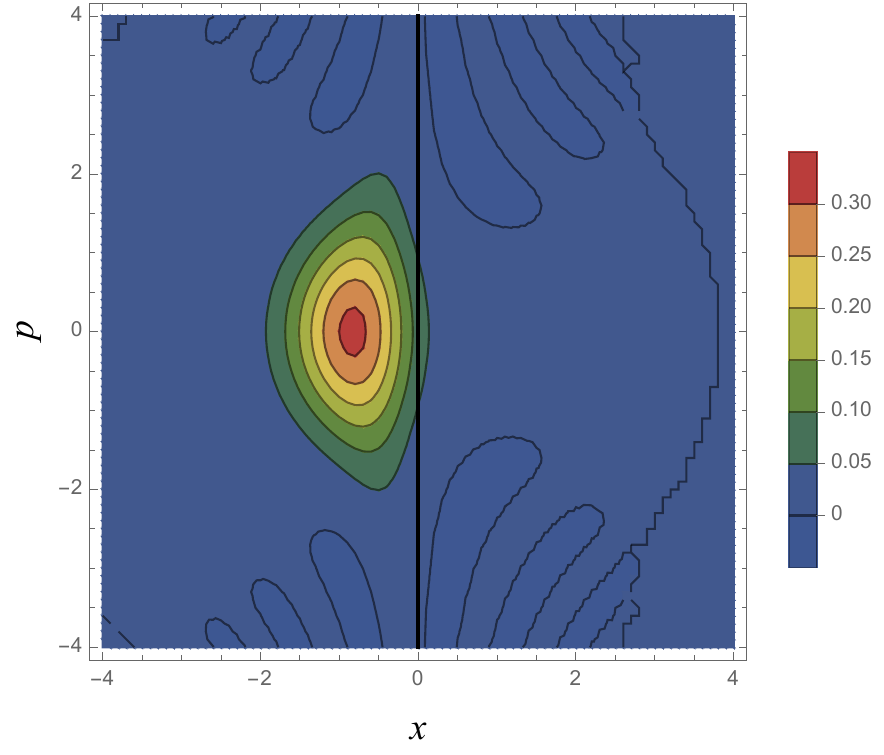}
	\subcaption{$\ell=2$.}
	\label{}
	\end{minipage}
	
	\medskip
	\begin{minipage}[t]{0.48\linewidth}
	\centering
	\includegraphics[width=\linewidth]{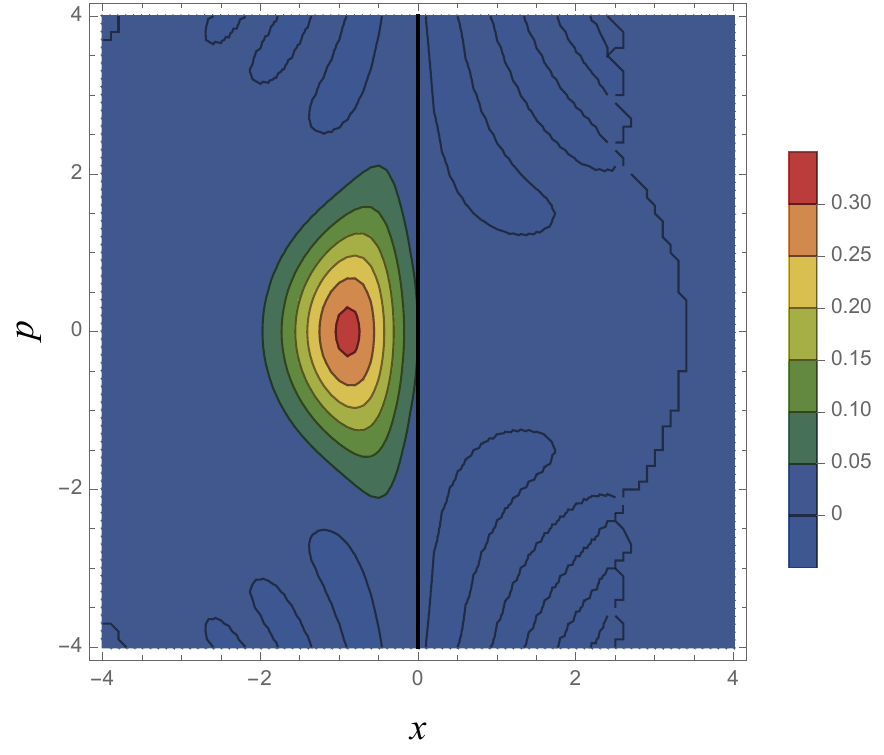}
	\subcaption{$\ell=3$.}
	\label{}
	\end{minipage}
	\quad
	\begin{minipage}[t]{0.48\linewidth}
	\centering
	\includegraphics[width=\linewidth]{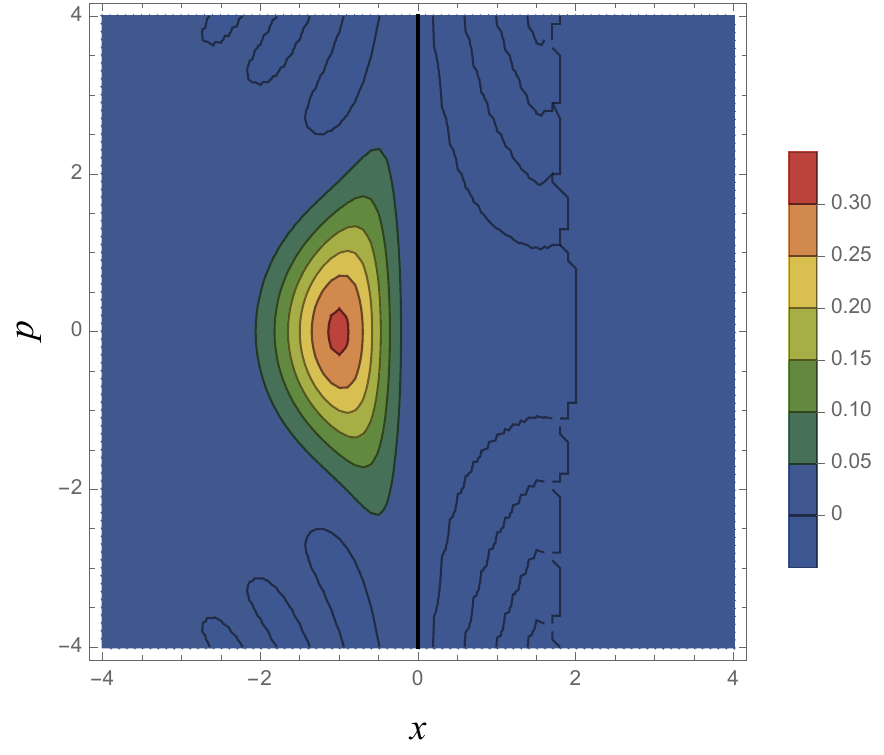}
	\subcaption{$\ell=10$.}
	\label{}
	\end{minipage}
\caption[Wigner distribution functions \eqref{eq:4-Wdf} for $n=0$ with several choices of $\ell$'s.]
	{Wigner distribution functions \eqref{eq:4-Wdf} for $n=0$ with several choices of $\ell$'s.
	These ground-states are not Hermite-polynomially solvable.}
\label{fig:4-Wdf_0}
\end{figure}

\begin{figure}[p]
\centering
	\begin{minipage}[t]{0.48\linewidth}
	\centering
	\includegraphics[width=\linewidth]{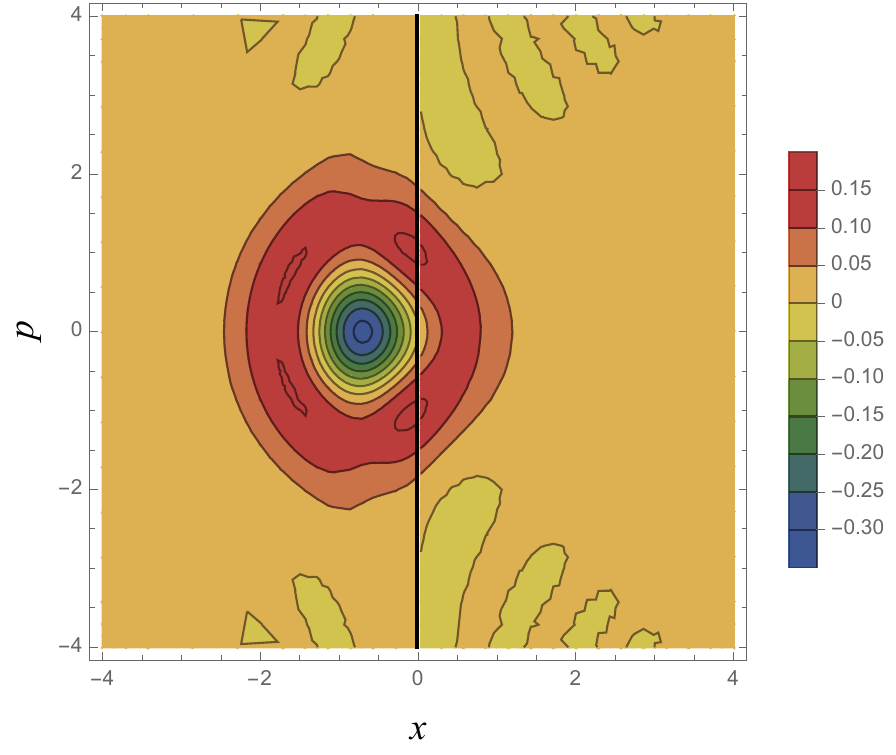}
	\subcaption{$\ell=1, n=1$.}
	\label{}
	\end{minipage}
	\quad
	\begin{minipage}[t]{0.48\linewidth}
	\centering
	\includegraphics[width=\linewidth]{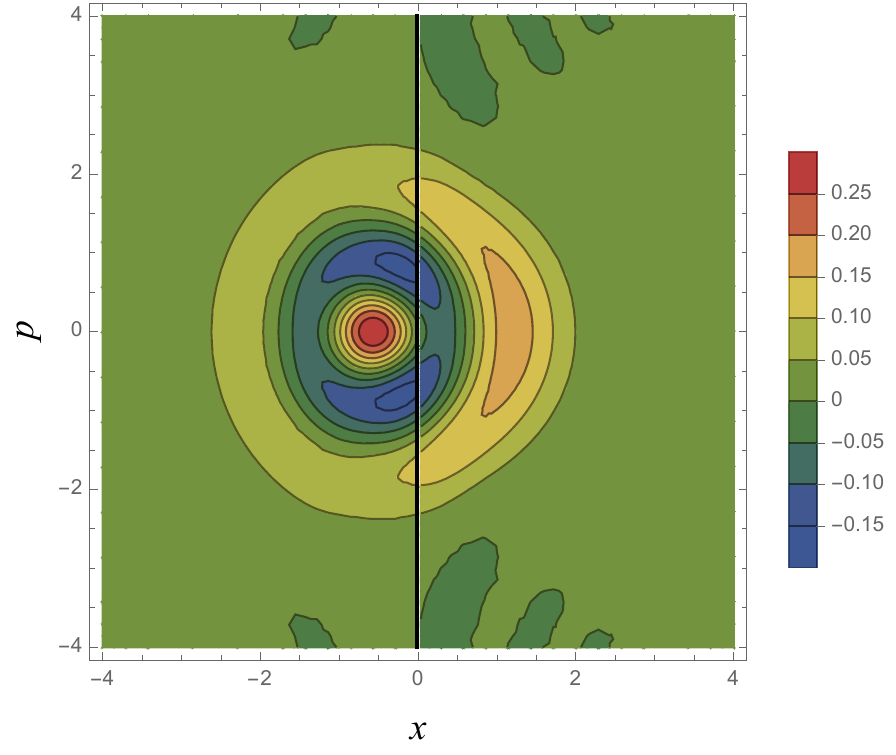}
	\subcaption{$\ell=1, n=2$.}
	\label{}
	\end{minipage}
	
	\medskip
	\begin{minipage}[t]{0.48\linewidth}
	\centering
	\includegraphics[width=\linewidth]{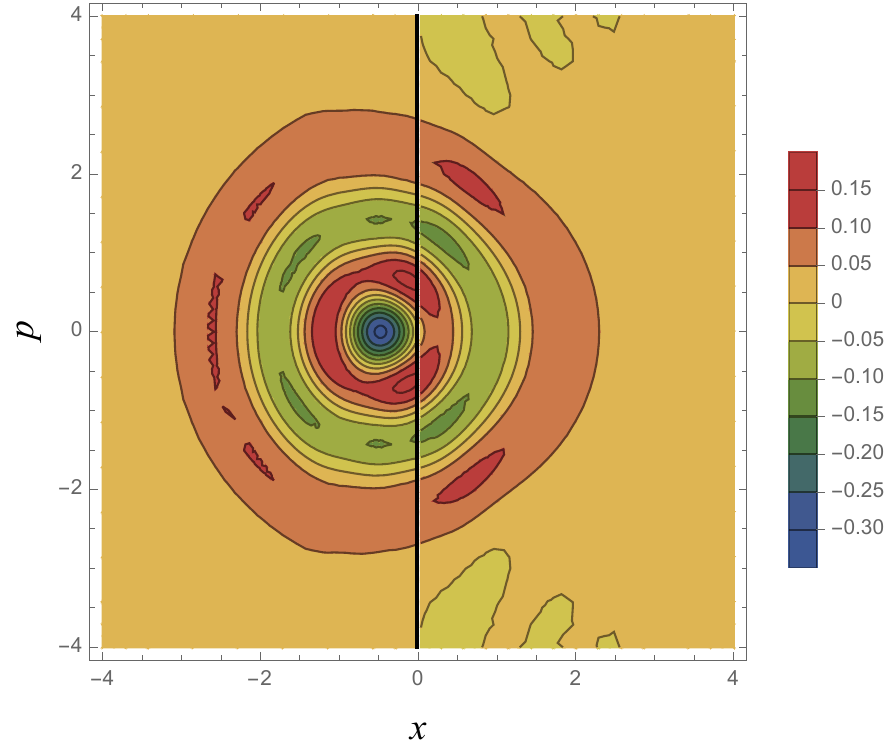}
	\subcaption{$\ell=1, n=3$.}
	\label{}
	\end{minipage}
	\quad
	\begin{minipage}[t]{0.48\linewidth}
	\centering
	\includegraphics[width=\linewidth]{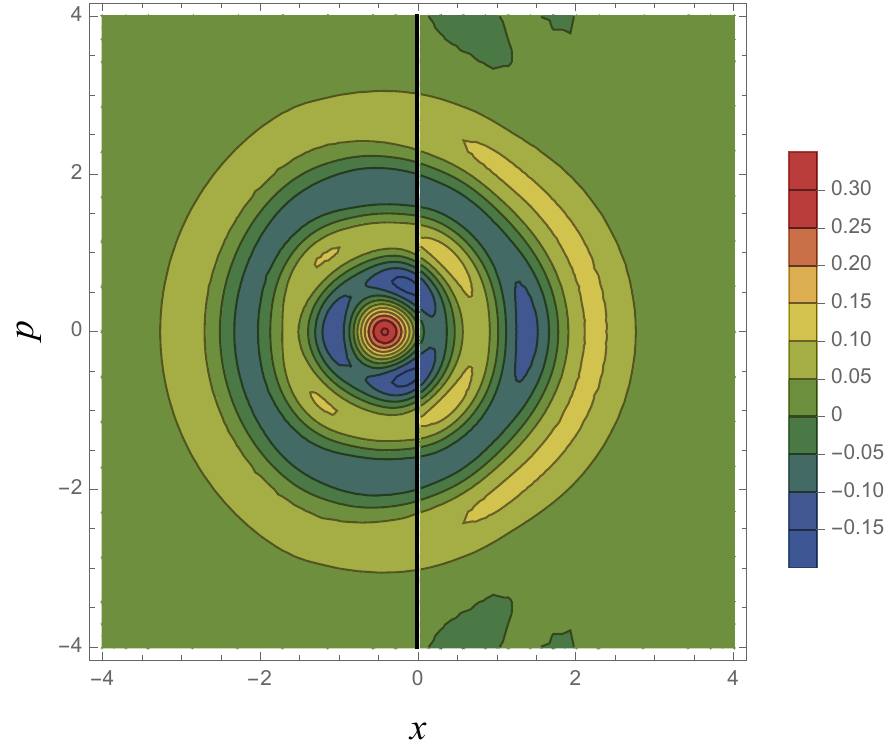}
	\subcaption{$\ell=1, n=4$.}
	\label{}
	\end{minipage}
	
	\medskip
	\begin{minipage}[t]{0.48\linewidth}
	\centering
	\includegraphics[width=\linewidth]{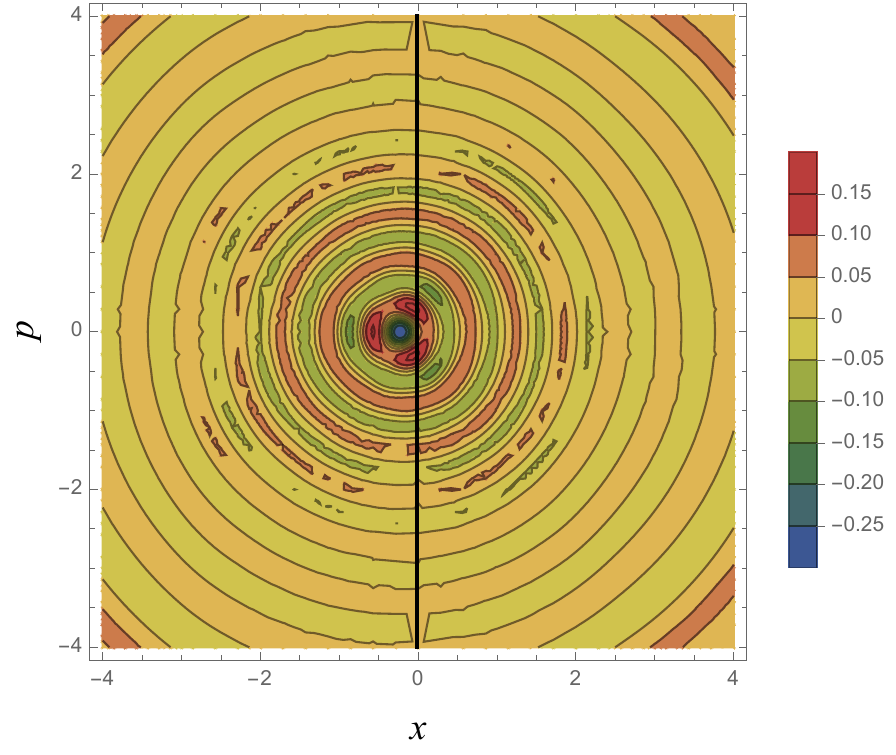}
	\subcaption{$\ell=1, n=15$.}
	\label{}
	\end{minipage}
	\quad
	\begin{minipage}[t]{0.48\linewidth}
	\centering
	\includegraphics[width=\linewidth]{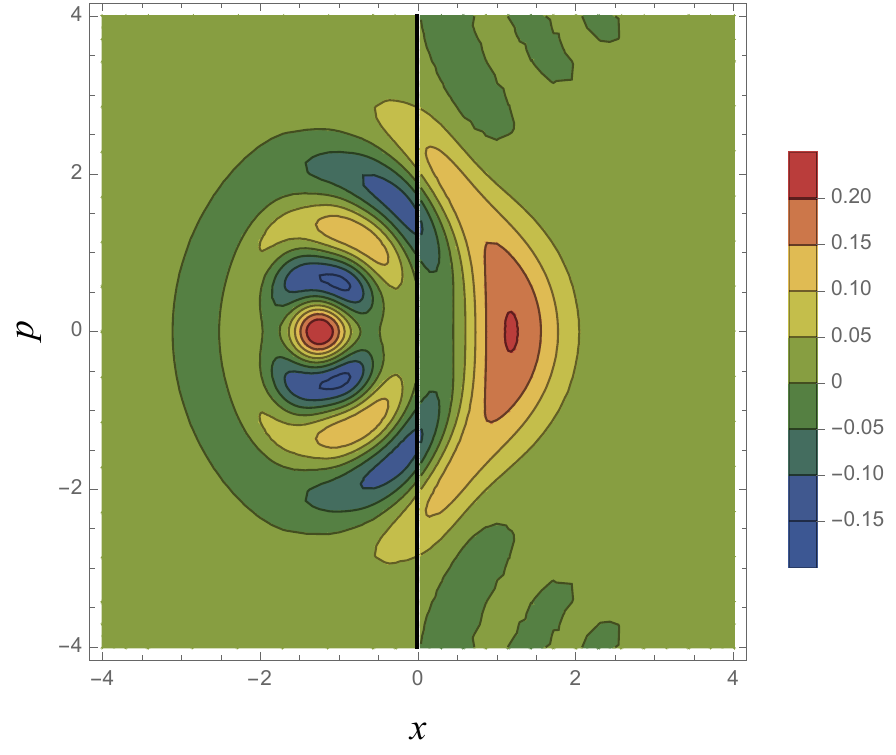}
	\subcaption{$\ell=3, n=4$.}
	\label{}
	\end{minipage}
\caption[Wigner distribution functions \eqref{eq:4-Wdf} for several choices of $(\ell,n)$.]
	{Wigner distribution functions \eqref{eq:4-Wdf} for several choices of $(\ell,n)$.
	Our choices of $n$ here are all corresponding to Hermite-polynomially solvable states.}
\label{fig:4-Wdf_H}
\end{figure}

On the other hand, we do not have the closed-form expression of the Wigner function for arbitrary $a$ (or $\ell$) and $n$ in our current problem~\eqref{eq:4-2-pot}.
With specific choices of $\ell$ and $n$, one could obtain the closed-form expressions, but it would be too complicated to show here.
Instead, we display several numerical calculations of the Wigner function in Figs. \ref{fig:4-Wdf_0} and \ref{fig:4-Wdf_H}.
Our analyses are exclusively focused on the cases with $a=4\ell$ for simplicity.

\begin{example}[$n=0$]
We show the Wigner functions of the ground states of this system~\eqref{eq:4-2-pot} with various $\ell$'s in Fig. \ref{fig:4-Wdf_0}.
\end{example}

\begin{example}[$n\geqslant\ell$]
Here, we show the Wigner function of the eigenstates whose wavefunctions are expressed in terms of the Hermite polynomials, in Fig. \ref{fig:4-Wdf_H}.
\end{example}

Our results show that, in comparison to the case of the ordinary harmonic oscillator, the Wigner functions remain symmetric in the $p$-direction, while the finite jump at $x=0$ of the potential~\eqref{eq:4-2-pot} causes asymmetry in the $x$-direction.
For $n=0$, the Wigner function tends to localize in $x<0$, and as $\ell$ grows, less and less probability density is seen in $x>0$, which is as expected because a larger $\ell$ means a deeper pocket of the potential well to trap a particle.

An interesting property of the Wigner function of the system is that the distribution is anisotropic in the phase space.
The local extrema appear in certain directions.
Although it seems that the Wigner functions are circular symmetric in a wide view for $n-\ell \gg 1$, they continue to be asymmetric on closer look around the origin of the phase space.
The circular symmetry will never be restored at $n-\ell\to\infty$.

We would like to further point out that the Wigner distribution functions of the system~\eqref{eq:4-2-pot} take on negative values for any $n$.
For $n \geqslant 1$, we have regions of negative values concentrically in the phase space, which is also observed in the case of the ordinary harmonic oscillator.
Our model has other negative-valued regions outside the concentric ones with larger $|p|$'s even for $n=0$.
This is also attributed to the finite jump of the potential energy.

\subsubsection{On the isospectral properties}
Let $\mathcal{H}_{\ell}^{[0]}$ denote the Hamiltonian of our one-dimensional quantum mechanical system with $a=4\ell$ \eqref{eq:4-SE4l}:
\begin{equation}
\mathcal{H}_{\ell}^{[0]} = -\frac{d^2}{dx^2} + V(x;4\ell) ~.
\end{equation}
Also, we write $E_{\ell,n}$ and $\psi_{\ell,n}^{[0]}$ ($n=0,1,2,\ldots$) for the energy eigenvalues and the corresponding eigenfunctions hereafter.

According to the Crum's theorem~\cite{10.1093/qmath/6.1.121}, there are infinitely many associated Hamiltonian systems $\mathcal{H}_{\ell}^{[M]}, M=1,2,\ldots$, which are essentially isospectral to $\mathcal{H}_{\ell}^{[0]}$.
They are
\begin{equation}
\mathcal{H}_{\ell}^{[M]} \coloneqq 
\mathcal{H}_{\ell}^{[0]} - 2\frac{d^2}{dx^2}\ln\mathrm{W}\left[ \psi_{\ell,0}^{[0]},\psi_{\ell,1}^{[0]},\psi_{\ell,2}^{[0]},\ldots,\psi_{\ell,M-1}^{[0]} \right](x) ~,
\label{eq:4-Ham_DC}
\end{equation}
in which $\mathrm{W}[f_1,\ldots,f_m](x)$ is the Wronskian defined as 
\begin{equation}
\mathrm{W}[f_1,\ldots,f_m](x) \coloneqq \det\left( \frac{d^{j-1}f_k(x)}{dx^{j-1}} \right)_{1\leqslant j,k\leqslant m} ~.
\end{equation}
$\mathcal{H}_{\ell}^{[M]}$ is such a system that the $M$ lowest eigenstates are deleted from $\mathcal{H}_{\ell}^{[0]}$, and shares all the eigenvalues above $E_{\ell,M}$ with $\mathcal{H}_{\ell}^{[0]}$.
The corresponding eigenfunctions $\{ \psi_{\ell,n}^{[M]} \}$ are related to $\{ \psi_{\ell,n}^{[0]} \}$ by the Dourboux--Crum transformation~\cite{darboux,10.1093/qmath/6.1.121}:
\begin{equation}
\psi_{\ell,n}^{[M]}(x) =
\frac{\mathrm{W}\left[ \psi_{\ell,0}^{[0]},\psi_{\ell,1}^{[0]},\ldots,\psi_{\ell,M-1}^{[0]},\psi_{\ell,n+M}^{[0]} \right](x)}{\mathrm{W}\left[ \psi_{\ell,0}^{[0]},\psi_{\ell,1}^{[0]},\ldots,\psi_{\ell,M-1}^{[0]} \right](x)} ~,
\end{equation}
which satisfies the following Schr\"{o}dinger equation: 
\begin{equation}
\mathcal{H}_{\ell}^{[M]}\psi_{\ell,n}^{[M]}(x) = E_{\ell,n+M}\psi_{\ell,n}^{[M]}(x) ~,~~~
n=0,1,2,\ldots ~.
\end{equation}

\begin{example}
Let us take $\ell=1$ and $M=1$ as an example.
In the context of supersymmetric quantum mechanics, the associated Hamiltonian with $M=1$ is often referred to as the (supersymmetric) partner.
Here, the Hamiltonian is 
\begin{equation}
\mathcal{H}_{1}^{[1]} = \mathcal{H}_{1}^{[0]} - 2\frac{d^2}{dx^2}\ln\psi_{1,0}^{[0]}(x) ~,~~~
\mathcal{H}_{1}^{[1]}\psi_{1,n}^{[1]}(x) = E_{1,n+1}\psi_{1,n}^{[1]}(x) ~,~~~
n=0,1,2,\ldots ~,
\end{equation}
where
\begin{equation}
\psi_{1,n}^{[1]}(x) =
\frac{\mathrm{W}\left[ \psi_{1,0}^{[0]},\psi_{1,n+1}^{[0]} \right](x)}{\psi_{1,0}^{[0]}(x)} ~.
\end{equation}
We plot them for the first several eigenstates in Fig. \ref{fig:4-sol_l1M1}.

Note from the explicit calculation that $\mathcal{H}_{1}^{[0]}$ and $\mathcal{H}_{1}^{[1]}$ are not shape invariant~\cite{gendenshtein1983derivation}.
The same can be applied to $\mathcal{H}_{\ell}^{[M]}$ and $\mathcal{H}_{\ell}^{[M+1]}$.
One might guess that the Hermite-polynomial solvability of our system is due to the shape invariance as in the case of the harmonic oscillator.
However, this is not the case.

\begin{figure}[t]
\centering
\includegraphics[scale=0.8]{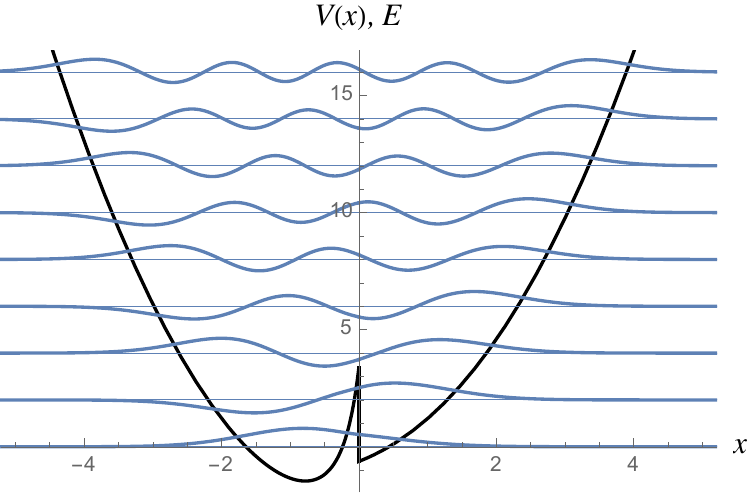}
\caption[The solutions of the eigenvalue problem \eqref{eq:4-Ham_DC} with $\ell=1$ and $M=1$.]
	{The solutions of the eigenvalue problem \eqref{eq:4-Ham_DC} with $\ell=1$ and $M=1$.
	Thin blue lines show the energy spectrum, and the blue curve on each line is the corresponding eigenfunction.
	The potential of this system is also plotted in this figure by a black curve.}
\label{fig:4-sol_l1M1}
\end{figure}
\end{example}

\begin{figure}[t]
\centering
\includegraphics[scale=0.8]{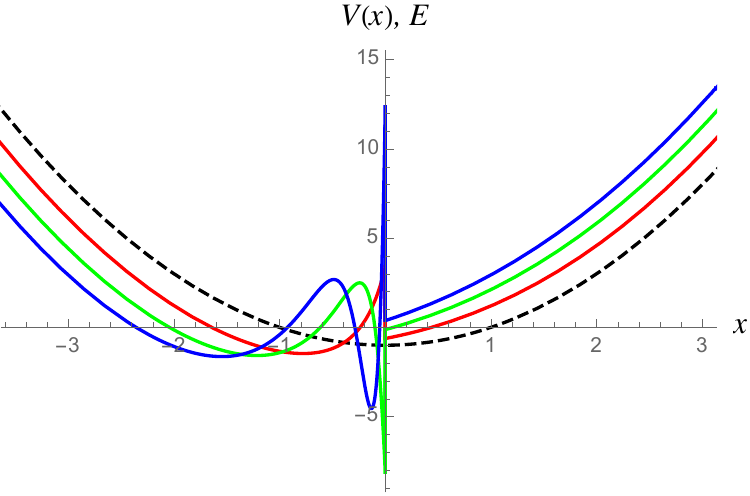}
\caption[The sequence $\{ \mathcal{H}_{\ell}^{[\ell]} \}$.]
	{The sequence $\{ \mathcal{H}_{\ell}^{[\ell]} \}$. 
	The potentials for $\ell=1,2,3$ are plotted in red, green and blue respectively.
	Those potentials are all isospectral to the 1-d harmonic oscillator potential (dashed black curve).
	$\mathcal{H}_{\ell}^{[\ell]}$'s are non-analytic at $x=0$, but never diverge.}
\label{fig:4-iso_sequence}
\end{figure}

We choose $M=\ell$ here.
The resulting Hamiltonian is 
\begin{equation}
\mathcal{H}_{\ell}^{[\ell]} =
\mathcal{H}_{\ell}^{[0]} - 2\frac{d^2}{dx^2}\ln\mathrm{W}\left[ \psi_{\ell,0}^{[0]},\psi_{\ell,1}^{[0]},\psi_{\ell,2}^{[0]},\ldots,\psi_{\ell,\ell-1}^{[0]} \right](x) ~,
\label{eq:ham_iso}
\end{equation}
which corresponds to the deletion of all negative-energy states of $\mathcal{H}_{\ell}^{[0]}$.
This Hamiltonian is strictly isospectral to the 1-dim. harmonic oscillator potential $\mathcal{H}_{\rm HO}(x)=x^2-1$.
Since $\ell$ can be any positive integer, we now have infinitely many isospectral potentials of the harmonic oscillator in our procedure above.
We plot the first several potentials of the sequence $\{ \mathcal{H}_{\ell}^{[\ell]} \}$ ($\ell = 1,2,\ldots$) in Fig. \ref{fig:4-iso_sequence}.

\begin{remark}[Krein--Adler transformation]
A further generalization of the isospectral deformation, \textit{i.e.}, deletions of the eigenstates from the original system, was formulated by Krein and Adler independently~\cite{krein1957continuous,adler1994modification}.
During this deformation, the eigenstates with the following indices are deleted:
\begin{equation}
\mathcal{D} \coloneqq \{ d_1,d_1+1 < d_2,d_2+1 < \cdots < d_N,d_N+1 \} ~,~~~
d_1,\ldots,d_N \in \mathbb{Z}_{\geqslant 0} ~,
\end{equation}
where taking $d_1=0$ and $d_{j+1}=d_j+2$ for all $j$ corresponds to the case of the Crum's theorem.
The Hamiltonian is 
\begin{equation}
\mathcal{H}_{\ell}^{\mathcal{D}} \coloneqq 
\mathcal{H}_{\ell}^{[0]} - 2\frac{d^2}{dx^2}\ln\mathrm{W}\left[ \psi_{\ell,d_1}^{[0]},\psi_{\ell,d_1+1}^{[0]},\ldots,\psi_{\ell,d_N+1}^{[0]} \right](x) ~,
\label{eq:4-Ham_KA}
\end{equation}
which shares all the energy spectrum with $\mathcal{H}_{\ell}^{[0]}$ except that those indexed by $\mathcal{D}$ are deleted.
The eigenfunctions are
\begin{equation}
\psi_{\ell,\tilde{n}}^{\mathcal{D}}(x) \coloneqq 
\frac{\mathrm{W}\left[ \psi_{\ell,d_1}^{[0]},\psi_{\ell,d_1+1}^{[0]},\ldots,\psi_{\ell,d_N+1}^{[0]},\psi_{\ell,\tilde{n}}^{[0]} \right](x)}{\mathrm{W}\left[  \psi_{\ell,d_1}^{[0]},\psi_{\ell,d_1+1}^{[0]},\ldots,\psi_{\ell,d_N+1}^{[0]} \right](x)} ~,~~~
\tilde{n} \in \mathbb{Z}_{\geqslant 0} \backslash \mathcal{D} ~,
\end{equation}
satisfying
\begin{equation}
\mathcal{H}_{\ell}^{\mathcal{D}}\psi_{\ell,\tilde{n}}^{\mathcal{D}}(x) = E_{\ell,\tilde{n}}\psi_{\ell,\tilde{n}}^{\mathcal{D}}(x) ~.
\end{equation}

\begin{example}
Taking $\ell=1$ and $\mathcal{D}=\{ 1,2 \}$, we get
\begin{equation}
\mathcal{H}_{1}^{\{ 1,2 \}} = \mathcal{H}_{1}^{[0]} - 2\frac{d^2}{dx^2}\ln\mathrm{W}\left[ \psi_{1,1}^{[0]},\psi_{1,2}^{[0]} \right](x) ~,~~~
\mathcal{H}_{1}^{\{ 1,2 \}}\psi_{1,\tilde{n}}^{\{ 1,2 \}}(x) = E_{1,\tilde{n}}\psi_{1,\tilde{n}}^{\{ 1,2 \}}(x) ~,~~~
\tilde{n}=0,3,4,\ldots ~,
\end{equation}
where
\begin{equation}
\psi_{1,\tilde{n}}^{\{ 1,2 \}}(x) =
\frac{\mathrm{W}\left[ \psi_{1,1}^{[0]},\psi_{1,2}^{[0]},\psi_{\ell,\tilde{n}}^{[0]} \right](x)}{\mathrm{W}\left[ \psi_{1,1}^{[0]},\psi_{1,2}^{[0]} \right](x)} ~,
\end{equation}
(See Fig.\,\ref{fig:4-sol_KA_12}).
\end{example}

\begin{figure}[t]
\centering
\includegraphics[scale=0.8]{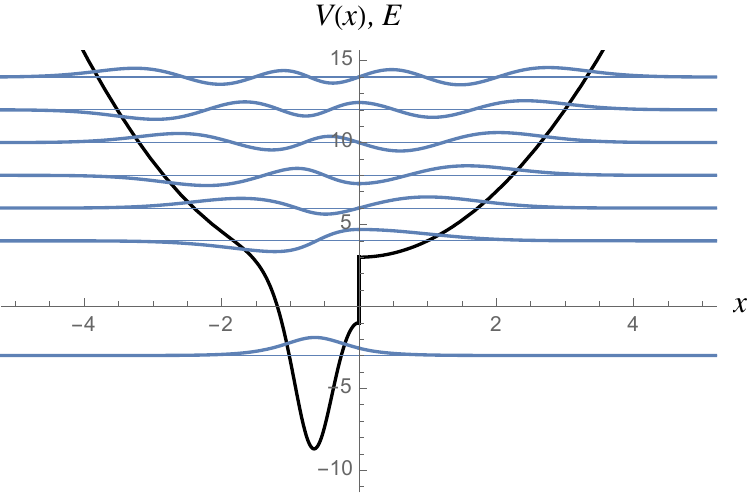}
\caption[The solutions of the eigenvalue problem \eqref{eq:4-Ham_KA} with $\ell=1$ and $\mathcal{D}=\{ 1,2 \}$.]
	{The solutions of the eigenvalue problem \eqref{eq:4-Ham_KA} with $\ell=1$ and $\mathcal{D}=\{ 1,2 \}$.
	Thin blue lines show the energy spectrum, and the blue curve on each line is the corresponding eigenfunction.
	The potential of this system is also plotted in this figure by a black curve.}
\label{fig:4-sol_KA_12}
\end{figure}
\end{remark}

\subsection{Harmonic Oscillator with a Step and a Ramp}
\begin{figure}[t]
\centering
\includegraphics[scale=0.7]{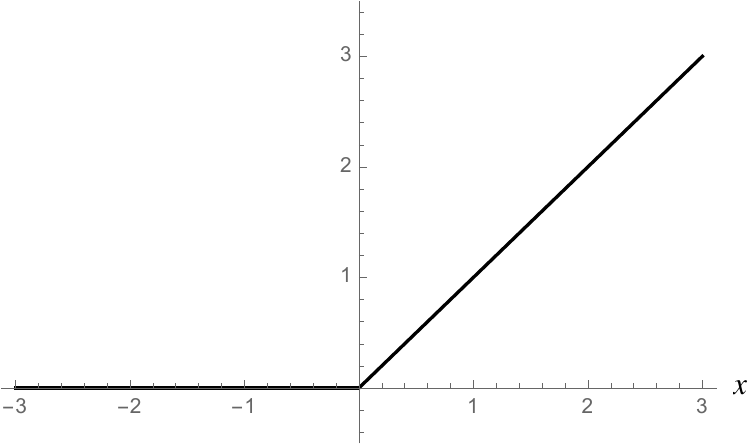}
\caption{The ramp function, $x\,\theta(x)$.}
\label{fig:4-RampF}
\end{figure}

We add a linear potential $-gx$ to the potential \eqref{eq:4-2-pot} for $x<0$, 
\begin{equation}
V(x) = \begin{cases}
	x^2 - 1 - a - gx = \left( x - \dfrac{g}{2} \right)^2 - 1 - a - \dfrac{g^2}{4} & (x<0) \\[1ex]
	x^2 - 1 & (x>0)
\end{cases} ~,
\label{eq:4-2-pot-ag}
\end{equation}
where $g$ is a real constant.
This is also a confining potential and has infinitely many discrete eigenvalues $\{ E_n \}$.
Taking $g=0$ coincides with the potential \eqref{eq:4-2-pot}.
Note that the function $x\,\theta(x)$ is often referred to as the ramp function, which is named after the shape of its graph (See Fig. \ref{fig:4-RampF}).

\begin{remark}
In this paper, we restrict ourselves to $a>0$ in the potential \eqref{eq:4-2-pot-ag}.
Note that, unlike the case of a harmonic oscillator with a step \eqref{eq:4-2-pot}, the constraint on $a$ breaks the generality.
For $a<0$, discussions are almost parallel to those for $a>0$ (See the following), but a kind of double-well potentials appear and they are more likely to be physically applicable.
\end{remark}

\subsubsection{The solutions}
One can construct the eigenfunctions for arbitrary $a$ and $g$:
\begin{equation}
\psi_n(x) =  \begin{cases}
	\mathrm{e}^{-\frac{(x-\frac{g}{2})^2}{2}}\left[ \alpha_- \,{}_1F_1\left( -\dfrac{E_n+a+\frac{g^2}{4}}{4};\dfrac{1}{2};\left( x-\dfrac{g}{2} \right)^2 \right) \right. & \\
	\left. \hspace{0.175\linewidth} + \beta_- \left( x-\dfrac{g}{2} \right)\, {}_1F_1\left( -\dfrac{E_n+a+\frac{g^2}{4}-2}{4};\dfrac{3}{2};\left( x-\dfrac{g}{2} \right)^2 \right)  \right] 
	& (x<0) \\[3ex]
	\mathrm{e}^{-\frac{x^2}{2}}\left[ \alpha_+ \,{}_1F_1\left( -\dfrac{E_n}{4};\dfrac{1}{2};x^2 \right) + \beta_+\, x\, {}_1F_1\left( -\dfrac{E_n-2}{4};\dfrac{3}{2};x^2 \right)  \right] 
	& (x>0)
\end{cases} ~,
\label{eq:4-wf-ag}
\end{equation}
in which $\alpha_{\pm},\beta_{\pm}$ are constants. 
From the boundary conditions at $x=0$ \eqref{eq:4-BC_origin}, $\alpha_{+}$ and $\beta_{+}$ are
\[
\alpha_{+} = \psi_n(0^-) ~,~~~
\beta_{+} = \frac{d\psi_n(0^-)}{dx} ~.
\]
On the other hand, those at $x\to\pm\infty$ \eqref{eq:4-BC_inf} yield the following simultaneous transcendental equations:
\begin{equation}
\frac{1}{\psi_n(0^-)}\frac{d\psi_n(0^-)}{dx} = -\frac{2\varGamma\left( -\frac{E_n-2}{4} \right)}{\varGamma\left( -\frac{E_n}{4} \right)} ~,~~~
\frac{\beta_-}{\alpha_-} = \frac{2\varGamma\Big( -\frac{E_n+a+\frac{g^2}{4}-2}{4} \Big)}{\varGamma\Big( -\frac{E_n+a+\frac{g^2}{4}}{4} \Big)} ~,
\label{eq:4-trans-ag}
\end{equation}
which are to be solved graphically, and determine the energy eigenvalues $\{ E_n \}$ as is shown in the following example.

\begin{figure}[t]
\centering
\includegraphics[scale=0.8]{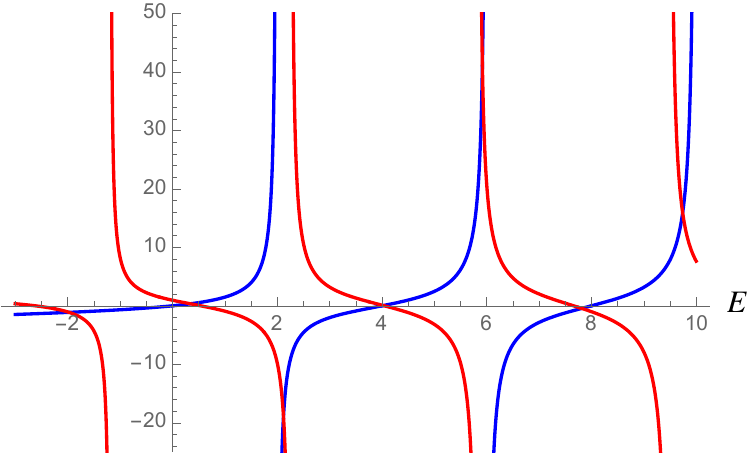}
\caption[Graphical solution of Eqs. \eqref{eq:4-trans-ag}.]
	{Graphical solution of Eqs. \eqref{eq:4-trans-ag}.
	The blue curves correspond to the left-hand side of the equation, while the red ones are the right-hand side.
	The intersections of these curves determine the energy eigenvalues.
	The numerical solutions are displayed in Tab. \ref{tab:4-Eigenvalues_a2g1}.}
\label{fig:4-GraphicalSol_a2g1}
\end{figure}

\begin{example}
\label{ex:4-a2g1}
We first solve equations \eqref{eq:4-trans-ag} with $a=2$ and $g=1$ to obtain the energy spectrum (See Fig. \ref{fig:4-GraphicalSol_a2g1}).
The first several energy eigenvalues are displayed in Tab. \ref{tab:4-Eigenvalues_a2g1} with six digits.
With the knowledge of the energy spectrum, one can determine the coefficients $\alpha_{\pm},\beta_{\pm}$ for each $n$, and therefore the eigenfunction $\psi_n(x)$.
The solution of the Schr\"{o}dinger equation with the potential \eqref{eq:4-2-pot-ag} with $a=2$ and $g=1$ is summarized in Fig. \ref{fig:4-sol_a2g1}.

\begin{table}[p]
\caption[First several energy eigenvalues for \eqref{eq:4-2-pot-ag} with $a=2$ and $g=1$ with six digits.]
	{First several energy eigenvalues for \eqref{eq:4-2-pot-ag} with $a=2$ and $g=1$ with six digits.
	These values are obtained by solving Eq. \eqref{eq:4-trans-ag} or finding the intersections in Fig.\ref{fig:4-GraphicalSol_a2g1} numerically.}
\label{tab:4-Eigenvalues_a2g1}
\centering
\begin{tabular}{cr}
\toprule
$n$ & $E_n$~~~ \\
\midrule
$0$ & $-1.97196$ \\
$1$ & $0.343665$ \\
$2$ & $2.12101$ \\
$3$ & $4.02740$ \\
$4$ & $5.91817$ \\
$5$ & $7.81348$ \\
\bottomrule
\end{tabular}
\end{table}

\begin{figure}[p]
\centering
\includegraphics[scale=0.8]{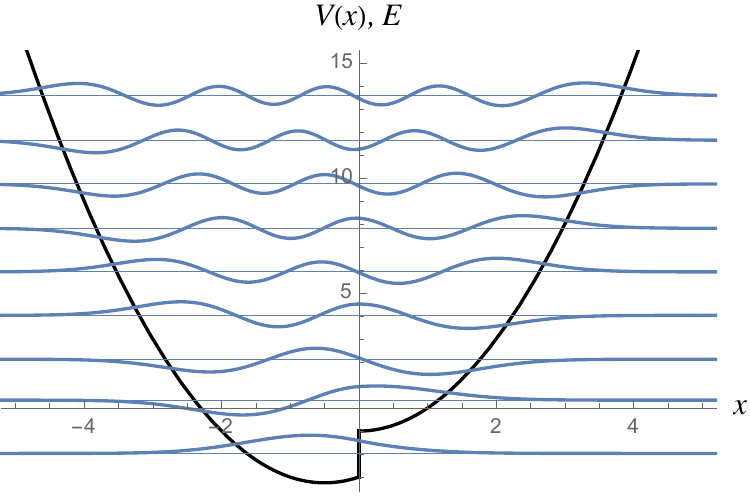}
\caption[The solution of the Schr\"{o}dinger equation with the potential \eqref{eq:4-2-pot-ag} with $a=2$ and $g=1$.]
	{The solution of the Schr\"{o}dinger equation with the potential \eqref{eq:4-2-pot-ag} with $a=2$ and $g=1$.
	Thin blue lines show the energy spectrum, and the blue curve on each line is the corresponding eigenfunction.
	The potential is also plotted in this figure by a black curve.}
\label{fig:4-sol_a2g1}
\end{figure}
\end{example}

\clearpage
\subsubsection{Case $a\to 0$: Harmonic oscillator with a ramp}
\begin{figure}[t]
\centering
\includegraphics[scale=1]{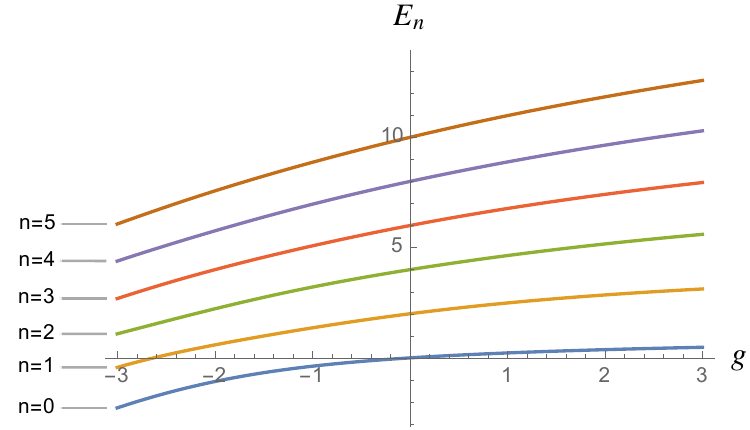}
\caption[The first six eigenvalues $E_n$ as functions of $g$ for $a=0$.]
	{The first six eigenvalues $E_n$ as functions of $g$ for $a=0$.
	They are all monotonically increasing in $g$, but never be equidistant for $g\neq 0$.}
\label{fig:4-Enes_a0g}
\end{figure}

Here, let us concentrate on the case $a\to 0$, where the potential consists of a harmonic oscillator plus a ramp function only,
\begin{equation}
V(x) = \begin{cases}
	x^2 - 1 - gx = \left( x - \dfrac{g}{2} \right)^2 - 1 - \dfrac{g^2}{4} & (x<0) \\[1ex]
	x^2 - 1 & (x>0)
\end{cases} ~.
\label{eq:4-2-pot-g}
\end{equation}
Here, we show how the energy spectrum changes as the external field is imposed. 
Remember that in the case of a harmonic oscillator plus a homogeneous external field, what happens is a constant shift of energies.
However, for our present case, Fig. \ref{fig:4-Enes_a0g} shows that that is not the case and the spectrum is never equidistant except for $g=0$.
For each $n$, the energy eigenvalue $E_n$ increases monotonically in $g$.

\subsubsection{Hermite-polynomial solutions}
As was mentioned, a potential is said to be \textit{quasi}-exactly solvable, when several eigenstates are explicitly obtained whereas the others are not~\cite{MR1329549,turbiner2016one}.
In our model \eqref{eq:4-2-pot-ag}, we can make only one state solvable via the Hermite polynomials, while for other states the wavefunctions are expressed only by the confluent hypergeometric functions and they are not reduced to any orthogonal polynomials.
This situation is similar to those in Refs. \cite{quantum4030022,10.1063/5.0127371}.

The construction is as follows.
First we choose $g$ such that $\phi^{\rm (H)}_m(x-g/2)$ has one of either zeros or extrema at $x=0$ (See Tab. \ref{tab:4-extrema-zeros}).
Suppose that $\phi^{\rm (H)}_{m}(-g/2)$ is the $j$-th zero [extremum] from the left.
Then, when the remaining model parameter $a$ is set to
\begin{equation}
a = 2k - \frac{g^2}{4} ~,
\label{eq:4-cond_on_a-QES}
\end{equation}
where $k \in\mathbb{Z}_{>0}$ is smaller than or equal to, and of the opposite parity to [same parity as] $m$, the $\left( j + \frac{m-k-1}{2} \right)$-th [$\left( j + \frac{m-k}{2} -1 \right)$-th] excited state is Hermite-polynomially solvable.
Such a state is of the energy eigenvalue
\begin{equation}
E_{j + \frac{m-k-1}{2}} = 2j + m-k-1
\quad
\left[ E_{j + \frac{m+k}{2} -1} = 2j + m-k -2 \right] ~,
\end{equation}
and the corresponding wavefunction is
\begin{equation}
\psi_{j + \frac{m-k-1}{2}}(x) \left[ \psi_{j + \frac{m+k}{2} -1}(x) \right] = \begin{cases}
	\mathcal{N}^{(-)}\mathrm{e}^{-\frac{(x-\frac{g}{2})^2}{2}}H_m\left( x-\frac{g}{2} \right) & (x<0) \\
	\mathcal{N}^{(+)}\mathrm{e}^{-\frac{x^2}{2}}H_{m-k}(x) & (x>0)
\end{cases}
\end{equation}
with $\mathcal{N}^{(\pm)}$ are constants to be determined from the boundary condition at $x=0$.

\begin{table}[t]
\caption{The zeros and extrema of $\phi^{\rm (H)}_n(x) = \mathrm{e}^{-x^2/2}H_n(x)$ for $n=0,1,2,3,4$.}
\label{tab:4-extrema-zeros}
\centering
\medskip
\begin{tabular}{cllllll}
\toprule
Order $n$ & $0$ & $1$ & $2$ & $3$ & $4$ & $\cdots$ \\
\midrule
Zeros & --- & $x=0$ & $x=\pm\dfrac{1}{\sqrt{2}}$ & $x=0,\pm\sqrt{\dfrac{3}{2}}$ & $x=\pm\sqrt{\dfrac{3}{2}\pm\sqrt{\dfrac{3}{2}}}$ \\[2ex]
Extrema & $x=0$ & $x=\pm 1$ & $x=0,\pm\sqrt{\dfrac{5}{2}}$ & $x=\pm\dfrac{\sqrt{9\pm\sqrt{57}}}{2}$ & $x=0,\pm\sqrt{\dfrac{7}{2}\pm\sqrt{\dfrac{11}{2}}}$ \\
\bottomrule
\end{tabular}
\end{table}

\begin{example}
Let us pick such $g$'s that $\phi^{\rm (H)}_2(x-g/2)$ has an extremum at $x=0$.
There are two extrema, $j=\{1,2\}$, and $-g/2=\pm 1/\sqrt{2}$.
Then, only $k=1$ is allowed, and $a$ is specified as $a=3/2$.

For $g=-\sqrt{2}$, the first excited-state wavefunction consists of Hermite polynomials,
\begin{equation}
\psi_1(x) = \begin{cases}
	2\mathrm{e}^{\frac{1}{4}}\mathrm{e}^{-\frac{\left(x+\frac{1}{\sqrt{2}}\right)^2}{2}}H_2\left( x+ \frac{1}{\sqrt{2}} \right) & (x<0) \\
	\hspace{1.6em}\mathrm{e}^{-\frac{x^2}{2}}H_1(x) & (x>0) 
\end{cases} ~,
\end{equation}
and the energy is $E_1=2$.
On the other hand, for $g=\sqrt{2}$, the Hermite polynomials constitute the second excited-state wavefunction with the energy $E_2=4$:
\begin{equation}
\psi_2(x) = \begin{cases}
	-2\mathrm{e}^{\frac{1}{4}}\mathrm{e}^{-\frac{\left(x-\frac{1}{\sqrt{2}}\right)^2}{2}}H_2\left( x- \frac{1}{\sqrt{2}} \right) & (x<0) \\
	\hspace{1.6em}-\mathrm{e}^{-\frac{x^2}{2}}H_1(x) & (x>0) 
\end{cases} ~.
\end{equation}
The solutions of the Schr\"{o}dinger equation with the potential \eqref{eq:4-2-pot-ag} with $g=\mp\sqrt{2}$ and $k=1$ are plotted in Fig. \ref{fig:4-sol_gmpsqrt2k1}.

\begin{figure}
\begin{minipage}[b]{.48\linewidth}
\centering
\includegraphics[width=\linewidth]{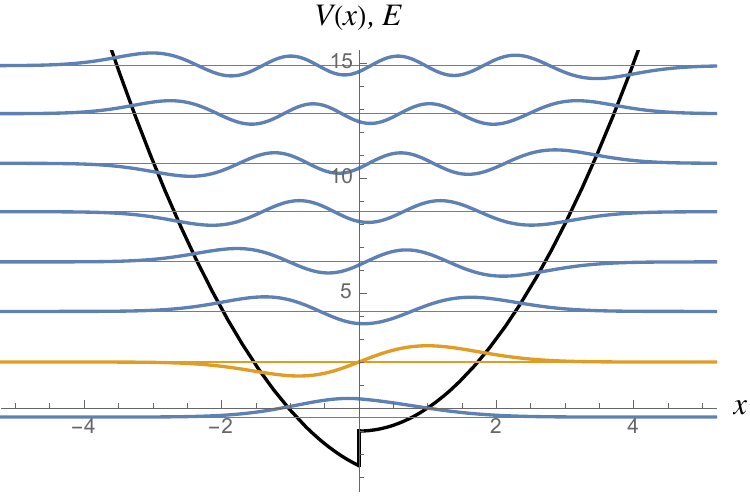}
\subcaption{$g=-\sqrt{2}$.}
\end{minipage}
\quad
\begin{minipage}[b]{.48\linewidth}
\centering
\includegraphics[width=\linewidth]{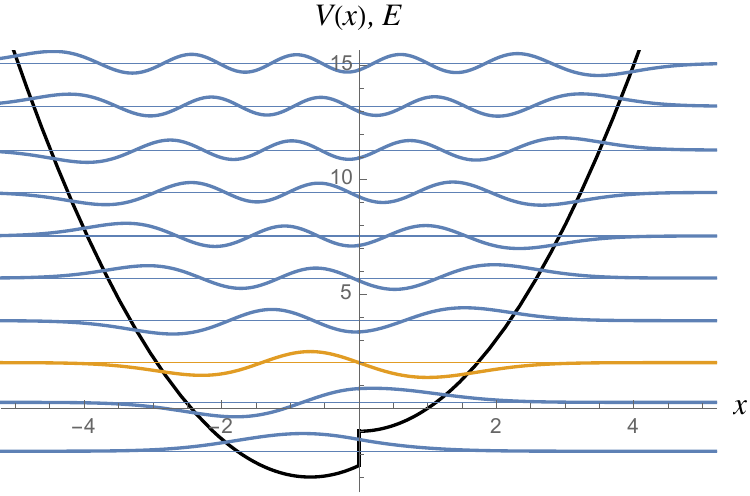}
\subcaption{$g=\sqrt{2}$.}
\end{minipage} 
\caption[The solutions for the potential \eqref{eq:4-2-pot-ag} with (a) $g=-\sqrt{2}$ and (b) $g=\sqrt{2}$.]
	{The solutions for the potential \eqref{eq:4-2-pot-ag} with (a) $g=-\sqrt{2}$ and (b) $g=\sqrt{2}$.
	The potential \eqref{eq:4-2-pot-ag} is displayed in each figure by a black curve.
	Thin lines show the energy spectrum, and the colored curve on each line is the corresponding eigenfunction.
	The states plotted in yellow possess the Hermite-polynomial solvability, while that colored in blue does not.}
\label{fig:4-sol_gmpsqrt2k1}
\end{figure}
\end{example}

\begin{example}
One application of our present work is to construct a sequence of the solvable potentials where only the ground-state wavefunction can be expressed by Hermite polynomials of different orders.

Such sequence is constructed as follows.
First we choose $g$ such that $\phi^{\rm (H)}_m(-g/2)$ is the first extremum from the left of $\phi^{\rm (H)}_m(x-g/2)$.
Here, $m$ can be any non-negative integer, and we choose $k=m$.
Then we identify the parameter $a$ using equation \eqref{eq:4-cond_on_a-QES}.
In this manner, one can construct infinitely many potentials whose ground-state wavefunctions are expressed by the Hermite polynomials but other eigenfunctions are not.
We show first several potentials $V_m(x)$, $m=1,2,3,4$, in Fig. \ref{fig:4-qes-sequence_pot} and the ground-state eigenfunctions $\psi^{(m)}_0(x)$:
\begin{equation}
\psi^{(m)}_0(x) = \begin{cases}
	\mathcal{N}_m\mathrm{e}^{-\frac{(x-\frac{g}{2})^2}{2}}H_m\left( x-\frac{g}{2} \right) & (x<0) \\
	\quad~\mathrm{e}^{-\frac{x^2}{2}} & (x>0)
\end{cases} ~,
\end{equation}
with $\mathcal{N}_m$ being a constant (See table \ref{tab:4-eg13_parameters}) in Fig. \ref{fig:4-qes-sequence_gswf}. 
They all have the energy $E_0=0$.
Taking $m=0$ means the ordinary harmonic oscillator.

Note that a similar procedure can be applied to construct a sequence of potentials such that only the $N$-th excited states can be expressed by Hermite polynomials of different orders.

\begin{table}[p]
\caption{Parameters of the wavefunction $\psi^{(m)}_0(x)$ for $m=1,2,3,4$.}
\label{tab:4-eg13_parameters}
\centering
\medskip
\begin{tabular}{lllll}
\toprule
$m$ & $1$ & $2$ & $3$ & $4$ \\
\midrule
$g$ & $2$ & $2\sqrt{\dfrac{5}{2}}$ & $\sqrt{9+\sqrt{57}}$ & $2\sqrt{\dfrac{7}{2}+\sqrt{\dfrac{11}{2}}}$ \\[2ex]
$\mathcal{N}_m$ & $-\dfrac{\sqrt{\mathrm{e}}}{2}$ & $\dfrac{\mathrm{e}^{5/4}}{8}$ & $-\dfrac{\mathrm{e}^{\frac{9+\sqrt{57}}{8}}}{2\sqrt{6(39+5\sqrt{57})}}$ & $\dfrac{\mathrm{e}^{\frac{7+\sqrt{22}}{4}}}{32(4+\sqrt{22})}$ \\
\bottomrule
\end{tabular}
\end{table}

\begin{figure}[p]
\begin{minipage}[b]{.48\linewidth}
\centering
\includegraphics[width=\linewidth]{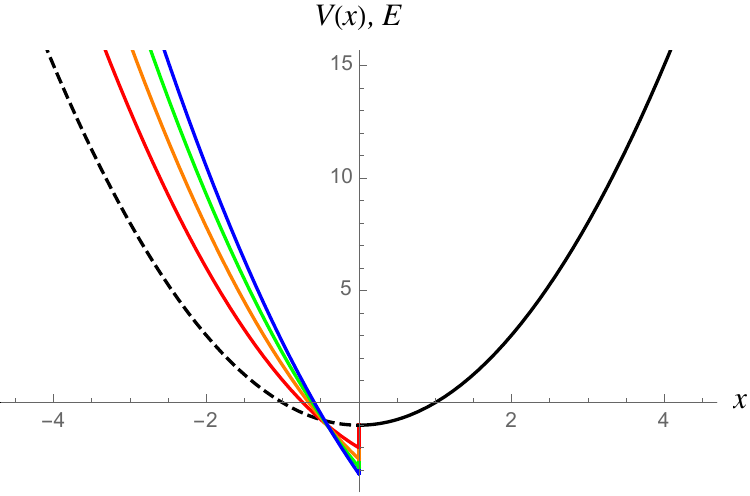}
\subcaption{Potentials $V_m(x)$.}
\label{fig:4-qes-sequence_pot}
\end{minipage}
\quad
\begin{minipage}[b]{.48\linewidth}
\centering
\includegraphics[width=\linewidth]{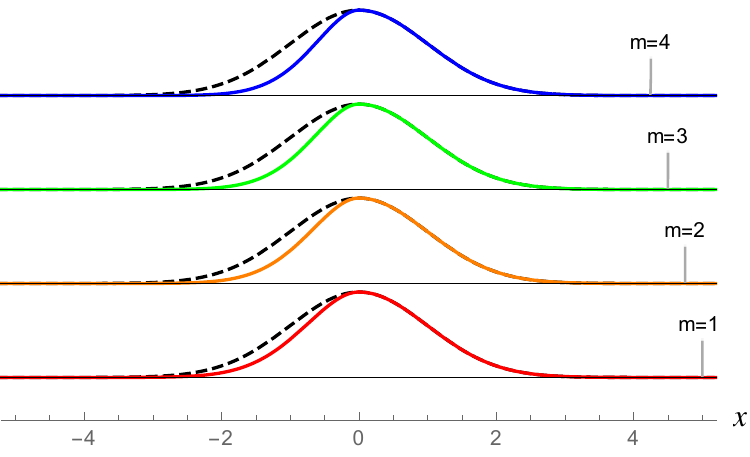}
\subcaption{The ground-state wavefunctions $\psi_0^{(m)}(x)$.}
\label{fig:4-qes-sequence_gswf}
\end{minipage} 
\caption[The sequence $\{ V_m(x) \}$, $\psi_0^{(m)}(x)$.]
	{The sequence $\{ V_m(x) \}$, $\psi_0^{(m)}(x)$.
	(a) The potentials for $m=1,2,3,4$ are plotted in red, orange, green and blue respectively, and $m=0$ (harmonic oscillator) by a black dashed curve.
	For $x>0$, they all share the same function, so we plotted them in the same color: black.
	(b) The ground-state wavefunctions of those potentials, whose energies are zero, are expressed in terms of Hermite polynomials with different orders.
	They are plotted in the same colors as the potentials.
	The black dashed curves are the ground-state wavefunctions of the harmonic oscillator.}
\label{fig:4-qes-sequence}
\end{figure}
\end{example}

\clearpage
\subsection{Harmonic Oscillator with Singularity Functions}
In this section, we have considered a harmonic oscillator with a step and/or a ramp.
The potentials \eqref{eq:4-2-pot}, \eqref{eq:4-2-pot-ag} and \eqref{eq:4-2-pot-g} can be abstracted as ``harmonic oscillators with singularity functions''.
For the singularity function, see Tab. \ref{tab:4-SingularityF}.

\begin{table}[t]
\caption{The Singularity Functions.}
\label{tab:4-SingularityF}
\centering
\begin{tabular}{ccc}
\toprule
order $n$ & $\langle x-x_0 \rangle^{n}$ & a.k.a. \\
\midrule
$n\leqslant -1$ & $\dfrac{d^{-n-1}}{dx^{-n-1}}\delta(x-x_0) \equiv \delta^{(-n-1)}(x-x_0)$ & \\
$\vdots$ & $\vdots$ & \\
$-2$ & $\dfrac{d}{dx}\delta(x-x_0) \equiv \delta^{(1)}(x-x_0)$ & \\
$-1$ & $\delta(x-x_0)$ & Dirac delta function\\
$0$ & $\theta(x-x_0)$ &Heaviside step function \\
$1$ & $(x-x_0)\theta(x-x_0)$ & Ramp function \\
$2$ & $(x-x_0)^2\theta(x-x_0)$ & \\
$\vdots$ & $\vdots$ & \\
$n\geqslant 0$ & $(x-x_0)^n\theta(x-x_0)$ & \\
\bottomrule
\end{tabular}
\end{table}

The case of a harmonic oscillator with a Dirac delta function has already been considered in Ref. \cite{Viana-Gomes_2011}.
The SWKB-induced quadratic oscillator \eqref{eq:4-SWKB_osc} we have dealt with in Sect. \ref{sec:4-2} is a special case of harmonic oscillators with a step and a parabolic ramp, while the potential \eqref{eq:3-sheared_osc} can be seen as a harmonic oscillator only with a parabolic ramp~\cite{Stillinger:1989aa,Chadzitaskos:2022kql}.
It would be quite a challenge to consider the general case of ``harmonic oscillators with singularity functions''.

\clearpage
\section{Summary of Chapter 4}
This chapter has been devoted to the exact solutions of the Schr\"{o}dinger equations with the classical-orthogonal-polynomially quasi-exactly solvable potentials defined by piecewise analytic functions.
General remarks on the solution methods are provided in Sect. \ref{sec:4-1}.

In Sect. \ref{sec:4-2}, we have solved the Schr\"{o}dinger equation for the novel solvable potential we have constructed in the previous chapter.
The potential is classical-orthogonal-polynomially quasi-exactly solvable.
We have identified the rule concerning which eigenstates are classical-orthogonal-polynomially solvable and which are not.

In order to dig deeper into the problem of the classical-orthogonal-polynomially quasi-exactly solvable potentials defined by piecewise analytic functions, we have considered simple modulations of the harmonic oscillator in Sect. \ref{sec:4-3}.
Here are our two illustrative examples: (1) a harmonic oscillator with a step, (2) a harmonic oscillator with a step and a ramp.
We note here that these problems, including the SWKB-induced one, are abstracted as harmonic oscillators with singularity functions, which is a new entry in the classical-orthogonal-polynomially (quasi-)exactly solvable potentials defined by piecewise analytic functions.

In the case of a harmonic oscillator with a step, we have shown that the energy spectra become isospectral, with several additional eigenstates, to the ordinary harmonic oscillator for special choices of a parameter.
We have further demonstrated that, using the Darboux--Crum transformation, one can systematically construct infinitely many potentials that are \textit{strictly} isospectral to the ordinary harmonic oscillator.

\subsubsection{Main statements of this chapter}
\begin{itemize}\setlength{\leftskip}{-1em}
\item The exact solutions of the Schr\"{o}dinger equations with a new entry of the classical-orthogonal-polynomially (quasi-)exactly solvable potentials defined by piecewise analytic functions: harmonic oscillator with singularity functions, are obtained (Sects. \ref{sec:4-2} and \ref{sec:4-3}).
\item Infinitely many potentials that are \textit{strictly} isospectral to the ordinary harmonic oscillator are constructed methodically (Sect. \ref{sec:4-3}).
\end{itemize}


\chapter{Conclusion}
\label{sec:5}

\section{Conclusion}
\label{sec:5-1}
In this thesis, we have studied the SWKB quantization condition and several solvable Schr\"{o}dinger equations.
First, we have applied the condition to various exactly solvable quantum mechanical systems, attempting to obtain a fundamental understanding of the quantization condition. 
It has turned out that the exactness of the SWKB quantization condition indicates that the system is exactly solvable via the classical orthogonal polynomials.
Moreover, we have formulated the inverse problem of the SWKB to construct (novel) classical-polynomially solvable superpotentials.
We have obtained the conventional shape-invariant potentials in our formulation and also an Hermite-polynomially quasi-exactly solvable potential, which is a member of the class of ``harmonic oscillators with singularity functions''.
The exact solutions of the Schr\"{o}dinger equations with the potentials in this class are also provided.

\medskip
The SWKB quantization condition is a quantization condition like the Bohr--Sommerfeld quantization condition and was proposed in the context of supersymmetric quantum mechanics.
Supersymmetric quantum mechanics, or exactly solvable quantum mechanics, has successfully been revealing various aspects of the exact solutions of Schr\"{o}dinger equation, such as Crum's theorem and shape invariance.
It is well-known that the condition amazingly reproduces exact bound-state spectra for all the conventional shape-invariant potentials.
However, it has recently turned out that it is not straightforward to put an interpretation on the condition, though there have been several attempts.
Moreover, the existing literature on the condition exclusively regarded it as a quantization condition of the energy, and few attempts have been made to apply the condition for different purposes.

Therefore, our first goal of the thesis was to understand the fundamental implication of the SWKB quantization condition, and then, based on this interpretation, we have explored how the condition can be applied to practical problems.

\medskip
To begin with, we have carried out extensive case studies to understand the physical meaning of the SWKB quantization condition. 
First, we have confirmed that this condition holds exactly for any conventional shape-invariant potentials, which have been well-known in the 1980s, by performing the integrals directly. 
Next, we have verified Bougie's argument that additive shape-invariant potentials do not always satisfy the condition equation.
We have further extended the argument in more general cases of the multi-indexed systems, and have found that the condition does not hold exactly, but it is still satisfied approximately.

In order to dig deeper into the approximate satisfaction of the condition, we have focused on the construction method of the multi-indexed systems.
They are constructed through the Darboux transformation of conventional shape-invariant potentials.
The other class of exactly solvable potentials that are also constructed through the Darboux transformation of conventional ones are the Krein--Adler systems.
We have applied the condition to those systems, and the result shows that the condition did not hold here, too.
A remarkable thing in the result is the following.
The condition estimates the energy with lower accuracy for $n$'s around the deleted levels.
This implies that the deviation is related to how the level structure of the system differs from that of the corresponding conventional shape-invariant system.

We then have worked out the way of evaluating the implications quantitatively.
Here, we have employed in our analysis the conditionally exactly solvable potentials by Juker and Roy, where conventional shape-invariant potentials are connected to Krein--Adler ones with a continuous parameter.
By varying this parameter, we have observed how the SWKB integrals change according to the shift of the level structure.
The potentials we employed also contain a parameter that realizes an isospectral deformation of conventional shape-invariant potentials, and we delivered a similar analysis with respect to this parameter.
These analyses have revealed that the SWKB condition indeed serves as an indicator of how much the level structure of the systems deviates from that of conventional shape-invariant ones, and the approximate satisfaction of the condition equation is guaranteed by the similarity between the level structures.
However, it has also become clear to us that it would be too difficult to make a mathematical statement regarding an interpretation of the SWKB condition.

In order to obtain a mathematical statement on an interpretation of the SWKB condition, we have reconsidered the exactness of the SWKB condition for conventional shape-invariant systems, particularly in a uniform manner.
All the conventional shape-invariant potentials are exactly solvable, and the solvability is guaranteed by the classical orthogonal polynomials.
We have focused on this aspect and demonstrated that the condition equations are actually reduced to three integral formulas, where the number three corresponds to the number of classical orthogonal polynomials that satisfy the precondition of Bochner's theorem and also have positive-definite weight functions. 
Furthermore, we have succeeded in verifying the statement with different classes of exactly solvable problems that are also solvable via classical orthogonal polynomials by considering a natural extension of the condition equation.
Thus, we have concluded that the SWKB condition signifies the solvability by the (three) classical orthogonal polynomials.

Based on the conclusion, we have then moved on to the matter of applying the SWKB quantization condition to some practical problems.
From the relation between the SWKB condition and the solvability through the classical orthogonal polynomials, we have come to the idea that the SWKB may have a meaning more than a mere quantization condition.
So far, our discussion was exclusively on the SWKB for obtaining the energy spectra from given superpotentials.
Now we would like to consider the other way around; we are to construct superpotentials from given energy spectra by means of the SWKB.
We have formulated this idea as an inverse problem concerning an integral equation. 
To ensure the uniqueness of the solutions and make it a well-posed problem, we have imposed an additional condition regarding the shape of the resulting superpotential.
We note that this inverse problem can be seen as an analogue of a classical problem of determining the potential energy from the period of oscillation in classical mechanics.

Our formulation successfully reconstructs all the shape-invariant potentials from given energy spectra.
We then turned to the problem of constructing novel solvable potentials by modifying the additional condition.
The resulting superpotentials are considered to be `modulations' of conventional shape-invariant potentials (In this thesis, we  have exclusively focused on modulations of the harmonic oscillator), and such a system is expected to exhibit solvability through the classical orthogonal polynomials, such as the Hermite polynomials.
In these systems, as they undergo modulations, not all the eigenfunctions can be expressed in terms of the classical orthogonal polynomials as in the case of the ordinary harmonic oscillator.
Instead, only a partial, but an infinite number of, eigenfunctions and their eigenvalues are equivalent to those of the undeformed one.
As a result, while in the case of the harmonic oscillator, the whole spectrum is equidistant, in the modulated problem, the spectra become equidistant with certain intervals. 
We have discussed in detail conditions on the appearance of solvability through the Hermite polynomials and the equidistant nature of the spectra.

The inverse problem of the SWKB, particularly the deformation of the harmonic oscillator, has provided us with a new perspective on solvable quantum-mechanical potentials.
That is, solvable quantum-mechanical potentials defined by piecewise analytic functions, especially polynomially solvable ones.
Such potentials have gained attention not only in the context of the SWKB but also in other areas. 
Historically, they were discussed in subatomic physics, describing confinement in nuclei or nucleons.
Recently, there has been a great interest in the context of non-polynomially solvable quantum-mechanical systems.
In this thesis, we have solved the Schr\"{o}dinger equations with potentials defined by piecewise quadratic functions and discussed their Hermite-polynomial (quasi-)solvability and their spectral properties.
We have also presented a systematic method of constructing an infinite number of potentials that are completely isospectral to the ordinary harmonic oscillator.

\section{Perspectives}
\label{sec:5-2}
The history of the problem of solving Schr\"{o}dinger equation goes back to the beginning of the 20th century, or if we extend it to a more general problem of the eigenvalue problem for a differential equation, it is much older.
Thus, one might think that there are few problems to solve in this context.
However, exactly solvable quantum mechanics is an ongoing research subject, and throughout this thesis, we have shown a new aspect of it regarding the SWKB formalism.

We would like to comment on another theme of exactly solvable quantum mechanics that has recently been discussed: discrete quantum mechanics~\cite{10.1063/1.2898695,Odake_2011,10.1143/PTP.125.851,Odake_2012,10.1063/1.3215983}.
It deals with second-order \textit{difference} equations, instead of Schr\"{o}dinger equation, which is a second-order differential equation.
We have richer problems with classical-polynomial solvability here.
It would be quite interesting to extend the SWKB formalism to discrete quantum mechanics.

Moreover, the profound understanding of the SWKB formalism could be a powerful tool for exploring new kinds of dualities in physics.
In the early days of quantum physics, quantization conditions played a significant role in understanding \textit{microscopic} phenomena by classical, \textit{macroscopic} theories.
In the case of WKB approximation, which is based on a perturbative treatment on $\hbar$, it was founded mathematically by A. Voros in exact WKB analysis~\cite{AIHPA_1983__39_3_211_0}.
This is now employed to understand non-perturbative phenomena in quantum physics.
Here, one finds a duality between perturbative and non-perturbative.
This idea is often referred to as resurgence theory~\cite{ecalle1985fonctions,10.1215/S0012-7094-98-09311-5,ANICETO20191,Sueishi:2020aa}.
Another example is called the ODE/IM correspondence~\cite{Patrick_Dorey_1999,Bazhanov:2001aa,Ito:2019aa}.
It has recently been discussed a relation between ordinary differential equations (ODE) and quantum integrable models (IM) through the exact WKB analysis.
A future study formulating exact `SWKB' analysis would be interesting.

Another possible area of future research would be to extend the ideas we have provided in the thesis to different theories with similar, but differential, equations.
As we have mentioned in the introduction, similar equations to Schr\"{o}dinger equation are everywhere in mathematical science.
When mathematical models of some phenomena are involved with a Schr\"{o}dinger-type eigenvalue problem, the knowledge of the solvability of Schr\"{o}dinger equations studied in this thesis can contribute to understanding phenomena and revealing the mathematical structures behind them.
Such phenomena can be found in soliton theories, spin systems, cosmology, optics, circuit theories, non-equilibrium statistical mechanics, theoretical/mathematical biology, information science, economics, financial engineering \textit{etc}.


\appendix
\chapter{Several Exactly Solvable Potentials}
\label{sec:B}

\section{Krein--Adler Systems}
\label{sec:B-1}
\subsection{The Construction Method}
The Krein--Adler transformation is a Darboux transformation with the choice of the seed solutions whose indices are 
\begin{equation}
\mathcal{D} = \{ d_1 ,d_2 ,\ldots, d_M \} ~,~~~
d_j \in \mathbb{Z}_{\geqslant 0} ~,
\end{equation}
in which the following conditions:
\begin{equation}
\forall n \in \mathbb{Z}_{\geqslant 0} ~,~~~
\prod_{j=1}^{M}(n - d_j) \geqslant 0 ~,
\label{eq:A-KA_D_conditions}
\end{equation}
are satisfied.
These condition mean that the set $\mathcal{D}$ is
\[
\mathcal{D} = \{ \ell_1, \ell_1+1 < \ell_2, \ell_2+1, \ldots, \ell_{M'},~\ell_{M'}+1 \} ~,~~~
\ell_j \in \mathbb{Z}_{\geqslant 0} ~,
\]
where the choice $\ell_1 = 0$, $\ell_{j+1} = \ell_j+2$ simply corresponds to the case of the Darboux--Crum transformation.
The conditions \eqref{eq:A-KA_D_conditions} are required so that the resulting potential has no singularity points in the domain.

The Krein--Adler transformation maps a Hamiltonian $\mathcal{H}$ to
\begin{equation}
\mathcal{H}_{\mathcal{D}}^{\mathrm{(KA)}} 
\coloneqq \mathcal{H} - 2\hbar^2\frac{d^2}{dx^2}\ln\left|\mathrm{W}\left[ \psi_{d_1}, \ldots, \psi_{d_M} \right](x) \right| ~,
\label{eq:A-KA_ham}
\end{equation}
whose eigenvalue equation is
\begin{align}
&\mathcal{H}_{\mathcal{D}}^{\mathrm{(KA)}} \psi_{\mathcal{D};n}^{\mathrm{(KA)}}(x) = \mathcal{E}_{\mathcal{D};n}^{\mathrm{(KA)}} \psi_{\mathcal{D};n}^{\mathrm{(KA)}}(x) ~,~~~
n = 0,1,2,\ldots ~,
\label{eq:A-KA_SE} \\
&\mathcal{E}_{\mathcal{D};n}^{\mathrm{(KA)}} = \mathcal{E}_{\breve{n}} ~,~~~
\psi_{\mathcal{D};n}^{\mathrm{(KA)}}(x) = \frac{\mathrm{W}\left[\psi_{d_1}, \ldots, \psi_{d_M}, \psi_{\breve{n}} \right](x)}{\mathrm{W}\left[\psi_{d_1}, \ldots, \psi_{d_M} \right](x)} ~,~~~
\breve{n} \in \mathbb{Z}_{\geqslant 0} \backslash \mathcal{D} ~.
\end{align}
It should be emphasized here that the transformation corresponds to the deletion of the eigenstates with the indices $\mathcal{D}$.

\subsection{Examples}
In the following, we restrict ourselves to the cases of $\mathcal{D} = \{ d, d+1 \}$ for simplicity.
Here, $n$ and $\breve{n}$ are related by
\begin{equation}
\breve{n} = \begin{cases}
	n & (0 \leqslant n \leqslant d-1) \\
	n+2 & (n \geqslant d)
\end{cases} ~.
\end{equation}

\subsubsection{1-dim. harmonic oscillator}
The Krein--Adler transformation of the 1-dim. harmonic oscillator:
\[
\mathcal{H}^{\rm (H)} = -\hbar^2\frac{d^2}{dx^2} + \omega^2x^2 - \hbar\omega ~,~~~
\mathcal{H}^{\rm (H)}\phi_n^{\rm (H)}(x) = \mathcal{E}_n^{\rm (H)}\phi_n^{\rm (H)}(x) ~,
\]
is
\begin{equation}
\mathcal{H}_{\mathcal{D}}^{\mathrm{(K,H)}} = \mathcal{H}^{\mathrm{(H)}} - 2\hbar^2\frac{d^2}{dx^2}\ln\left|\mathrm{W}\left[ \phi_{d}^{\mathrm{(H)}}, \phi_{d+1}^{\mathrm{(H)}} \right](x) \right| ~,
\end{equation}
whose eigenvalues and the corresponding eigenfunctions are
\begin{equation}
\mathcal{E}_{\mathcal{D};n}^{\mathrm{(K,H)}} = 2\breve{n}\hbar\omega ~,~~~
\phi_{\mathcal{D};n}^{\mathrm{(K,H)}}(x) = \frac{\mathrm{W}\left[\phi_{d}^{\mathrm{(H)}}, \phi_{d+1}^{\mathrm{(H)}}, \phi_{\breve{n}}^{\mathrm{(H)}} \right](x)}{\mathrm{W}\left[\phi_{d}^{\mathrm{(H)}}, \phi_{d+1}^{\mathrm{(H)}} \right](x)} ~.
\label{eq:A-KAH2_phi}
\end{equation}
With the formulae for the Wronskian \eqref{eq:A-Wron_23} and \eqref{eq:A-Wron_24}, the eigenfunction are reduced to
\begin{equation}
\phi_{\mathcal{D};n}^{\mathrm{(K,H)}}(x) = \mathrm{e}^{-\frac{\xi^2}{2}}\frac{\mathrm{W}\left[ H_{d}, H_{d+1}, H_{\breve{n}} \right](\xi)}{\mathrm{W}\left[ H_{d}, H_{d+1} \right](\xi)} ~,~~~
\xi \equiv \sqrt{\frac{\omega}{\hbar}}\,x ~.
\end{equation}
Especially, the ground-state wavefunction is 
\begin{equation}
\phi_{\mathcal{D};0}^{\mathrm{(K,H)}}(x) = \mathrm{e}^{-\frac{\xi^2}{2}}\frac{\mathrm{W}\left[ H_{d}, H_{d+1}, 1 \right](\xi)}{\mathrm{W}\left[ H_{d}, H_{d+1} \right](\xi)} ~.
\end{equation}

\subsubsection{Radial oscillator}
The Krein--Adler transformation of the radial oscillator:
\[
\mathcal{H}^{\rm (L)} = -\hbar^2\frac{d^2}{dx^2} + \omega^2x^2 + \frac{\hbar^2g(g-1)}{x^2} - \hbar\omega(2g+1) ~,~~~
\mathcal{H}^{\rm (L)}\phi_n^{\rm (L)}(x) = \mathcal{E}_n^{\rm (L)}\phi_n^{\rm (L)}(x) ~,
\]
is
\begin{equation}
\mathcal{H}_{\mathcal{D}}^{\mathrm{(K,L)}} = \mathcal{H}^{\mathrm{(L)}} - 2\hbar^2\frac{d^2}{dx^2}\ln\left|\mathrm{W}\left[ \phi_{d}^{\mathrm{(L)}}, \phi_{d+1}^{\mathrm{(L)}} \right](x) \right| ~,
\end{equation}
whose eigenvalues and the corresponding eigenfunctions are
\begin{equation}
\mathcal{E}_{\mathcal{D};n}^{\mathrm{(K,L)}} = 4\breve{n}\hbar\omega ~,~~~
\phi_{\mathcal{D};n}^{\mathrm{(K,L)}}(x) = \frac{\mathrm{W}\left[\phi_{d}^{\mathrm{(L)}}, \phi_{d+1}^{\mathrm{(L)}}, \phi_{\breve{n}}^{\mathrm{(L)}} \right](x)}{\mathrm{W}\left[\phi_{d}^{\mathrm{(L)}}, \phi_{d+1}^{\mathrm{(L)}} \right](x)} ~.
\label{eq:A-KAL2_phi}
\end{equation}
With the formulae for the Wronskian \eqref{eq:A-Wron_23} and \eqref{eq:A-Wron_24}, the eigenfunction are reduced to
\begin{equation}
\phi_{\mathcal{D},\breve{n}}^{\mathrm{(K,L)}}(x) = \mathrm{e}^{-\frac{z}{2}}z^{\frac{g+2}{2}}\frac{\mathrm{W}\left[ L_{d}^{(g-\frac{1}{2})}, L_{d+1}^{(g-\frac{1}{2})}, L_{\breve{n}}^{(g-\frac{1}{2})} \right](z)}{\mathrm{W}\left[ L_{d}^{(g-\frac{1}{2})}, L_{d+1}^{(g-\frac{1}{2})} \right](z)} ~,~~~
z \equiv \frac{\omega}{\hbar}\,x^2 ~.
\end{equation}
Especially, the ground-state wavefunction is 
\begin{equation}
\phi_{\mathcal{D};0}^{\mathrm{(K,L)}}(x) = \mathrm{e}^{-\frac{z}{2}}z^{\frac{g+2}{2}}\frac{\mathrm{W}\left[ L_{d}^{(g-\frac{1}{2})}, L_{d+1}^{(g-\frac{1}{2})}, 1 \right](z)}{\mathrm{W}\left[ L_{d}^{(g-\frac{1}{2})}, L_{d+1}^{(g-\frac{1}{2})} \right](z)} ~.
\end{equation}

\subsubsection{P\"{o}schl--Teller potential}
The Krein--Adler transformation of the P\"{o}schl--Teller potential:
\[
\mathcal{H}^{\rm (J)} = -\hbar^2\frac{d^2}{dx^2} + \frac{\hbar^2 g(g-1)}{\sin^2x} + \frac{\hbar^2 h(h-1)}{\cos^2x} - \hbar^2(g+h)^2  ~,~~~
\mathcal{H}^{\rm (J)}\phi_n^{\rm (J)}(x) = \mathcal{E}_n^{\rm (J)}\phi_n^{\rm (J)}(x) ~,
\]
is
\begin{equation}
\mathcal{H}_{\mathcal{D}}^{\mathrm{(K,J)}} = \mathcal{H}^{\mathrm{(J)}} - 2\hbar^2\frac{d^2}{dx^2}\ln\left|\mathrm{W}\left[ \phi_{d}^{\mathrm{(J)}}, \phi_{d+1}^{\mathrm{(J)}} \right](x) \right| ~,
\end{equation}
whose eigenvalues and the corresponding eigenfunctions are
\begin{equation}
\mathcal{E}_{\mathcal{D};n}^{\mathrm{(K,J)}} = 4\hbar^2\breve{n}(\breve{n}+g+h) ~,~~~
\phi_{\mathcal{D};n}^{\mathrm{(K,J)}}(x) = \frac{\mathrm{W}\left[\phi_{d}^{\mathrm{(J)}}, \phi_{d+1}^{\mathrm{(J)}}, \phi_{\breve{n}}^{\mathrm{(J)}} \right](x)}{\mathrm{W}\left[\phi_{d}^{\mathrm{(J)}}, \phi_{d+1}^{\mathrm{(J)}} \right](x)} ~.
\label{eq:A-KAJ2_phi}
\end{equation}
With the formulae for the Wronskian \eqref{eq:A-Wron_23} and \eqref{eq:A-Wron_24}, the eigenfunction are reduced to
\begin{equation}
\phi_{\mathcal{D};n}^{\mathrm{(K,J)}}(x) = (1-y)^{\frac{g+2}{2}}(1+y)^{\frac{h+2}{2}}\frac{\mathrm{W}\left[ P_{d}^{(g-\frac{1}{2},h-\frac{1}{2})}, P_{d+1}^{(g-\frac{1}{2},h-\frac{1}{2})}, P_{\breve{n}}^{(g-\frac{1}{2},h-\frac{1}{2})} \right](y)}{\mathrm{W}\left[ P_{d}^{(g-\frac{1}{2},h-\frac{1}{2})}, P_{d+1}^{(g-\frac{1}{2},h-\frac{1}{2})} \right](y)} ~.
\end{equation}
where $y \equiv \cos 2x$.
Especially, the ground-state wavefunction is 
\begin{equation}
\phi_{\mathcal{D};0}^{\mathrm{(K,J)}}(x) = (1-y)^{\frac{g+2}{2}}(1+y)^{\frac{h+2}{2}}\frac{\mathrm{W}\left[ P_{d}^{(g-\frac{1}{2},h-\frac{1}{2})}, P_{d+1}^{(g-\frac{1}{2},h-\frac{1}{2})}, 1 \right](y)}{\mathrm{W}\left[ P_{d}^{(g-\frac{1}{2},h-\frac{1}{2})}, P_{d+1}^{(g-\frac{1}{2},h-\frac{1}{2})} \right](y)} ~.
\end{equation}

\section{Multi-indexed Systems}
\label{sec:B-2}
The eigenfunctions of the multi-indexed systems are expressed in terms of the multi-indexed polynomials.
The modifier ``multi-indexed'' reflects that the symbol has many superscripts and subficies as we shall show in Eqs. \eqref{eq:A-MI_SE} and \eqref{eq:A-MI_phi} \textit{etc}.

\subsection{The Construction Method}
\subsubsection{Virtual-state wavefunctions}
Let $\mathcal{H}_{\rm L}(x;g)$ and $\mathcal{H}_{\rm J}(x;g,h)$ denote the Hamiltonians $\mathcal{H}^{\rm (L)}$ and $\mathcal{H}^{\rm (J)}$ without the constant terms.
They are invariant under the following discrete symmetry transformations:
\begin{equation}
\mathrm{L}:~ g\to 1-g ~;\qquad
\mathrm{J}:~ g\to 1-g ~,~~~ h\to 1-h ~,
\end{equation}
and also $\mathcal{H}_{\rm L}(x;g)$ changes its sign under the discrete transformation of the coordinate: $x\to \mathrm{i}x$,
\begin{align}
\mathrm{L}&: & 
\mathcal{H}_{\rm L}(\mathrm{i}x;g) &= ~~\,\hbar^2\frac{d^2}{dx^2} - \omega^2x^2 - \frac{\hbar^2g(g-1)}{x^2} 
\equiv -\mathcal{H}_{\rm L}(x;g) ~,
\label{eq:A-Ham_L_ix} \\
&&
\mathcal{H}_{\rm L}(x;1-g) &= -\hbar^2\frac{d^2}{dx^2} + \omega^2x^2 + \frac{\hbar^2(1-g)(-g)}{x^2} 
\equiv \mathcal{H}_{\rm L}(x;g) ~, 
\label{eq:A-Ham_L_1-g} \\
\mathrm{J}&: & 
\mathcal{H}_{\rm J}(x;1-g,h) &= -\hbar^2\frac{d^2}{dx^2} + \frac{\hbar^2(1-g)(-g)}{\sin^2x} + \frac{\hbar^2h(h-1)}{\cos^2x} 
\equiv \mathcal{H}_{\rm J}(x;g,h) ~, 
\label{eq:A-Ham_J_1-g} \\
&& 
\mathcal{H}_{\rm J}(x;g,1-h) &= -\hbar^2\frac{d^2}{dx^2} + \frac{\hbar^2g(g-1)}{\sin^2x} + \frac{\hbar^2(1-h)(-h)}{\cos^2x} 
\equiv \mathcal{H}_{\rm J}(x;g,h) ~. 
\label{eq:A-Ham_J_1-h} 
\end{align}
Note that these transformations do not hold the differential equations \eqref{} and \eqref{} invariant.
Moreover, the eigenfunctions transformed by these transformations are no longer square-integrable.
However, they remain solutions of the Sch\"{o}dinger equations; the discrete transformations map a solution of the Schr\"{o}dinger equation to another. 

We solve the eigenvalue problems for the Hamiltonians \eqref{eq:A-Ham_L_ix}--\eqref{eq:A-Ham_J_1-h} by virtue of the discrete transformations.
Here we add constants to the Hamiltonians to make them positive semi-definite, and also we write $z\equiv \omega x^2/\hbar$ and $y\equiv\cos 2y$ in the following.
For Eq. \eqref{eq:A-Ham_L_ix}, the eigenvalues and the corresponding eigenfunctions are
\begin{equation}
\mathcal{E}_{\rm v}^{\rm (L),I}/\hbar\omega = -4\left( g + \mathrm{v} + \frac{1}{2} \right) ~,~~~
\varphi_{\mathrm{v}}^{\mathrm{(L),I}}(x) = \mathrm{e}^{\frac{z}{2}} z^{\frac{g}{2}} L_{\rm v}^{(g-\frac{1}{2})}(-z) ~,~~~
\mathrm{v} = 0,1,2,\ldots ~,
\end{equation}
and for Eq. \eqref{eq:A-Ham_L_1-g},
\begin{multline}
\quad
\mathcal{E}_{\rm v}^{\rm (L),II}/\hbar\omega = -4\left( g - \mathrm{v} - \frac{1}{2} \right) ~,~~~
\varphi_{\rm v}^{\rm (L),II}(x) = \mathrm{e}^{-\frac{z}{2}} z^{\frac{1-g}{2}} L_{\rm v}^{(\frac{1}{2}-g)}(z) ~, \\
\mathrm{v} = 0,1,\ldots, \left\lfloor g-\frac{1}{2} \right\rfloor' ~.
\end{multline}
On the other hand, for Eq. \eqref{eq:A-Ham_J_1-h},
\begin{multline}
\hspace{-.5em}
\mathcal{E}_{\mathrm{v}}^{\mathrm{(J),I}}/\hbar^2 = -4\left( g + \mathrm{v} + \frac{1}{2} \right)\left( h - \mathrm{v} - \frac{1}{2} \right) ,~~
\varphi_{\mathrm{v}}^{\mathrm{(J),I}}(x) = \left( \frac{1-y}{2} \right)^{\frac{g}{2}} \left( \frac{1+y}{2} \right)^{\frac{1-h}{2}} P_{\mathrm{v}}^{(g-\frac{1}{2},\frac{1}{2}-h)}(y) ~, \\
\mathrm{v} = 0,1,\ldots, \left\lfloor h-\frac{1}{2} \right\rfloor' ~,
\end{multline}
and for Eq. \eqref{eq:A-Ham_J_1-g},
\begin{multline}
\hspace{-.5em}
\mathcal{E}_{\mathrm{v}}^{\mathrm{(J),II}}/\hbar^2 = -4\left( g - \mathrm{v} - \frac{1}{2} \right)\left( h + \mathrm{v} + \frac{1}{2} \right) ,~~
\varphi_{\mathrm{v}}^{\mathrm{(J),II}}(x) = \left( \frac{1-y}{2} \right)^{\frac{1-g}{2}} \left( \frac{1+y}{2} \right)^{\frac{h}{2}} P_{\mathrm{v}}^{(\frac{1}{2}-g,h-\frac{1}{2})}(y) ~, \\
\mathrm{v} = 0,1,\ldots, \left\lfloor g-\frac{1}{2} \right\rfloor' ~.
\end{multline}
The eigenfunctions $\varphi_{\rm v}(x)$'s are often referred to as \textit{virtual-state wavefunctions}, satisfying the following five conditions:
\begin{itemize}
\item $\varphi_{\rm v}(x)$ has no zeros in the domain $(x_1,x_2)$;
\item $\varphi_{\rm v}(x)$ has negative energy;
\item $\varphi_{\rm v}(x)$ is a polynomial;
\item $\varphi_{\rm v}(x)$ is not square-integrable;
\item $\varphi_{\rm v}^{-1}(x)$ is not square-integrable, too.
\end{itemize}

\subsubsection{Construction of the systems}
In what follows, we choose the virtual-state wavefunctions as the seed solutions whose indices are
\begin{equation}
\mathcal{D} = \mathcal{D}^{\mathrm{I}}\cup\mathcal{D}^{\mathrm{II}}
= \{ d_1^{\mathrm{I}},\ldots,d_M^{\mathrm{I}} \}\cup\{ d_1^{\mathrm{II}},\ldots,d_N^{\mathrm{II}} \}
\end{equation}
with
\begin{equation}
d_1^{\mathrm{I}}<\cdots<d_M^{\mathrm{I}} \in\mathbb{Z}_{>0} ~,~~~
d_1^{\mathrm{II}}<\cdots<d_N^{\mathrm{II}} \in\mathbb{Z}_{>0} ~.
\end{equation}
Moreover, 
\begin{align}
\mathrm{L}:\qquad g &> \mathrm{max}\left\{ N+\frac{3}{2}, d_j^{\mathrm{II}}+\frac{1}{2} \right\} ~, \\
\mathrm{J}:\qquad g &> \mathrm{max}\left\{ N+2, d_j^{\mathrm{II}}+\frac{1}{2} \right\} ~,~~~
h > \mathrm{max}\left\{ M+2, d_j^{\mathrm{I}}+\frac{1}{2} \right\} ~.
\end{align}
so that all the virtual-state wavefunctions are accommodated.
Note that $\mathcal{D} = \mathcal{D}^{\rm I} \cup \mathcal{D}^{\rm II} = \{ \ell \} \cup \emptyset$ and $\mathcal{D} = \mathcal{D}^{\rm I} \cup \mathcal{D}^{\rm II} = \emptyset \cup \{ \ell \}$ correspond to type I/II exceptional $X_{\ell}$ systems.

A multi-indexed system is constructed by the Darboux transformation of the Hamiltonian $\mathcal{H}^{(\ast)}$ with the seed solutions above,
\begin{equation}
\mathcal{H}_{\mathcal{D}}^{\rm (MI\ast)} \coloneqq \mathcal{H}^{\rm (\ast)} - 2\hbar^2\frac{d^2}{dx^2}\ln\left|\mathrm{W}\left[\varphi_{d_1^{\rm I}}^{\rm (\ast),I},\ldots,\varphi_{d_M^{\rm I}}^{\rm (\ast),I},\varphi_{d_1^{\rm II}}^{\rm (\ast),II},\ldots,\varphi_{d_N^{\rm II}}^{\rm (\ast),II}\right](x)\right| ~,~~~
\ast = \mathrm{L,J} ~,
\label{eq:A-MI_Ham}
\end{equation}
whose eigenvalue equation is
\begin{align}
&\mathcal{H}_{\mathcal{D}}^{\rm (MI\ast)} \phi_{\mathcal{D};n}^{\rm (MI\ast)}(x) = \mathcal{E}_{\mathcal{D};n}^{\rm (MI\ast)} \phi_{\mathcal{D};n}^{\rm (MI\ast)}(x) ~,~~~
n = 0,1,2,\ldots ~,
\label{eq:A-MI_SE} \\
&\mathcal{E}_{\mathcal{D};n}^{\rm (MI\ast)} = \mathcal{E}_{n} ~,~~~
\phi_{\mathcal{D},n}^{\rm (MI\ast)}(x) = \frac{\mathrm{W}\left[\varphi_{d_1^{\mathrm{I}}}^{\mathrm{(\ast),I}},
\ldots,\varphi_{d_M^{\mathrm{I}}}^{\mathrm{(\ast),I}},\varphi_{d_1^{\mathrm{II}}}^{\mathrm{(\ast),II}},
\ldots,\varphi_{d_N^{\mathrm{II}}}^{\mathrm{(\ast),II}},\phi_n^{(\ast)}\right](x)}{\mathrm{W}\left[\varphi_{d_1^{\mathrm{I}}}^{\mathrm{(\ast),I}},
\ldots,\varphi_{d_M^{\mathrm{I}}}^{\mathrm{(\ast),I}},\varphi_{d_1^{\mathrm{II}}}^{\mathrm{(\ast),II}},\ldots,\varphi_{d_N^{\mathrm{II}}}^{\mathrm{(\ast),II}}
\right](x)} ~.
\label{eq:A-MI_phi}
\end{align}
Note that the resulting Hamiltonians are also shape invariant.

\begin{example}[Type I $X_1$-Laguerre]
The Darboux transformation of $\mathcal{H}^{\rm (L)}$ with the seed solution:
\[
\varphi_1^{\rm I}(x) = \mathrm{e}^{\frac{z}{2}} z^{\frac{g}{2}}\left( z + g + \dfrac{1}{2} \right) ~,
\]
yields 
\begin{align}
\mathcal{H}_1^{\rm I} &= -\hbar^2\frac{d^2}{dx^2} + \omega^2x^2 + \frac{\hbar^2g(g-1)}{x^2} - 2\hbar^2\frac{d^2}{dx^2}\ln\left| \varphi_1^{\rm I}(x) \right| \nonumber \\
&= -\hbar^2\frac{d^2}{dx^2} + \omega^2x^2 + \frac{\hbar^2g(g+1)}{x^2} - \frac{16\hbar^3\omega (2g+1)}{(2\omega x + 2\hbar g + \hbar)^2} + \frac{8\hbar^2\omega}{2\omega x + 2g +1} - \hbar\omega (2g + 3) ~.
\end{align}
The partner potential is computed as
\begin{equation}
\omega^2x^2 + \frac{\hbar^2(g+1)(g+2)}{x^2} - \frac{16\hbar^3\omega (2g+3)}{(2\omega x + 2\hbar g + 3\hbar)^2} + \frac{8\hbar^2\omega}{2\omega x + 2g + 3} - \hbar\omega (2g + 1) ~,
\end{equation}
therefore one can see that the system is shape invariant under $g\to g+1$.
The energy eigenvalues are turned out to be $\mathcal{E}_{1;n}^{\rm (MIL),I} = 4n\hbar\omega$
\end{example}

\section{Conditionally Exactly Solvable Systems by Junker and Roy}
\subsection{The Construction Method}
Let $W_0(x)$ be a conventional shape-invariant superpotential (it can be any exactly solvable superpotential in general).
We assume that the superpotential of the conditionally exactly solvable systems is of the following form:
\begin{equation}
W^{\rm (CES)}(x) \coloneqq W_0(x) + W_1(x) ~,
\label{eq:A-CES_spot}
\end{equation}
where $W_1(x)$ is a function of $x$ to be determined later.

The Hamiltonian reads
\begin{align}
\mathcal{H}_{\rm CES}^{[0]}
&= \left( -\hbar\frac{d}{dx} + W^{\rm (CES)}(x) \right) \left( \hbar\frac{d}{dx} + W^{\rm (CES)}(x) \right) \nonumber \\
&= -\hbar^2\frac{d^2}{dx^2} + W_0(x)^2 - \hbar\frac{dW_0}{dx} + 2W_0(x)W_1(x) + W_1(x)^2 - \hbar\frac{dW_1}{dx} ~,
\label{eq:A-CES_Ham0}
\end{align}
while the partner Hamiltonian is
\begin{align}
\mathcal{H}_{\rm CES}^{[1]}
&= \left( \hbar\frac{d}{dx} + W^{\rm (CES)}(x) \right) \left( -\hbar\frac{d}{dx} + W^{\rm (CES)}(x) \right) \nonumber \\
&= -\hbar^2\frac{d^2}{dx^2} + W_0(x)^2 + \hbar\frac{dW_0}{dx} + 2W_0(x)W_1(x) + W_1(x)^2 + \hbar\frac{dW_1}{dx} ~.
\label{eq:A-CES_Ham1}
\end{align}
In each equation, the first three terms are a conventional shape-invariant Hamiltonian.

Now let us assume that the last three terms in Eq. \eqref{eq:A-CES_Ham1} correspond to a constant shift of energy, say $b$,
\begin{equation}
2W_0(x)W_1(x) + W_1(x)^2 + \hbar\frac{dW_1}{dx} = b ~,
\label{eq:A-CES_b}
\end{equation}
where $b$ has to satisfy
\begin{equation}
b > \mathcal{E}_0 - \mathcal{E}_1 = -\mathcal{E}_1 ~,
\label{eq:A-CES_cond-b}
\end{equation}
because the lowest energy of $\mathcal{H}_{\rm CES}^{[1]}$ cannot be smaller than, or equal to, that of $\mathcal{H}_{\rm CES}^{[0]}$.
We call this condition \underline{\sc Condition I}.
In order to determine $W_1(x)$, we solve the Ricatti-type differential equation \eqref{eq:A-CES_b}.
Substituting 
\begin{equation}
W_1(x) = \hbar\frac{\partial_x u(x)}{u(x)} = \hbar\frac{d}{dx}\ln u(x) ~,
\end{equation}
it follows that $u(x)$ satisfies the following second-order linear differential equation:
\begin{equation}
\hbar^2\frac{d^2u}{dx^2} + 2\hbar W_0(x)\frac{du}{dx} - bu(x) = 0 ~,
\label{eq:A-CES_u}
\end{equation}
whose general solution is a two-parameter family, say $\alpha$ and $\beta$.
We are to specify those parameters so that $u(x)$ has no zero, and therefore $W_1(x)$ is not singular, in the domain.
We call this condition \underline{\sc Condition II}.

\subsection{Example}
\subsubsection{1-dim. harmonic oscillator}
Take $W_0(x) = \omega x$ as an example.
Then, the partner Hamiltonian $\mathcal{H}_{\rm CES}^{[1]}$ is
\begin{equation}
\mathcal{H}_{\rm CES}^{[1]} = -\hbar^2\frac{d^2}{dx^2} + \omega^2x^2 + \hbar\omega + b'\hbar\omega ~,~~~
b \equiv b'\hbar\omega ~,
\end{equation}
whose eigenvalues and the corresponding eigenfunctions are
\begin{equation}
\mathcal{E}^{[1]}_n = [2(n+1)+b']\hbar\omega ~,~~~
\phi^{[1]}_n(x) = \mathrm{e}^{-\frac{\omega x^2}{2\hbar}}H_n\left(\sqrt{\frac{\omega}{\hbar}}x\right) ~.
\end{equation}
From \underline{\sc Condition I}, the parameter $b'$ satisfies
\begin{equation}
b' > -2 ~.
\label{eq:A-CESH_cond1}
\end{equation}

Eq. \eqref{eq:A-CES_u} reads
\begin{equation}
\hbar^2\frac{d^2u}{dx^2} + 2\hbar\omega x\frac{du}{dx} - b'\hbar\omega u(x) = 0 ~,
\end{equation}
whose general solution is
\begin{equation}
u(x) = \alpha{}_1F_1\left( -\frac{b'}{4},\frac{1}{2};-\frac{\omega}{\hbar}x^2 \right) + \beta \sqrt{\frac{\omega}{\hbar}}\,x\,{}_1F_1\left( \frac{1}{2}-\frac{b'}{4},\frac{3}{2};-\frac{\omega}{\hbar}x^2 \right) 
\eqqcolon u^{\rm (H)}(x) ~,
\label{eq:A-CESH_u}
\end{equation}
which has no zero if
\begin{equation}
\qquad
\left| \frac{\beta}{\alpha} \right| < \frac{2\varGamma\left(\frac{b}{4}+1\right)}{\varGamma\left(\frac{b}{4}+\frac{1}{2}\right)} 
\qquad\text{: \underline{\sc Condition II}} ~.
\label{eq:A-CESH_cond2}
\end{equation}
We fix $\alpha = 1$ without loss of generality.

Hence, the conditionally exactly solvable Hamiltonian $\mathcal{H}_{\rm CES}^{[0]}$ is obtained as
\begin{align}
\mathcal{H}_{\rm CES}^{[0]}
&= -\hbar^2\frac{d^2}{dx^2} + \omega^2x^2 + \hbar\omega + b'\hbar\omega - 2\hbar^2\left[ \frac{\partial_x^2u^{\rm (H)}(x)}{u^{\rm (H)}(x)} - \left( \frac{\partial_xu^{\rm (H)}(x)}{u^{\rm (H)}(x)} \right)^2 \right] \nonumber \\
&= -\hbar^2\frac{d^2}{dx^2} + \omega^2x^2 + \hbar\omega - b'\hbar\omega + 2\hbar\frac{\partial_xu^{\rm (H)}(x)}{u^{\rm (H)}(x)}  \left[ 2\omega x + \frac{\partial_xu^{\rm (H)}(x)}{u^{\rm (H)}(x)}  \right] ~.
\end{align}
The eigenvalues and the corresponding eigenfunctions are
\begin{align}
\mathcal{E}_0^{\rm (C,H)} &= 0 ~,~~~
\phi_0^{\rm (C,H)}(x) = \frac{\mathrm{e}^{-\frac{\omega x^2}{2\hbar}}}{u^{\rm (H)}(x)} ~, \\
\mathcal{E}_n^{\rm (C,H)} &= \mathcal{E}^{[1]}_{n-1} = (2n+b)\hbar\omega ~,~~~
\phi_n^{\rm (C,H)}(x) = \left( -\hbar\frac{d}{dx} + \omega x + \hbar\frac{\partial_xu^{\rm (H)}(x)}{u^{\rm (H)}(x)} \right) \phi^{[1]}_{n-1}(x) \nonumber \\ 
&\hspace{.675\linewidth}\text{for $n\geqslant 1$} ~.
\end{align}

\section{Deformed Shape-invariant Systems}
\subsection{The Construction Method}
\subsubsection{Position-dependent effective mass}
We consider a situation where the mass of a particle has a position dependency, $m\to m(x)$, whose idea was originally introduced to explain the tunneling of electrons in superconductor~\cite{PhysRevLett.5.147,PhysRevLett.5.464}.
A naive guessing would tell you that the kinetic term, say $T$, of a Hamiltonian is replaced as
\begin{equation}
T = -\frac{\hbar^2}{2m}\frac{d^2}{dx^2}
\quad\to\quad
-\frac{\hbar^2}{2m(x)}\frac{d^2}{dx^2} ~.
\end{equation}
However, this operator is not Harmitian.
The Hermitian kinetic term with a position-dependent effective mass proposed by von Roos~\cite{PhysRevB.27.7547} reads
\begin{equation}
T = -\frac{\hbar^2}{4}\left[ m(x)^{\alpha}\frac{d}{dx}m(x)^{\beta}\frac{d}{dx}m(x)^{\gamma} + m(x)^{\gamma}\frac{d}{dx}m(x)^{\beta}\frac{d}{dx}m(x)^{\alpha} \right] ~.
\label{eq:A-T_hamitian}
\end{equation}
Since $[ \bullet ]$ has the dimension of $\mathsf{M}^{-1}\mathsf{L}^{-2}$, the parameters $\alpha,\beta,\gamma$ satisfy
\begin{equation}
\alpha + \beta + \gamma = -1 ~.
\end{equation}
For example, BenDaniel--Duke kinetic energy operator is a kinetic energy operator with a position-dependent effective mass proposed by BenDaniel and Duke in 1966~\cite{PhysRev.152.683,PhysRevB.39.13434}, which corresponds to the case $\alpha=\gamma=0,\beta=-1$,
\begin{equation}
T = -\frac{\hbar^2}{2}\frac{d}{dx}\frac{1}{m(x)}\frac{d}{dx} ~.
\end{equation}

When dealing with the factorization of a Hamiltonian:
\[
\mathcal{H} = T + V(x) = \big( -\hat{\pi} + W(x) \big) \big( \hat{\pi} + W(x) \big) ~,
\]
it is convenient to write
\begin{equation}
\frac{1}{2m(x)} \equiv \eta(x)^2 ~,~~~
\eta(x) > 0 ~.
\end{equation}
Then, Eq. \eqref{eq:A-T_hamitian} changes into
\begin{equation}
T = -\frac{\hbar^2}{2} \left[ \eta(x)^{\rho}\frac{d}{dx}\eta(x)^{\sigma}\frac{d}{dx}\eta(x)^{\tau} + \eta(x)^{\tau}\frac{d}{dx}\eta(x)^{\sigma}\frac{d}{dx}\eta(x)^{\rho} \right] ~,~~~
\rho + \sigma + \tau = 2 ~,
\end{equation}
which is simplified to
\begin{align}
T 
&=  -\frac{\hbar^2}{2} \left[ 2\eta^2\frac{d^2}{dx^2} + 4\eta\partial_x\eta\frac{d}{dx} + (\rho+\tau) \eta\partial_x^2\eta + (\rho+\tau-2\rho\tau) (\partial_x\eta)^2 \right] \nonumber \\
&= -\hbar^2 \left[ \left( \sqrt{\eta}\frac{d}{dx}\sqrt{\eta} \right)^2 - \frac{1-\rho-\tau}{2}\eta\partial_x^2\eta - \left( \frac{1}{2}- \rho \right) \left( \frac{1}{2} - \tau \right)(\partial_x\eta)^2 \right] ~.
\end{align}
Thereby the Hamiltonian with a position-dependent effective mass is
\begin{align}
\mathcal{H} &= T + V(x)
= -\hbar^2 \left( \sqrt{\eta(x)}\frac{d}{dx}\sqrt{\eta(x)} \right)^2 + V_{\rm eff}(x)
\equiv -\hat{\pi}^2 + V_{\mathrm{eff}}(x) \nonumber \\
&= -\hbar^2\eta(x)^2\frac{d^2}{dx^2} - 2\hbar^2\eta(x)\partial_x\eta(x)\frac{d}{dx} - \frac{\hbar^2}{4}\left( 2\eta(x)\partial_x^2\eta(x) + [\partial_x\eta(x)]^2 \right) + V_{\mathrm{eff}}(x) ~,
\end{align}
where the effective potential $V_{\rm eff}(x) \in \mathbb{R}$ is
\begin{equation}
V_{\rm eff}(x) \equiv \hbar^2 \left[ \frac{1-\rho-\tau}{2}\eta(x)\partial_x^2\eta(x) + \left( \frac{1}{2}- \rho \right) \left( \frac{1}{2} - \tau \right)[\partial_x\eta(x)]^2 \right] + V(x) ~.
\end{equation}

\subsubsection{Deformed SUSY QM and deformed shape invariance}
We define a Hamiltonian $\mathcal{H}^{[0]}$ as a product of two operators $\mathcal{A}^{\pm}$,
\begin{equation}
\mathcal{H}^{[0]} \coloneqq \mathcal{A}^{+}\mathcal{A}^{-} ~,~~~
\mathcal{A}^{\pm} \coloneqq \mp\hbar\sqrt{\eta(x)}\frac{d}{dx}\sqrt{\eta(x)} + W(x) ~,
\end{equation}
where those operators are defined in $x \in (x_1,x_2) $, and $W(x)$ is the superpotential of this system.
This Hamiltonian has the lowest energy zero.
The Hamiltonian $\mathcal{H}^{[0]}$ is 
\begin{align}
\mathcal{H}^{[0]} &= \left( -\hbar\sqrt{\eta(x)}\frac{d}{dx}\sqrt{\eta(x)} + W(x) \right) \left( \hbar\sqrt{\eta(x)}\frac{d}{dx}\sqrt{\eta(x)} + W(x) \right) \nonumber\\
&= -\hbar^2\left( \sqrt{\eta(x)}\frac{d}{dx}\sqrt{\eta(x)} \right)^2 + W(x)^2 - \hbar\eta(x)\frac{dW(x)}{dx} ~.
\end{align}
The effective potential $V_{\rm eff}^{[0]}(x)$ is identified as
\begin{equation}
V_{\rm eff}^{[0]}(x) \equiv W(x)^2 - \hbar\eta(x)\frac{dW(x)}{dx} ~.
\end{equation}
The Schr\"{o}dinger equation is
\[
\mathcal{H}^{[0]}\phi_n^{[0]}(x) = \mathcal{E}_n\phi_n^{[0]}(x) ~,~~~
n = 0,1,2 \ldots ~,
\]
where the ground-state wavefunction is determined by $\mathcal{A}^{-}\phi_0^{[0]} = 0$, 
\begin{equation}
\phi_0^{[0]} \propto \frac{1}{\sqrt{\eta(x)}}\exp\left[ -\frac{1}{\hbar}\int^x\frac{W(\bar{x})}{\eta(\bar{x})}\,d\bar{x} \right] ~.
\end{equation}
Note that $\eta(x)\equiv\mathrm{const.}$ corresponds to the ordinary quantum mechanical system discussed in Chap. \ref{sec:2}.
Also, this system can be seen as a quantum mechanical system on the curved space whose metric function is
\begin{equation}
g(x) \equiv \frac{1}{\eta(x)^2} 
\end{equation}
(See also Refs. \cite{CQuesne_2004,quesne-SIGMA2007}).

As in the ordinary quantum mechanics, we require that $\phi_n^{[0]}(x)$ is
\begin{enumerate}\setlength{\leftskip}{2em}
\renewcommand{\labelenumi}{(\roman{enumi})}
\item \quad$\displaystyle\int_{x_1}^{x_2} \left| \phi_n^{[i]}(x) \right|^2 \,dx < \infty$\qquad : square integrability,
\end{enumerate}
and also
\begin{enumerate}\setlength{\leftskip}{2em}
\renewcommand{\labelenumi}{(\roman{enumi})}
\setcounter{enumi}{1}
\item \quad$\displaystyle\left| \phi_n^{[i]}(x) \right|^2 \eta(x) \to 0 \quad \text{as}\quad x \to x_{1,2}$\qquad : Hermicity of Hamiltonian.
\end{enumerate}
The second requirement is to ensure that the Hermicity of the `deformed' momentum operator $\pi$:
\begin{equation}
\int_{x_1}^{x_2} \psi(x) \sqrt{\eta(x)} \left( -\mathrm{i}\frac{d}{dx} \right) \sqrt{\eta(x)} \phi(x) \,dx
= \left[ \int_{x_1}^{x_2} \phi(x) \sqrt{\eta(x)} \left( -\mathrm{i}\frac{d}{dx} \right) \sqrt{\eta(x)} \psi(x) \,dx \right]^{\ast} ~.
\end{equation}

Furthermore, the partner Hamiltonian of $\mathcal{H}^{[0]}$ is defined as
\begin{align}
\mathcal{H}^{[1]} \coloneqq \mathcal{A}^{-}\mathcal{A}^{+} 
&= -\hbar^2\left( \sqrt{\eta(x)}\frac{d}{dx}\sqrt{\eta(x)} \right)^2 + W(x)^2 + \hbar\eta(x)\frac{dW(x)}{dx} \nonumber \\
&\equiv -\hbar^2\left( \sqrt{\eta(x)}\frac{d}{dx}\sqrt{\eta(x)} \right)^2 + V_{\mathrm{eff}}^{[1]}(x) - \mathcal{E}_1 ~,
\end{align}
satisfying
\[
\mathcal{H}^{[1]}\phi_n^{[1]}(x) = \mathcal{E}_{n+1}\phi_n^{[1]}(x) ~,~~~
n = 0,1,2 \ldots ~,
\]
The partner Hamiltonians are connected by the following intertwining relations
\begin{equation}
\mathcal{A}^{-}\mathcal{H}^{[0]} = \mathcal{H}^{[1]}\mathcal{A}^{-} ~,~~~
\mathcal{A}^{+}\mathcal{H}^{[1]} = \mathcal{H}^{[0]}\mathcal{A}^{+} ~.
\end{equation}
Thus, a similar argument to the case in ordinary quantum mechanics tells that the two Hamiltonians are isospectral except for the ground state of $\mathcal{H}^{[0]}$.
It is important to note here that the eigenstates are never degenerated,
\[
0 = \mathcal{E}_0 < \mathcal{E}_1 < \mathcal{E}_2 < \cdots ~.
\]

Now, let us write the dependency of the model parameters explicitly to discuss the shape-invariant property in the above formalism with a position-dependent effective mass, $W(x)\equiv W(x;\bm{a})$, $\mathcal{A}^{\pm}\equiv\mathcal{A}^{\pm}(\bm{a})$ and $\phi_n^{[i]}(x)\equiv\phi_n^{[i]}(x;\bm{a})$.
The system possesses the \textit{deformed} shape invariance, or is \textit{deformed} shape-invariant, when
\begin{equation}
W(x;\bm{a})^2 + \hbar\eta(x)\frac{dW(x;\bm{a})}{dx} = W(x;f(\bm{a}))^2 - \hbar\eta(x)\frac{dW(x;f(\bm{a}))}{dx} + R(\bm{a}) ~.
\label{eq:A-DSIcond}
\end{equation}
That is, we impose the deformed shape-invariant condition on the effective potential $V_{\mathrm{eff}}^{[i]}(x)$, not on the potential $V(x)$.

For the deformed shape-invariant system, the energy eigenvalues and the corresponding eigenfunctions are expressed as
\begin{equation}
\mathcal{E}_n = \sum_{k=0}^{n-1}R^k(\bm{a}) ~,~~~
\phi_n^{[0]}(x;a_0) \propto \mathcal{A}^{+}(\bm{a})\mathcal{A}^{+}(f(\bm{a}))\cdots\mathcal{A}^{+}(f^{n-1}(\bm{a})) \phi_0^{[0]}(x;f^n(\bm{a})) ~.
\end{equation}

\subsection{Example}
\subsubsection{Deformed harmonic oscillator}
For a superpotential
\begin{equation}
W(x) = \omega x ~,~~~
x \in (-\infty,\infty) ~,
\end{equation}
a characteristic function $\eta(x)$ of the following form realizes deformed shape invariance:
\begin{equation}
\eta(x) = 1 + \alpha x^2 ~.
\end{equation}
$\alpha=0$ corresponds to the ordinary harmonic oscillator.
The effective potentials $V_{\mathrm{eff}}^{[0]}(x)$, $V_{\mathrm{eff}}^{[1]}(x)$ are
\begin{equation}
V_{\mathrm{eff}}^{[0]}(x) = \omega^2x^2 - \hbar\omega (1+\alpha x^2)  ~,~~~
V_{\mathrm{eff}}^{[1]}(x) = \omega^2x^2 + \hbar\omega (1+\alpha x^2) ~.
\end{equation}
One can see that the system is shape invariant under $\omega \to \omega+\hbar\alpha$.
The energy eigenvalues turned out to be $\mathcal{E}_n = 2\hbar\omega n + \hbar^2\alpha n^2$.

The Schr\"{o}dinger equation for this system is
\begin{multline}
-\hbar^2(1+\alpha x^2)^2\frac{d^2\phi_n^{[0]}(x)}{dx^2} - 4\hbar^2\alpha x(1+\alpha x^2)\frac{d\phi_n^{[0]}(x)}{dx} + \left[ - \hbar^2\alpha ( 2\alpha x^2 + 1 ) \right. \\ 
\left. + V_{\mathrm{eff}}^{[0]}(x) \right] \phi_n^{[0]}(x)
= (2\hbar\omega n + \hbar^2\alpha n^2) \phi_n^{[0]}(x) ~,
\end{multline}
which transforms to that for the $\dfrac{1}{\sin^2x}$-potential under $x \mapsto z = z(x) = \cot^{-1}\sqrt{\alpha}\,x$ with the identification $g\equiv \omega/\hbar\alpha$.
Thus, the eigenfunction $\phi_n^{[0]}(x)$ is turned out to be
\begin{equation}
\phi_n^{[0]}(x) = \left( \frac{1}{\sqrt{1+\alpha x^2}} \right)^{\frac{\omega}{\hbar\alpha}+1} P_n^{(\frac{\omega}{\hbar\alpha}+\frac{1}{2}),(\frac{\omega}{\hbar\alpha}+\frac{1}{2})}\left( \frac{\sqrt{\alpha}\,x}{\sqrt{1+\alpha x^2}} \right) ~.
\end{equation}

\begin{remark}[Semi-confined harmonic oscillator]
We emphasize here that there is another kind of exactly solvable, especially classical-orthogonal-polynomially solvable, quantum mechanical system with a position-dependent effective mass other than the deformed shape-invariant systems.
The Hamiltonian is 
\begin{equation}
\mathcal{H} = -\hbar^2\frac{d}{dx}\frac{1}{M(x)}\frac{d}{dx} + M(x)\omega^2x^2 - \hbar\omega 
\end{equation}
with
\begin{equation}
M(x) = \begin{cases}
	\dfrac{a}{x+a} & (x > -a) \\
	\infty & (x \leqslant -a) 
\end{cases} ~.
\end{equation}
This Hamiltonian is completely isospectral to the ordinary harmonic oscillator:
\[
\mathcal{E}_n = 2\hbar\omega n ~,
\]
and the eigenfunctions are
\begin{equation}
\phi_n (x) = \begin{cases}
	\displaystyle \left( 1+\frac{x}{a} \right)^{\frac{\omega a^2}{\hbar}}\mathrm{e}^{-\frac{\omega a}{\hbar}(x+a)} L_n^{(\frac{2\omega a^2}{\hbar})}\left( \frac{2\omega a}{\hbar}(x+a) \right) & (x > -a) \\
	0 & (x \leqslant -a)
\end{cases} ~.
\end{equation}
For more details, see Refs. \cite{Jafarov-2021,doi:10.1142/S0217979222502277,Jafarov:2022aa,nagiyev2023wigner,jafarov2023dynamical}.
\end{remark}

\chapter{Definitions, Theorems and Formulae}
\label{sec:Formulae}

\section{Wronskian}
\label{sec:A.3}
\subsubsection{Definition}
\begin{equation}
\mathrm{W}\left[ f_1,\ldots,f_m \right](x) \coloneqq
\mathrm{det}\left( \frac{\mathrm{d}^{j-1}f_k(x)}{\mathrm{d}x^{j-1}} \right)_{1\leqslant j,k\leqslant m}
\end{equation}

\subsubsection{Formulae}
\begin{align}
&\mathrm{W}\left[ gf_1,gf_2,\ldots,gf_n \right](x)
= g(x)^n \mathrm{W}\left[ f_1,f_2,\ldots,f_n \right](x) 
\label{eq:A-Wron_23} \\[5pt]
&\mathrm{W}\left[ f_1(y),f_2(y),\ldots,f_n(y) \right](x)
= y'(x)^{\frac{n(n-1)}{2}} \mathrm{W}\left[ f_1,f_2,\ldots,f_n \right](y) 
\label{eq:A-Wron_24} \\[5pt]
&\mathrm{W}\big[ \mathrm{W}\left[ f_1,f_2,\ldots,f_n,g \right],\mathrm{W}\left[ f_1,f_2,\ldots,f_n,h \right] \big](x) \nonumber \\
&\hspace{0.2\paperwidth}= \mathrm{W}\left[ f_1,f_2,\ldots,f_n \right](x)\mathrm{W}\left[ f_1,f_2,\ldots,f_n,g,h \right](x)
\label{eq:A-Wron_25}
\end{align}

\section{Maximum and Minimum}
\subsubsection{Maximum}
\begin{equation}
\max\{ x,y \} = \begin{cases}
	x & (x \geqslant y) \\
	y & (x < y) \\
\end{cases}
\end{equation}

\subsubsection{Minimum}
\begin{equation}
\min\{ x,y \} = \begin{cases}
	x & (x \leqslant y) \\
	y & (x > y) \\
\end{cases}
\end{equation}

\section{Floor and Ceiling Function}
\subsubsection{Floor function}
\begin{align}
&\lfloor x \rfloor\hspace{3pt} \coloneqq \max\{ n \in \mathbb{Z}; n \leqslant x \} \\
&\lfloor x \rfloor' \coloneqq \max\{ n \in \mathbb{Z}; n < x \}
\end{align}

\subsubsection{Ceiling function}
\begin{align}
&\lceil x \rceil\hspace{3pt} \coloneqq \min\{ n \in \mathbb{Z}; n \geqslant x \} \\
&\lceil x \rceil' \coloneqq \min\{ n \in \mathbb{Z}; n > x \}
\end{align}

\section{Sign Function}
\subsubsection{Definition}
\begin{equation}
\sgn (x) \coloneqq \begin{cases}
	\dfrac{x}{|x|} & (x \neq 0) \\[1ex]
	\;\,0 & (x = 0)
\end{cases}
\end{equation}

\section{Hypergeometric Function}
\subsection{Hypergeometric Function}
\subsubsection{Definition}
\begin{equation}
{}_pF_q\left(
	\begin{tabular}{c}
	$\alpha_1,\cdots,\alpha_p$ \\
	$\beta_1,\cdots,\beta_q$
	\end{tabular}	 
\bigg|~z \right) \equiv {}_pF_q(\alpha_1,\ldots,\alpha_r;\beta_1,\ldots,\beta_s;z) \coloneqq
\sum_{n=0}^{\infty} \frac{(\alpha_1)_n\cdots(\alpha_p)_n}{(\beta_1)_n\ldots(\beta_q)_n}\frac{z^n}{n!}
\end{equation}
where $(a)_n$ is the Pochhammer's symbol:
\begin{align}
(a)_n \coloneqq \left\{
	\begin{array}{ll}
	1 & \text{(for $n=0$)} \\[5pt]
	\displaystyle\prod_{k=1}^n (a+k-1) = a(a+1)\cdots(a+n-1) & \text{(for $n\geqslant 1$)}
	\end{array}
\right. ~.
\end{align}

Especially, for $p=2, q=1$,
\begin{equation}
{}_2F_1 \left(
	\begin{tabular}{c}
	$\alpha,\beta$ \\
	$\gamma$
	\end{tabular}	 
\bigg|~z \right) \equiv {}_2F_1(\alpha,\beta;\gamma;z) \equiv {}_2F_1(\alpha,\beta,\gamma;z) =
\sum_{n=0}^{\infty} \frac{(\alpha)_n(\beta)_n}{(\gamma)_n}\frac{z^n}{n!}
\end{equation}
is the Gaussian hypergeometric function, and for $p=q=1$,
\begin{equation}
{}_1F_1 \left(
	\begin{tabular}{c}
	$\alpha$ \\
	$\gamma$
	\end{tabular}	 
\bigg|~z \right) \equiv {}_1F_1(\alpha;\gamma;z) \equiv {}_2F_1(\alpha,\gamma;z) =
\sum_{n=0}^{\infty} \frac{(\alpha)_n}{(\gamma)_n}\frac{z^n}{n!}
\end{equation}
is the confuluent hypergeometric function (Kummer's function).

\subsection{Confluent Hypergeometric Function (Kummer's Function)}
\subsubsection{Definition}
Confluent hypergeometric function (Kummer's function):
\begin{equation}
{}_1F_1(\alpha,\gamma;z) \coloneqq \sum_{n=0}^{\infty} \frac{(\alpha)_n}{(\gamma)_n}\frac{z^n}{n!} ~,
\end{equation}
satisfies the following Kummer's equation:
\begin{equation}
z\frac{\mathrm{d}^2u}{\mathrm{d}z^2} + (\gamma-z)\frac{\mathrm{d}u}{\mathrm{d}z} - \alpha u = 0 ~.
\end{equation}

\subsubsection{Asymptotic expansion}
\begin{equation}
{}_1F_1(\alpha,\gamma;z) \sim
\left( \frac{\varGamma(\gamma)}{\varGamma(\gamma-\alpha)}z^{-\alpha} + \frac{\varGamma(\gamma)}{\varGamma(\alpha)}\mathrm{e}^z z^{\alpha-\gamma} \right) \left( 1+\mathcal{O}(z^{-1}) \right) ~,~~~
|z|\to\infty
\end{equation}

\section{Classical Orthogonal Polynomials}
\subsection{Hermite Polynomials}
\label{sec:A.5.1}
\subsubsection{Differential equation}
\begin{equation}
y''(x) - 2xy'(x) + 2ny(x) = 0
\label{eq:H_deq}
\end{equation}

\subsubsection{Rodrigues' formula}
\begin{equation}
H_n(x) = (-1)^n\mathrm{e}^{x^2}\frac{d^n}{dx^n}\mathrm{e}^{-x^2}
\label{eq:H_Rodrigues}
\end{equation}

\subsubsection{Recurrence relations}
\begin{gather}
H_{n+1}(x) = 2xH_n(x) - 2nH_{n-1}(x) \\
H'_n(x) = 2nH_{n-1}(x)
\end{gather}

\subsubsection{Orthogonality}
\begin{equation}
\int_{-\infty}^{\infty} \mathrm{e}^{-x^2}H_m(x)H_n(x) \,dx = 2^n\pi^{1/2}n!\delta_{m,n}
\end{equation}

\subsection{Laguerre Polynomials}
\subsubsection{Differential equation}
\begin{equation}
xy''(x) + (\alpha + 1 - x)y'(x) + ny(x) = 0
\label{eq:L_deq}
\end{equation}

\subsubsection{Rodrigues' formula}
\begin{equation}
L_n^{(\alpha)}(x) = \frac{\mathrm{e}^xx^{-\alpha}}{n!}\frac{d^n}{dx^n}(\mathrm{e}^{-x}x^{n+\alpha})
\label{eq:L_Rodrigues}
\end{equation}

\subsubsection{Recurrence relation}
\begin{equation}
(n+1)L_{n+1}^{(\alpha)}(x) - (2n + \alpha + 1 - x)L_n^{(\alpha)} + (n+\alpha)L_{n-1}^{(\alpha)} = 0
\end{equation}

\subsubsection{Orthogonality}
\begin{equation}
\int_0^{\infty} \mathrm{e}^{-x}x^{\alpha}L_m^{(\alpha)}(x)L_n^{(\alpha)}(x) \,dx = \frac{(n+\alpha)!}{n!}\delta_{m,n}
\end{equation}

\subsection{Jacobi Polynomials}
\subsubsection{Differential equation}
\begin{equation}
(1-x^2)y''(x) - [\beta - \alpha (\alpha + \beta + 2)x]y'(x) + n(n + \alpha + \beta + 1)y(x) = 0
\label{eq:J_deq}
\end{equation}

\subsubsection{Rodrigues' formula}
\begin{equation}
P_n^{(\alpha,\beta)}(x) = \frac{(-1)^n}{2^nn!}\frac{1}{(1-x)^{\alpha}(1+x)^{\beta}}\frac{d^n}{dx^n}\left[ (1-x)^{n+\alpha}(1+x)^{n+\beta} \right]
\label{eq:J_Rodrigues}
\end{equation}

\subsubsection{Recurrence relation}
\begin{multline}
2(n+1)(n + \alpha + \beta + 1)(2n + \alpha + \beta)P_{n+1}^{(\alpha,\beta)}(x) \\
= (2n + \alpha + \beta + 1)\left[ (2n + \alpha + \beta + 2)(2 + \alpha + \beta)x + \alpha^2 - \beta^2 \right]P_n^{(\alpha,\beta)}(x) \\
- 2(n+\alpha)(n+\beta)(2n + \alpha + \beta + 2)P_{n-1}^{(\alpha,\beta)}(x)
\end{multline}

\subsubsection{Orthogonality}
\begin{equation}
\int_{-1}^1 (1-x)^{\alpha}(1+x)^{\beta} P_m^{(\alpha,\beta)}(x)P_n^{(\alpha,\beta)}(x) \,dx = \frac{2^{\alpha+\beta+1}}{2n + \alpha + \beta + 1}\frac{(n+\alpha)!(n+\beta)!}{(n + \alpha + \beta)!n!}\delta_{m,n}
\end{equation}

\section{Integral Formulae}
\subsection[Ref. {[42]}]{Ref. \cite{hruska1997accuracy}}
\begin{itemize}\setlength{\leftskip}{-1em}
\item For $0<a<b$,
	\begin{equation}
	\int_a^b \sqrt{(y-a)(b-y)} \,\frac{dy}{y} = \frac{\pi}{2}(a+b)-\pi\sqrt{ab} ~.
	\label{eq:int_formula1}
	\end{equation}
\item For $a<b$,
	\begin{equation}
	\int_a^b \sqrt{(y-a)(b-y)} \,\frac{dy}{y^2+1} = \frac{\pi}{\sqrt{2}}\sqrt{\sqrt{(1+a^2)(1+b^2)}-ab+1}-\pi ~.
	\label{eq:int_formula2}
	\end{equation}
\item For $-1<a<b<1$,
	\begin{equation}
	\int_a^b \sqrt{(y-a)(b-y)} \,\frac{dy}{1-y^2} = \frac{\pi}{2}\left[ 2-\sqrt{(1-a)(1-b)}-\sqrt{(1+a)(1+b)} \right] ~.
	\label{eq:int_formula3}
	\end{equation}
\item For $1<a<b$,
	\begin{equation}
	\int_a^b \sqrt{(y-a)(b-y)} \,\frac{dy}{y^2-1} = \frac{\pi}{2}\left[ \sqrt{(a+1)(b+1)}-\sqrt{(a-1)(b-1)}-2 \right] ~.
	\label{eq:int_formula4}
	\end{equation}
\end{itemize}

\subsection[Ref. {[83]}]{Ref. \cite{gradshteyn2014tables}}

\begin{multline}
\int \frac{\cos^2x}{a\cos^2 + c\sin^2x} \,dx 
= \frac{1}{4(a-c)^2} \left[ (a-c)x + -\frac{c(a-c)}{\sqrt{ac-b^2}}\arctan\frac{c\tan x+b}{\sqrt{ac-b^2}} \right] \\
\text{for $ac>0$}
\label{eq:int_dSIH}
\end{multline}

\begin{align}
&\int \frac{\sqrt{a+b x + cx^2}}{x(x+p)} \,dx
= \frac{1}{p}\left[ \int \frac{\sqrt{a+bx + cx^2}}{x} \,dx - \int \frac{\sqrt{a+b x + cx^2}}{x+p} \,dx \right] 
\label{eq:int_dSIL1b} \\
&\int \frac{\sqrt{a+bx + cx^2}}{x} \,dx 
= \sqrt{a+bx + cx^2} + a\int \frac{\mathrm{d}x}{x\sqrt{a+b x + cx^2}} + \frac{b}{2}\int \frac{\mathrm{d}x}{\sqrt{a+bx + cx^2}}  \\
&\int \frac{\sqrt{a+b x + cx^2}}{x+p} \,dx = c\int \frac{x}{\sqrt{a+bx + cx^2}} \,dx + (b - cp)\int \frac{\mathrm{d}x}{\sqrt{a+bx + cx^2}} \nonumber \\
&\hspace*{0.3\linewidth} + (a - bp + cp^2)\int \frac{\mathrm{d}x}{(x+p)\sqrt{a+b x + cx^2}} \qquad\text{for $x+p>0$}
\label{eq:int_dSIL1e}
\end{align}

\begin{align}
&\int \frac{dx}{x\sqrt{a+bx + cx^2}} = \frac{1}{\sqrt{-a}}\arctan\frac{2a+b x}{2\sqrt{-a}\sqrt{a+b x + cx^2}} \qquad \text{for $a<0$} 
\label{eq:int_dSIL2b} \\
&\int \frac{dx}{\sqrt{a+b x + cx^2}} = -\frac{1}{\sqrt{-c}}\arcsin\frac{2cx+b}{\sqrt{b^2-4ac}} \qquad \text{for $c<0$ and $b^2-4ac>0$} \\
&\int \frac{dx}{(x+p)\sqrt{a+b x + cx^2}} = \frac{1}{\sqrt{a+bx + cx^2}}\arcsin\frac{-2(a+bx + cx^2) + b + 2p}{\sqrt{b^2-4ac}} \nonumber \\
&\hspace*{0.55\linewidth} \text{for $\frac{1}{x+p}>0$ and $b^2-4ac>0$}
\label{eq:int_dSIL2e}
\end{align}

\subsection{Others}
\begin{multline}
\int \frac{\sqrt{a^2 - x^2}}{(p + qx)(x+a+1)(x+a-1)} \,dx \\
= \frac{\sqrt{1-2a}}{2(p + q - aq)}\arctan\frac{a^2+ax-x}{\sqrt{1-2a}\sqrt{a^2-x^2}}
- \frac{\sqrt{1+2a}}{2(p - q - aq)}\arctan\frac{a^2+ax+x}{\sqrt{1+2a}\sqrt{a^2-x^2}} \\
+ \frac{\sqrt{p^2-a^2q^2}}{(p + q - aq)(p - q - aq)}\arctan\frac{a^2q+px}{\sqrt{p^2-a^2q^2}\sqrt{a^2-x^2}}
\label{eq:int_dSIJ}
\end{multline}

\section{Darboux Transformation}
Let us consider a Schr\"{o}dinger-type differential equation:
\begin{equation}
\mathcal{H}\psi(x) = \mathcal{E}\psi(x) ~,~~~
\mathcal{H} = -\hbar^2\frac{d^2}{dx^2} + V(x) ~.
\label{eq:A-SE_gene}
\end{equation}
Here, the Darboux transformation can be applicable to generic equations of the above form, and $\mathcal{E}, V(x) \in \mathbb{C}$.

\subsubsection{General formulation}
Let $\{ \varphi_j(x), \widetilde{\mathcal{E}}_j \}$ ($j=1,2,\ldots$) be distinct solutions of Eq. \eqref{eq:A-SE_gene}:
\begin{equation}
\mathcal{H}\varphi_j(x) = \widetilde{\mathcal{E}}_j\varphi_j(x) ~,~~~ 
j = 1,2,\ldots,M ~.
\end{equation}
Here, $\varphi_j(x)$ does not have to be square-integrable, and $\widetilde{\mathcal{E}}_j$ can take complex values.
They are called seed solutions.

First, we do \textit{formal} factorization of the Hamiltonian $\mathcal{H}$ in terms of a seed solution $\varphi_1(x)$,
\begin{equation}
\mathcal{H} = \hbar^2\left( -\frac{d}{dx} - \frac{d}{dx}\ln|\varphi_1(x)| \right) \left( \frac{d}{dx} - \frac{d}{dx}\ln|\varphi_1(x)| \right) + \widetilde{\mathcal{E}}_1 
\equiv \hbar^2\mathcal{A}_{\varphi}^{\dag}\mathcal{A}_{\varphi} + \widetilde{\mathcal{E}}_1 ~.
\end{equation}
We define a new Hamiltonian $\mathcal{H}^{(1)}$ by
\begin{align}
\mathcal{H}^{(1)}
&\coloneqq \hbar^2\mathcal{A}_{\varphi}\mathcal{A}_{\varphi}^{\dag} + \widetilde{\mathcal{E}}_1
= \hbar^2\left( \frac{d}{dx} - \frac{d}{dx}\ln|\varphi_1(x)| \right)\left( -\frac{d}{dx} - \frac{d}{dx}\ln|\varphi_1(x)| \right) + \widetilde{\mathcal{E}} \nonumber \\
&= \mathcal{H} - 2\hbar^2\frac{d^2}{dx^2}\ln|\varphi_1(x)| ~.
\end{align}
These two Hamiltonian are related via the following intertwining relations:
\begin{equation}
\mathcal{H}^{(1)}\mathcal{A}_{\varphi} = \mathcal{A}_{\varphi}\mathcal{H} ~,~~~
\mathcal{A}^{\dag}_{\varphi}\mathcal{H}^{(1)} = \mathcal{H}\mathcal{A}^{\dag}_{\varphi} ~.
\end{equation}
Thus 
\begin{align}
\psi^{(1)}(x) &\coloneqq \mathcal{A}_{\varphi}\psi(x) = \frac{\mathrm{W}[\varphi_1,\psi](x)}{\varphi_1(x)} \\
\varphi_j^{(1)}(x) &\coloneqq \mathcal{A}_{\varphi}\varphi_j(x) = \frac{\mathrm{W}[\varphi_1,\varphi_j](x)}{\varphi_1(x)} ~,~~~
j = 2,3,\ldots ~,
\end{align}
are solutions of the Schr\"{o}dinger equation for $\mathcal{H}^{(1)}$.
Moreover, 
\begin{equation}
\mathcal{A}_{\varphi}^{\dag}\varphi_{1}^{(1)}(x) = 0
\quad\Longrightarrow\quad
\varphi_{1}^{(1)}(x) = \frac{1}{\varphi_1(x)} ~,
\end{equation}
is also a solution.
Namely,
\begin{align}
&\mathcal{H}^{(1)}\psi^{(1)}(x) = \mathcal{E}\psi^{(1)}(x) ~, \\
&\mathcal{H}^{(1)}\varphi_j^{(1)}(x) = \widetilde{\mathcal{E}}_j\varphi_j^{(1)}(x) ~,~~~
j=1,2,\ldots, M ~.
\end{align}

By repeating the above Darboux transformation $M$ times, we arrive at the following theorem:

\begin{theorem}[Darboux, 1882~\cite{darboux}]
Let $\psi(x)$ be a solution of the original Schr\"{o}dinger equation:
\[
\mathcal{H}\psi(x) = \mathcal{E}\psi(x) ~.
\]
Then the functions 
\begin{align*}
\psi^{(M)}(x) &\coloneqq \frac{\mathrm{W}[\varphi_1,\ldots,\varphi_M,\psi](x)}{\mathrm{W}[\varphi_1,\ldots,\varphi_M](x)} \\
\varphi_j^{(M)}(x) &\coloneqq \frac{\mathrm{W}[\varphi_1,\ldots,\varphi_{j-1},\varphi_{j+1},\ldots,\varphi_M](x)}{\mathrm{W}[\varphi_1,\ldots,\varphi_M](x)} ~,~~~
j=1,2,\ldots,M ~,
\end{align*}
where
\[
\mathcal{H}\varphi_j(x) = \widetilde{\mathcal{E}}_j\varphi_j(x) ~,~~~
j=j=1,2,\ldots, M ~,
\]
satisfy the following Schr\"{o}dinger equation with the same energy:
\begin{align*}
&\mathcal{H}^{(M)}\psi^{(M)}(x) = \mathcal{E}\psi^{(M)}(x) ~, \\
&\mathcal{H}^{(M)}\varphi_j^{(M)}(x) = \widetilde{\mathcal{E}}_j\varphi_j^{(M)}(x) ~,~~~
j=1,2,\ldots, M ~,
\end{align*}
with
\[
\mathcal{H}^{(M)} \coloneqq \mathcal{H} - 2\hbar^2\frac{d^2}{dx^2}\ln\left| \mathrm{W}[\varphi_1,\ldots,\varphi_M](x) \right| ~.
\]
\end{theorem}

\begin{remark}
In Sect. \ref{sec:B}, we have constructed two solvable systems through Darboux transformations.
Let us summarize which choices of seed solutions $\{ \varphi_{d_j}(x) \}$ correspond to which constructions below.
\begin{itemize}\setlength{\leftskip}{-1em}
\item $\varphi_{d_j}(x)$'s are the eigenfunctions of the original Hamiltonian $\mathcal{H}$.

	\hfill{$\Rightarrow$ Krein--Adler systems (Sect. \ref{sec:B-1}).}
\item $\varphi_{d_j}(x)$'s are the virtual-state wavefunctions of the original Hamiltonian $\mathcal{H}$.

	\hfill{$\Rightarrow$ Multi-indexed systems (Sect. \ref{sec:B-2}).}
\end{itemize}
\end{remark}

\clearpage
\thispagestyle{empty}
~

\clearpage
\pagestyle{fancy}
\fancyhead{}
\fancyhead[LE]{\quad\thepage}
\fancyhead[RO]{\thepage\quad}
\fancyfoot{}

\addcontentsline{toc}{chapter}{\bibname}
\bibliographystyle{naturemag}
\bibliography{PhDThesis_bib.bib}


\end{document}